\documentclass[a4paper,11pt]{article}
\pdfoutput=1 
\usepackage{verbatim}

\usepackage{jheppub}
\usepackage{amsmath,amsfonts,amssymb,amsthm,color}
\usepackage[T1]{fontenc} 
\usepackage{enumerate}
\usepackage{tikz, tikz-3dplot, pgfplots}
\usepackage[utf8]{inputenc}
\usepackage{slashed}
\usepackage{booktabs}
\usepackage{caption}
\usepackage{subcaption}
\usepackage{footmisc}
\usepackage{multirow,graphicx}

\usepackage{float}
\usepackage{changepage}

\usepackage{array}

\usepackage{changepage}
\usepackage{float}
\usepackage{tikz}
\usetikzlibrary{decorations.text}
\usepackage{enumitem}
\setlist{itemsep=0pt}
\usetikzlibrary{math}
\usepackage{tikz}
\usepackage{pgfplots}
\usetikzlibrary{decorations.text,intersections, pgfplots.fillbetween}
\usetikzlibrary{math}
 \usetikzlibrary{snakes,3d,shapes.geometric,shadows.blur}
\usepackage{tikz-3dplot}

\definecolor{amaranthred}{rgb}{0.83,0.13,0.18}
\definecolor{amazon}{rgb}{0.23,0.48,0.34}
\definecolor{bdazzledblue}{rgb}{0.18,0.35,0.58}
\definecolor{absolutezero}{rgb}{0.0,0.28,0.73}
\definecolor{bitterlemon}{rgb}{0.79,0.88,0.05}
\definecolor{byzantine}{rgb}{0.74,0.2,0.64}
\definecolor{turquoise}{rgb}{0.19, 0.84, 0.78}

\newcolumntype{C}{>{\centering\arraybackslash}m{20em}}
\newcolumntype{M}[1]{>{\centering\arraybackslash}m{#1}}

\newcommand{\comm}[1]{} %for commenting out blocks of text

\def\({\left(}
\def\){\right)}
\def\[{\left[}
\def\]{\right]}

\def\coeff#1#2{{\textstyle \frac{#1}{#2}}}

\def\One{{\hbox{ 1\kern-.8mm l}}}

\def\barray{\begin{array}}
\def\earray{\end{array}}
\def\be{\begin{equation}}
\def\ee{\end{equation}}
\def\bea{\begin{eqnarray}}
\def\eea{\end{eqnarray}}
\def\bal{\begin{align}}
\def\eal{\end{align}}
\def\nn{\nonumber}

\def\-{\,-\,}
\def\={\,=\,}
\def\+{\,+\,}
\def\equi{\,\equiv\,}

\numberwithin{equation}{section} % replaces the hack below

\definecolor{cardinal}{rgb}{0.6,0,0}
\definecolor{darkgreen}{rgb}{0,0.4,0}
\definecolor{golden}{rgb}{0.92, 0.7, 0}
\definecolor{midnight}{rgb}{0, 0, 0.5}
\definecolor{darkblue}{rgb}{0, 0, 0.7}
\definecolor{purple}{rgb}{0.5, 0, 0.5}

\def\IR{\mathbb{R}}

\def\cA{{\cal A}}

\def\cC{{\cal C}}

\def\cF{{\cal F}}
\def\cK{{\cal K}}
\def\cG{{\cal G}}

\def\cN{{\cal N}}

\def\cO{{\cal O}}

\makeatletter
\newcommand\footnoteref[1]{\protected@xdef\@thefnmark{\ref{#1}}\@footnotemark}
\makeatother

\usetikzlibrary{positioning,calc,fadings,decorations.pathreplacing}

\tikzset{
 diffuse color/.initial = black,                       % (1) The color value does not carry through to the pre- and post- actions
}

\tikzset{
 linear opacity/.initial=0.5,                          %     Initial value must be defined here to define the next key
 linear stroke/.style = {                              %     Define a style to draw a diffused sstroke
   preaction={                                         %     This uses preactions to draw the gradiens
     draw=\pgfkeysvalueof{/tikz/diffuse color},        %     Draw and colour the path (See 1). This is disabled by `/tikz/diffuse gradient' (See 2)
     line width = (2.0-#1)*\pgflinewidth,              %     Vary the Line width for each stroke, change 2.0 to 1.0 to normalize scaling
     opacity=\pgfkeysvalueof{/tikz/linear opacity},white}},  %     Keep resetting the opacity for each line, there is probably a cleaner means of setting this (See 3)
 diffuse gradient/.style={                             %     This style executes `/tikz/diffuse' stroke multiple times to achieve the gradient effect
   draw = none,                                        % (2) Disable the default draw operation
   linear opacity=#1,                                  % (3) Set a consistent value for opacity. In retrospect one could have simply assigned this to /tikz/opacity
   linear stroke/.list={0.0,#1,...,1.0}},              %     Draw the line multiple times. Ideally we would use (1.0-1/#1) as the final value and (1/#1) as the stepsize
 diffuse gradient/.default=1,                          %     Set an initial step size
}

\tikzset{
 non-linear stroke/.style = {                          %     Define a style to draw a diffused sstroke
   preaction={                                         %     This uses preactions to draw the gradiens
     draw=\pgfkeysvalueof{/tikz/diffuse color},        %     Draw and colour the path (See 1). This is disabled by `/tikz/diffuse falloff' (See 4)
     line width = (2.0-#1)*\pgflinewidth,              %     Vary the Line width for each stroke, change 2.0 to 1.0 to normalize scaling
     opacity=#1,white}},                                     %     Vary the opacity for each stroke
 diffuse falloff/.style={                              %     This style executes `/tikz/diffuse' stroke multiple times to achieve the gradient effect
   draw = none,                                        % (4) Disable the default draw operation
   non-linear stroke/.list={0.0,#1,...,1.0}},          %     Draw the line multiple times. Ideally we would use (1.0-1/#1) as the final value and (1/#1) as the stepsize
 diffuse falloff/.default=1,                           %     Set an initial step size
}
 
\newcommand{\AdSBTZTS}{
    \begin{tikzpicture}[remember picture]
\def\deb{-10} 
\def\inter{0.45} 
\def\ha{2.8} 
\def\zaxisline{4} 
\def\rodsize{1.1} 
\def\numrod{1.5} 

\def\fin{\deb+1+2*\rodsize+\numrod*\rodsize}

\draw (\deb+5.75,\ha-5*\inter) node{$\Longleftrightarrow$};

\draw[black,thin] (\deb+1,\ha) -- (\fin,\ha);
\draw[black,thin] (\deb,\ha-\inter) -- (\fin-1,\ha-\inter);
\draw[black,thin] (\deb,\ha-2*\inter) -- (\fin,\ha-2*\inter);
\draw[black,thin] (\deb,\ha-3*\inter) -- (\fin,\ha-3*\inter);
\draw[black,thin] (\deb,\ha-4*\inter) -- (\fin,\ha-4*\inter);
\draw[black,thin] (\deb,\ha-5*\inter) -- (\fin,\ha-5*\inter);
\draw[black,thin] (\deb,\ha-6*\inter) -- (\fin,\ha-6*\inter);
\draw[black,thin] (\deb,\ha-7*\inter) -- (\fin,\ha-7*\inter);
\draw[black,thin] (\deb,\ha-8*\inter) -- (\fin,\ha-8*\inter);

\draw [decorate, 
    decoration = {brace,
        raise=5pt,
        amplitude=5pt},line width=0.2mm,gray] (\deb-0.6,\ha-2.5*\inter+0.05) --  (\deb-0.6,\ha+0.5*\inter-0.05);
\draw [decorate, 
    decoration = {brace,
        raise=5pt,
        amplitude=5pt},line width=0.2mm,gray] (\deb-0.6,\ha-3.5*\inter+0.05) --  (\deb-0.6,\ha-2.5*\inter-0.05);
\draw [decorate, 
    decoration = {brace,
        raise=5pt,
        amplitude=5pt},line width=0.2mm,gray] (\deb-0.6,\ha-7.5*\inter+0.05) --  (\deb-0.6,\ha-3.5*\inter-0.05);
        
\draw[gray] (\deb-1.2,\ha-1*\inter) node{S$^3$};
\draw[gray] (\deb-1.2,\ha-3*\inter) node{S$^1$};
\draw[gray] (\deb-1.2,\ha-5.5*\inter) node{T$^4$};
    
\draw (\deb-0.5,\ha) node{$\varphi_1$};
\draw (\deb-0.5,\ha-\inter) node{$\varphi_2$};
\draw (\deb-0.5,\ha-2*\inter) node{$\psi$};
\draw (\deb-0.5,\ha-3*\inter) node{$y$};
\draw (\deb-0.5,\ha-4*\inter) node{$x_1$};
\draw (\deb-0.5,\ha-5*\inter) node{$x_2$};
\draw (\deb-0.5,\ha-6*\inter) node{$x_3$};
\draw (\deb-0.5,\ha-7*\inter) node{$x_4$};
\draw (\deb-0.5,\ha-8*\inter) node{$t$};

\draw[black, dotted, line width=1mm] (\deb,\ha) -- (\deb+0.5,\ha);
\draw[black,line width=1mm] (\deb+0.5,\ha) -- (\deb+0.5+0.5*\rodsize,\ha);
\draw[black,line width=1mm] (\fin-0.5-0.5*\rodsize,\ha-\inter) -- (\fin-0.55,\ha-\inter);
\draw[black, dotted,line width=1mm] (\fin-0.5,\ha-\inter) -- (\fin,\ha-\inter);

\draw[amazon,line width=1mm] (\deb+0.5+0.5*\rodsize,\ha-4*\inter) -- (\deb+0.5+1*\rodsize,\ha-4*\inter);
\draw[amaranthred,line width=1mm] (\deb+0.5+1*\rodsize,\ha-3*\inter) -- (\deb+0.5+1.5*\rodsize,\ha-3*\inter);
\draw[gray,line width=1mm] (\deb+0.5+1.5*\rodsize,\ha-8*\inter) -- (\deb+0.5+2*\rodsize,\ha-8*\inter);
\draw[amaranthred,line width=1mm] (\deb+0.5+2*\rodsize,\ha-3*\inter) -- (\deb+0.5+2.5*\rodsize,\ha-3*\inter);
\draw[bdazzledblue,line width=1mm] (\deb+0.5+2.5*\rodsize,\ha-2*\inter) -- (\deb+0.5+3*\rodsize,\ha-2*\inter);

\draw[gray,dotted,line width=0.2mm] (\deb+0.5+0.5*\rodsize,\ha) -- (\deb+0.5+0.5*\rodsize,\ha-8*\inter);
\draw[gray,dotted,line width=0.2mm] (\deb+0.5+1*\rodsize,\ha) -- (\deb+0.5+1*\rodsize,\ha-8*\inter);
\draw[gray,dotted,line width=0.2mm] (\deb+0.5+1.5*\rodsize,\ha) -- (\deb+0.5+1.5*\rodsize,\ha-8*\inter);
\draw[gray,dotted,line width=0.2mm] (\deb+0.5+2*\rodsize,\ha) -- (\deb+0.5+2*\rodsize,\ha-8*\inter);
\draw[gray,dotted,line width=0.2mm] (\deb+0.5+2.5*\rodsize,\ha) -- (\deb+0.5+2.5*\rodsize,\ha-8*\inter);
\draw[gray,dotted,line width=0.2mm] (\deb+0.5+3*\rodsize,\ha) -- (\deb+0.5+3*\rodsize,\ha-8*\inter);
\end{tikzpicture}}

\newcommand{\AdSBTZ}{
    \begin{tikzpicture}[remember picture]
\def\deb{-10} 
\def\inter{0.45} 
\def\ha{2.8} 
\def\zaxisline{4} 
\def\rodsize{1.1} 
\def\numrod{1.5} 

\def\fin{\deb+1+2*\rodsize+\numrod*\rodsize}

\draw (\deb+5.75,\ha-5*\inter) node{$\Longleftrightarrow$};

\draw[black,thin] (\deb+1,\ha) -- (\fin,\ha);
\draw[black,thin] (\deb,\ha-\inter) -- (\fin-1,\ha-\inter);
\draw[black,thin] (\deb,\ha-2*\inter) -- (\fin,\ha-2*\inter);
\draw[black,thin] (\deb,\ha-3*\inter) -- (\fin,\ha-3*\inter);
\draw[black,thin] (\deb,\ha-4*\inter) -- (\fin,\ha-4*\inter);
\draw[black,thin] (\deb,\ha-5*\inter) -- (\fin,\ha-5*\inter);
\draw[black,thin] (\deb,\ha-6*\inter) -- (\fin,\ha-6*\inter);
\draw[black,thin] (\deb,\ha-7*\inter) -- (\fin,\ha-7*\inter);
\draw[black,thin] (\deb,\ha-8*\inter) -- (\fin,\ha-8*\inter);

\draw [decorate, 
    decoration = {brace,
        raise=5pt,
        amplitude=5pt},line width=0.2mm,gray] (\deb-0.6,\ha-2.5*\inter+0.05) --  (\deb-0.6,\ha+0.5*\inter-0.05);
\draw [decorate, 
    decoration = {brace,
        raise=5pt,
        amplitude=5pt},line width=0.2mm,gray] (\deb-0.6,\ha-3.5*\inter+0.05) --  (\deb-0.6,\ha-2.5*\inter-0.05);
\draw [decorate, 
    decoration = {brace,
        raise=5pt,
        amplitude=5pt},line width=0.2mm,gray] (\deb-0.6,\ha-7.5*\inter+0.05) --  (\deb-0.6,\ha-3.5*\inter-0.05);
        
\draw[gray] (\deb-1.2,\ha-1*\inter) node{S$^3$};
\draw[gray] (\deb-1.2,\ha-3*\inter) node{S$^1$};
\draw[gray] (\deb-1.2,\ha-5.5*\inter) node{T$^4$};
    
\draw (\deb-0.5,\ha) node{$\varphi_1$};
\draw (\deb-0.5,\ha-\inter) node{$\varphi_2$};
\draw (\deb-0.5,\ha-2*\inter) node{$\psi$};
\draw (\deb-0.5,\ha-3*\inter) node{$y$};
\draw (\deb-0.5,\ha-4*\inter) node{$x_1$};
\draw (\deb-0.5,\ha-5*\inter) node{$x_2$};
\draw (\deb-0.5,\ha-6*\inter) node{$x_3$};
\draw (\deb-0.5,\ha-7*\inter) node{$x_4$};
\draw (\deb-0.5,\ha-8*\inter) node{$t$};

\draw[black, dotted, line width=1mm] (\deb,\ha) -- (\deb+0.5,\ha);
\draw[black,line width=1mm] (\deb+0.5,\ha) -- (\deb+0.5+0.5*\rodsize,\ha);
\draw[black,line width=1mm] (\fin-0.5-0.5*\rodsize,\ha-\inter) -- (\fin-0.55,\ha-\inter);
\draw[black, dotted,line width=1mm] (\fin-0.5,\ha-\inter) -- (\fin,\ha-\inter);

\draw[gray,line width=1mm] (\deb+0.5+0.5*\rodsize,\ha-8*\inter) -- (\deb+0.5+1*\rodsize,\ha-8*\inter);
\draw[amaranthred,line width=1mm] (\deb+0.5+1*\rodsize,\ha-3*\inter) -- (\deb+0.5+1.5*\rodsize,\ha-3*\inter);
\draw[gray,line width=1mm] (\deb+0.5+1.5*\rodsize,\ha-8*\inter) -- (\deb+0.5+2*\rodsize,\ha-8*\inter);
\draw[amaranthred,line width=1mm] (\deb+0.5+2*\rodsize,\ha-3*\inter) -- (\deb+0.5+2.5*\rodsize,\ha-3*\inter);
\draw[gray,line width=1mm] (\deb+0.5+2.5*\rodsize,\ha-8*\inter) -- (\deb+0.5+3*\rodsize,\ha-8*\inter);

\draw[gray,dotted,line width=0.2mm] (\deb+0.5+0.5*\rodsize,\ha) -- (\deb+0.5+0.5*\rodsize,\ha-8*\inter);
\draw[gray,dotted,line width=0.2mm] (\deb+0.5+1*\rodsize,\ha) -- (\deb+0.5+1*\rodsize,\ha-8*\inter);
\draw[gray,dotted,line width=0.2mm] (\deb+0.5+1.5*\rodsize,\ha) -- (\deb+0.5+1.5*\rodsize,\ha-8*\inter);
\draw[gray,dotted,line width=0.2mm] (\deb+0.5+2*\rodsize,\ha) -- (\deb+0.5+2*\rodsize,\ha-8*\inter);
\draw[gray,dotted,line width=0.2mm] (\deb+0.5+2.5*\rodsize,\ha) -- (\deb+0.5+2.5*\rodsize,\ha-8*\inter);
\draw[gray,dotted,line width=0.2mm] (\deb+0.5+3*\rodsize,\ha) -- (\deb+0.5+3*\rodsize,\ha-8*\inter);
\end{tikzpicture}}

\newcommand{\AdSTS}{
    \begin{tikzpicture}[remember picture]
\def\deb{-10} 
\def\inter{0.45} 
\def\ha{2.8} 
\def\zaxisline{4} 
\def\rodsize{1.1} 
\def\numrod{1.5} 

\def\fin{\deb+1+2*\rodsize+\numrod*\rodsize}

\draw (\deb+5.75,\ha-5*\inter) node{$\Longleftrightarrow$};

\draw[black,thin] (\deb+1,\ha) -- (\fin,\ha);
\draw[black,thin] (\deb,\ha-\inter) -- (\fin-1,\ha-\inter);
\draw[black,thin] (\deb,\ha-2*\inter) -- (\fin,\ha-2*\inter);
\draw[black,thin] (\deb,\ha-3*\inter) -- (\fin,\ha-3*\inter);
\draw[black,thin] (\deb,\ha-4*\inter) -- (\fin,\ha-4*\inter);
\draw[black,thin] (\deb,\ha-5*\inter) -- (\fin,\ha-5*\inter);
\draw[black,thin] (\deb,\ha-6*\inter) -- (\fin,\ha-6*\inter);
\draw[black,thin] (\deb,\ha-7*\inter) -- (\fin,\ha-7*\inter);
\draw[black,thin] (\deb,\ha-8*\inter) -- (\fin,\ha-8*\inter);

\draw [decorate, 
    decoration = {brace,
        raise=5pt,
        amplitude=5pt},line width=0.2mm,gray] (\deb-0.6,\ha-2.5*\inter+0.05) --  (\deb-0.6,\ha+0.5*\inter-0.05);
\draw [decorate, 
    decoration = {brace,
        raise=5pt,
        amplitude=5pt},line width=0.2mm,gray] (\deb-0.6,\ha-3.5*\inter+0.05) --  (\deb-0.6,\ha-2.5*\inter-0.05);
\draw [decorate, 
    decoration = {brace,
        raise=5pt,
        amplitude=5pt},line width=0.2mm,gray] (\deb-0.6,\ha-7.5*\inter+0.05) --  (\deb-0.6,\ha-3.5*\inter-0.05);
        
\draw[gray] (\deb-1.2,\ha-1*\inter) node{S$^3$};
\draw[gray] (\deb-1.2,\ha-3*\inter) node{S$^1$};
\draw[gray] (\deb-1.2,\ha-5.5*\inter) node{T$^4$};
    
\draw (\deb-0.5,\ha) node{$\varphi_1$};
\draw (\deb-0.5,\ha-\inter) node{$\varphi_2$};
\draw (\deb-0.5,\ha-2*\inter) node{$\psi$};
\draw (\deb-0.5,\ha-3*\inter) node{$y$};
\draw (\deb-0.5,\ha-4*\inter) node{$x_1$};
\draw (\deb-0.5,\ha-5*\inter) node{$x_2$};
\draw (\deb-0.5,\ha-6*\inter) node{$x_3$};
\draw (\deb-0.5,\ha-7*\inter) node{$x_4$};
\draw (\deb-0.5,\ha-8*\inter) node{$t$};

\draw[black, dotted, line width=1mm] (\deb,\ha) -- (\deb+0.5,\ha);
\draw[black,line width=1mm] (\deb+0.5,\ha) -- (\deb+0.5+0.5*\rodsize,\ha);
\draw[black,line width=1mm] (\fin-0.5-0.5*\rodsize,\ha-\inter) -- (\fin-0.55,\ha-\inter);
\draw[black, dotted,line width=1mm] (\fin-0.5,\ha-\inter) -- (\fin,\ha-\inter);

\draw[amazon,line width=1mm] (\deb+0.5+0.5*\rodsize,\ha-4*\inter) -- (\deb+0.5+1*\rodsize,\ha-4*\inter);
\draw[amaranthred,line width=1mm] (\deb+0.5+1*\rodsize,\ha-3*\inter) -- (\deb+0.5+1.5*\rodsize,\ha-3*\inter);
\draw[bdazzledblue,line width=1mm] (\deb+0.5+1.5*\rodsize,\ha-2*\inter) -- (\deb+0.5+2*\rodsize,\ha-2*\inter);
\draw[amaranthred,line width=1mm] (\deb+0.5+2*\rodsize,\ha-3*\inter) -- (\deb+0.5+2.5*\rodsize,\ha-3*\inter);
\draw[byzantine,line width=1mm] (\deb+0.5+2.5*\rodsize,\ha-5*\inter) -- (\deb+0.5+3*\rodsize,\ha-5*\inter);

\draw[gray,dotted,line width=0.2mm] (\deb+0.5+0.5*\rodsize,\ha) -- (\deb+0.5+0.5*\rodsize,\ha-8*\inter);
\draw[gray,dotted,line width=0.2mm] (\deb+0.5+1*\rodsize,\ha) -- (\deb+0.5+1*\rodsize,\ha-8*\inter);
\draw[gray,dotted,line width=0.2mm] (\deb+0.5+1.5*\rodsize,\ha) -- (\deb+0.5+1.5*\rodsize,\ha-8*\inter);
\draw[gray,dotted,line width=0.2mm] (\deb+0.5+2*\rodsize,\ha) -- (\deb+0.5+2*\rodsize,\ha-8*\inter);
\draw[gray,dotted,line width=0.2mm] (\deb+0.5+2.5*\rodsize,\ha) -- (\deb+0.5+2.5*\rodsize,\ha-8*\inter);
\draw[gray,dotted,line width=0.2mm] (\deb+0.5+3*\rodsize,\ha) -- (\deb+0.5+3*\rodsize,\ha-8*\inter);
\end{tikzpicture}}

\newcommand{\AdSS}{
    \begin{tikzpicture}[remember picture]
\def\deb{-10} 
\def\inter{0.45} 
\def\ha{2.8} 
\def\zaxisline{4} 
\def\rodsize{1.1} 
\def\numrod{1.5} 

\def\fin{\deb+1+2*\rodsize+\numrod*\rodsize}

\draw (\deb+5.75,\ha-5*\inter) node{$\Longleftrightarrow$};

\draw[black,thin] (\deb+1,\ha) -- (\fin,\ha);
\draw[black,thin] (\deb,\ha-\inter) -- (\fin-1,\ha-\inter);
\draw[black,thin] (\deb,\ha-2*\inter) -- (\fin,\ha-2*\inter);
\draw[black,thin] (\deb,\ha-3*\inter) -- (\fin,\ha-3*\inter);
\draw[black,thin] (\deb,\ha-4*\inter) -- (\fin,\ha-4*\inter);
\draw[black,thin] (\deb,\ha-5*\inter) -- (\fin,\ha-5*\inter);
\draw[black,thin] (\deb,\ha-6*\inter) -- (\fin,\ha-6*\inter);
\draw[black,thin] (\deb,\ha-7*\inter) -- (\fin,\ha-7*\inter);
\draw[black,thin] (\deb,\ha-8*\inter) -- (\fin,\ha-8*\inter);

\draw [decorate, 
    decoration = {brace,
        raise=5pt,
        amplitude=5pt},line width=0.2mm,gray] (\deb-0.6,\ha-2.5*\inter+0.05) --  (\deb-0.6,\ha+0.5*\inter-0.05);
\draw [decorate, 
    decoration = {brace,
        raise=5pt,
        amplitude=5pt},line width=0.2mm,gray] (\deb-0.6,\ha-3.5*\inter+0.05) --  (\deb-0.6,\ha-2.5*\inter-0.05);
\draw [decorate, 
    decoration = {brace,
        raise=5pt,
        amplitude=5pt},line width=0.2mm,gray] (\deb-0.6,\ha-7.5*\inter+0.05) --  (\deb-0.6,\ha-3.5*\inter-0.05);
        
\draw[gray] (\deb-1.2,\ha-1*\inter) node{S$^3$};
\draw[gray] (\deb-1.2,\ha-3*\inter) node{S$^1$};
\draw[gray] (\deb-1.2,\ha-5.5*\inter) node{T$^4$};
    
\draw (\deb-0.5,\ha) node{$\varphi_1$};
\draw (\deb-0.5,\ha-\inter) node{$\varphi_2$};
\draw (\deb-0.5,\ha-2*\inter) node{$\psi$};
\draw (\deb-0.5,\ha-3*\inter) node{$y$};
\draw (\deb-0.5,\ha-4*\inter) node{$x_1$};
\draw (\deb-0.5,\ha-5*\inter) node{$x_2$};
\draw (\deb-0.5,\ha-6*\inter) node{$x_3$};
\draw (\deb-0.5,\ha-7*\inter) node{$x_4$};
\draw (\deb-0.5,\ha-8*\inter) node{$t$};

\draw[black, dotted, line width=1mm] (\deb,\ha) -- (\deb+0.5,\ha);
\draw[black,line width=1mm] (\deb+0.5,\ha) -- (\deb+0.5+0.5*\rodsize,\ha);
\draw[black,line width=1mm] (\fin-0.5-0.5*\rodsize,\ha-\inter) -- (\fin-0.55,\ha-\inter);
\draw[black, dotted,line width=1mm] (\fin-0.5,\ha-\inter) -- (\fin,\ha-\inter);

\draw[amaranthred,line width=1mm] (\deb+0.5+0.5*\rodsize,\ha-3*\inter) -- (\deb+0.5+2*\rodsize,\ha-3*\inter);
\draw[bdazzledblue,line width=1mm] (\deb+0.5+2*\rodsize,\ha-2*\inter) -- (\deb+0.5+3*\rodsize,\ha-2*\inter);

\draw[gray,dotted,line width=0.2mm] (\deb+0.5+0.5*\rodsize,\ha) -- (\deb+0.5+0.5*\rodsize,\ha-8*\inter);
\draw[gray,dotted,line width=0.2mm] (\deb+0.5+2*\rodsize,\ha) -- (\deb+0.5+2*\rodsize,\ha-8*\inter);
\draw[gray,dotted,line width=0.2mm] (\deb+0.5+3*\rodsize,\ha) -- (\deb+0.5+3*\rodsize,\ha-8*\inter);
\end{tikzpicture}}

\newcommand{\AdST}{
    \begin{tikzpicture}[remember picture]
\def\deb{-10} 
\def\inter{0.45} 
\def\ha{2.8} 
\def\zaxisline{4} 
\def\rodsize{1.1} 
\def\numrod{1.5} 

\def\fin{\deb+1+2*\rodsize+\numrod*\rodsize}

\draw (\deb+5.75,\ha-5*\inter) node{$\Longleftrightarrow$};

\draw[black,thin] (\deb+1,\ha) -- (\fin,\ha);
\draw[black,thin] (\deb,\ha-\inter) -- (\fin-1,\ha-\inter);
\draw[black,thin] (\deb,\ha-2*\inter) -- (\fin,\ha-2*\inter);
\draw[black,thin] (\deb,\ha-3*\inter) -- (\fin,\ha-3*\inter);
\draw[black,thin] (\deb,\ha-4*\inter) -- (\fin,\ha-4*\inter);
\draw[black,thin] (\deb,\ha-5*\inter) -- (\fin,\ha-5*\inter);
\draw[black,thin] (\deb,\ha-6*\inter) -- (\fin,\ha-6*\inter);
\draw[black,thin] (\deb,\ha-7*\inter) -- (\fin,\ha-7*\inter);
\draw[black,thin] (\deb,\ha-8*\inter) -- (\fin,\ha-8*\inter);

\draw [decorate, 
    decoration = {brace,
        raise=5pt,
        amplitude=5pt},line width=0.2mm,gray] (\deb-0.6,\ha-2.5*\inter+0.05) --  (\deb-0.6,\ha+0.5*\inter-0.05);
\draw [decorate, 
    decoration = {brace,
        raise=5pt,
        amplitude=5pt},line width=0.2mm,gray] (\deb-0.6,\ha-3.5*\inter+0.05) --  (\deb-0.6,\ha-2.5*\inter-0.05);
\draw [decorate, 
    decoration = {brace,
        raise=5pt,
        amplitude=5pt},line width=0.2mm,gray] (\deb-0.6,\ha-7.5*\inter+0.05) --  (\deb-0.6,\ha-3.5*\inter-0.05);
        
\draw[gray] (\deb-1.2,\ha-1*\inter) node{S$^3$};
\draw[gray] (\deb-1.2,\ha-3*\inter) node{S$^1$};
\draw[gray] (\deb-1.2,\ha-5.5*\inter) node{T$^4$};
    
\draw (\deb-0.5,\ha) node{$\varphi_1$};
\draw (\deb-0.5,\ha-\inter) node{$\varphi_2$};
\draw (\deb-0.5,\ha-2*\inter) node{$\psi$};
\draw (\deb-0.5,\ha-3*\inter) node{$y$};
\draw (\deb-0.5,\ha-4*\inter) node{$x_1$};
\draw (\deb-0.5,\ha-5*\inter) node{$x_2$};
\draw (\deb-0.5,\ha-6*\inter) node{$x_3$};
\draw (\deb-0.5,\ha-7*\inter) node{$x_4$};
\draw (\deb-0.5,\ha-8*\inter) node{$t$};

\draw[black, dotted, line width=1mm] (\deb,\ha) -- (\deb+0.5,\ha);
\draw[black,line width=1mm] (\deb+0.5,\ha) -- (\deb+0.5+0.5*\rodsize,\ha);
\draw[black,line width=1mm] (\fin-0.5-0.5*\rodsize,\ha-\inter) -- (\fin-0.55,\ha-\inter);
\draw[black, dotted,line width=1mm] (\fin-0.5,\ha-\inter) -- (\fin,\ha-\inter);

\draw[amaranthred,line width=1mm] (\deb+0.5+0.5*\rodsize,\ha-3*\inter) -- (\deb+0.5+2*\rodsize,\ha-3*\inter);
\draw[amazon,line width=1mm] (\deb+0.5+2*\rodsize,\ha-4*\inter) -- (\deb+0.5+3*\rodsize,\ha-4*\inter);

\draw[gray,dotted,line width=0.2mm] (\deb+0.5+0.5*\rodsize,\ha) -- (\deb+0.5+0.5*\rodsize,\ha-8*\inter);
\draw[gray,dotted,line width=0.2mm] (\deb+0.5+2*\rodsize,\ha) -- (\deb+0.5+2*\rodsize,\ha-8*\inter);
\draw[gray,dotted,line width=0.2mm] (\deb+0.5+3*\rodsize,\ha) -- (\deb+0.5+3*\rodsize,\ha-8*\inter);
\end{tikzpicture}}

\newcommand{\AdS}{
    \begin{tikzpicture}[remember picture]
\def\deb{-10} 
\def\inter{0.45} 
\def\ha{2.8} 
\def\zaxisline{4} 
\def\rodsize{1.1} 
\def\numrod{1.5} 

\def\fin{\deb+1+2*\rodsize+\numrod*\rodsize}

\draw (\deb+5.75,\ha-5*\inter) node{$\Longleftrightarrow$};

\draw[black,thin] (\deb+1,\ha) -- (\fin,\ha);
\draw[black,thin] (\deb,\ha-\inter) -- (\fin-1,\ha-\inter);
\draw[black,thin] (\deb,\ha-2*\inter) -- (\fin,\ha-2*\inter);
\draw[black,thin] (\deb,\ha-3*\inter) -- (\fin,\ha-3*\inter);
\draw[black,thin] (\deb,\ha-4*\inter) -- (\fin,\ha-4*\inter);
\draw[black,thin] (\deb,\ha-5*\inter) -- (\fin,\ha-5*\inter);
\draw[black,thin] (\deb,\ha-6*\inter) -- (\fin,\ha-6*\inter);
\draw[black,thin] (\deb,\ha-7*\inter) -- (\fin,\ha-7*\inter);
\draw[black,thin] (\deb,\ha-8*\inter) -- (\fin,\ha-8*\inter);

\draw [decorate, 
    decoration = {brace,
        raise=5pt,
        amplitude=5pt},line width=0.2mm,gray] (\deb-0.6,\ha-2.5*\inter+0.05) --  (\deb-0.6,\ha+0.5*\inter-0.05);
\draw [decorate, 
    decoration = {brace,
        raise=5pt,
        amplitude=5pt},line width=0.2mm,gray] (\deb-0.6,\ha-3.5*\inter+0.05) --  (\deb-0.6,\ha-2.5*\inter-0.05);
\draw [decorate, 
    decoration = {brace,
        raise=5pt,
        amplitude=5pt},line width=0.2mm,gray] (\deb-0.6,\ha-7.5*\inter+0.05) --  (\deb-0.6,\ha-3.5*\inter-0.05);
        
\draw[gray] (\deb-1.2,\ha-1*\inter) node{S$^3$};
\draw[gray] (\deb-1.2,\ha-3*\inter) node{S$^1$};
\draw[gray] (\deb-1.2,\ha-5.5*\inter) node{T$^4$};
    
\draw (\deb-0.5,\ha) node{$\varphi_1$};
\draw (\deb-0.5,\ha-\inter) node{$\varphi_2$};
\draw (\deb-0.5,\ha-2*\inter) node{$\psi$};
\draw (\deb-0.5,\ha-3*\inter) node{$y$};
\draw (\deb-0.5,\ha-4*\inter) node{$x_1$};
\draw (\deb-0.5,\ha-5*\inter) node{$x_2$};
\draw (\deb-0.5,\ha-6*\inter) node{$x_3$};
\draw (\deb-0.5,\ha-7*\inter) node{$x_4$};
\draw (\deb-0.5,\ha-8*\inter) node{$t$};

\draw[black, dotted, line width=1mm] (\deb,\ha) -- (\deb+0.5,\ha);
\draw[black,line width=1mm] (\deb+0.5,\ha) -- (\deb+0.5+0.5*\rodsize,\ha);
\draw[black,line width=1mm] (\fin-0.5-0.5*\rodsize,\ha-\inter) -- (\fin-0.55,\ha-\inter);
\draw[black, dotted,line width=1mm] (\fin-0.5,\ha-\inter) -- (\fin,\ha-\inter);

\draw[amaranthred,line width=1mm] (\deb+0.5+0.5*\rodsize,\ha-3*\inter) -- (\deb+0.5+3*\rodsize,\ha-3*\inter);

\draw[gray,dotted,line width=0.2mm] (\deb+0.5+0.5*\rodsize,\ha) -- (\deb+0.5+0.5*\rodsize,\ha-8*\inter);
\draw[gray,dotted,line width=0.2mm] (\deb+0.5+3*\rodsize,\ha) -- (\deb+0.5+3*\rodsize,\ha-8*\inter);
\end{tikzpicture}}

\newcommand{\BTZ}{
    \begin{tikzpicture}[remember picture]
\def\deb{-10} 
\def\inter{0.45} 
\def\ha{2.8} 
\def\zaxisline{4} 
\def\rodsize{1.1} 
\def\numrod{1.5} 

\def\fin{\deb+1+2*\rodsize+\numrod*\rodsize}

\draw (\deb+5.75,\ha-5*\inter) node{$\Longleftrightarrow$};

\draw[black,thin] (\deb+1,\ha) -- (\fin,\ha);
\draw[black,thin] (\deb,\ha-\inter) -- (\fin-1,\ha-\inter);
\draw[black,thin] (\deb,\ha-2*\inter) -- (\fin,\ha-2*\inter);
\draw[black,thin] (\deb,\ha-3*\inter) -- (\fin,\ha-3*\inter);
\draw[black,thin] (\deb,\ha-4*\inter) -- (\fin,\ha-4*\inter);
\draw[black,thin] (\deb,\ha-5*\inter) -- (\fin,\ha-5*\inter);
\draw[black,thin] (\deb,\ha-6*\inter) -- (\fin,\ha-6*\inter);
\draw[black,thin] (\deb,\ha-7*\inter) -- (\fin,\ha-7*\inter);
\draw[black,thin] (\deb,\ha-8*\inter) -- (\fin,\ha-8*\inter);

\draw [decorate, 
    decoration = {brace,
        raise=5pt,
        amplitude=5pt},line width=0.2mm,gray] (\deb-0.6,\ha-2.5*\inter+0.05) --  (\deb-0.6,\ha+0.5*\inter-0.05);
\draw [decorate, 
    decoration = {brace,
        raise=5pt,
        amplitude=5pt},line width=0.2mm,gray] (\deb-0.6,\ha-3.5*\inter+0.05) --  (\deb-0.6,\ha-2.5*\inter-0.05);
\draw [decorate, 
    decoration = {brace,
        raise=5pt,
        amplitude=5pt},line width=0.2mm,gray] (\deb-0.6,\ha-7.5*\inter+0.05) --  (\deb-0.6,\ha-3.5*\inter-0.05);
        
\draw[gray] (\deb-1.2,\ha-1*\inter) node{S$^3$};
\draw[gray] (\deb-1.2,\ha-3*\inter) node{S$^1$};
\draw[gray] (\deb-1.2,\ha-5.5*\inter) node{T$^4$};
    
\draw (\deb-0.5,\ha) node{$\varphi_1$};
\draw (\deb-0.5,\ha-\inter) node{$\varphi_2$};
\draw (\deb-0.5,\ha-2*\inter) node{$\psi$};
\draw (\deb-0.5,\ha-3*\inter) node{$y$};
\draw (\deb-0.5,\ha-4*\inter) node{$x_1$};
\draw (\deb-0.5,\ha-5*\inter) node{$x_2$};
\draw (\deb-0.5,\ha-6*\inter) node{$x_3$};
\draw (\deb-0.5,\ha-7*\inter) node{$x_4$};
\draw (\deb-0.5,\ha-8*\inter) node{$t$};

\draw[black, dotted, line width=1mm] (\deb,\ha) -- (\deb+0.5,\ha);
\draw[black,line width=1mm] (\deb+0.5,\ha) -- (\deb+0.5+0.5*\rodsize,\ha);
\draw[black,line width=1mm] (\fin-0.5-0.5*\rodsize,\ha-\inter) -- (\fin-0.55,\ha-\inter);
\draw[black, dotted,line width=1mm] (\fin-0.5,\ha-\inter) -- (\fin,\ha-\inter);

\draw[gray,line width=1mm] (\deb+0.5+0.5*\rodsize,\ha-8*\inter) -- (\deb+0.5+3*\rodsize,\ha-8*\inter);

\draw[gray,dotted,line width=0.2mm] (\deb+0.5+0.5*\rodsize,\ha) -- (\deb+0.5+0.5*\rodsize,\ha-8*\inter);
\draw[gray,dotted,line width=0.2mm] (\deb+0.5+3*\rodsize,\ha) -- (\deb+0.5+3*\rodsize,\ha-8*\inter);
\end{tikzpicture}}

\tikzset{%
  >=latex, % option for nice arrows
  inner sep=0pt,%
  outer sep=2pt,%
  mark coordinate/.style={inner sep=0pt,outer sep=0pt,minimum size=3pt,
    fill=black,circle}%
}

\title{\boldmath Non-BPS Bubbling Geometries in AdS$_3$
}

\author{Ibrahima Bah and}
\author{Pierre Heidmann} 
\affiliation{Department of Physics and Astronomy, Johns Hopkins University, 3400 North Charles Street, Baltimore, MD 21218, USA}

\emailAdd{iboubah@jhu.edu}
\emailAdd{pheidma1@jhu.edu}

\abstract{We construct large classes of non-BPS smooth horizonless geometries that are asymptotic to AdS$_3\times$S$^3\times$T$^4$ in type IIB supergravity.  These geometries are supported by electromagnetic flux corresponding to D1-D5 charges.  We show that Einstein equations for systems with eight commuting Killing vectors decompose into a set of Ernst equations, thereby admitting an integrable structure. This feature, which can a priori be applied to other AdS$_D\times\cC$ settings in supergravity, allows us to use solution-generating techniques associated with the Ernst formalism.  We explicitly derive solutions by applying the charged Weyl formalism that we have previously developed.  These are sourced internally by a chain of bolts that correspond to regions where the orbits of the commuting Killing vectors collapse smoothly.  We show that these geometries can be interpreted as non-BPS T$^4$ and S$^3$ deformations on global AdS$_3\times$S$^3\times$T$^4$ that are located at the center of AdS$_3$.  These non-BPS deformations can be made arbitrarily small and should therefore correspond to non-supersymmetric operators in the D1-D5 CFT.  Finally, we also construct interesting bound states of non-extremal BTZ black holes connected by regular bolts.  }

\preprint{}
\begin{document}

\maketitle
\flushbottom

%%%%%%%%%%%%%%%%%%%%%%%%%%%%%%%%%%%%%
\section{Introduction}
\label{sec:Intro}
%%%%%%%%%%%%%%%%%%%%%%%%%%%%%%%%%%%%%

An intriguing question in cosmological physics is the existence of gravitational solitons that can describe various dark and ultra-compact objects which, from afar, are indistinguishable from black holes in general relativity.  Recently, the authors have developed a framework for constructing and studying large classes of gravitational solitons that are described by smooth horizonless solutions with arbitrary mass and charges in supergravity \cite{Bah:2020ogh,Bah:2020pdz,Bah:2021owp,Bah:2021rki,Heidmann:2021cms,Bah:2022yji}.  These are thought of as coherent states of quantum gravity that admit classical descriptions.  The main novelty of our work is that it provides systematic tools for obtaining generically non-supersymmetric solutions, thereby allowing for a compelling case for gravitational solitons in the real world.

There are various interesting questions relating to gravitational solitons.\footnote{The necessary conditions for their existence in theories of gravity were discussed in \cite{Gibbons:2013tqa}.  There are questions on their production mechanics and their stability,  whose study has been initiated in \cite{Bah:2021irr}. }  The main question that motivates this paper is how to understand them more precisely as coherent states of quantum gravity.  Holography and the AdS/CFT correspondence provide the best definition of a quantum theory of gravity via conformal field theories.

The main objective of this paper is to derive a construction mechanism that allows us to systematically build non-supersymmetric asymptotically-AdS solutions that contain various gravitational solitons in the bulk.  In this point of view,  the solitons can be characterized by non-supersymmetric CFT operators under the AdS/CFT dictionary.  This provides a more precise definition of the solitons as states of a quantum theory.  This perspective has been very fruitful in the context of BPS states in various AdS/CFT contexts.  For instance,  the classification and matching of $\coeff{1}{2}$-BPS states in AdS$_5$/CFT$_4$ by the Lin-Lunin-Maldacena (LLM) geometries with their CFT duals has allowed for a precision test of the duality in $\cN=4$ Super-Yang-Mills \cite{Lin:2004nb}.  Similar success can be reported for the fuzzball and microstate geometry program where $\coeff{1}{4}$- and $\coeff{1}{8}$-BPS states have been classified,  and precision holography tests have been carried out for the AdS$_3$/CFT$_2$ duality in the context of the D1-D5 CFT \cite{Lunin:2002iz,Kanitscheider:2007wq,Giusto:2012yz,Giusto2015,Bena:2015bea,Shigemori:2020yuo}, and for the less understood AdS$_2$/CFT$_1$ duality \cite{Bena:2018bbd}.

However, much less is known beyond the comfortable frontiers of supersymmetry and about the holographic descriptions of non-BPS states.  The first challenge is to construct, from Einstein's equations in supergravity frameworks, large families of smooth non-BPS geometries that are asymptotic to AdS$_D\times\cC$ where $\cC$ describes a compact space.  The second challenge is to develop holographic dictionaries beyond supersymmetry.  In the former case,  only a few atypical sets of non-BPS smooth geometries \cite{Jejjala:2005yu,Bena:2009en,DallAgata:2010srl,Bossard:2014yta,Bossard:2014ola,Bena:2015drs,Bossard:2017vii} were known until recently.\footnote{There are other interesting constructions as \cite{Edery:2020kof,Edery:2022crs} in AdS$_3$,  but being constructed from Einstein gravity in three dimensions,  they do not admit UV description within a quantum gravity theory.} Some promising breakthroughs have been achieved first from a consistent S$^3$ truncation of six-dimensional supergravity \cite{Mayerson:2020tcl,Houppe:2020oqp,Ganchev:2021pgs} and second from the ``charged Weyl formalism'' established by the authors \cite{Bah:2020ogh,Bah:2020pdz,Bah:2021owp,Bah:2021rki,Heidmann:2021cms,Bah:2022yji}.  In all of these solutions,  only a small subset admits field theory descriptions in the context of AdS$_3$/CFT$_2$ in type IIB supergravity: the JMaRT geometry \cite{Chakrabarty:2015foa} and the microstrata \cite{Ganchev:2021ewa}.  This was done by describing them as non-BPS descendants of well-understood BPS configurations, using spectral flow for the former and relating them perturbatively to known BPS solutions for the latter.

This article contributes to bridging the gap towards non-BPS holography. The initial goal is to adapt the formalism of \cite{Bah:2020ogh,Bah:2020pdz,Bah:2021owp,Bah:2021rki,Heidmann:2021cms,Bah:2022yji} for a systematic construction of regular non-BPS geometries that are asymptotic to AdS$_D\times\cC$ and that can be considered as non-BPS deformations on known BPS backgrounds.  These deformations regularly backreact by inducing geometric transitions where compact directions in $\cC$ degenerate in the spacetime.  These form a chain of bolts along which the geometries smoothly cap off.   In this paper,  we focus on AdS$_3\times$S$^3\times$T$^4$ in type IIB supergravity which is dual to the D1-D5 CFT,\footnote{Note that some solutions in this paper require a KKm charge, $k$, such that the asymptotic is AdS$_3\times$S$^3/\mathbb{Z}_k\times$T$^4$.  The holographic dictionary of such a system is much less understood than that of the D1-D5 system already at the level of BPS states. }  but the formalism can be generalized to other AdS frameworks as AdS$_2\times$S$^3\times$T$^6$ in M-theory \cite{NonBPSAdS2}.

Another motivation for this paper is the development of solution-generating techniques for asymptotically-AdS spacetimes.  In asymptotically-flat spaces,  numerous approaches from the Ernst formalism, inverse scattering,  B\"{a}cklund transformations, or monodromy methods have been successfully applied to generate a variety of geometries.\footnote{see \cite{Belinski:2001ph,Belinsky:1971nt,Belinsky:1979mh,PhysRevLett.41.1197,Alekseev:1999kj,Alekseev:1999bv,Stephani:2003tm} as a non-exhaustive list of techniques used to generate solutions in four-dimensional General Relativity in vacuum of with gauge fields.\label{noteint} } These usually rely on the action of the Geroch group in gravity \cite{Geroch:1970nt,Geroch:1972yt} which arises from geometries with $d-2$ commuting Killing vectors, where $d$ is the dimension of the spacetime.  This latter assumption reduces Einstein equations for the $d$-dimensional spacetime to an integrable system on a two-dimensional plane,  usually parametrized by $(\rho,z)$ in the cylindrical Weyl coordinate system.  Generic solutions are generated by rod sources, which are segments on the $z$-axis,  and that induce spacelike or timelike coordinate degeneracies in the spacetime.

In backgrounds with cosmological constants, these powerful solution-generating techniques fail to apply since Einstein equations no longer reduce to a two-dimensional problem.  We overcome this issue by considering geometries of the form AdS$_D\times\cC$ with suitable flux in supergravity.  The curvature of the AdS spacetime can be balanced off with that of the internal space $\cC$,  and the action of the Geroch group still applies on the overall space when assuming $d-2$ commuting Killing vector.  Thus, we show for the first time how powerful solution-generating techniques in general relativity can be used to generate AdS solutions in supergravity.  This has been an open problem for many generations.  

In this paper, we focus on the D1-D5 system of type IIB on T$^4\times$S$^1$, and highlight the integrable structure of the equations of motion for backgrounds with eight commuting Killing vectors.  We show that inverse scattering or monodromy methods can be generically used to construct regular geometries, supported by D1-D5 flux, that are asymptotic to AdS$_3\times$S$^3\times$T$^4$ and admit a chain of regular rod sources in the interior.  These rods generate regular horizons of non-BPS D1-D5 black holes or smooth bolts that correspond to regions where circles from any of the three asymptotic components of AdS$_3\times$S$^3\times$T$^4$ degenerate.

In the charged Weyl formalism \cite{Bah:2020pdz,Bah:2021owp,Bah:2021rki,Heidmann:2021cms}, the integrable structure allows to restrict to a class of solutions that relies on a linear structure of the cylindrical axially-symmetric Laplace problem with non-BPS sources.  Moreover,  we reparametrize the two-dimensional system by considering the AdS$_3$ radial distance and the backbone angle of the S$^3$, $(r,\theta)$.  In this perspective, global AdS$_3$ is generated by a single rod source where the S$^1$ in AdS$_3$ degenerates at $r=0$.  This geometry on its own preserves supersymmetry.  Our construction allows for additional rod sources where circles in the T$^4$ or S$^3$ can also degenerate.  These deformations explicitly break supersymmetry and thereby correspond to non-BPS solutions.  They induce bolts that all sit at the center of AdS$_3$,  $r=0$,  and along segments on the sphere as depicted in Table \ref{tab:IntroTab}.  The regularity conditions at the bolts lead to ``non-BPS bubble equations'' which fix their size in terms of the asymptotic data: the radii of the T$^4$ and S$^1$, and the D1-D5 charges.  

We have the freedom to independently dial the total charges with respect to the sizes of the T$^4$ and S$^1$.  Therefore,  we can explore a multi-parameter family of solutions where the sizes of the extra rod sources can be small and treated as non-BPS perturbations on a global AdS$_3\times$S$^3\times$T$^4$ spacetime.  This is essential for deriving dual descriptions of these states and for understanding the dual operators that lead to these geometries.   Unlike any previously known asymptotically-AdS$_3$ smooth geometries,  these new deformations are the first ones that break the rigidity and symmetry of the T$^4$ or the S$^3$,  and that do not rely on a four-dimensional hyper-Kh\"aler base.

We can also consider rod sources where the timelike Killing vector degenerates, these will induce horizons of non-extremal D1-D5 black holes,  which are BTZ black holes.  With that regard,  we construct static bound states of non-extremal BTZ black holes in type IIB that are separated by smooth bolts where either the S$^1$,  S$^3$ or T$^4$ degenerates (see Table  \ref{tab:IntroTab2}).  They could be interesting solutions for the study of final states of black strings in Gregory-Laflamme instability in a similar fashion as in \cite{Lehner:2011wc,Emparan:2021ewh}.  

Our constructions provide a new perspective for exploring non-BPS smooth deformations of BPS black-hole microstates in string theory.  In the specific case of this paper, these are smooth non-BPS geometries in the D1-D5 system.  An important success of the microstate geometry program in supersymmetric settings has been that many black-hole microstates are coherent enough to admit classical descriptions. The result in this paper also demonstrates that large classes of non-BPS excitations and their associated degrees of freedom can generate coherent configurations that admit descriptions in terms of smooth geometries in AdS.  This is surprising and unexpected since it is usually believed that supersymmetry is a crucial ingredient to allow for smooth topological structure and to prevent gravitational collapse. Characterizing these states as CFT operators, and the mechanism by which they can decay will be important in understanding the fate of black-hole microstates and their associated geometries as they radiate away various excitations.  We hope to initiate a program to explore non-BPS structures in holography and AdS/CFT by providing explicit examples in string theory and supergravity.  

In this paper, we focus on the construction of non-BPS AdS$_3$ solutions and discuss their bulk description.  In upcoming work,  we would like to study their physical aspects in holography and provide systematic methods that allow for a study in D1-D5 CFT.

Before proceeding, we provide a summary of results and a roadmap for the paper.  The sections  \ref{sec:EOMGen} and \ref{sec:linearbranch} consist in the derivation of the solution-generating technique to systematically construct non-BPS states in AdS$_3\times$S$^3\times$T$^4$ in type IIB,  and which can a priori apply to other AdS$_D\times\cC$ frameworks. The reader particularly interested in the smooth non-BPS geometries in AdS$_3$ can jump to the self-contained sections \ref{sec:AdS3} and \ref{sec:AdS3Gen}.  The reader interested in the non-extremal BTZ bound states can directly go to section \ref{sec:BTZBS}. 

\subsection{Summary of results}

In section \ref{sec:EOMGen},  we derive the equations of motion for static D1-D5 solutions with eight commuting Killing vectors and discuss their integrable structure and relation to Ernst equations.  We derive generically the internal and asymptotic boundary conditions to impose for the construction of geometries that are asymptotic to AdS$_3\times$S$^3\times$T$^4$ and generated by rod sources.  These correspond to regions where a spacelike or timelike coordinate degenerates.

In section \ref{sec:linearbranch},  we summarize the linear branch of solutions obtained from the charged Weyl formalism \cite{Bah:2020pdz,Bah:2021owp,Bah:2021rki,Heidmann:2021cms},  and we adapt to asymptotically-AdS$_3$ solutions in type IIB.  We show that one can construct a large variety of smooth bubbling geometries or D1-D5 black hole bound states in this ansatz.  They are asymptotic to AdS$_3\times$S$^3\times$T$^4$ and terminate as a chain of rod sources where either a spacelike Killing vector degenerates, defining a smooth bolt, or the timelike Killing vector inducing a horizon.  The degenerating circle can either come from the T$^4$ direction,  parametrized by $(x_1,x_2,x_3,x_4)$,  or from the Hopf fiber of the S$^3$,  $\psi$,  or the S$^1$ inside the AdS$_3$ part,  denoted in this paper as the $y$-circle.  The T$^4$ bolts carry a D5 charge,  the S$^1$ bolts and the horizons induce D1 and D5 charges,  while the S$^3$ bolts have no charges. Thus,  since D1 and D5 charges are required to have asymptotically-AdS$_3$ solutions in type IIB, smooth horizonless solutions require at least one-rod source that forces the S$^1$ to degenerate.

In section \ref{sec:AdS3},  we construct simple smooth geometries obtained with our solution-generating technique,  and we discuss their physics.  The solutions depend on two variables that we choose to be the radial distance in AdS$_3$ and the angular position on the S$^3$,  $(r,\theta)$.  The simplest solution is sourced by a single rod that forces the degeneracy of the S$^1$ (see the first line of Table \ref{tab:IntroTab}).  We show that it corresponds to a rigid S$^3\times$T$^4$ fibration over a global AdS$_3$ spacetime as depicted on the right-hand side of the figure.  Thus,  our ansatz contains the BPS bulk dual of the NS-NS ground state of the D1-D5 CFT.  We then decorate this solution by adding a rod that forces either a circle in T$^4$ to degenerate (see the second line of Table \ref{tab:IntroTab}) or the S$^3$ Hopf fiber (see third line).  The spacetimes still cap off at $r=0$ where the rod sources are localized,  but the S$^3$ splits now into two regions given by two ranges of $\theta$: a first region where the S$^1$ shrinks as for the global AdS$_3$ solution and a second region that corresponds to the smooth T$^4$ or S$^3$ degeneracy.  We argue that these deformations break all supersymmetry and that their sizes are fixed by regularity in terms of the asymptotic quantities: the total D1-D5 charges and the radii of the internal directions.  Moreover,  the S$^3$ deformation requires imposing a smooth orbifold action on the S$^3$ such that the solution is asymptotic to AdS$_3\times$S$^3/\mathbb{Z}_k\times$T$^4$.  Finally,  we show that by either considering the volume of T$^4$ much smaller than the total D5-brane charge or by imposing $k\gg 1$,  these deformations are much smaller than the S$^1$ bolt such that they can be considered as non-BPS perturbations on global AdS$_3\times$S$^3\times$T$^4$ that have induced non-trivial smooth topological transformations at the center of AdS$_3$.

In section \ref{sec:AdS3Gen},  we derive more generic smooth geometries obtained from an arbitrary number of rod sources as depicted in the last line of Table \ref{tab:IntroTab2}.  They consist of an arbitrary
\begin{table}[H]\sffamily
\begin{adjustwidth}{-0.06\textwidth}{-0.06\textwidth}
\begin{tabular}{l|C@{\hspace{2pt}} CC}
\toprule
Sol. & Rod-source diagram & Geometry and topology  \\ 
\midrule
\rotatebox{90}{\hspace{-1.25cm} {\scriptsize Global AdS$_3\times$S$^3\times$T$^4$}}  & \vspace{0.2cm}\AdS & \includegraphics[width=20em]{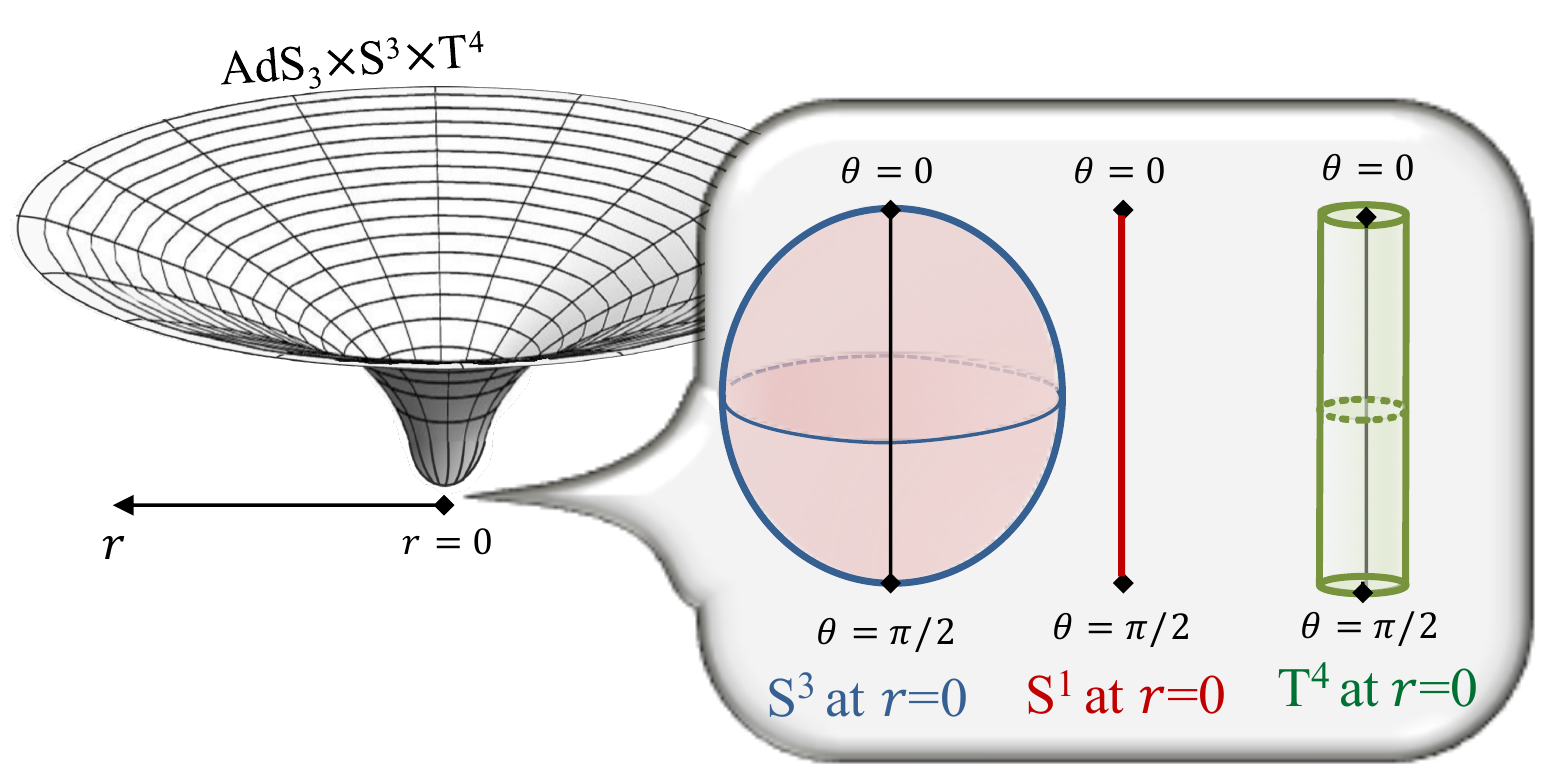}   \\ 
\hline
\rotatebox{90}{\hspace{-1.1cm} {\scriptsize A T$^4$deformation}}& \vspace{0.2cm}\AdST & \includegraphics[width=20em]{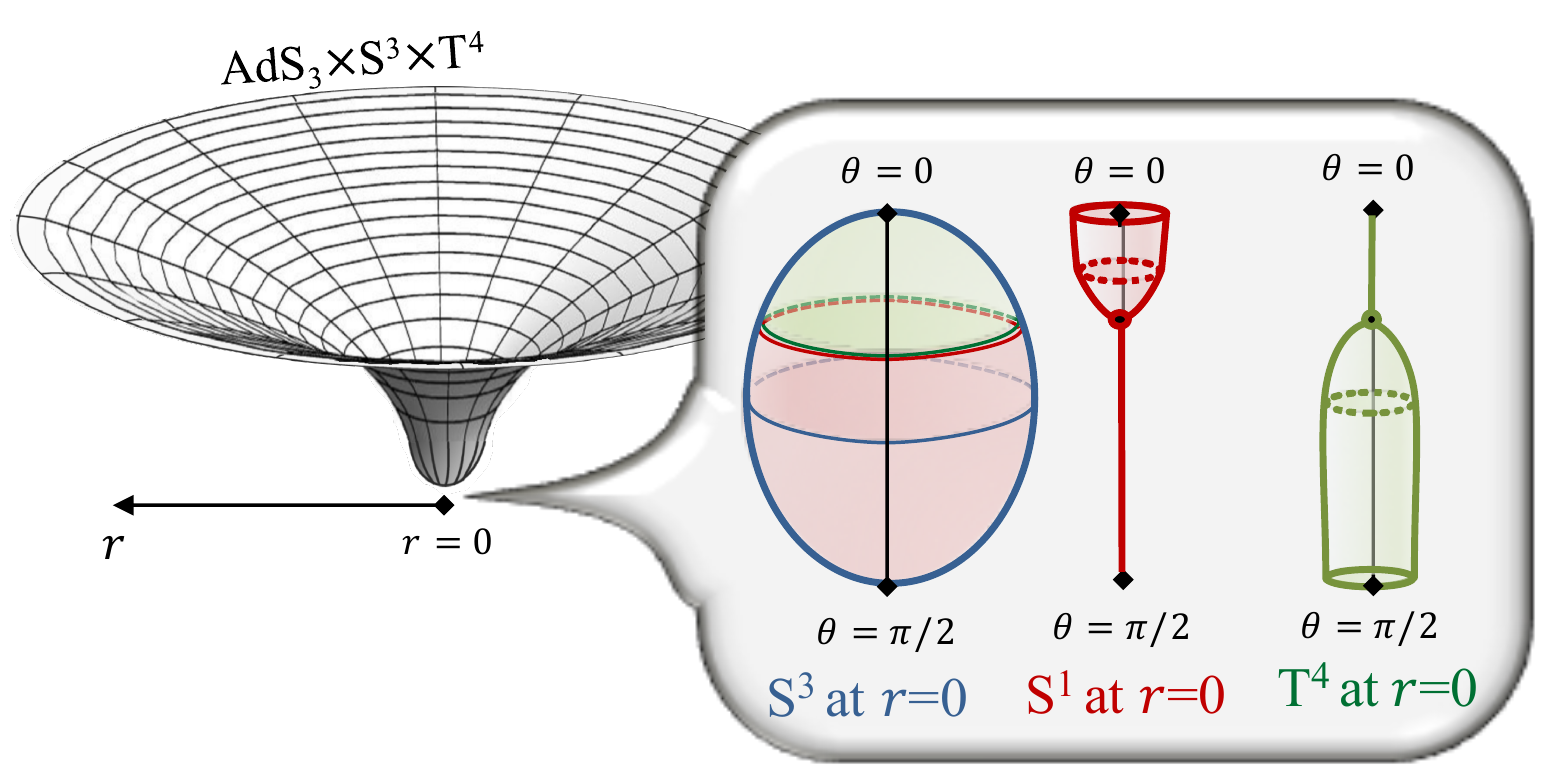}   \\ 
\hline
\rotatebox{90}{\hspace{-1.1cm} {\scriptsize A S$^3$deformation}} & \vspace{0.2cm}\AdSS & \includegraphics[width=20em]{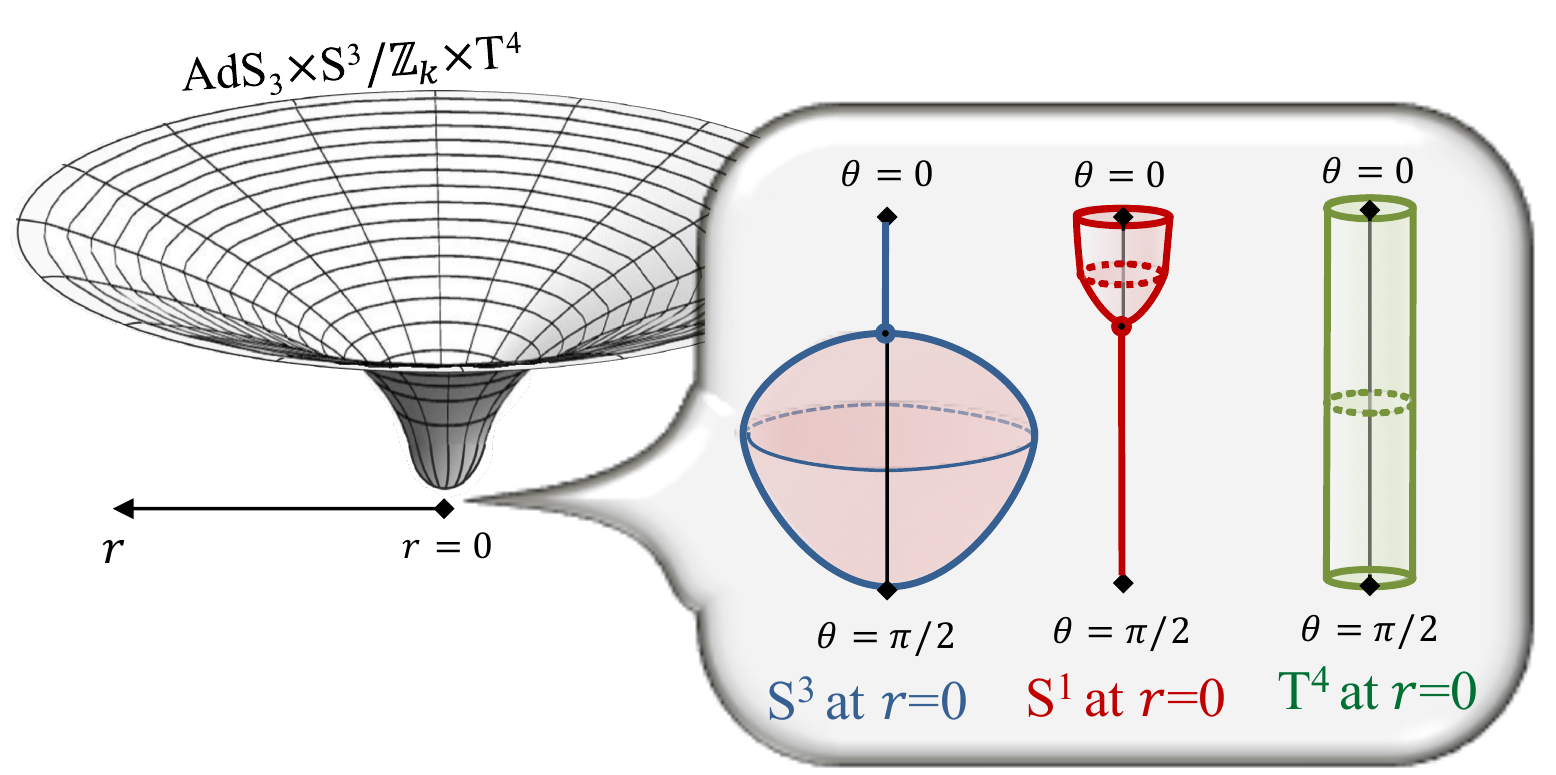}   \\ 
\hline
\rotatebox{90}{\hspace{-2cm} {\scriptsize Chain of T$^4$ \& S$^3$ deformations}} &\vspace{0.2cm}\AdSTS  & \includegraphics[width=20em]{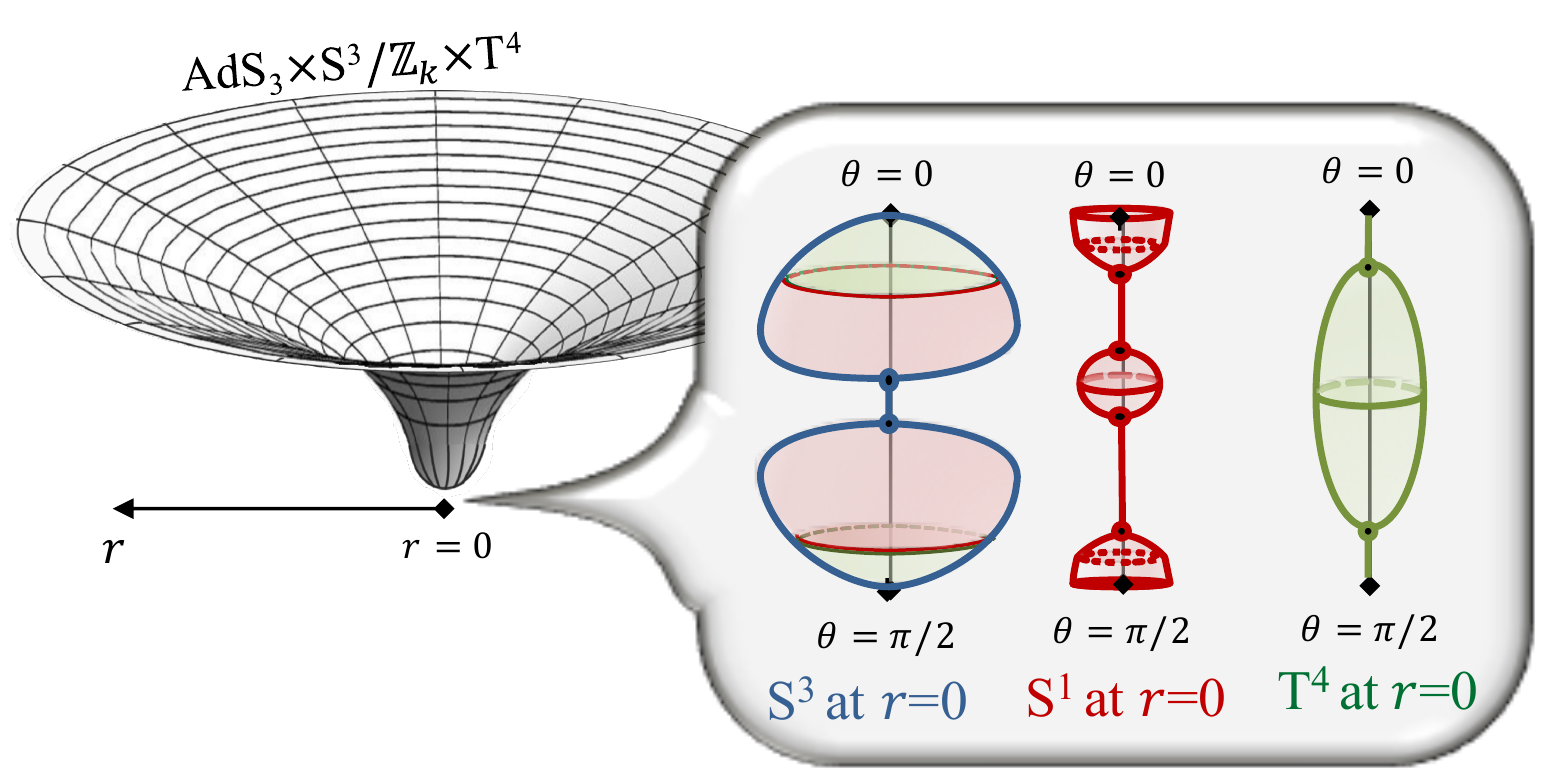}   \\ 
\bottomrule 
\end{tabular}
\caption{\label{tab:IntroTab} Description of the static axially-symmetric smooth  solutions constructed in this paper that are asymptotic to AdS$_3\times$S$^3\times$T$^4$ in type IIB with a potential orbifold on the S$^3$.  The left-hand sides depict the rod sources that force a direction to degenerate on the symmetry axis.  The S$^1$ ($y$) is the internal circle of the AdS$_3$ region,  $\varphi_1$ and $\varphi_2$ are spherical angles of the S$^3$ while $\psi$ is its Hopf fibration angle,  and the T$^4$ is parametrized by $x_a$.  The right-hand sides show the geometries as a function of $r$ and $\theta$ giving the position on the AdS$_3$ and S$^3$ respectively.  We depicted the topology of the S$^3$, S$^1$, and T$^4$ at the end-to-spacetime locus, $r=0$,  where the S$^3$ splits into different regions corresponding to the locus of each rod source.}
\end{adjustwidth}
\end{table} 
\noindent number of non-BPS T$^4$ and S$^3$ deformations such that the spacetimes terminate smoothly at $r=0$ as a chain of bolts that split the S$^3$ into different regions with non-trivial topology.   Similarly,  all rod sizes are fixed in terms of the asymptotic quantities and, for the same regime of parameters,  the deformations can be treated as smooth perturbations on global AdS$_3$.  Thus,  the only free parameters that are not asymptotic quantities are the number of rods which can be understood as the ``quantum bits'' to be included in the geometries,  and the nature of the rods,  which would correspond to the exact nature of the ``bits'' in this analogy.  The absence of moduli is related to the non-supersymmetric nature of these solutions which will be important to understand in the dual CFT.

In section \ref{sec:BTZBS},  we allow the geometries to have rod sources that force the timelike coordinate to degenerate and induce horizons.  We first consider the solution that consists of a single rod of this kind (see the first line of Table \ref{tab:IntroTab2}).  We show that it corresponds to a rigid 
\begin{table}[H]\sffamily
\begin{adjustwidth}{-0.06\textwidth}{-0.06\textwidth}
\begin{tabular}{l|C@{\hspace{2pt}} CC}
\toprule
Sol. & Rod-source diagram & Geometry and Topology  \\ 
\midrule
\rotatebox{90}{\hspace{-1.65cm} {\scriptsize Non-extreme BTZ$\times$S$^3\times$T$^4$}}  & \vspace{0.2cm}\BTZ & \includegraphics[width=20em]{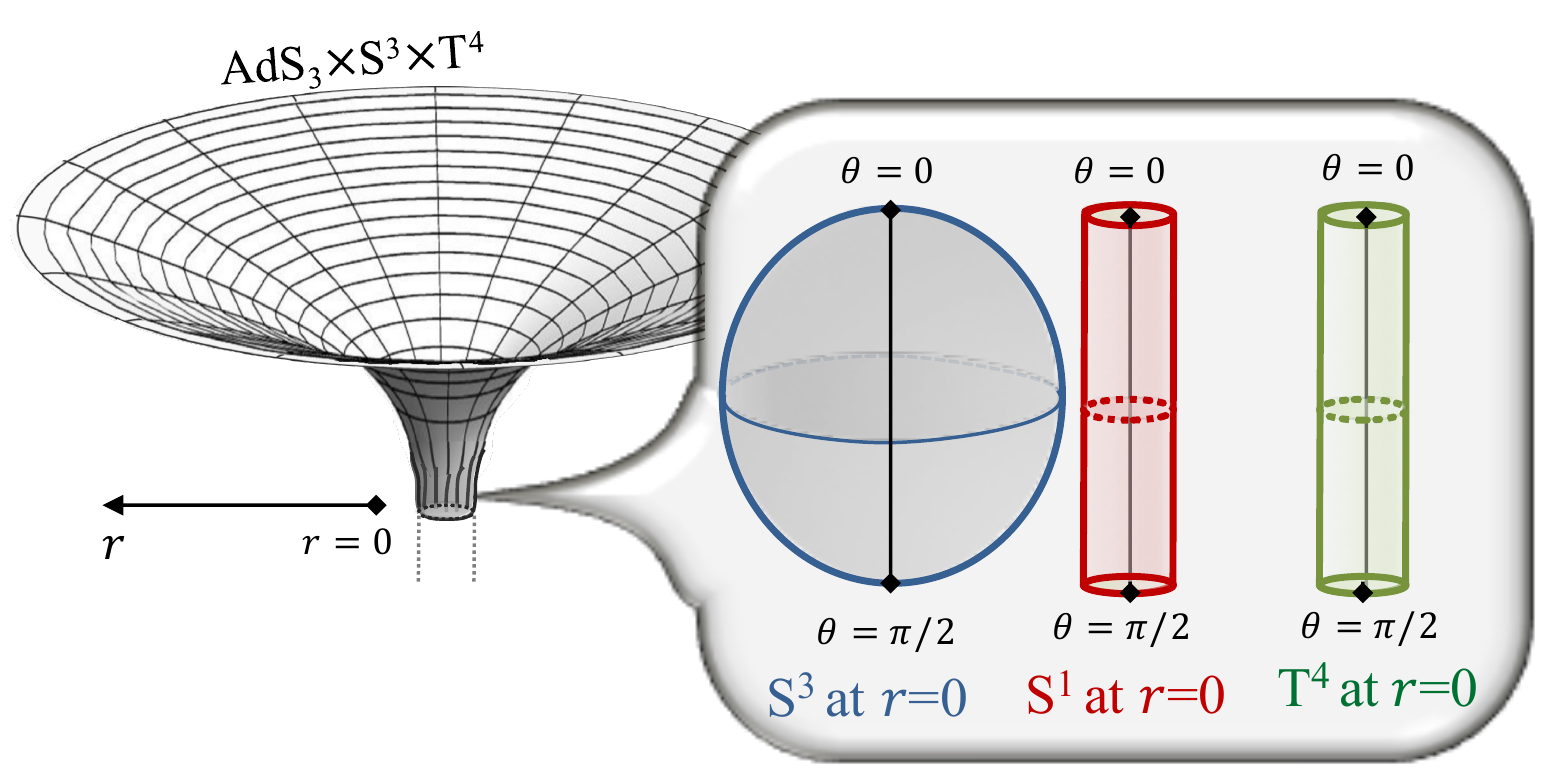}   \\ 
\hline
\rotatebox{90}{\hspace{-1.7cm} {\scriptsize Chain of BTZ and S$^1$ bolts}}& \vspace{0.2cm}\AdSBTZ & \includegraphics[width=20em]{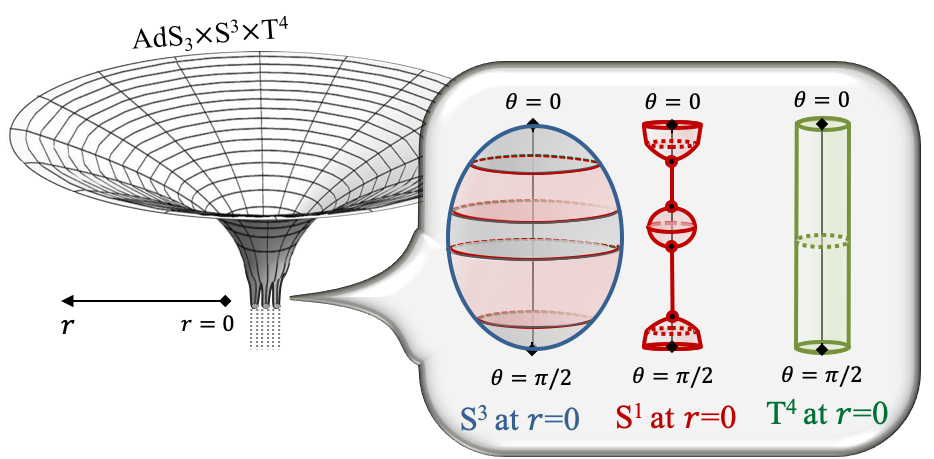}   \\ 
\hline
\rotatebox{90}{\hspace{-1.3cm} {\scriptsize Generic bound state}} & \vspace{0.2cm}\AdSBTZTS & \includegraphics[width=20em]{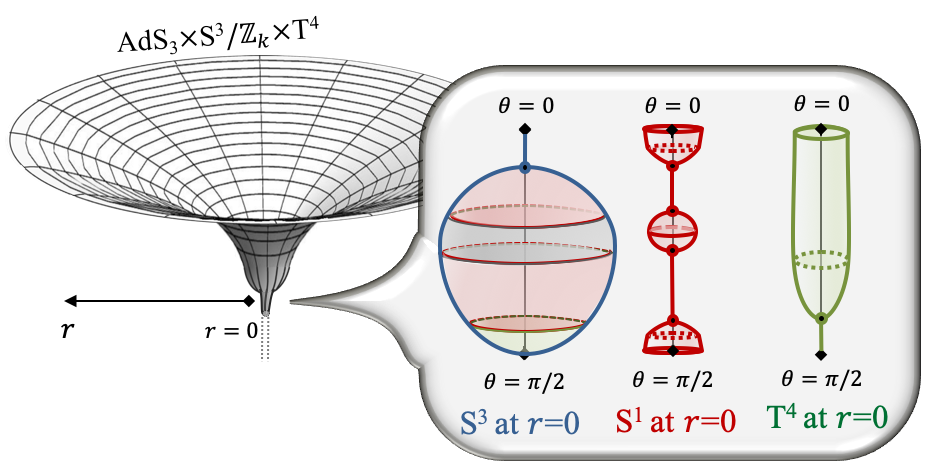}   \\ 
\bottomrule 
\end{tabular}
\caption{\label{tab:IntroTab2} Description of the axially-symmetric static non-extremal BTZ bound states constructed in this paper.  We used the same conventions as in Table \ref{tab:IntroTab}.}
\end{adjustwidth}
\end{table}  
\noindent S$^3\times$T$^4$ fibration over a static non-extremal BTZ black hole.  We then consider bound states of non-extremal black holes by having chains of such rods with bolts where the S$^1$ shrinks (see the second line of Table \ref{tab:IntroTab2}) or chains with generic bolts  (see the third line of Table \ref{tab:IntroTab2}).   The former has the advantage to involve AdS$_3$ directions only such that the S$^3\times$T$^4$ do not change topology in the IR.  For both types of solutions,  the spacetimes are regular for $r>0$,  and $r=0$ corresponds to the bound state of non-extremal black holes separated by bubbles.  One moves along these different objects by shifting $\theta$,  that is by moving along the S$^3$.  The black holes are in thermal equilibrium and the bolts are regular if all sizes are fixed in terms of the asymptotic quantities and the temperature.  Each type of rod can be considered as small perturbations by assuming some hierarchy of scales in between different asymptotic quantities and temperature.

%%%%%%%%%%%%%%%%%%%%%%%%%%%%%%%%%%%%%
\section{Integrable structure for non-BPS geometries in AdS$_3$}
\label{sec:EOMGen}
%%%%%%%%%%%%%%%%%%%%%%%%%%%%%%%%%%%%%

In this section, we derive the equations of motion obtained from type IIB supergravity and analyze their integrable structure.  We restrict to axially-symmetric and static backgrounds on T$^4\times$S$^1\times$S$^1$ with D1-D5 flux \cite{Bah:2020pdz,Bah:2021owp,Bah:2021rki,Heidmann:2021cms}. More precisely, we consider geometries that depend on two variables and have seven U(1) isometries and a time translation symmetry. Finally, we will have specific attention to boundary conditions that lead to regular geometries that are asymptotic to AdS$_3\times$S$^3\times$T$^4$.

\subsection{Einstein equations for the D1-D5 system}
\label{sec:EOM}

We consider static and axially-symmetric solutions of type IIB supergravity that depend on two variables. In the Weyl formalism, we can freely choose these coordinates, denoted as $(\rho,z)$, such that the induced metric on the two-dimensional space is conformally flat and that the induced metric on the remaining eight-dimensional spacetime satisfies $\det h_E \= -\rho^2$, where $h_E$ is the metric in the Einstein frame \cite{Weyl:book,Emparan:2001wk}. Moreover, one can consider one of the U(1) isometry, denoted as $\phi$, to have a metric coefficient proportional to $\rho^2$ such that the $(\rho,z,\phi)$ space defines a three-dimensional base in the Weyl cylindrical coordinate system, and $z$ plays the role of the axis of symmetry \cite{Emparan:2001wk,Bah:2020pdz,Heidmann:2021cms}.

The solutions are constructed on a T$^4\times$S$^1\times$S$^1$ and are supported by D1-D5 flux. The common S$^1$ direction of the D1 and D5 branes is parametrized by $y$, while the T$^4$ wrapped by the D5 branes is parametrized by $(x_1,x_2,x_3,x_4)$.  Finally,  we consider the remaining S$^1$, parametrized by an angle $\psi$,  as a Hopf fibration over the $(\rho,z,\phi)$ base.

An ansatz of metric and fields that suits the spacetime symmetries and flux is given, in the string frame,\footnote{The relation between the metric in the string frame and Einstein frame is $g=e^{\frac{\Phi}{2}}g_E$.} by
\begin{align}
ds_{10}^2 \= & \sqrt{\frac{W_0}{Z_1 Z_5}} \,\left[ - \frac{dt^2}{W_1}+ W_1\,dy^2\right]  \+ \sqrt{\frac{Z_1}{Z_5}}\,\sum_{a=1}^{4} W_{a+1}\,dx_a^2 \nn \\
&\+ \sqrt{W_0\,Z_1 Z_5} \left[\frac{1}{Z_0} \left(d\psi+H_0 \,d\phi\right)^2 +Z_0 \,\left(e^{2\nu} \left( d\rho^2 +dz^2\right) + \rho^2d\phi^2 \right) \right] \,, \label{eq:TypeIIBAnsatz} 
\end{align}
\begin{align}
C^{(2)} \= &H_5 \,d\phi \wedge d\psi -T_1 \,dt\wedge dy \,,\qquad  e^\Phi \= \sqrt{\frac{Z_1}{Z_5}\,W_0}\,,\qquad C^{(0)}=C^{(4)}=B_2=0\,. \nn
\end{align}
where  $C^{(p)}$, $B_2$, and $\Phi$ are the R-R gauge fields, the NS-NS gauge field, and dilaton respectively.  The warp factors and gauge potentials are functions of $(\rho,z)$. Moreover,  $\det h_E=-\rho^2$ requires
\begin{equation}
\prod_{i=2}^5 W_i \=1 \qquad \Longrightarrow \qquad W_5 \= \frac{1}{W_2 W_3 W_4}\,.
\label{eq:constraintW}
\end{equation}

We refer the reader interested in the derivation of the equations of motion to \cite{Heidmann:2021cms} or to Appendix \ref{App:EOMtypeIIB} for a more direct calculation. 

There are an electric gauge potential, $T_1$, induced by the D1 branes and two magnetic gauge potentials, $H_0$ for the KKm vector and $H_5$ for the D5 branes. We have introduced three warp factors, $\{Z_I\}_{I=0,1,5}$, which couple naturally with each gauge potential.  In addition, we have five independent warp factors, $\{W_\Lambda\}_{\Lambda=0,1,2,3,4}$, which are associated with the T$^4\times$S$^1$ deformations. Finally, $e^{2\nu}$ determines the nature of the three-dimensional base.  We introduce the cylindrical Laplacian operator of a flat three-dimensional base for axisymmetric functions:
\begin{equation}
\Delta \equi \frac{1}{\rho}\,\partial_\rho \left( \rho \,\partial_\rho \right) \+ \partial_z^2\,.
\label{eq:Laplacian}
\end{equation}
The Einstein equations can be written down in a uniform way if one defines electric duals of the magnetic D5 and KK gauge potentials and decompose $\nu$ such that 
\begin{equation}
dT_5 \equi \frac{-1}{\rho Z_5^2} \,\star_2 dH_5 \,, \qquad dT_0 \equi \frac{-1}{\rho Z_0^2} \,\star_2 dH_0 \,, \qquad \nu = \nu_{Z_1}+\nu_{Z_5} +\nu_{Z_0}+\sum_{i=0}^5 \nu_{W_i}\,,
\label{eq:ElecDual}
\end{equation}
where $\star_2$ is the Hodge star operator in the $(\rho,z)$ flat space and $\nu_X$ are the individual contributions of the warp factors in $\nu$. The equations of motion decompose into 9 sectors:
\begin{itemize}
\item[•] \underline{Six vacuum sectors for $ \Lambda=0,1,2,3,4,5$:}
\begin{align}
&\Delta \log W_\Lambda  \= 0\,,\label{eq:EOMVac}\\
&\frac{2}{\rho}\, \partial_z \nu_{W_\Lambda} \= \epsilon_\Lambda\,\partial_\rho \log W_\Lambda \,\partial_z \log W_\Lambda\,, \quad \frac{4}{\rho}\, \partial_\rho \nu_{W_\Lambda} \= \epsilon_\Lambda\left[\left( \partial_\rho \log W_\Lambda\right)^2 \- \left(\partial_z \log W_\Lambda\right)^2\right],\nn
\end{align}
where $\epsilon_\Lambda=1$ for $\Lambda=1$ and $\epsilon_\Lambda=
1/2$ otherwise.  Note that the equation for $\Lambda=5$ is not independent of the others due to the constraint \eqref{eq:constraintW}. However, it still gives a non-trivial contribution $\nu_{W_5}$.
\item[•] \underline{Three Maxwell sectors for $ I=0,1,5$:}
\begin{align}
& \Delta \log Z_I \= - Z_I^2  \left[ (\partial_\rho T_I)^2 + (\partial_z T_I)^2 \right]\,, \quad  \partial_\rho \left(\rho Z_I^2 \,\partial_\rho T_I\right)\+\partial_z \left( \rho Z_I^2 \,\partial_z T_I\right)  \=0 \,,\nn \\
&\frac{2}{\rho}\, \partial_z \nu_{Z_I} \= \partial_\rho \log Z_I \,\partial_z \log Z_I - Z_I^2 \,\partial_\rho  T_I\partial_z  T_I\,, \label{eq:EOMMaxwell}\\
& \frac{4}{\rho}\, \partial_\rho \nu_{Z_I} \=\left( \partial_\rho \log Z_I\right)^2 \- \left(\partial_z \log Z_I\right)^2 - Z_I^2 \, \left((\partial_\rho T_I)^2-(\partial_z T_I)^2 \right)\,. \nn
\end{align}
\end{itemize}
The equations for the T$^4\times$S$^1$ deformations,  $W_\Lambda$,  and their associated $\nu_W$ form a linear system of equations. They are identical to vacuum Weyl equations \cite{Weyl:book,Emparan:2001wk}. The logarithms of $W_\Lambda$ are harmonic functions for which solutions sourced by segments, i.e. rods,  on the $z$-axis are explicitly known.

The equations in the Maxwell sectors are coupled non-linear equations and admit an interesting structure which we discuss next.  

\subsection{Integrable structure and linear solutions}
\label{sec:integrableStruc}

A remarkable feature of the three Maxwell sectors \eqref{eq:EOMMaxwell} is that they are identical to equations obtained from four-dimensional axially-symmetric static geometries with a single one-form gauge field.  More precisely,  a background given by
\begin{equation} \label{eq:4dsys}
ds_4^2 \= - \frac{dt^2}{Z_I^2} + Z_I^2 \left[ e^{8 \nu_{Z_I}} \left( d\rho^2 +dz^2 \right) +\rho^2 d\phi^2 \right]\,,\qquad F = -2 dT_I \wedge dt\,,
\end{equation}
leads to the exact same equations as for the three Maxwell sectors \eqref{eq:EOMMaxwell}.\footnote{We have considered the four-dimensional Einstein-Maxwell action $$(16 \pi G_4)\,S_4 =\int d^4x \sqrt{g} \left(R - \frac{1}{4} F_{\mu\nu}F^{\mu\nu} \right).$$}

The equations of motion of this system admit integrable structures that are well established from the Ernst formalism and inverse scattering.  These integrable structures follow the fact that the ansatz in \eqref{eq:4dsys} admit an action from the Geroch group \cite{Geroch:1970nt,Geroch:1972yt}.  Our general system above will inherit all of these structures that can allow for a large phase of solutions.  Indeed, monodromy methods and B\"acklund transformations can be used to extract solutions \cite{Belinsky:1979mh,PhysRevLett.41.1197,Alekseev:1999kj,Alekseev:1999bv,Stephani:2003tm}.  In particular, they can be used as solution-generating methods for non-BPS AdS solutions.  

In this paper,  we focus on a specific linear class of solutions of the Maxwell system which can be obtained from the integrable structure of the Ernst formalism.\footnote{See section 18.6.3 of \cite{Stephani:2003tm}.} The gravitational potential of the spacetime in \eqref{eq:4dsys} is the redshift factor $Z_I$.  We can consider an ansatz where the electric potential is a function of the gravitational potential  $T_I(Z_I)$.  By plugging into the equations of motion of $(Z_I,T_I)$ in \eqref{eq:EOMMaxwell},  we found that both potentials are expressed in terms of a function for which the logarithm is harmonic and three complex constants:
\begin{equation}\label{eq:soluX}
Z \= \frac{e^{b} L - e^{-b}L^{-1}}{2 a}\,,\qquad T \= \frac{\sqrt{1+a^2 Z^2}}{Z}+c\,, \qquad \Delta \log L \=0
\end{equation} where $(a,b,c) \in \mathbb{C}$ and we have dropped the $I$ index for clarity.   

The key ingredient is the potential $L$ for which the logarithm satisfies the three-dimensional axially-symmetric Laplace equation.  Arbitrary solutions can be obtained by considering arbitrary sources to this linear equation in a similar fashion as for vacuum Weyl solutions \cite{Weyl:book,Emparan:2001wk},  but with now non-trivial electromagnetic flux turned on.  This is the reason why this branch of solutions has been denoted as ``the charged Weyl formalism'' in \cite{Bah:2020ogh,Bah:2020pdz,Bah:2021owp,Heidmann:2021cms,Bah:2021rki}. This is an explicit realization of the integrable structure which exists for the four-dimensional system in \eqref{eq:4dsys},  and thus inherited by the type IIB D1-D5 system in \eqref{eq:TypeIIBAnsatz}. 

Note that the electric potential $T$ does not simply reduce to the BPS branch where it counterbalances the gravitational potential $T=\frac{1}{Z}$. Thus, the structure in \eqref{eq:soluX} can be taken as a non-BPS but still linear generalization of BPS multicenter solutions \cite{Gauntlett:2002nw,Bena:2005va,Bena:2007kg,Heidmann:2017cxt,Bena:2017fvm,Bena:2018bbd,Heidmann:2018vky}.  This has composed much of the recent progress in constructing asymptotically-flat non-BPS smooth horizonless solutions in \cite{Bah:2020ogh,Bah:2020pdz,Bah:2021owp,Heidmann:2021cms,Bah:2021rki}, while \cite{Bah:2022yji} exploits inverse scattering methods.  One of the main goals of this paper is to show how this linear structure can be adapted to construct large families of non-BPS asymptotically-AdS$_3$ smooth geometries.

\subsection{Boundary conditions}
\label{sec:BC}

In \cite{Bah:2020ogh,Bah:2020pdz,Bah:2021owp,Bah:2021rki,Heidmann:2021cms,Bah:2022yji}, the ansatz \eqref{eq:TypeIIBAnsatz} has been used to generate non-BPS bubbling geometries that are either asymptotic to $\IR^{1,3}\times$S$^1\times$S$^1\times$T$^4$ or $\IR^{1,4}\times$S$^1\times$T$^4$.   In this section,  we introduce new boundary conditions for asymptotically AdS$_3\times$S$^3\times$T$^4$ geometries.  Moreover,  several solutions will require that the asymptotic S$^3$ has a smooth orbifold action. Thus,  we will more generically consider boundary conditions leading to AdS$_3\times$S$^3/\mathbb{Z}_k\times$T$^4$.

For the solutions constructed in \cite{Bah:2020ogh,Bah:2020pdz,Bah:2021owp,Bah:2021rki,Heidmann:2021cms,Bah:2022yji},  the spacetimes end as a chain of smooth bubbles. The internal bubbles are induced by rods on the $z$-axis, which are finite segments where a spacelike Killing vector degenerates smoothly.  They correspond to bolts where one of the compact circles degenerates on the symmetry axis.  The local geometry at each bolt is $\IR^2\times\mathcal{C}_\text{Bubble}$ where $\mathcal{C}_\text{Bubble}$ is a compact space defining the topology of the bubble.  Such a geometric transition can be produced by imposing appropriate singular behaviors on the warp factors $(W_\Lambda,Z_I,\nu)$ at the rods so that all metric components but one are finite.

\subsubsection{Asymptotic boundary conditions}
\label{sec:asympBC}

We introduce the asymptotic spherical coordinates 
\begin{equation}
\rho \equi \frac{r^2}{4}\, \sin 2\theta \,,\qquad z \equi \frac{r^2}{4}\, \cos 2\theta\,.
\label{eq:AsympCoord}
\end{equation}
We consider the following asymptotic behaviors at large $r$,  that are compatible with the equations of motion, 
\begin{align}
&W_\Lambda,\, e^{2\nu} \,\sim \,1\,,\qquad Z_1 \,\sim\, \frac{Q_1}{r^2} \,, \qquad Z_5 \,\sim\, \frac{Q_5}{r^2} \,, \qquad Z_0 \,\sim\, \frac{4k}{r^2} \,, \nn \\
&  H_0   \,\sim \,k\cos 2\theta\,,  \qquad H_5  \,\sim \, \frac{Q_5}{4} \cos 2\theta\,,  \qquad T_1 \,\sim \, \frac{r^2}{Q_1}\,.
\label{eq:AsympBehav}
\end{align}
The metric and fields \eqref{eq:TypeIIBAnsatz} are asymptotic to
\begin{equation}
\begin{split}
ds_{10}^2& \,\sim \,  k\sqrt{Q_1 Q_5} \left[\frac{r^2}{k Q_1 Q_5}\,(-dt^2+dy^2)+ \frac{dr^2}{r^2} \+ d\Omega_3^2\right]  +\sqrt{\frac{Q_1}{Q_5}}\sum_{a=1}^4 dx_a^2 \,, \\ \label{eq:AdS3Asymp}
C^{(2)} &\,\sim \, \frac{Q_5}{4} \cos 2\theta \, d\phi \wedge d\psi - \frac{r^2}{Q_1} dt\wedge dy\,, \qquad e^{\Phi } \sim \sqrt{\frac{Q_1}{Q_5}}\,,
\end{split}
\end{equation}
where $Q_1$ and $Q_5$ are the supergravity D1 and D5 brane charges and $d\Omega_3^2$ is the line element of the S$^3$,  
\begin{equation}
d\Omega_3^2 =d\theta^2+\cos^2 \theta\,d\varphi_1^2+\sin^2 \theta\,d\varphi_2^2 \,.
\end{equation}
We have defined \emph{the spherical angles} of the S$^3$ from the Hopf fibration angles such as
\begin{equation}
\varphi_1 \equi \frac{1}{2}\left(\phi +\frac{\psi}{k}\right)\,,\qquad \varphi_2 \equi  \frac{1}{2}\left(\phi -\frac{\psi}{k}\right)\quad \Leftrightarrow \quad \phi \= \varphi_1 +\varphi_2 \,,\qquad \psi= k \,(\varphi_1 -\varphi_2)\,.
\label{eq:DefHyperspher}
\end{equation}
We define the periodicity of the compact directions such that
\begin{equation}
\begin{split}
(\psi,\phi) &\= (\psi,\phi) \+ (4\pi,0)\,,\qquad (\psi,\phi) \= (\psi,\phi) \+ (2\pi,2\pi)\,,\\
 y &\= y \+2\pi  R_y \,,\qquad x_a \= x_a \+ 2\pi R_{x_a}\,, \quad a=1,2,3,4\,,
\end{split}
\label{eq:psi&phiPerio}
\end{equation}
where $R_{y}$ and $R_{x_a}$ correspond to the radii of the S$^1$ and the T$^4$ directions.  Thus, the geometries are asymptotic to AdS$_3\times$S$^3/\mathbb{Z}_k\times$T$^4$ for which the AdS$_3$ and S$^3$ radii are equal to $(k^2 Q_1 Q_5)^\frac{1}{4}$.  One can restrict to solutions without orbifold asymptotically by simply considering $k=1$ in the above expressions.

To conclude, one can obtain asymptotically AdS$_3$ solutions if the warp factors $Z_I$ vanish at a large distance as $r^{-2}$.  This requires sourcing them internally, such that they have a singular behavior.  These singularities must be carefully tuned to correspond to regular coordinate degeneracies.

\subsubsection{Internal boundary conditions}

As previously argued by the author in \cite{Bah:2020ogh,Bah:2020pdz,Bah:2021owp,Bah:2021rki,Heidmann:2021cms,Bah:2022yji},  the type IIB ansatz \eqref{eq:TypeIIBAnsatz} allows for the construction of non-BPS smooth bubbling geometries by generating bolts on the $z$-axis.  These are obtained when the warp factors are sourced at segments of the $z$-axis and have suitable singular behaviors.  We consider a source in between $z_- \leq z \leq z_+$ at $\rho=0$ and introduce the following behavior as we approach $\rho \to 0$, 
\begin{equation}
Z_I \,\propto \, \rho^{-2\alpha_{Z_I}}\,,\qquad W_\Lambda \,\propto \, \rho^{-2\alpha_{W_\Lambda}}\,,\qquad e^{2\nu} \,\propto \, \rho^{2\alpha_{\nu}}\,,
\label{eq:InternalBehaviors}
\end{equation}
where $\alpha_X$ are constants.  

Therefore,  there are only 7 non-trivial combinations for which the rod corresponds to a regular coordinate degeneracy on the $z$-axis such that the local metric behaves as
\begin{equation}
ds_{10}^2 \,\propto\, d\rho^2 - \frac{\rho^2}{\kappa_t^2} dt^2 +ds(\text{horizon})^2\,,\quad \text{or}\quad ds_{10}^2 \,\propto\, d\rho^2 + \frac{\rho^2}{\kappa_x^2}  dx^2 - g_{tt} dt^2 +ds(\text{bubble})^2\,,
\label{eq:boltDef}
\end{equation}
where $x$ is one of the compact direction $(\psi,y,x_1,x_2,x_3,x_4)$, and $\kappa$ is a constant. Moreover, $ds(\text{horizon})$ or $ds(\text{bubble})$ is the line element of the compact space that corresponds to either a horizon if the rod induces the degeneracy of the timelike direction or a bubble if it is a spacelike direction.  In addition,  $\kappa$ must be fixed by regularity in terms of the periodicity of the compact direction,  generically denoted as $x\to x+2\pi R_x$, or the black hole temperature, $T$:
\begin{equation}
\kappa_t \= \frac{1}{2\pi T}\,,\quad \text{or}\quad \kappa_x \= R_x\,.
\label{eq:ConstraintInternal}
\end{equation}

The 7 values of $\alpha_X$ that lead to these local geometries are summarized in Table \ref{tab:internalBCGen}.
\begin{table}[h]
  \centering
  \begin{tabular}{|c||c!{\vrule width 1.4pt}c|c|c!{\vrule width 1.4pt}c|c|c|c|c|}
\hline 
 & $\alpha_\nu$ & $\alpha_{Z_0}$ & $\alpha_{Z_1}$& $\alpha_{Z_5}$&  $\alpha_{W_0}$ &$\alpha_{W_1}$& $\alpha_{W_2}$& $\alpha_{W_3}$& $\alpha_{W_4}$ \\ \hline \hline
Horizon & $1$ &$\frac{1}{2}$ &$\frac{1}{2}$ &$\frac{1}{2}$ &$0$ &$\frac{1}{2}$ &$0$ &$0$ &$0$ \\ \noalign{\hrule height 0.8pt}
$\psi$ degeneracy & $1$ &$1$ &$0$ &$0$ &$0$ &$0$ &$0$ &$0$ &$0$ \\ \hline
$y$ degeneracy & $1$ &$\frac{1}{2}$ &$\frac{1}{2}$ &$\frac{1}{2}$  &$0$ &$-\frac{1}{2}$&$0$ &$0$ &$0$ \\ \hline
$x_1$ degeneracy & $1$ &$\frac{1}{2}$ &$0$ &$\frac{1}{2}$ &$\frac{1}{2}$ &$0$ &$-\frac{3}{4}$ &$\frac{1}{4}$ &$\frac{1}{4}$ \\ \hline
$x_2$ degeneracy & $1$ &$\frac{1}{2}$ &$0$ &$\frac{1}{2}$  &$\frac{1}{2}$ &$0$ &$\frac{1}{4}$ &$-\frac{3}{4}$ &$\frac{1}{4}$ \\ \hline
$x_3$ degeneracy & $1$ &$\frac{1}{2}$ &$0$ &$\frac{1}{2}$ &$\frac{1}{2}$ &$0$ &$\frac{1}{4}$ &$\frac{1}{4}$ &$-\frac{3}{4}$ \\ \hline
$x_4$ degeneracy & $1$ &$\frac{1}{2}$ &$0$ &$\frac{1}{2}$  &$\frac{1}{2}$ &$0$ &$\frac{1}{4}$ &$\frac{1}{4}$ &$\frac{1}{4}$ \\ \hline
  \end{tabular}
  \caption{\label{tab:internalBCGen} The seven choices of boundary conditions at a rod source \eqref{eq:InternalBehaviors} leading to a regular coordinate degeneracy of the timelike direction or a compact spacelike direction.}
\end{table}
Note that $\alpha_\nu$ is not an independent parameter a priori since the first derivatives of $\nu$ are quadratically sourced by $Z_I$ and $W_\Lambda$. However, by studying the local behavior of the equations for $\nu$, \eqref{eq:EOMVac} and \eqref{eq:EOMMaxwell}, one can show that 
\begin{equation}
\alpha_\nu \= \alpha_{Z_0}^2+\alpha_{Z_1}^2+\alpha_{Z_5}^2+\sum_{i=1}^4 \alpha_{W_i}^2 + \frac{\alpha_{W_0}^2}{2} + \alpha_{W_2} \alpha_{W_3}+ \alpha_{W_2} \alpha_{W_4}+ \alpha_{W_3} \alpha_{W_4},
\end{equation}
and the internal boundary conditions in Table \ref{tab:internalBCGen} are consistent.

If $\alpha_X =0$ then the corresponding warp factor is not sourced at the rod.  Moreover,  if $\alpha_{Z_1}$ or $\alpha_{Z_5}\neq 0$,  the associated gauge potential, $T_1$ or $H_5$,  is also sourced at the rod and carries a charge.  More precisely,  a rod corresponding to a horizon leads to a D1-D5 black hole,  a rod corresponding to the degeneracy of the $\psi$-circle induces a bolt without D1 and D5 charges,  a rod obtained from the degeneracy of a T$^4$ direction carries a D5 charge while a rod making the $y$-circle degenerate corresponds to a D1-D5 bolt.

%%%%%%%%%%%%%%%%%%%%%%%%%%%%%%%%%%%%%
\section{A linear branch of solutions}
\label{sec:linearbranch}
%%%%%%%%%%%%%%%%%%%%%%%%%%%%%%%%%%%%%

In this section,  we summarize the charged Weyl formalism that has been used in \cite{Bah:2020pdz,Bah:2021owp,Bah:2021rki,Heidmann:2021cms} to construct smooth asymptotically-flat non-BPS geometries in various supergravity frameworks.  They satisfy the same equations as the ones derived in section \ref{sec:EOM} and are based on the linear branch of solutions that we introduced in section \ref{sec:integrableStruc}.  We will adapt the formalism in the context of building asymptotically-AdS$_3$ non-BPS geometries in type IIB supergravity.

\subsection{Charged Weyl formalism}

As introduced in section \ref{sec:integrableStruc},  the eight independent sectors of coupled differential equations \eqref{eq:EOMVac} and \eqref{eq:EOMMaxwell} can be solved by considering \emph{eight functions} for which their logarithms are harmonic functions:
\begin{equation}
\Delta \log L_I \= \Delta \log W_\Lambda = 0 \,,\qquad  I = 0,5,p\,,\quad \Lambda=0,1,..,4,
\end{equation}
where $\Delta$ is the flat Laplacian \eqref{eq:Laplacian}.  Then,  the type IIB fields of \eqref{eq:TypeIIBAnsatz} are given by \eqref{eq:soluX}
\begin{equation}
Z_I \= \frac{e^{b_I} \, L_I - e^{-b_I} \,L_I^{-1}}{2 a_I} \,,\qquad T_I \= \frac{\sqrt{1+a_I^2 Z_I^2}}{Z_I}\,,\qquad \star_2 dH_I \= \frac{\rho}{a_I}\, d(\log L_I)\,,
\label{eq:LinearSolGen}
\end{equation}
where $a_I$ and $b_I$ are positive arbitrary constants, and the base warp factor $\nu$ can be obtained by integrating \eqref{eq:EOMVac} and \eqref{eq:EOMMaxwell}.\footnote{The equation for the $\nu_{Z_I}$ simplifies in terms of the $L_I$ such that it takes the same form as the vacuum equations for $\nu_{W_\Lambda}$:
$$
\frac{2}{\rho}\, \partial_z \nu_{Z_I} = \partial_\rho \log L_I \,\partial_z \log L_I\,, \quad \frac{4}{\rho}\, \partial_\rho \nu_{Z_I} =\left( \partial_\rho \log L_I \right)^2 - \left(\partial_z \log L_I\right)^2.
$$}

\begin{figure}[h]
\centering
\includegraphics[height=75mm]{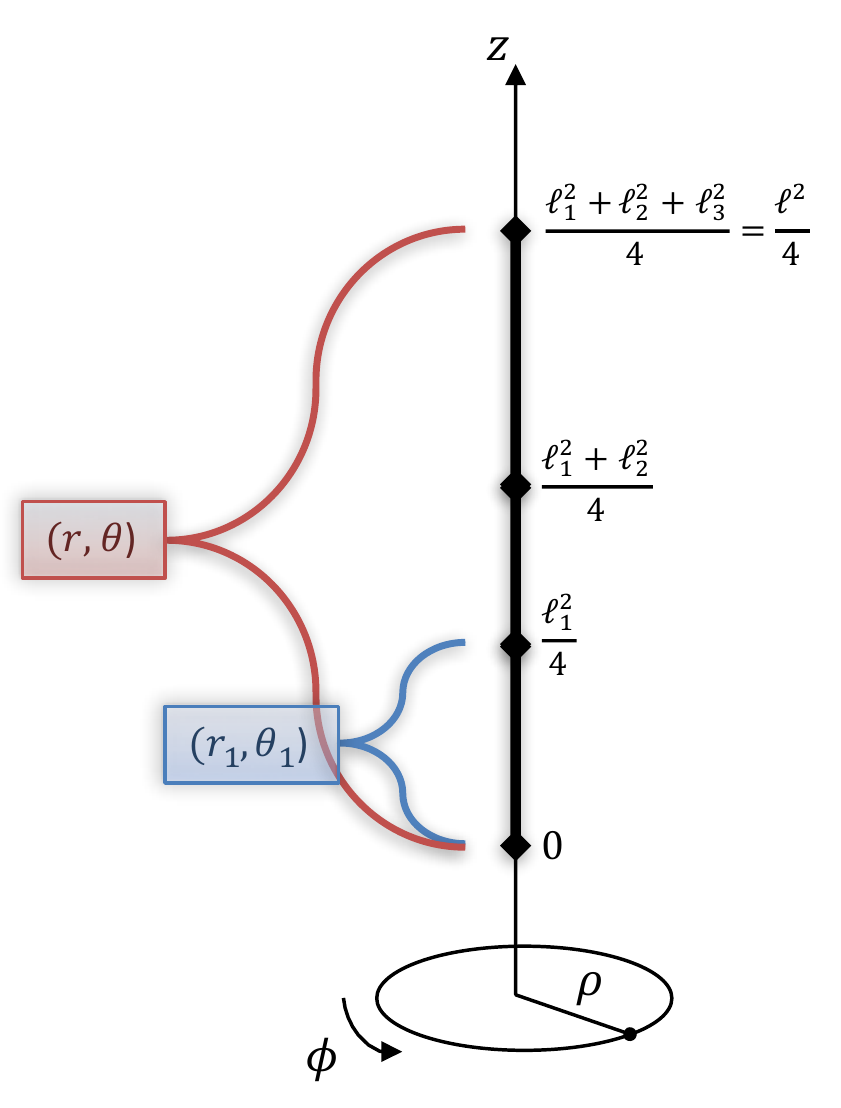}
\caption{Schematic description of connected rod sources on the $z$-axis. We depict the local spherical coordinates on the first rod,  $(r_1,\theta_1)$ \eqref{eq:DefDistance}, and the global spherical coordinates on the whole configuration, $(r,\theta)$ \eqref{eq:DefDistanceglobal}.}
\label{fig:rodsources}
\end{figure}

By axisymmetry,  the harmonic functions can be sourced on the $z$-axis by \emph{an arbitrary number of rods}.  We assume for the scope of this paper that they have a \emph{finite length} and are \emph{connected}.\footnote{Inspired by the results of \cite{Elvang:2002br,Bah:2020pdz,Bah:2021owp,Bah:2021rki,Heidmann:2021cms,Bah:2022yji},  we assume that the rod sources are connected to prevent from struts in between two disconnected rods.  A strut is a string with negative tension that manifests itself as a conical excess along a segment where a compact coordinate degenerates and that cannot be resolved classically in supergravity.} Thus,  we consider $n$ connected rod sources such that the origin of the $z$-axis is located at the extremity of the first rod.  We depicted a generic rod configuration in Fig.\ref{fig:rodsources}.  The rod lengths are denoted as $\ell_i^2/4$,  $i=1,\ldots,n$,  while the overall length is $\ell^2/4$ such that
\begin{equation}
\ell^2 \equi \sum_{i=1}^n \ell_i^2\,.
\end{equation}

We introduce the \emph{local spherical coordinates around the $i^\text{th}$ rod}, $(r_i,\theta_i)$,  given by
\begin{equation}
\begin{split}
r_i^2 &\equi 2\left[\sqrt{\rho^{2}+\left(z-\frac{1}{4} \sum_{j=1}^i \ell_j^2\right)^{2}}+\sqrt{\rho^{2}+\left(z-\frac{1}{4} \sum_{j=1}^{i-1} \ell_j^2\right)^{2}} -\frac{\ell_i^2}{4}\right]\,, \\
\cos 2\theta_i &\equi \frac{4}{\ell_i^2} \left[\sqrt{\rho^{2}+\left(z-\frac{1}{4} \sum_{j=1}^{i-1} \ell_j^2\right)^{2}}- \sqrt{\rho^{2}+\left(z-\frac{1}{4} \sum_{j=1}^{i} \ell_j^2\right)^{2}} \right]\,,
\end{split}
\label{eq:DefDistance}
\end{equation}
where $0\leq \theta_i\leq \frac{\pi}{2}$ and $r_i \geq 0$.  The coordinate $r_i$ measures the radial distance to the rod.  Indeed,  taking $r_i=0$ and varying $\theta_i$ from $0$ to $\pi/2$ is equivalent to a shift along the $i^\text{th}$ rod such that $\rho=0$ with $z$ varying from $\coeff{1}{4}\sum_{j=1}^{i-1}\ell_j^2$ to $\coeff{1}{4}\sum_{j=1}^{i}\ell_j^2$.

Moreover,  it will be convenient to express the solutions in terms of the \emph{global spherical coordinates} $(r,\theta)$,
\begin{equation}
\begin{split}
r^2 &\equi 2\left[\sqrt{\rho^{2}+\left(z-\frac{\ell^2}{4} \right)^{2}}+\sqrt{\rho^{2}+z^{2}} -\frac{\ell^2}{4}\right]\,, \\
\cos 2\theta &\equi \frac{4}{\ell^2} \left[\sqrt{\rho^{2}+z^{2}}-\sqrt{\rho^{2}+\left(z-\frac{\ell^2}{4} \right)^{2}}\right]\,,
\end{split}
\label{eq:DefDistanceglobal}
\end{equation}
which implies
\begin{equation}
\begin{split}
\rho= \frac{r\sqrt{r+\ell^2}}{4} \sin 2\theta \,,\qquad z = \frac{2 r^2+\ell^2}{8} \cos 2\theta+  \frac{\ell^2}{8}\,.
\end{split}
\end{equation}
They are the spherical coordinates centered on the whole rod configuration such that the rod sources are located at $r^2=0$ and varying $\theta$ from $0$ to $\pi/2$ moves from the first rod to the very last.  In this coordinate system,  the local spherical coordinates $(r_i,\theta_i)$, given in terms of $\rho$ and $z$ in \eqref{eq:DefDistance},  are given by
\begin{align}
4r_i^2 \= & \sqrt{\left((2r^2+\ell^2) \,\cos 2\theta+\ell^2 -2\sum_{j=1}^{i-1}\ell_j^2 \right)^2+4r^2(r^2+\ell^2)\sin^22\theta}\nn\\ 
& +   \sqrt{\left((2r^2+\ell^2) \,\cos 2\theta+\ell^2 -2\sum_{j=1}^{i}\ell_j^2 \right)^2+4r^2(r^2+\ell^2)\sin^22\theta} \-2\ell_i^2\,, \nn\\
4\ell_i^2\cos^2 \theta_i \=&\sqrt{\left((2r^2+\ell^2) \,\cos 2\theta+\ell^2 -2\sum_{j=1}^{i-1}\ell_j^2 \right)^2+4r^2(r^2+\ell^2)\sin^22\theta}\label{eq:ri&thetaidef} \\ 
& -  \sqrt{\left((2r^2+\ell^2) \,\cos 2\theta+\ell^2 -2\sum_{j=1}^{i}\ell_j^2 \right)^2+4r^2(r^2+\ell^2)\sin^22\theta}\+2\ell_i^2 \,, .\nn
\end{align}

The eight functions $(L_I,W_\Lambda)$ are \emph{sourced at the rods with specific weights} $(P_i^{(I)},G_i^{(\Lambda)})$:
\begin{equation}
L_I \= \prod_{i=1}^{n} \left(1+ \frac{\ell_i^2}{r_i^2} \right)^{P_i^{(I)}}\,,\quad W_\Lambda \=  \prod_{i=1}^{n} \left(1+ \frac{\ell_i^2}{r_i^2} \right)^{G_i^{(\Lambda)}}\,,\quad I = 0,5,p\,,\quad \Lambda=0,1,..,4.
\label{eq:HarmFunc2}
\end{equation}
The warp factors $(Z_0,Z_1,Z_5)$ and the gauge potential $T_1$ can be directly derived from \eqref{eq:LinearSolGen}, and we have in addition
\begin{equation}
\begin{split}
H_0  &\= \frac{1}{4a_0}\sum_{i=1}^{n} \ell_i^2 P_i^{(0)}\,\cos 2\theta_i\,,\qquad  H_5 \= \frac{1}{4a_5}\sum_{i=1}^{n} \ell_i^2 P_i^{(5)}\,\cos 2\theta_i\,,\\
e^{2\nu} &\=   \prod_{i, j=1}^{n} \left( \frac{\left( \left(r_i^2+\ell_i^2 \right) \cos^2\theta_i +  \left(r_j^2+\ell_j^2\right) \sin^2\theta_j \right)\left(r_i^2 \cos^2\theta_i +  r_j^2 \sin^2\theta_j \right)}{\left( \left(r_i^2+\ell_i^2 \right) \cos^2\theta_i + r_j^2 \sin^2\theta_j \right)\left(r_i^2 \cos^2\theta_i +  \left(r_j^2+\ell_j^2\right) \sin^2\theta_j \right)}\right)^{\alpha_{ij}}\,, 
\end{split}
\end{equation}
where we have defined
\begin{equation}
\alpha_{ij}\equi \sum_{I=0,1,5} P_i^{(I)} P_j^{(I)} \+ \frac{1}{2}\left[\sum_{\Lambda=0}^4 G_i^{(\Lambda)}G_j^{(\Lambda)}  \+\sum_{\Lambda,\Sigma=2}^4 G_i^{(\Lambda)}G_j^{(\Sigma)} +G_i^{(1)} G_j^{(1)}\right]\,.
\label{eq:AlphaDef}
\end{equation}
The magnetic dual of the electric D1 gauge potential has exactly the same form as $H_5$ by replacing $a_5 \to a_1$ and $P_i^{(5)} \to P_i^{(1)}$.  By considering the asymptotic spherical coordinates \eqref{eq:AsympCoord},  we have $\cos 2\theta_i \sim \cos 2\theta$ at large distance.  Therefore,  the solutions correspond to D1-D5 geometries in type IIB with potential KKm charges, $k$, along $\psi$ such that the supergravity charges are given, in unit of volume, by
\begin{equation}
Q_1 \=  \frac{1}{a_1}\sum_{i=1}^{n} \ell_i^2 P_i^{(1)}\,,\qquad Q_5 \=  \frac{1}{a_5}\sum_{i=1}^{n} \ell_i^2 P_i^{(5)}\,,\qquad k \=  \frac{1}{4a_0}\sum_{i=1}^{n} \ell_i^2 P_i^{(0)}\,. 
\label{eq:D1D5KKmchargesGen}
\end{equation}
One can use these expressions to directly fix the constants $a_I$ in terms of the net charges and the rod parameters.

The charged Weyl formalism allows to extract large families of geometries that solve Einstein equations of the D1-D5 system in type IIB \eqref{eq:TypeIIBAnsatz}.  They are given by an arbitrary number,  $n$,  of rod sources on the $z$-axis. Each rod has nine associated parameters: eight weights $(P_i^{(I)},G_i^{(\Lambda)})$ and a length parameter $\ell_i^2$.  Moreover,  we have six other independent parameters that fix the asymptotic of the solutions which are the D1-D5 charges,  the orbifold parameter $k$, and the constant parameter $b_I$. 

In \cite{Bah:2021owp,Bah:2021rki,Heidmann:2021cms},  asymptotically-flat regular geometries have been extracted from these solutions.  All weights are fixed such that they correspond to \emph{bolts}  or \emph{horizons}.  By computing the ADM mass and comparing it to the D1-D5 charges, the solutions have been shown to be non-BPS as soon as the rods do not degenerate to point sources $\ell_i^2\neq 0$.

We will adapt the solutions to construct geometries asymptotic to AdS$_3\times$S$^3/\mathbb{Z}_k\times$T$^4$ in type IIB by applying the internal and asymptotic boundary conditions introduced in section \ref{sec:BC}.  As their asymptotically-flat cousins,  the solutions will be internally sourced by rods leading to a chain of regular bolts and/or black holes.  The difference in the asymptotic constraints however will change the constants $b_I$ and modify the geometries globally.  

Before doing so, we point out some useful expressions between the local and global spherical coordinates:
\begin{align}
&r_{i}^2 \cos^2 \theta_i \= (r_{i+1}^2+\ell_{i+1}^2) \cos^2\theta_{i+1}\,,\qquad  r_{i+1}^2 \sin^2 \theta_{i+1} \= (r_i^2+\ell_i^2) \sin^2\theta_i\,, \nn \\
&\prod_{i=1}^{n} \left(1+ \frac{\ell_i^2}{r_i^2} \right) \= 1+ \frac{\ell^2}{r^2}\,,\qquad \sum_{i=1}^{n} \ell_i^2 \cos 2\theta_i \= \ell^2 \cos 2\theta\,,\nn \\
& \prod_{i, j=1}^{n}\frac{\left( \left(r_i^2+\ell_i^2 \right) \cos^2\theta_i +  \left(r_j^2+\ell_j^2\right) \sin^2\theta_j \right)\left(r_i^2 \cos^2\theta_i +  r_j^2 \sin^2\theta_j \right)}{\left( \left(r_i^2+\ell_i^2 \right) \cos^2\theta_i + r_j^2 \sin^2\theta_j \right)\left(r_i^2 \cos^2\theta_i +  \left(r_j^2+\ell_j^2\right) \sin^2\theta_j \right)} \label{eq:SimplRelations2}\\
& \hspace{8.5cm} \= \frac{r^2(r^2+\ell^2)}{(r^2+\ell^2\sin^2\theta)(r^2+\ell^2\cos^2\theta)}\,. \nn
\end{align}

\subsection{Asymptotically-AdS$_3$ solutions}

We first derive the constraints on the asymptotics before discussing the internal boundary conditions at the rods.

\subsubsection{Asymptotic boundary conditions}

We expand the warp factors and gauge potentials at large distance by considering the asymptotic spherical coordinates \eqref{eq:AsympCoord}. We find that $W_\Lambda$,  $e^{2\nu}$, $H_0$ and $H_5$ have already the right behavior given by \eqref{eq:AsympBehav} and we have
\begin{equation}
Z_0 \,\sim\, \frac{4k \sinh b_0}{\sum_{i=1}^{n} \ell_i^2 P_i^{(0)}} + \frac{4k \cosh b_0}{r^2}\,,\qquad Z_{1,5} \,\sim\, \frac{Q_{1,5} \sinh b_{1,5}}{\sum_{i=1}^n \ell_i^2 P_i^{(1,5)}} + \frac{Q_{1,5} \cosh b_{1,5}}{r^2}\,.
\end{equation}
Thus,  from the result in section \ref{sec:asympBC},  we find that the solutions are asymptotic to AdS$_3\times$S$^3/\mathbb{Z}_k$ $\times$T$^4$ if  one imposes
\begin{equation}
b_0 \= b_1 \= b_5 \= 0\,.
\end{equation}
Moreover,  one can consider solutions that are asymptotic to AdS$_3\times$S$^3\times$T$^4$ without orbifold action on the S$^3$ by simply considering $k=1$.

\subsubsection{Internal boundary conditions}
\label{sec:InternalBCLinear}

Each rod locus,  $\rho=0$ and $\coeff{1}{4}\sum_{j=1}^{i-1}\ell_j^2\leq z \leq \coeff{1}{4}\sum_{j=1}^{i}\ell_j^2$,  corresponds to $r_i=0$ and $0\leq \theta_i \leq \pi/2$ in the local spherical coordinates \eqref{eq:DefDistance}.  Thus,  the eight functions $(L_I, W_\Lambda)$  \eqref{eq:HarmFunc2} are either blowing or vanishing if $P_i^{(I)}$ or $G_i^{(\Lambda)}$ are non-zero.  The warp factors have the same behavior as the generic one given in \eqref{eq:InternalBehaviors},  and the exponents $\alpha_{Z_I}$ and $\alpha_{W_\Lambda}$ can be related to the weights at the $i^\text{th}$ rod such as
\begin{equation}
\alpha_{Z_I} \=  |P^{(I)}_i|\,,\qquad \alpha_{W_\Lambda} \= G^{(\Lambda)}_i\,,
\end{equation}

Therefore,  we transpose the seven possible choices of regular internal boundary conditions summarized in Table \ref{tab:internalBCGen} in terms of rod weights in Table \ref{tab:internalBC}.
\begin{table}[h]
  \centering
  \begin{tabular}{|c||c|c|c!{\vrule width 1.4pt}c|c|c|c|c|}
\hline 
 &  $P_i^{(0)}$ & $P_i^{(1)}$& $P_i^{(5)}$&  $G_i^{(0)}$&$G_i^{(1)}$& $G_i^{(2)}$& $G_i^{(3)}$& $G_i^{(4)}$ \\ \hline \hline
Horizon & $\frac{1}{2}$ &$\frac{1}{2}$ &$\frac{1}{2}$ &$0$ &$\frac{1}{2}$ &$0$ &$0$ &$0$ \\ \noalign{\hrule height 0.8pt}
$\psi$ degeneracy & $1$ &$0$ &$0$ &$0$ &$0$ &$0$ &$0$ &$0$ \\ \hline
$y$ degeneracy & $\frac{1}{2}$ &$\frac{1}{2}$ &$\frac{1}{2}$  &$0$ &$-\frac{1}{2}$&$0$ &$0$ &$0$ \\ \hline
$x_1$ degeneracy  &$\frac{1}{2}$ &$0$ &$\frac{1}{2}$ &$\frac{1}{2}$ &$0$ &$-\frac{3}{4}$ &$\frac{1}{4}$ &$\frac{1}{4}$ \\ \hline
$x_2$ degeneracy &$\frac{1}{2}$ &$0$ &$\frac{1}{2}$  &$\frac{1}{2}$ &$0$ &$\frac{1}{4}$ &$-\frac{3}{4}$ &$\frac{1}{4}$ \\ \hline
$x_3$ degeneracy &$\frac{1}{2}$ &$0$ &$\frac{1}{2}$ &$\frac{1}{2}$ &$0$ &$\frac{1}{4}$ &$\frac{1}{4}$ &$-\frac{3}{4}$ \\ \hline
$x_4$ degeneracy &$\frac{1}{2}$ &$0$ &$\frac{1}{2}$  &$\frac{1}{2}$ &$0$ &$\frac{1}{4}$ &$\frac{1}{4}$ &$\frac{1}{4}$ \\ \hline
  \end{tabular}
  \caption{\label{tab:internalBC} The seven possible weights at the $i^\text{th}$ rod leading to a regular coordinate degeneracy of the timelike direction or a compact spacelike direction.}
\end{table}

Moreover,  there are additional constraints given by \eqref{eq:ConstraintInternal} such that the rods define smooth bolts or horizons.  These will give a set of $n$ algebraic equations that constrain the rod lengths in terms of the charges, temperature, and radii of the compact dimensions. For smooth solutions without horizons,  these equations will be denoted as \emph{bubble equations}.  

Interestingly,  for regular sources,  the exponents $\alpha_{ij}$ \eqref{eq:AlphaDef} simplify to
\begin{equation}
\alpha_{ij} \= \begin{cases} 1 \quad \text{if the $i^\text{th}$ and $j^\text{th}$ rods are of the same nature,} \\
\frac{1}{2} \quad \text{otherwise,}
\end{cases}
\label{eq:AlphaSimple}
\end{equation}
where ``same nature'' means that the same coordinate degenerates at both rods.

Note that a necessary condition for having asymptotically-AdS$_3$ solutions is to have $Q_1 , Q_5 , k \neq 0$ \eqref{eq:D1D5KKmchargesGen}.  Moreover, a non-zero charge $Q_1$ requires at least one $P_i^{(1)}$ turned on.   However,  only two types of rods can have $P_i^{(1)}\neq 0$: the ones corresponding to a horizon or to a degeneracy of the  S$^1$ parametrized by $y$.  Thus,  horizonless configurations with D1-brane charge necessarily require rods where the S$^1$ shrinks.  In other words, asymptotically-AdS$_3$ smooth horizonless geometries must force the common S$^1$ of the D1 and D5 branes to degenerate somewhere in the spacetime.

\subsection{Final form of the solutions}
\label{sec:linearbranchSum}

We remind that the type IIB fields are
\begin{align}
ds_{10}^2 \= & \sqrt{\frac{W_0}{Z_1 Z_5}} \,\left[ - \frac{dt^2}{W_1}+ W_1\,dy^2\right]  \+ \sqrt{\frac{Z_1}{Z_5}}\,\left( W_2 dx_2^2+W_3 dx_2^2 +W_4 dx_3^2 +\frac{dx_4^2}{W_2 W_3 W_4}\right)\nn \\
&\+ \sqrt{W_0\,Z_1 Z_5} \left[\frac{1}{Z_0} \left(d\psi+H_0 \,d\phi\right)^2 +Z_0 \,\left(e^{2\nu} \left( d\rho^2 +dz^2\right) + \rho^2d\phi^2 \right) \right] \,,\label{eq:TypeIIBAnsatz2} \\
C^{(2)} \= &H_5 \,d\phi \wedge d\psi -T_1 \,dt\wedge dy \,,\qquad  e^\Phi \= \sqrt{\frac{Z_1}{Z_5}\,W_0}\,,\qquad C^{(0)}=C^{(4)}=B_2=0\,.\nn 
\end{align}
The geometries obtained from the linear branch of solutions of the equations \eqref{eq:EOMVac} and \eqref{eq:EOMMaxwell} that are asymptotic to AdS$_3\times$S$^3/\mathbb{Z}_k\times$T$^4$ are sourced by $n$ connected rods on the $z$-axis of length $\ell_i^2/4$.  The main fields are given by eight functions such that their logarithms are harmonic functions sourced at the rods
\begin{equation}
L_I \= \prod_{i=1}^{n} \left(1+ \frac{\ell_i^2}{r_i^2} \right)^{P_i^{(I)}}\,,\quad W_\Lambda \=  \prod_{i=1}^{n} \left(1+ \frac{\ell_i^2}{r_i^2} \right)^{G_i^{(\Lambda)}}\,,\quad I = 0,5,p\,,\quad \Lambda=0,1,..,4,
\label{eq:HarmFunc}
\end{equation}
and we have
\begin{align}
Z_1 &\= Q_1\,\frac{L_1-L_1^{-1}}{2 \sum_{i=1}^{n} \ell_i^2 P_i^{(1)}}\,,\qquad T_1 \= \frac{\sum_{i=1}^{n} \ell_i^2 P_i^{(1)}}{Q_1}\,\frac{L_1^2+1}{L_1^2-1}\,,\nn \\
 Z_5& \=Q_5\, \frac{L_5-L_5^{-1}}{2 \sum_{i=1}^{n} \ell_i^2 P_i^{(5)}}\,,\qquad H_5 \= \frac{Q_5}{4\sum_{i=1}^{n} \ell_i^2 P_i^{(5)}}\,\, \sum_{i=1}^{n} \ell_i^2 P_i^{(5)}\,\cos 2\theta_i\,,\nn \\
Z_0 &\= 2k\,\frac{L_0 -L_0^{-1}}{\sum_{i=1}^{n} \ell_i^2 \,P_i^{(0)}}\,, \qquad H_0 \= \frac{k}{\sum_{i=1}^n \ell_i^2 \,P_i^{(0)}}\,\,\sum_{i=1}^{n}\ell_i^2 P_i^{(0)}\,\cos 2\theta_i\,,  \label{eq:LinearAdS3} \\
e^{2\nu} &\=   \prod_{i, j=1}^{n} \left( \frac{\left( \left(r_i^2+\ell_i^2 \right) \cos^2\theta_i +  \left(r_j^2+\ell_j^2\right) \sin^2\theta_j \right)\left(r_i^2 \cos^2\theta_i +  r_j^2 \sin^2\theta_j \right)}{\left( \left(r_i^2+\ell_i^2 \right) \cos^2\theta_i + r_j^2 \sin^2\theta_j \right)\left(r_i^2 \cos^2\theta_i +  \left(r_j^2+\ell_j^2\right) \sin^2\theta_j \right)}\right)^{\alpha_{ij}}\,, \nn
\end{align}
where the local spherical coordinates at each rod $(r_i,\theta_i)$ are given in terms of Weyl cylindrical coordinates,  $(\rho,z)$, in \eqref{eq:DefDistance} and in terms of the global spherical coordinates, $(r,\theta)$, in \eqref{eq:ri&thetaidef}.  The exponents $\alpha_{ij}$ are given in \eqref{eq:AlphaSimple}.

The weights $(P_i^{(I)},G_i^{(\Lambda)})$ at each rod take one of the seven possible values in Table \ref{tab:internalBC} depending on the coordinate that degenerates at this location.  The weights $(P_i^{(1)},P_i^{(5)})$ are associated to the local D1-D5 brane charges at the rod $(q_{D1}^{(i)},q_{D5}^{(i)})$ given by
\begin{equation}
q_{D1}^{(i)} \= \frac{\ell_i^2 \,P_i^{(1)}}{\sum_{j=1}^n \ell_j^2 \,P_j^{(1)}}\,Q_1\,,\qquad q_{D5}^{(i)} \= \frac{\ell_i^2 \,P_i^{(5)}}{\sum_{j=1}^n \ell_j^2 \,P_j^{(5)}}\,Q_5\,.
\label{eq:ChargeAtRodGen}
\end{equation}

Moreover,  the solutions are constrained by $n$ regularity equations that must be derived in a case-by-case manner \eqref{eq:ConstraintInternal}.  They will fix all rod lengths, $\ell_i^2$, in terms of the asymptotic quantities that are the D1 and D5 charges,  the radii of the compact directions,  the asymptotic orbifold parameter $k$, and possibly the temperature if the solutions have horizons.  Thus,  the solutions have no moduli after regularity.  The only free parameters that are not asymptotic quantities are the number of rods,  $n$, which can be understood as the ``quantum bits'' to be included in the geometries, and the nature of the rods, which would correspond to the exact nature of the ``bits'' in this analogy.

%%%%%%%%%%%%%%%%%%%%%%%%%%%%%%%%%%%%%
\section{Examples of non-BPS bubbling deformations in AdS$_3\times$S$^3\times$T$^4$}
\label{sec:AdS3}
%%%%%%%%%%%%%%%%%%%%%%%%%%%%%%%%%%%%%

In this section, we construct geometries that are asymptotic to AdS$_3\times$S$^3\times$T$^4$ or AdS$_3\times$S$^3/\mathbb{Z}_k$ $\times$T$^4$ using the linear branch of solutions.  We restrict to simple examples with the least number of rods to illustrate the physics of the solutions,  and we focus on smooth bubbling geometries without horizons. 

As previously argued in section \ref{sec:InternalBCLinear},  one needs at least one rod that forces the degeneracy of the S$^1$ (the $y$-circle).  This generates the necessary D1 and D5 brane charges to be asymptotic to  AdS$_3$ in type IIB.  Thus,  we first construct the solution obtained from such a single rod.  We obtain a global AdS$_3\times$S$^3\times$T$^4$ spacetime with a conical defect on the S$^3$ that can be tuned.  The rod inducing the degeneracy of the S$^1$ is located at the center of AdS$_3$, that is at $r=0$ in the global spherical coordinates \eqref{eq:DefDistanceglobal}.

Then,  we show that the linear branch of solutions allows decorating this  solution by rods that lead to smooth non-BPS T$^4$ or S$^3$ deformations in type IIB.  We focus on two examples with different physics:
\begin{itemize}
\item[•] First, we construct solutions that correspond to global AdS$_3 \times$S$^3\times$T$^4$ but with an extra rod that forces a T$^4$ coordinate to degenerate.  The spacetime still caps off smoothly at $r=0$ but,  the S$^3$ splits into two regions there: a region where the S$^1$ degenerates and a region where the T$^4$ direction pinches off.  We will show that these new smooth bubbling solutions break supersymmetry, they break the rigidity of the T$^4$,  and the symmetry of the S$^3$ and AdS$_3$ parts.  Moreover, we will show that when $R_{x_1} \ll \sqrt{Q_5}$,  where $R_{x_1}$ is the radius of the T$^4$ direction that shrinks,  the solutions can be seen as a small non-BPS perturbation on a global AdS$_3$ background in type IIB.  The backreaction has forced the T$^4$ to degenerate smoothly at the center of AdS$_3$ and at a specific locus on the S$^3$.  
\item[•] Second,  we do the exact same analysis with a rod that now forces the Hopf fibration angle of the S$^3$,  $\psi$,  to degenerate.  We will show that the smoothness of the solutions requires to impose an asymptotic conical defect $k$ on the S$^3$ such that the solutions are asymptotic to AdS$_3 \times$S$^3/\mathbb{Z}_k\times$T$^4$.  Otherwise, the physics is relatively similar,  the solutions cap off smoothly at $r=0$ as a degeneracy of the S$^1$ or the S$^3$ depending on the position on the S$^3$.  We show that such deformation breaks supersymmetry of the unperturbed global AdS$_3$ spacetime by breaking the symmetry on the S$^3$ and AdS$_3$ parts. However, the solutions still preserve the rigidity of the T$^4$.  Furthermore,  we argue that, in the large orbifold limit $k \gg 1$,  the extra rod becomes a small non-BPS perturbation on a global AdS$_3$ background for which the backreaction has forced the S$^3$ to degenerate smoothly at the center of AdS$_3$.
\end{itemize}

\subsection{Global AdS$_3$ as a single rod solution}
\label{sec:GlobalAdS3}

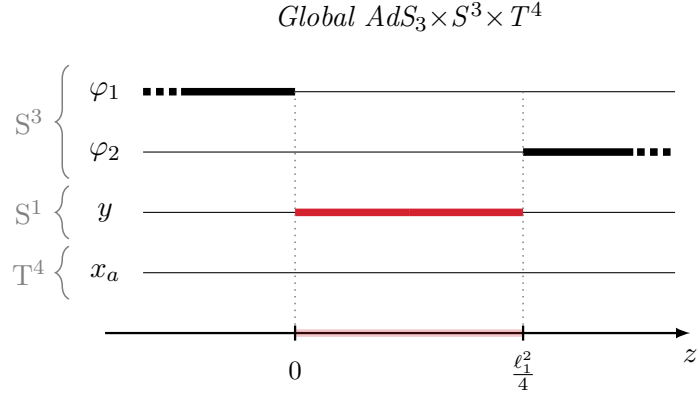
\begin{figure}[h]
\centering
    \begin{tikzpicture}
%% some definitions

\def\deb{-10} 
\def\inter{0.8} 
\def\ha{2.8} 
\def\zaxisline{4} 
\def\rodsize{1.5} 
\def\numrod{2} 

\def\fin{\deb+1+2*\rodsize+\numrod*\rodsize} 

%% 

%\draw (\deb-2,\ha-\zaxisline*0.5*\inter) node{$\Longrightarrow$};

%% Pic title

\draw (\deb+0.5+\rodsize+0.5*\numrod*\rodsize,\ha+1) node{{{\it Global AdS$_3\times$S$^{\,3}\times$T$^{\,4}$}}}; 

%% Each line black line and names

\draw[black,thin] (\deb+1,\ha) -- (\fin,\ha);
\draw[black,thin] (\deb,\ha-\inter) -- (\fin-1,\ha-\inter);
\draw[black,thin] (\deb,\ha-2*\inter) -- (\fin,\ha-2*\inter);
\draw[black,thin] (\deb,\ha-3*\inter) -- (\fin,\ha-3*\inter);
\draw[black,->,line width=0.3mm] (\deb-0.5,\ha-\zaxisline*\inter) -- (\fin+0.2,\ha-\zaxisline*\inter);

\draw [decorate, 
    decoration = {brace,
        raise=5pt,
        amplitude=5pt},line width=0.2mm,gray] (\deb-0.8,\ha-1.5*\inter+0.05) --  (\deb-0.8,\ha+0.5*\inter-0.05);
\draw [decorate, 
    decoration = {brace,
        raise=5pt,
        amplitude=5pt},line width=0.2mm,gray] (\deb-0.8,\ha-2.5*\inter+0.05) --  (\deb-0.8,\ha-1.5*\inter-0.05);
\draw [decorate, 
    decoration = {brace,
        raise=5pt,
        amplitude=5pt},line width=0.2mm,gray] (\deb-0.8,\ha-3.5*\inter+0.05) --  (\deb-0.8,\ha-2.5*\inter-0.05);
        
\draw[gray] (\deb-1.5,\ha-0.5*\inter) node{S$^3$};
\draw[gray] (\deb-1.5,\ha-2*\inter) node{S$^1$};
\draw[gray] (\deb-1.5,\ha-3*\inter) node{T$^4$};
    
\draw (\deb-0.5,\ha) node{$\varphi_1$};
\draw (\deb-0.5,\ha-\inter) node{$\varphi_2$};
\draw (\deb-0.5,\ha-2*\inter) node{$y$};
\draw (\deb-0.5,\ha-3*\inter) node{$x_a$};

\draw (\fin+0.2,\ha-\zaxisline*\inter-0.3) node{$z$};

%% First two line and their rods

\draw[black, dotted, line width=1mm] (\deb,\ha) -- (\deb+0.5,\ha);
\draw[black,line width=1mm] (\deb+0.5,\ha) -- (\deb+0.5+\rodsize,\ha);
\draw[black,line width=1mm] (\fin-0.5-\rodsize,\ha-\inter) -- (\fin-0.55,\ha-\inter);
\draw[black, dotted,line width=1mm] (\fin-0.5,\ha-\inter) -- (\fin,\ha-\inter);

%% Next lines and their rods

\draw[amaranthred,line width=1mm] (\deb+0.5+\rodsize,\ha-2*\inter) -- (\deb+0.5+2*\rodsize,\ha-2*\inter);
\draw[amaranthred,line width=1mm] (\deb+0.5+2*\rodsize,\ha-2*\inter) -- (\deb+0.5+3*\rodsize,\ha-2*\inter);

\draw[amaranthred,line width=1mm,opacity=0.25] (\deb+0.5+\rodsize,\ha-\zaxisline*\inter) -- (\deb+0.5+3*\rodsize,\ha-\zaxisline*\inter);

%% Vertical lines and coordinates

\draw[gray,dotted,line width=0.2mm] (\deb+0.5+\rodsize,\ha) -- (\deb+0.5+\rodsize,\ha-\zaxisline*\inter);
\draw[gray,dotted,line width=0.2mm] (\deb+0.5+3*\rodsize,\ha) -- (\deb+0.5+3*\rodsize,\ha-\zaxisline*\inter);

\draw[line width=0.3mm] (\deb+0.5+3*\rodsize,\ha-\zaxisline*\inter+0.1) -- (\deb+0.5+3*\rodsize,\ha-\zaxisline*\inter-0.1);
\draw[line width=0.3mm] (\deb+0.5+\rodsize,\ha-\zaxisline*\inter+0.1) -- (\deb+0.5+\rodsize,\ha-\zaxisline*\inter-0.1);

\draw (\deb+0.5+\rodsize,\ha-\zaxisline*\inter-0.5) node{{\small $0$}};
\draw (\deb+0.5+3*\rodsize,\ha-\zaxisline*\inter-0.5) node{{\small $\frac{\ell_1^2}{4}$}};

\end{tikzpicture}
\caption{Rod diagram of the shrinking directions on the $z$-axis after sourcing the solutions with one rod that forces the degeneracy of the $y$-circle.  We took the figures of \cite{Astorino:2022fge} as models.}
\label{fig:rodsourceAdS3}
\end{figure}  

We consider a single rod source, $n=1$,  such that it forces the S$^1$ ($y$-circle) to degenerate,  i.e.  $P_1^{(0)}=P_1^{(1)}= P_1^{(5)}= -G_1^{(1)} =1/2$ while all other weights are taken to be zero (see Table \ref{tab:internalBC}).  The rod profile has been depicted in Fig.\ref{fig:rodsourceAdS3}.

The metric warp factors are \eqref{eq:LinearAdS3}
\begin{align}
Z_0 &\= \frac{4k}{r_1\sqrt{r_1^2+\ell_1^2}},\qquad Z_1 \= \frac{Q_1}{ r_1\sqrt{r_1^2+\ell_1^2}},\qquad Z_5 \= \frac{Q_5}{r_1\sqrt{r_1^2+\ell_1^2}},\label{eq:LinearAdS32} \\
W_1 &\=\left(1+\frac{\ell_1^2}{r_1^2}\right)^{-\frac{1}{2}}\,,\quad W_0=W_2=W_3=W_4=1\,,\quad e^{2\nu} \= \frac{r_1^2(r_1^2+\ell_1^2)}{(r_1^2+\ell_1^2\sin^2\theta_1)(r_1^2+\ell_1^2\cos^2\theta_1)}\,,\nn
\end{align}
while the gauge potentials give
\begin{equation}
 H_0 \=k \cos 2\theta_1,\qquad T_1 \= \frac{\frac{\ell_1^2}{2}+r_1^2}{Q_1},\qquad H_5 \= \frac{Q_5}{4}\cos 2\theta_1. 
\end{equation}

With a single rod in the configuration, the global spherical coordinates,  $(r,\theta)$ \eqref{eq:DefDistanceglobal}, are identical to the coordinates centered on the rod, $(r_1,\theta_1)$, and we have $\ell^2= \ell_1^2$.  Thus, it is more convenient to change coordinates from the Weyl coordinates to these spherical coordinates and from the Hopf coordinates of the S$^3$ to the spherical coordinates \eqref{eq:DefHyperspher}.  The type IIB solution \eqref{eq:TypeIIBAnsatz2} gives\footnote{We have also performed a global gauge transformation on $C^{(2)}$,  and use,  as a consequence of the change of coordinates \eqref{eq:DefDistanceglobal},  \begin{equation}
d\rho^2 +dz^2 = \frac{\left( r^2+\ell^2\cos^2\theta\right)\left( r^2+\ell^2\sin^2\theta\right)}{4} \, \left(\frac{dr^2}{r^2+\ell^2}+ d\theta^2 \right).
\label{eq:ChangeMetWeylSp}
\end{equation}}
\begin{align}
ds_{10}^2 \= &\frac{1}{\sqrt{Q_1 Q_5}} \left[-(r^2+\ell^2) \,dt^2 +r^2\,dy^2 \right] \+ \sqrt{\frac{Q_1}{Q_5}}\,\sum_{a=1}^4 dx_a^2 \label{eq:metGlobalAdS3}\\ 
& \+ k\sqrt{Q_1 Q_5} \left[\frac{dr^2}{r^2+\ell^2} + d\theta^2 + \cos^2 \theta \,d\varphi_1^2+ \sin^2 \theta\,d\varphi_2^2 \right],\nn\\
C^{(2)} \= &k \,Q_5 \cos^2 \theta \,d\varphi_2 \wedge d\varphi_1 -\frac{r^2+\ell^2}{Q_1} \,dt\wedge dy \,,\quad  e^\Phi \= \sqrt{\frac{Q_1}{Q_5}}\,.\nn
\end{align}
Therefore,  it corresponds to \emph{global AdS$_3\times$S$^{\,3}/\mathbb{Z}_k\times$T$^{\,4}$}, with $Q_1$ and $Q_5$ D1 and D5 charges.\footnote{Even if the gauge field has a magnetic contribution,  $k Q_5 \cos^2\theta d\varphi_2 \wedge d\varphi_1$,  the net D5 charge is still $Q_5$ since the periodicity of $(\varphi_1,\varphi_2)$ is $2\pi$ and $2\pi/k$ \eqref{eq:DefHyperspher}.  One can simply use the Hopf coordinates, $C^{(2)} = \frac{Q_5}{2} \cos^2\theta d\phi \wedge d\psi$, and use the periodicities \eqref{eq:psi&phiPerio} for the integration.}

%%%%%%%%%%%%%%%%%%%%%%%%%%%%%%%%
\begin{figure}[h]
\centering
\includegraphics[scale=0.65]{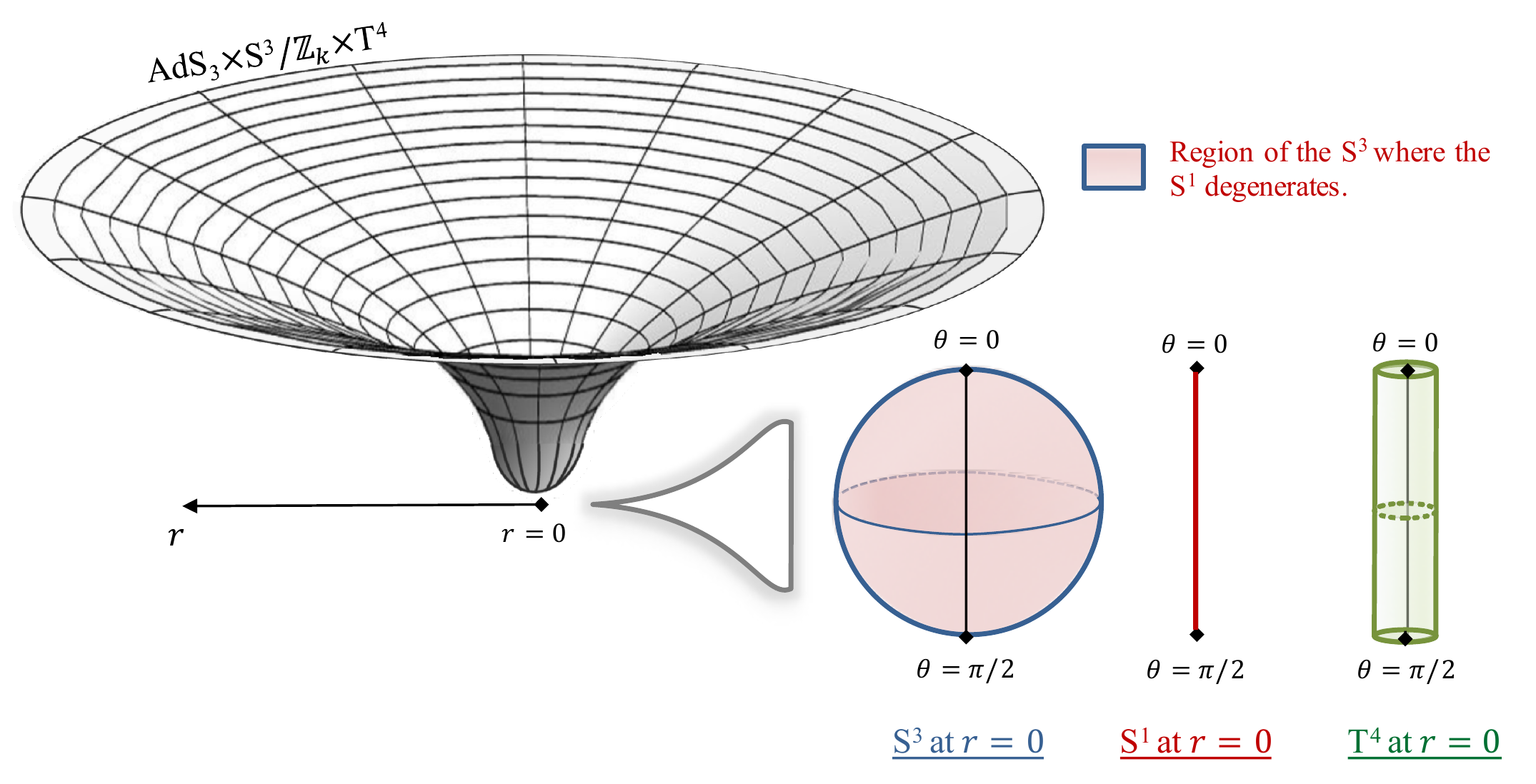}
\caption{Schematic description of the spacetime built from a single rod inducing the degeneracy of the S$^1$.  On the left-hand side,  we depict the overall geometry in terms of the radius $r$.  On the right-hand side,  we describe the behavior of the S$^3$,  S$^1$, and T$^4$ at $r=0$ and as a function of $\theta$,  the S$^3$ coordinate.  Global AdS$_3\times$S$^3\times$T$^4$ ends smoothly as a coordinate degeneracy of the S$^1$ at $r=0$ and the T$^4$ is rigid.}
\label{fig:AdS3pic}
\end{figure}
%%%%%%%%%%%%%%%%%%%%%%%%%%%%%%%%

At the unique rod, that is at the center of the global AdS$_3$ spacetime $r=0$,  the $y$-circle degenerates.  It corresponds to a smooth origin of $\IR^2$ if and only if
\begin{equation}
R_y \= \frac{\sqrt{k \,Q_1 Q_5}}{\ell}\,,
\label{eq:RegCondAdS3}
\end{equation}
where $R_y$ is the radius of the $y$-circle \eqref{eq:psi&phiPerio}.  We retrieve the usual regularity condition for a global AdS$_3 \times$S$^3\times$T$^4$ without orbifold by considering $k=1$.

In Fig.\ref{fig:AdS3pic},  we introduce our conventions for illustrating the geometries by applying them to the present global AdS$_3\times$S$^3/\mathbb{Z}_k\times$T$^4$ spacetime.  On the left-hand side,  we describe the geometries in terms of the radius $r$. Then, on the right-hand side, we specify the topology of the S$^3$,  S$^1$, and T$^4$ at $r=0$ as a function of $\theta$,  giving the position on the S$^3$.  The spacetime ends at $r=0$ where the S$^1$ shrinks for arbitrary $\theta$ while the T$^4$ is rigid.

\subsection{T$^4$ deformation at the center of AdS$_3$ and at a pole of the S$^3$}
\label{sec:AdS3+T4}

\begin{figure}[h]
\centering
    \begin{tikzpicture}
%% some definitions

\def\deb{-10} 
\def\inter{0.8} 
\def\ha{2.8} 
\def\zaxisline{5} 
\def\rodsize{1.5} 
\def\numrod{2.5} 

\def\fin{\deb+1+2*\rodsize+\numrod*\rodsize} 

%% 

%\draw (\deb-2,\ha-\zaxisline*0.5*\inter) node{$\Longrightarrow$};

%% Pic title

\draw (\deb+0.5+\rodsize+0.5*\numrod*\rodsize,\ha+1) node{{{\it Global AdS$_3\times$S$^{\,3}\times$T$^{\,4}$ with a T$^{\,4}$ deformation}}}; 

%% Each line black line and names

\draw[black,thin] (\deb+1,\ha) -- (\fin,\ha);
\draw[black,thin] (\deb,\ha-\inter) -- (\fin-1,\ha-\inter);
\draw[black,thin] (\deb,\ha-2*\inter) -- (\fin,\ha-2*\inter);
\draw[black,thin] (\deb,\ha-3*\inter) -- (\fin,\ha-3*\inter);
\draw[black,thin] (\deb,\ha-4*\inter) -- (\fin,\ha-4*\inter);
\draw[black,->, line width=0.3mm] (\deb-0.5,\ha-\zaxisline*\inter) -- (\fin+0.2,\ha-\zaxisline*\inter);

\draw [decorate, 
    decoration = {brace,
        raise=5pt,
        amplitude=5pt},line width=0.2mm,gray] (\deb-0.8,\ha-1.5*\inter+0.05) --  (\deb-0.8,\ha+0.5*\inter-0.05);
\draw [decorate, 
    decoration = {brace,
        raise=5pt,
        amplitude=5pt},line width=0.2mm,gray] (\deb-0.8,\ha-2.5*\inter+0.05) --  (\deb-0.8,\ha-1.5*\inter-0.05);
\draw [decorate, 
    decoration = {brace,
        raise=5pt,
        amplitude=5pt},line width=0.2mm,gray] (\deb-0.8,\ha-4.5*\inter+0.05) --  (\deb-0.8,\ha-2.5*\inter-0.05);
        
\draw[gray] (\deb-1.5,\ha-0.5*\inter) node{S$^3$};
\draw[gray] (\deb-1.5,\ha-2*\inter) node{S$^1$};
\draw[gray] (\deb-1.5,\ha-3.5*\inter) node{T$^4$};

\draw (\deb-0.5,\ha) node{$\varphi_1$};
\draw (\deb-0.5,\ha-\inter) node{$\varphi_2$};
\draw (\deb-0.5,\ha-2*\inter) node{$y$};
\draw (\deb-0.5,\ha-3*\inter) node{$x_1$};
\draw (\deb-0.5,\ha-4*\inter) node{$x_a$};

\draw (\fin+0.2,\ha-\zaxisline*\inter-0.3) node{$z$};

%% First two line and their rods

\draw[black, dotted, line width=1mm] (\deb,\ha) -- (\deb+0.5,\ha);
\draw[black,line width=1mm] (\deb+0.5,\ha) -- (\deb+0.5+\rodsize,\ha);
\draw[black,line width=1mm] (\fin-0.5-\rodsize,\ha-\inter) -- (\fin-0.55,\ha-\inter);
\draw[black, dotted,line width=1mm] (\fin-0.5,\ha-\inter) -- (\fin,\ha-\inter);

%% Next lines and their rods

\draw[amaranthred,line width=1mm] (\deb+0.5+\rodsize,\ha-2*\inter) -- (\deb+0.5+2*\rodsize,\ha-2*\inter);
\draw[amaranthred,line width=1mm] (\deb+0.5+2*\rodsize,\ha-2*\inter) -- (\deb+0.5+3*\rodsize,\ha-2*\inter);
\draw[amazon,line width=1mm] (\deb+0.5+3*\rodsize,\ha-3*\inter) -- (\deb+0.5+3.5*\rodsize,\ha-3*\inter);

\draw[amaranthred,line width=1mm,opacity=0.25] (\deb+0.5+\rodsize,\ha-\zaxisline*\inter) -- (\deb+0.5+3*\rodsize,\ha-\zaxisline*\inter);
\draw[amazon,line width=1mm,opacity=0.25] (\deb+0.5+3*\rodsize,\ha-\zaxisline*\inter) -- (\deb+0.5+3.5*\rodsize,\ha-\zaxisline*\inter);

%% Vertical lines and coordinates

\draw[gray,dotted,line width=0.2mm] (\deb+0.5+\rodsize,\ha) -- (\deb+0.5+\rodsize,\ha-\zaxisline*\inter);
\draw[gray,dotted,line width=0.2mm] (\deb+0.5+3*\rodsize,\ha) -- (\deb+0.5+3*\rodsize,\ha-\zaxisline*\inter);
\draw[gray,dotted,line width=0.2mm] (\deb+0.5+3.5*\rodsize,\ha) -- (\deb+0.5+3.5*\rodsize,\ha-\zaxisline*\inter);

\draw[line width=0.3mm] (\deb+0.5+3*\rodsize,\ha-\zaxisline*\inter+0.1) -- (\deb+0.5+3*\rodsize,\ha-\zaxisline*\inter-0.1);
\draw[line width=0.3mm] (\deb+0.5+\rodsize,\ha-\zaxisline*\inter+0.1) -- (\deb+0.5+\rodsize,\ha-\zaxisline*\inter-0.1);
\draw[line width=0.3mm] (\deb+0.5+3.5*\rodsize,\ha-\zaxisline*\inter+0.1) -- (\deb+0.5+3.5*\rodsize,\ha-\zaxisline*\inter-0.1);

\draw (\deb+0.5+\rodsize,\ha-\zaxisline*\inter-0.5) node{{\small $0$}};
\draw (\deb+0.5+3*\rodsize,\ha-\zaxisline*\inter-0.5) node{{\small $\frac{\ell_1^2}{4}$}};
\draw (\deb+0.9+3.5*\rodsize,\ha-\zaxisline*\inter-0.5) node{{\small $\frac{\ell_1^2+\ell_2^2}{4}=\frac{\ell^2}{4}$}};

\end{tikzpicture}
\caption{Rod diagram of the shrinking directions on the $z$-axis after sourcing the solutions with two connected rods that force the degeneracy of the $y$ and $x_1$ circles respectively. }
\label{fig:rodsourceAdS3+T4}
\end{figure}
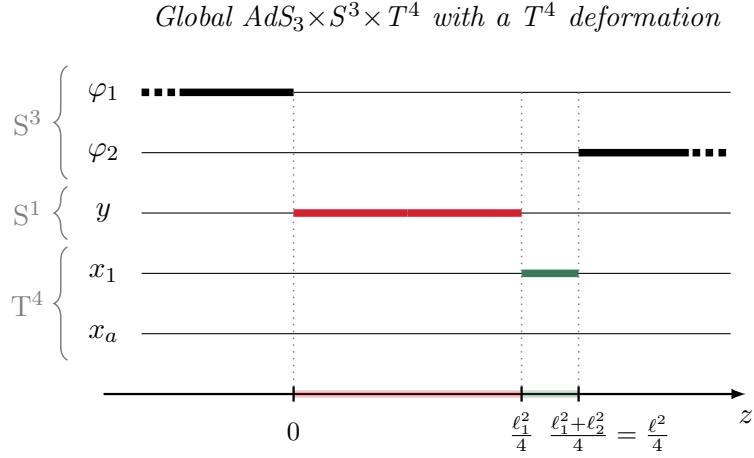  

We consider two connected rod sources (see Fig.\ref{fig:rodsourceAdS3+T4}).  The first one is identical to the previous section while the second one induces the degeneracy of $x_1$,  a T$^4$ direction.  From Table  \ref{tab:internalBC},  the weights at the second rod are fixed such that $P_2^{(0)}=P_2^{(5)}=G_2^{(0)} = -\frac{2}{3}G_2^{(2)}=2G_2^{(3)}=2G_2^{(4)} =1/2$ while all its other weights are zero.  Moreover, we will assume that the S$^3$ has no conical defect asymptotically: $k=1$.

\subsubsection{The solution}

We refer the reader interested in the derivation of the type IIB fields to Appendix \ref{App:GlobalAdS3+T4}. The solution \eqref{eq:TypeIIBAnsatz2}, obtained from \eqref{eq:LinearAdS3} with the rod configuration considered in this section,  gives
\begin{align}
ds_{10}^2 \= & \sqrt{Q_1 Q_5\,\cF_1 }\,\Biggl[  \cF_2\,\frac{4\,(d\rho^2+dz^2)}{\left( r^2+\ell^2\cos^2\theta\right)\left( r^2+\ell^2\sin^2\theta\right)} + \cos^2 \theta \,d\varphi_1^2 + \sin^2 \theta\, d\varphi_2^2 \Biggr] \nn \\
&+\frac{1}{\sqrt{Q_1 Q_5\,\cF_1}} \left[-(r^2+\ell^2) \,dt^2 +r^2 \,\cF_3\,dy^2 \right] + \sqrt{\frac{Q_1}{Q_5}\,\cF_1}\,\left( \frac{dx_1^2}{\cF_3} + \sum_{a=2}^4 dx_a^2 \right) , \label{eq:met1AdS3+T4}\\ 
C^{(2)} \= &Q_5\cos^2\theta\,d\varphi_2 \wedge d\varphi_1 -\frac{r^2+\ell^2}{Q_1\,\cF_1} \,dt\wedge dy \,,\qquad  e^\Phi \=\sqrt{\frac{Q_1}{Q_5}\,\cF_1}\,, \nn 
\end{align}
where we have defined \emph{three deformation factors}
\begin{equation}
\cF_1 \equi\frac{\ell^2}{\ell^2-\ell_2^2}\left(1-\frac{\ell_2^2 \,r^2}{\ell^2\,r_2^2} \right)\,,\quad \cF_2 \equi \frac{r_2^2(r_2^2+\ell_2^2)}{(r_2^2+\ell_2^2\cos^2 \theta_2)(r_2^2+\ell_2^2\sin^2 \theta_1)} \,,\quad \cF_3 \equi 1+\frac{\ell_2^2 }{r_2^2} \,, \label{eq:FcalDef}
\end{equation}
that are trivial in the limit where the extra rod vanishes $\ell_2 \to 0$.  We remind that $(r,\theta)$ are the global spherical coordinates of the two-rod configuration \eqref{eq:DefDistanceglobal}, while $(r_i,\theta_i)$ are the local spherical coordinates centered at the $i^\text{th}$ rod, given in terms of $(\rho,z)$ in \eqref{eq:DefDistance} and of $(r,\theta)$ in \eqref{eq:ri&thetaidef}.

The role of the second rod as a deformation on top of a global AdS$_3\times$S$^3\times$T$^4$ background can be highlighted by changing from the Weyl coordinates $(\rho,z)$ to the global spherical coordinates $(r,\theta)$:
\begin{align}
ds_{10}^2 \= & \sqrt{Q_1 Q_5\,\cF_1} \,\Biggl[  \cF_2\,\left(\frac{dr^2}{r^2+\ell^2}+d\theta^2 \right)+ \cos^2 \theta \,d\varphi_1^2 + \sin^2 \theta\, d\varphi_2^2 \Biggr] \nn \\
&+\frac{1}{\sqrt{Q_1 Q_5\,\cF_1}} \left[-(r^2+\ell^2) \,dt^2 +r^2 \,\cF_3\,dy^2 \right] + \sqrt{\frac{Q_1}{Q_5}\,\cF_1}\,\left( \frac{dx_1^2}{\cF_3} + \sum_{a=2}^4 dx_a^2 \right) , \label{eq:met2AdS3+T4}
\end{align}

The solutions are asymptotic to AdS$_3\times$S$^3\times$T$^4$ as in \eqref{eq:AdS3Asymp} since all $\cF_I$ goes to $1$ at large $r$.  The second rod does not only break the rigidity of the T$^4$ but also deforms the S$^3$ and AdS$_3$ spaces.  This deformation can be made regular as a smooth coordinate degeneracy of the $x_1$-circle at the second rod. We analyze the topology at the rod sources and derive the regularity conditions in the upcoming section.  We will also show that the T$^4$ deformation breaks the supersymmetry of the unperturbed global AdS$_3$ background.

\subsubsection{Regularity conditions and topology}
\label{sec:RegCondAdS3+T4}

The rod sources are located at $\rho=0$ and $0\leq z \leq \ell^2/4$.  In the $(r,\theta)$ coordinate system \eqref{eq:DefDistanceglobal},  they are at $r=0$ and 
\begin{equation}
0 \leq \theta \leq \theta_c \quad \Rightarrow\text{ Locus of the $2^\text{nd}$ rod} \quad \bigl| \quad  \theta_c \leq \theta \leq \frac{\pi}{2} \quad \Rightarrow\text{ Locus of the $1^\text{st}$ rod,}\label{eq:RodLocTheta1}
\end{equation} 
where
\begin{equation}
\cos^2 \theta_c \equi \frac{\ell_1^2}{\ell_1^2+\ell_2^2} \equi 1 -\frac{\ell_2^2}{\ell^2}\,.
\label{eq:DefThetaCri}
\end{equation}
First,  at $r>0$,  one can check that $r_i>0$ and all $\cF_I$ are finite and positive.  Therefore,  all metric components \eqref{eq:met2AdS3+T4} are finite and the geometries are regular there for $\theta\neq 0,\pi/2$.  The loci $r>0$ and $\theta=0,\pi/2$ correspond to the two semi-infinite segments above and below the rod sources on the $z$-axis depicted in Fig.\ref{fig:rodsourceAdS3+T4}.  They define the North and South poles of the S$^3$ where $\varphi_2$ and $\varphi_1$ degenerate respectively.  One can check that $\cF_2=1$ at these loci and the angles degenerate smoothly without conical singularities at the poles: $ds(S^3) \sim d\theta^2 + \cos^2 \theta d\varphi_1^2 +\sin^2 \theta d\varphi_2^2$.  The spacetime is therefore regular outside the rod sources at $r>0$ and has a S$^3\times$S$^1\times$T$^4$ topology.

At the sources, $r=0$,  the $y$-circle degenerates at the first rod, $\theta_c \leq \theta \leq \pi/2$ where $r_1=0$ and $r_2>0$, while the $x_1$-circle degenerates at the second rod,  $0 \leq \theta \leq \theta_c$ where $r_2=0$ and $r_1>0$.  The local geometries are better described in terms of the local spherical coordinates,  $ i=1\text{ or }2$, 
\begin{equation}
\rho = \frac{r_i \sqrt{r_i^2+\ell_i^2}}{4}\,\sin 2\theta_i \,,\qquad z \= \frac{2r_i^2+\ell_i^2}{8} \,\cos 2\theta_i + \frac{1}{4} \sum_{j=1}^i \ell_j^2 - \frac{\ell_i^2}{8}\,, 
\label{eq:LocalSpher1}
\end{equation}
which implies
\begin{equation}
d\rho^2 +dz^2 \= \frac{\left( r_i^2+\ell_i^2\cos^2\theta_i\right)\left( r_i^2+\ell_i^2\sin^2\theta_i\right)}{4} \, \left(\frac{dr_i^2}{r_i^2+\ell_i^2}+ d\theta_i^2 \right)\,.
\label{eq:LocalSpher2}
\end{equation}
Therefore, at the first rod $r_1 \to 0$,\footnotemark  $\,$ the time slices of the metric \eqref{eq:met1AdS3+T4} give
\begin{align}
ds_{10} \,\sim \,  &\frac{\sqrt{Q_1 Q_5}}{\ell_1\,\ell} \,\left[\left( dr_1^2 +\frac{\ell^2}{Q_1 Q_5}\,r_1^2 \,dy^2\right)+ (\ell_2^2+\ell_1^2\sin^2\theta_1) \,d\varphi_2^2 \right]+ \sqrt{\frac{Q_1}{Q_5}} \sum_{a=2}^4 dx_a^2\nn \\
& \hspace{-0.2cm}+\frac{\ell_1 \sqrt{Q_1 Q_5}}{\ell} \left[d\theta_1^2+ \cos^2\theta_1 \,d\varphi_1^2+\frac{\ell^2}{Q_5(\ell_2^2+\ell_1^2 \sin^2 \theta_1)}\sin^2 \theta_1 \,dx_1^2 \right]\,.\label{eq:Met1stRodAdS3+T4}
\end{align}

\footnotetext{When $r_1\to 0$,  we have $$ r_2^2 \sim \ell_1^2 \sin^2 \theta_1\,,\quad \cos \theta_2 \sim \frac{r_1 \cos \theta_1}{\sqrt{\ell_2^2+\ell_1^2 \sin^2\theta_1}}\,,\quad r^2\sim  \frac{\ell^2 \sin^2 \theta_1}{\ell_2^2+\ell_1^2 \sin^2 \theta_1}\,r_1^2\,,\quad \ell^2 \cos^2\theta \sim \ell_1^2 \cos^2 \theta_1.$$} 

\noindent At the second rod,  $r_2 \to 0$,\footnote{When $r_2\to 0$,  we have $$r_1^2 \sim \ell_2^2 \cos^2 \theta_2\,,\quad \sin \theta_1 \sim \frac{r_2 \sin \theta_2}{\sqrt{\ell_1^2+\ell_2^2 \cos^2\theta_2}}\,,\quad r^2 \sim \frac{\ell^2 \cos^2 \theta_2}{\ell_1^2+\ell_2^2 \cos^2 \theta_2}\,r_2^2\,,\quad \ell^2\sin^2 \theta \sim \ell_2^2 \sin^2 \theta_2\,.$$} we have
\begin{align}
ds_{10} \,\sim \,  &\frac{\ell_1}{\sqrt{\ell_1^2+\ell_2^2 \cos^2\theta_2}}\Biggl[\frac{\sqrt{Q_1 Q_5}}{\ell_1\,\ell} \left[ \left( dr_2^2 +\frac{\ell^2}{Q_5\,\ell_2^2}\,r_2^2 \,dx_1^2\right)+ (\ell_1^2+\ell_2^2 \cos^2 \theta_2)^2 d\varphi_1^2 \right] \\ 
&\hspace{2.3cm}+ \sqrt{\frac{Q_1}{Q_5}} \sum_{a=2}^4 dx_a^2  + \frac{\ell_2^2\sqrt{Q_1 Q_5}}{\ell_1\,\ell}\left[d\theta_2^2+\frac{\ell^2}{Q_1 Q_5}\cos^2 \theta_2 \,dy^2+ \sin^2\theta_2 \,d\varphi_2^2 \right] . \nn
\end{align}

%%%%%%%%%%%%%%%%%%%%%%%%%%%%%%%%
\begin{figure}[t]
\centering
\includegraphics[scale=0.65]{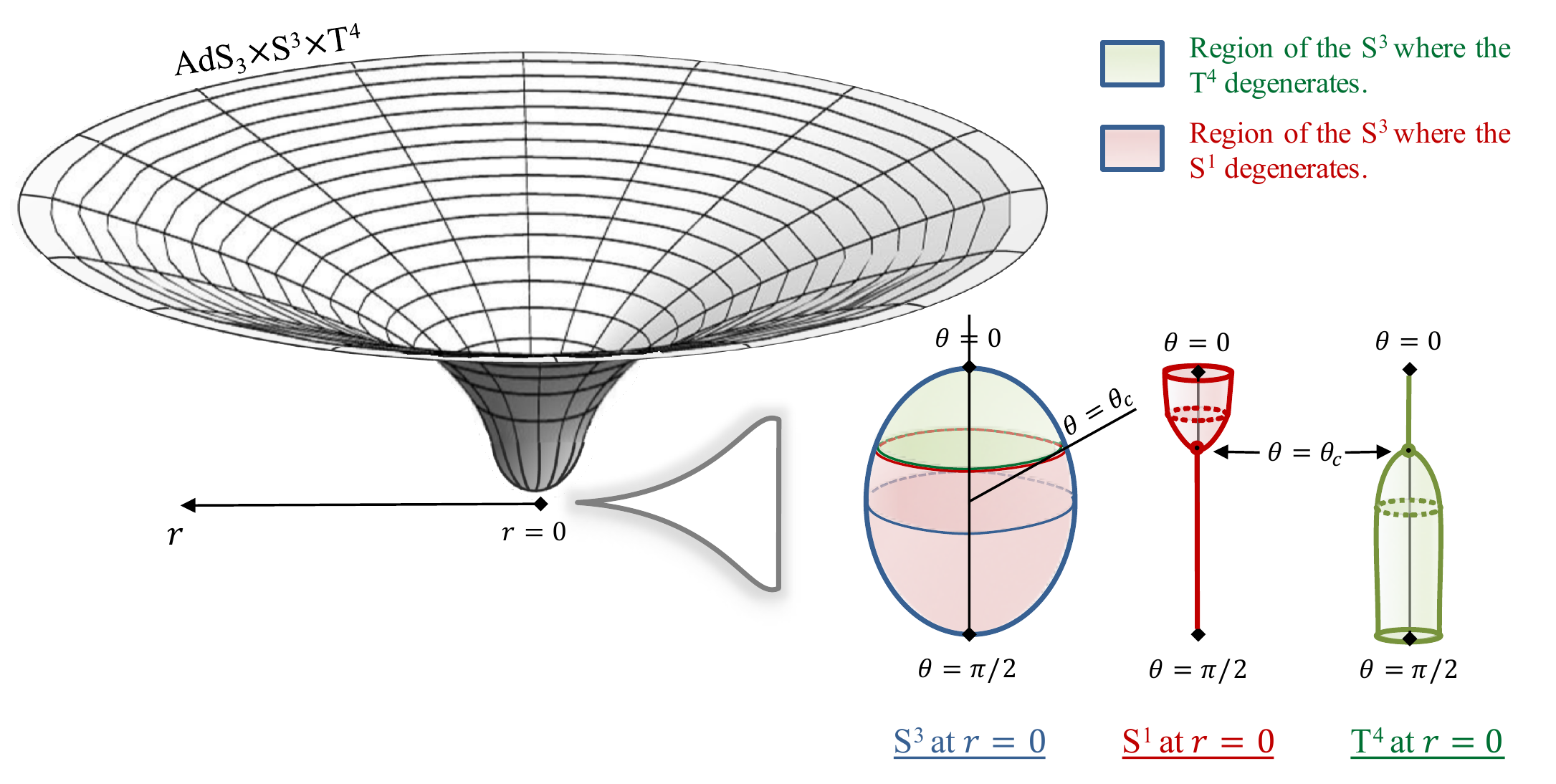}
\caption{Schematic description of the spacetime induced by a chain of two rods,  inducing the degeneracy of the S$^1$ ($y$) and a T$^4$ direction ($x_1$).  On the left-hand side, we depict the overall geometry in terms of the radius $r$.  On the right-hand side,  we describe the behavior of the S$^3$,  S$^1$, and T$^4$ at $r=0$ and as a function of $\theta$,  giving the position on the S$^3$.  At $r=0$,  the spacetime ends smoothly as a coordinate degeneracy of either the S$^1$ or the T$^4$ depending on the position on the S$^3$. }
\label{fig:AdS3+T41pic}
\end{figure}
%%%%%%%%%%%%%%%%%%%%%%%%%%%%%%%%

\noindent Both geometries correspond to regular S$^3\times$T$^4$ fibrations over an origin of a $\IR^2$ space if\footnote{We remind that $R_y$ and $R_{x_1}$ are the radius of the $y$ and $x_1$ circles, defined by their periodicities $y= y+2\pi R_y$ and $x_1 = x_1+2\pi R_{x_1}$ \eqref{eq:psi&phiPerio}.}
\begin{equation}
R_y \= \frac{\sqrt{Q_1 Q_5}}{\ell}\,,\qquad R_{x_1} \= \frac{\ell_2\,\sqrt{Q_5}}{\ell} \,,
\label{eq:RegCondAdS3+T4}
\end{equation}
which implies
\begin{equation}
\ell_1 = \frac{\sqrt{Q_1 (Q_5- R_{x_1}^2)}}{R_y}\,,\qquad \ell_2 = \frac{ R_{x_1 }\sqrt{Q_1}}{R_y}\quad \Rightarrow \quad \ell^2\= \ell_1^2+\ell_2^2 \= \frac{Q_1 Q_5}{R_y^2}.
\label{eq:RegCondAdS3+T4bis}
\end{equation}

The three-form flux, $F_3=dC^{(2)}$,  is regular at the rods since the components along the shrinking directions vanish. Moreover, the first rod carries D1 and D5 brane charges while the second rod carries a D5 brane charge given by \eqref{eq:ChargeAtRodGen}
\begin{equation}
q_{D1}^{(1)}\= Q_1 \,,\quad q_{D5}^{(1)}  \=Q_5-R_{x_1}^2\,,\qquad \text{and}\qquad \,\, q_{D1}^{(2)}  \=0\,,\quad  q_{D5}^{(2)}  \=R_{x_1}^2\,.
\end{equation}

The presence of the second rod has modified the regularity condition of the global AdS$_3\times$S$^3\times$T$^4$ background \eqref{eq:RegCondAdS3}.  The total length of the sources, $\ell^2$, is still given by $Q_1 Q_5/R_{y}^2$ but it has been now distributed on both sources. 

Moreover,  the deformation has also drastically changed the topology at the center of the spacetime $r=0$.  Indeed,  the S$^3$ is parametrized by $(\theta,\varphi_1,\varphi_2)$ in the unperturbed geometry \eqref{eq:metGlobalAdS3},  and the $(x_1,x_2,x_3,x_4)$ define a rigid T$^4$.  The role of $\varphi_2$ and $x_1$ has been interchanged after perturbation at $r_1 = 0$ \eqref{eq:Met1stRodAdS3+T4}, i.e. $r=0$ and $\theta_c\leq \theta\leq\pi/2$,  and the local S$^3$ is now parametrized by $(\theta_1,\varphi_1,x_1)$.  At the second rod, $r=0$ and $0\leq \theta\leq \theta_c$,  the local S$^3$ is given by $(\theta_2,y,\varphi_2)$, and since both rods and S$^3$ are connected we have an overall S$^3$ given by $(\theta,\varphi_1,\varphi_2)$ at $r=0$.

Thus, we obtained regular geometries that are asymptotic to AdS$_3\times$S$^3\times$T$^4$.  The spacetime ends smoothly at $r=0$ as a chain of two bolts where the S$^1$ and one of the T$^4$ coordinates degenerate alternately.  We have depicted the geometries and the behavior of the S$^3$, S$^1$, and T$^4$ at the end-of-spacetime point in Fig.\ref{fig:AdS3+T41pic}.  At this locus,  the two bolts split the S$^3$ into two regions such that the T$^4$ shrinks in the Northern Hemisphere while the S$^1$ degenerates in the Southern Hemisphere.  The intersection is set to $\theta= \theta_c$ \eqref{eq:DefThetaCri} such that 
\begin{equation}
\cos^2\theta_c \= 1-\frac{R_{x_1}^2}{Q_5}.
\label{eq:DefThetaCribis}
\end{equation} 
Therefore,  one of the regions can be very small relative to the other if there is a hierarchy of scales between $\ell_1^2$ and $\ell_2^2$,  i.e.  $R_{x_1}^2$ and $Q_5$.

\subsubsection{Supersymmetry breaking}
\label{sec:nonBPSness}

We now give arguments suggesting that the solution does not preserve any supersymmetry.

\noindent First,  the solution necessarily breaks the supersymmetry of the undeformed global AdS$_3$ spacetime in type IIB.  Supersymmetric solutions in ten-dimensional supergravity are characterized by the existence of a Killing vector that can be either time-like or null \cite{Tomasiello:2011eb,Giusto:2013rxa}.  Global AdS$_3$ and all superstratum excitations \cite{Bena:2015bea,Bena:2016agb,Bena:2017xbt,Bena:2018bbd,Ceplak:2018pws,Heidmann:2019zws,Heidmann:2019xrd} are $\frac{1}{4}$- and $\frac{1}{8}$- BPS solutions in type IIB respectively that are based on a null Killing spinor defining a null direction in the spacetime.  Thus,  they can be generically decomposed into a $(u,v)=(t-y,t+y)$ null fibration over an eight-dimensional space, which, in the case of those spacetimes,  is composed of a rigid T$^4$ fibration over a four-dimensional almost hyper-Kh\"aler  base \cite{Giusto:2013rxa}.  This structure can be made manifest for the global AdS$_3$ metric \eqref{eq:metGlobalAdS3} by performing a spectral flow from the NS-NS sector to the R-R sector,
\begin{equation}
\varphi_1 \to \varphi_1 - \frac{y}{R_y}\,,\qquad \varphi_2 \to \varphi_2 - \frac{t}{R_y}\,,
\end{equation}
such that a null direction appears and the four-dimensional base is the flat metric. 
However,  there are no spectral flows,  boosts along $y$ and shifts of coordinates with the T$^4$ directions that can produce a globally null direction in the T$^4$ deformation metric \eqref{eq:met1AdS3+T4}.  This is a consequence of the non-trivial deformation factors,  $\cF_1$, and $\cF_3$, that cannot be compensated by constant shifts.  Thus,  the deformed solution cannot preserve the same Killing spinors as the global AdS$_3\times$S$^3\times$T$^4$ spacetime in type IIB and necessarily breaks all its supersymmetry.

Second,  it is still possible that new supersymmetries emerge when the T$^4$ deformation is included.   This necessarily requires the existence of a timelike Killing spinor.  To disprove that,  one should a priori derive Killing spinor equations for our generic ansatz \eqref{eq:TypeIIBAnsatz2} and show that the solution at hand does not satisfy some of them.  Since this computation is rather tedious and requires a project on its own we postpone it for future work.  We just give a few arguments that suggest that the solution does not preserve any supersymmetries:
\begin{itemize}
\item[•] First,  if one reduces to five dimensions along the T$^4$ and a generic direction, $y+\alpha t +\beta \varphi_1+\gamma \varphi_2$,  one can show that the solution does not satisfy the supersymmetric conditions of $\cN=1$ five-dimensional supergravity derived in \cite{Bellorin:2006yr}.  Indeed,  one cannot generate a four-dimensional hyper-Kh\"aler base in five dimensions by performing a change of variables and boosts on the S$^3$ angles $(\varphi_1,\varphi_2)$.
\item[•] Second,  one can construct the ``unbounded'' solution of our bound state of two rods by considering the same rod configuration as in Fig.\ref{fig:rodsourceAdS3+T4} but with disconnected rods.  The rods will then be separated by a segment that does not source the warp factors but still induces the degeneracy of a compact direction given as a linear combination of the $\varphi_1$ and $\varphi_2$ angles.  However,  this degeneracy has necessarily a conical excess which means that the sources are separated by a strut.  A strut corresponds to a string with negative tension that accounts for the repulsive force needed to compensate for the self-attraction  between both sources \cite{Costa:2000kf,Regge:1961px,Bah:2021owp}.  Thus,  both sources are not in equilibrium at a finite distance as it is usually the case between BPS sources.  In asymptotically-flat spacetimes,  the rods are non-supersymmetric such that their inherent mass was larger than their charges and the electromagnetic force does not compensate for their gravitational attraction.  A similar phenomenon appears here which greatly suggests that the solution is non-supersymmetric.
\end{itemize}

It would be interesting to have a precise idea on how the T$^4$ deformation breaks all supersymmetry as it has been done for other new smooth type IIB geometries in AdS$_3$ \cite{Ganchev:2021pgs,Ganchev:2021ewa}.  Those ``microstrata'' are non-BPS extensions of superstrata where supersymmetry is broken by superposing both left-moving and right-moving excitations in AdS$_3$ while keeping a rigid T$^4$. 

\subsubsection{Two interesting limits}

The regularity condition \eqref{eq:RegCondAdS3+T4bis} requires 
\begin{equation}
R_{x_1}^2 \leq Q_5\,,
\end{equation}
and there are two interesting limits whether $R_{x_1}^2 \ll Q_5$ or $R_{x_1}^2 \sim Q_5$.   

\begin{itemize}
\item[•] \underline{A regular non-BPS perturbation on global AdS$_3\times$S$^3\times$T$^4$:}
\end{itemize}

We first assume that $R_{x_1}^2 \ll Q_5$ which makes the second rod infinitesimally small compared to the first,  $\ell_2 \ll \ell_1$.  Moreover,  $R_{x_1}$ can be related to the volume of T$^4$,  $\text{Vol}(\text{T}^4)= (2\pi)^4 V_4$,  by assuming that all T$^4$ directions have the same radii: $R_{x_1}= V_4^{1/4}$.  Therefore,  we have to assume
\begin{equation}
\sqrt{V_4} \ll Q_5 \qquad \Rightarrow \qquad \ell_2 \ll \ell_1.
\end{equation}

All terms proportional to $\ell_2^2$ are then perturbations.  Since $\theta_c \sim 0$ \eqref{eq:DefThetaCribis},  this is valid up to a small region around the North pole of the S$^3$ and around the center of spacetime,  $r\sim 0$ and $\theta\sim0$, where $r_2 \sim 0$.   The perturbation is localized there and has a large effect such that the $x_1$ coordinate degenerates.  Outside this small region,  $\cF_I = 1 +\cO(\sqrt{V_4}/Q_5)$ and the solution corresponds to a small non-BPS excitation on a global AdS$_3\times$S$^3\times$T$^4$ background.  It is localized at the North pole of the S$^3$ and at the center of AdS$_3$ such that it forces the smooth degeneracy of a T$^4$ circle.  

\begin{itemize}
\item[•] \underline{Singular D1 branes on a D5 bubble:}
\end{itemize}

At the other side of the parameter space,  $\ell_1$ can be made strictly zero by considering
\begin{equation}
Q_5 \= R_{x_1}^2 \= \sqrt{V_4} \qquad \Rightarrow \qquad \ell_1 = 0\,.
\end{equation}
This does not eliminate all effects of the rod as it shrinks to a point source that has zero size but still carries a D1 brane charge.  From Fig.\ref{fig:AdS3+T41pic},  the point source is located at the South pole of the S$^3$,  $\theta=\pi/2$,  at $r=0$.  The type IIB solution \eqref{eq:met2AdS3+T4} becomes
\begin{align}
ds_{10}^2 \= & \sqrt{Q_1 Q_5}\,\sqrt{\frac{r^2+\ell^2}{r^2+\ell^2 \cos^2\theta}} \,\Biggl[  \frac{dr^2}{r^2+\ell^2}+d\theta^2+ \cos^2 \theta \,d\varphi_1^2 + \sin^2 \theta\, d\varphi_2^2 \Biggr] \label{eq:D1SingonD5Bu}  \\
&\hspace{-0.45cm} +\sqrt{\frac{(r^2+\ell^2)(r^2+\ell^2\cos^2\theta)}{Q_1 Q_5}} \left[-dt^2 +dy^2 \right] + \sqrt{\frac{Q_1(r^2+\ell^2)}{Q_5(r^2+\ell^2 \cos^2\theta)}} \,\left( \frac{r^2dx_1^2}{r^2+\ell^2} + \sum_{a=2}^4 dx_a^2 \right) ,  \nn\\ 
C^{(2)} \= &Q_5\cos^2\theta\,d\varphi_2 \wedge d\varphi_1 -\frac{r^2+\ell^2 \cos^2\theta}{Q_1} \,dt\wedge dy \,,\qquad  e^\Phi \=\sqrt{\frac{Q_1}{Q_5}}\,\sqrt{\frac{r^2+\ell^2}{r^2+\ell^2 \cos^2\theta}}\,.\nn 
\end{align}
The solutions are regular for $r>0$.  Moreover,  there is still a bolt where $x_1$ degenerates smoothly at $r=0$ and $\theta\neq \pi/2$,  which is the locus of the second rod,  and it carries $Q_5$ D5 brane charge.  However,  the first rod has now degenerated to a singular horizon at $r=0$ and $\theta=\pi/2$.  At this locus,  we have a blowing T$^4$ parametrized by $(x_2,x_3,x_4,\varphi_2)$ while $(\theta_1,\varphi_1,x_1)$ describes a shrinking S$^3$ and the time and $y$ fiber degenerate.  Thus,  the solutions correspond to singular D1 branes at the South pole of a D5 bubble.

Moreover,  the singularity is resolved by considering $Q_5 \lesssim R_{x_1}^2$,  i.e. $\ell_1\ll \ell_2$.  A geometric transition occurs that transforms the singular point source into a small bolt where the $y$ circle degenerates smoothly at the vicinity of the South pole of the S$^3$ and at the center of spacetime $r=0$.

\subsection{T$^4$ deformation at the center of AdS$_3$}
\label{sec:AdS3+T42}

In the previous section,  the non-BPS T$^4$ deformation was naturally centered around the North pole of the S$^3$ and at the center of the AdS$_3$.   In this section, we show that the deformation can be localized elsewhere on the S$^3$.

\begin{figure}[h]
\centering
    \begin{tikzpicture}
%% some definitions

\def\deb{-10} 
\def\inter{0.8} 
\def\ha{2.8} 
\def\zaxisline{5} 
\def\rodsize{1.5} 
\def\numrod{2.5} 

\def\fin{\deb+1+2*\rodsize+\numrod*\rodsize} 

%% 

%\draw (\deb-2,\ha-\zaxisline*0.5*\inter) node{$\Longrightarrow$};

%% Pic title

\draw (\deb+0.5+\rodsize+0.5*\numrod*\rodsize,\ha+1) node{{{\it Global AdS$_3\times$S$^{\,3}\times$T$^{\,4}$ with a T$^{\,4}$ deformation}}}; 

%% Each line black line and names

\draw[black,thin] (\deb+1,\ha) -- (\fin,\ha);
\draw[black,thin] (\deb,\ha-\inter) -- (\fin-1,\ha-\inter);
\draw[black,thin] (\deb,\ha-2*\inter) -- (\fin,\ha-2*\inter);
\draw[black,thin] (\deb,\ha-3*\inter) -- (\fin,\ha-3*\inter);
\draw[black,thin] (\deb,\ha-4*\inter) -- (\fin,\ha-4*\inter);
\draw[black,->, line width=0.3mm] (\deb-0.5,\ha-\zaxisline*\inter) -- (\fin+0.2,\ha-\zaxisline*\inter);

\draw [decorate, 
    decoration = {brace,
        raise=5pt,
        amplitude=5pt},line width=0.2mm,gray] (\deb-0.8,\ha-1.5*\inter+0.05) --  (\deb-0.8,\ha+0.5*\inter-0.05);
\draw [decorate, 
    decoration = {brace,
        raise=5pt,
        amplitude=5pt},line width=0.2mm,gray] (\deb-0.8,\ha-2.5*\inter+0.05) --  (\deb-0.8,\ha-1.5*\inter-0.05);
\draw [decorate, 
    decoration = {brace,
        raise=5pt,
        amplitude=5pt},line width=0.2mm,gray] (\deb-0.8,\ha-4.5*\inter+0.05) --  (\deb-0.8,\ha-2.5*\inter-0.05);
        
\draw[gray] (\deb-1.5,\ha-0.5*\inter) node{S$^3$};
\draw[gray] (\deb-1.5,\ha-2*\inter) node{S$^1$};
\draw[gray] (\deb-1.5,\ha-3.5*\inter) node{T$^4$};

\draw (\deb-0.5,\ha) node{$\varphi_1$};
\draw (\deb-0.5,\ha-\inter) node{$\varphi_2$};
\draw (\deb-0.5,\ha-2*\inter) node{$y$};
\draw (\deb-0.5,\ha-3*\inter) node{$x_1$};
\draw (\deb-0.5,\ha-4*\inter) node{$x_a$};

\draw (\fin+0.2,\ha-\zaxisline*\inter-0.3) node{$z$};

%% First two line and their rods

\draw[black, dotted, line width=1mm] (\deb,\ha) -- (\deb+0.5,\ha);
\draw[black,line width=1mm] (\deb+0.5,\ha) -- (\deb+0.5+\rodsize,\ha);
\draw[black,line width=1mm] (\fin-0.5-\rodsize,\ha-\inter) -- (\fin-0.55,\ha-\inter);
\draw[black, dotted,line width=1mm] (\fin-0.5,\ha-\inter) -- (\fin,\ha-\inter);

%% Next lines and their rods

\draw[amaranthred,line width=1mm] (\deb+0.5+\rodsize,\ha-2*\inter) -- (\deb+0.5+2*\rodsize,\ha-2*\inter);
\draw[amaranthred,line width=1mm] (\deb+0.5+2.5*\rodsize,\ha-2*\inter) -- (\deb+0.5+3.5*\rodsize,\ha-2*\inter);
\draw[amazon,line width=1mm] (\deb+0.5+2*\rodsize,\ha-3*\inter) -- (\deb+0.5+2.5*\rodsize,\ha-3*\inter);

\draw[amaranthred,line width=1mm,opacity=0.25] (\deb+0.5+\rodsize,\ha-\zaxisline*\inter) -- (\deb+0.5+2*\rodsize,\ha-\zaxisline*\inter);
\draw[amazon,line width=1mm,opacity=0.25] (\deb+0.5+2*\rodsize,\ha-\zaxisline*\inter) -- (\deb+0.5+2.5*\rodsize,\ha-\zaxisline*\inter);
\draw[amaranthred,line width=1mm,opacity=0.25] (\deb+0.5+2.5*\rodsize,\ha-\zaxisline*\inter) -- (\deb+0.5+3.5*\rodsize,\ha-\zaxisline*\inter);

%% Vertical lines and coordinates

\draw[gray,dotted,line width=0.2mm] (\deb+0.5+\rodsize,\ha) -- (\deb+0.5+\rodsize,\ha-\zaxisline*\inter);
\draw[gray,dotted,line width=0.2mm] (\deb+0.5+2*\rodsize,\ha) -- (\deb+0.5+2*\rodsize,\ha-\zaxisline*\inter);
\draw[gray,dotted,line width=0.2mm] (\deb+0.5+2.5*\rodsize,\ha) -- (\deb+0.5+2.5*\rodsize,\ha-\zaxisline*\inter);
\draw[gray,dotted,line width=0.2mm] (\deb+0.5+3.5*\rodsize,\ha) -- (\deb+0.5+3.5*\rodsize,\ha-\zaxisline*\inter);

\draw[line width=0.3mm] (\deb+0.5+\rodsize,\ha-\zaxisline*\inter+0.1) -- (\deb+0.5+\rodsize,\ha-\zaxisline*\inter-0.1);
\draw[line width=0.3mm] (\deb+0.5+2*\rodsize,\ha-\zaxisline*\inter+0.1) -- (\deb+0.5+2*\rodsize,\ha-\zaxisline*\inter-0.1);
\draw[line width=0.3mm] (\deb+0.5+2.5*\rodsize,\ha-\zaxisline*\inter+0.1) -- (\deb+0.5+2.5*\rodsize,\ha-\zaxisline*\inter-0.1);
\draw[line width=0.3mm] (\deb+0.5+3.5*\rodsize,\ha-\zaxisline*\inter+0.1) -- (\deb+0.5+3.5*\rodsize,\ha-\zaxisline*\inter-0.1);

\draw (\deb+0.5+\rodsize,\ha-\zaxisline*\inter-0.5) node{{\small $0$}};
\draw (\deb+0.5+2*\rodsize,\ha-\zaxisline*\inter-0.5) node{{\small $\frac{\ell_1^2}{4}$}};
\draw (\deb+0.5+2.5*\rodsize,\ha-\zaxisline*\inter-0.5) node{{\small $\frac{\ell_1^2+\ell_2^2}{4}$}};
\draw (\deb+0.9+3.5*\rodsize,\ha-\zaxisline*\inter-0.5) node{{\small $\frac{\ell_1^2+\ell_2^2+\ell_3^2}{4}=\frac{\ell^2}{4}$}};

\end{tikzpicture}
\caption{Rod diagram of the shrinking directions on the $z$-axis after sourcing the solutions with three connected rods that force the degeneracy of the $y$ and $x_1$ circles.}
\label{fig:rodsourceAdS3+T42}
\end{figure}
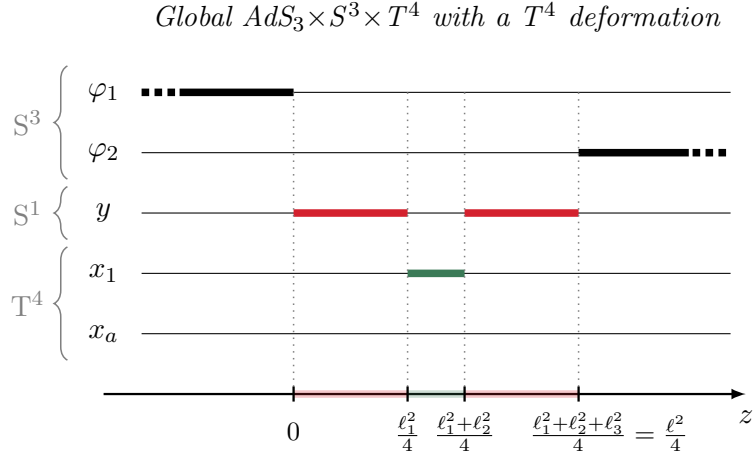  

We consider three rod sources (see Fig.\ref{fig:rodsourceAdS3+T42}).  The first two are identical to the previous section while the last one is chosen such that it induces the degeneracy of the $y$ coordinate.  From Table  \ref{tab:internalBC}, it requires fixing the weights at the third rod such that $P_3^{(0)}=P_3^{(1)}= P_3^{(5)}= -G_3^{(1)} =1/2$ while all its other weights are taken to be zero.  We still consider that $k=1$ such that the S$^3$ has no conical defect asymptotically.

\subsubsection{The solution}

We refer the reader interested in the derivation of the type IIB fields to Appendix \ref{App:GlobalAdS3+T42}. The solution \eqref{eq:TypeIIBAnsatz2}, obtained from \eqref{eq:LinearAdS3} with the rod configuration considered in this section,  gives
\begin{align}
ds_{10}^2 \= & \sqrt{Q_1 Q_5\,\cF_1} \,\Biggl[  \widetilde{\cF}_2\,\left(\frac{dr^2}{r^2+\ell^2}+d\theta^2 \right)+ \cos^2 \theta \,d\varphi_1^2 + \sin^2 \theta\, d\varphi_2^2 \Biggr] \nn \\
&+\frac{1}{\sqrt{Q_1 Q_5\,\cF_1}} \left[-(r^2+\ell^2) \,dt^2 +r^2 \,\cF_3\,dy^2 \right] + \sqrt{\frac{Q_1}{Q_5}\,\cF_1}\,\left( \frac{dx_1^2}{\cF_3} + \sum_{a=2}^4 dx_a^2 \right) , \label{eq:metAdS3+T42}\\ 
C^{(2)} \= &Q_5\cos^2\theta\,d\varphi_2 \wedge d\varphi_1 -\frac{r^2+\ell^2}{Q_1\,\cF_1} \,dt\wedge dy \,,\qquad  e^\Phi \=\sqrt{\frac{Q_1}{Q_5}\,\cF_1}\,, \nn 
\end{align}
where we have defined,  in addition to the deformation factors introduced in \eqref{eq:FcalDef},\footnote{The deformation factors $\cF_1$ and $\cF_3$, given in \eqref{eq:FcalDef},  give different values for the present configuration since $\ell^2=\ell_1^2+\ell_2^2+\ell_3^2$ and $(r_2,\theta_2)$ are different.}
\begin{equation}
\begin{split}
\widetilde{\cF}_2 &\equi \frac{r_2^4(r_2^2+\ell_2^2)^2}{(r_2^2+\ell_2^2\cos^2 \theta_2)(r_2^2+\ell_2^2\sin^2 \theta_2)(r_2^2+\ell_2^2\sin^2 \theta_1)(r_2^2+\ell_2^2\cos^2 \theta_3)} \,.
\end{split}\label{eq:FcalDefbis}
\end{equation}

We remind that $(r,\theta)$ are the global spherical coordinates of the three-rod configuration \eqref{eq:DefDistanceglobal}, while $(r_i,\theta_i)$ are the local spherical coordinates centered at the $i^\text{th}$ rod, given in terms of $(\rho,z)$ in \eqref{eq:DefDistance} and of $(r,\theta)$ in \eqref{eq:ri&thetaidef}.

%%%%%%%%%%%%%%%%%%%%%%%%%%%%%%%%
\begin{figure}[t]
\centering
\includegraphics[scale=0.7]{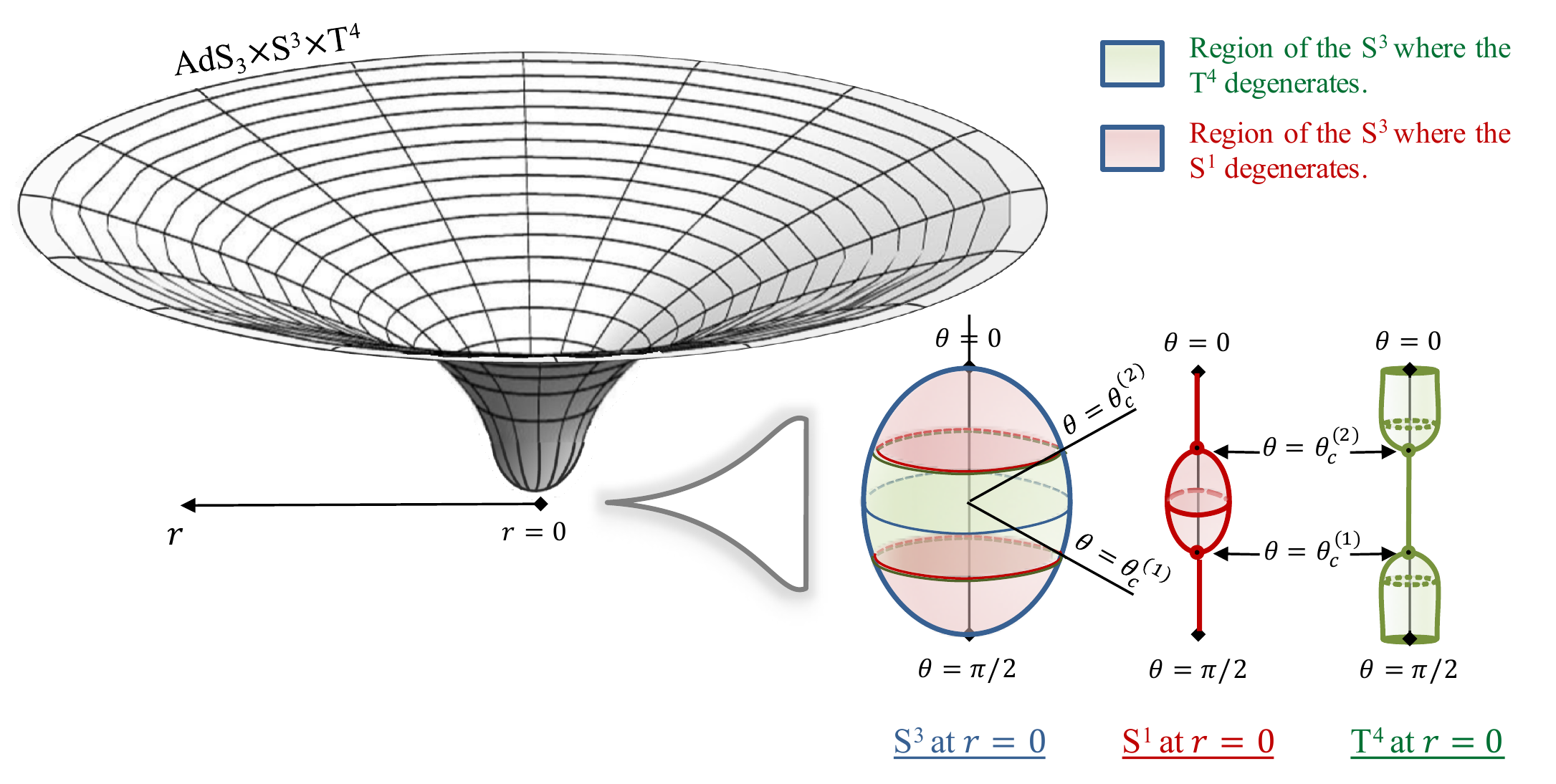}
\caption{Schematic description of the spacetime induced by a chain of three rods,  two inducing the degeneracy of the S$^1$ ($y$) with one in the middle inducing the degeneracy of a T$^4$ direction ($x_1$).  On the left-hand side, we depict the overall geometry in terms of $r$.  On the right-hand side,  we describe the behavior of the S$^3$,  S$^1$, and T$^4$ at $r=0$ and as a function of $\theta$.  }
\label{fig:AdS3+T42pic}
\end{figure}
%%%%%%%%%%%%%%%%%%%%%%%%%%%%%%%%

The details of the regularity analysis and the description of the topology can be found in Appendix \ref{App:GlobalAdS3+T42} which we summarize here.  The solutions are regular for $r>0$ with a S$^1\times$S$^3\times$T$^4$ topology and are asymptotic to AdS$_3\times$S$^3\times$T$^4$ since $\cF_I \sim \widetilde{\cF}_2 \sim 1$ at large $r$.  The rod sources are located at $r=0$ with $\theta$ from $0$ to $\pi/2$ such that
\begin{equation}
0 \leq \theta \leq \theta_c^{(2)} \,\,\, \Rightarrow\text{ $3^\text{rd}$ rod} \,\,\, \bigl| \,\,\, \theta_c^{(2)} \leq \theta \leq \theta_c^{(1)} \,\,\, \Rightarrow\text{ $2^\text{nd}$ rod} \,\,\, \bigl|  \,\,\,  \theta_c^{(1)} \leq \theta \leq \frac{\pi}{2} \,\,\, \Rightarrow\text{ $1^\text{st}$ rod,}
\label{eq:RodLocTheta2}
\end{equation} 
where we have defined
\begin{equation}
\cos^2 \theta_c^{(1)} \equi  \frac{\ell_1^2}{\ell^2}\,,\qquad \cos^2 \theta_c^{(2)} \equi  \frac{\ell_1^2+\ell_2^2}{\ell^2}\,.
\label{eq:DefThetaCri2}
\end{equation}
The rods correspond to bolts where either $y$ or $x_1$ degenerates defining an origin of a $\IR^2$ space.  The transverse spaces are S$^3_{\theta_1 x_1 \varphi_1}\times$T$^4_{\varphi_2 x_2 x_3 x_4}$ at the first rod,  S$^2_{\theta_2 y}\times$T$^5_{\varphi_1 \varphi_2 x_2 x_3 x_4}$ at the second rod and S$^3_{\theta_3 x_1 \varphi_2}\times$T$^4_{\varphi_1 x_2 x_3 x_4}$ at the third rod.  The bolts are regular if 
\begin{equation}
\ell_1^2 \= \ell_3^2 \= \frac{Q_1 Q_5}{2R_y^2} \left(1-\frac{R_{x_1}}{ \sqrt{Q_5}}\right)\,,\quad \ell_2^2 \= \frac{Q_1 \sqrt{Q_5 }\,R_{x_1}}{2R_y^2} \,\quad \Rightarrow \quad \ell^2 \= \frac{Q_1 Q_5}{R_y^2}  \left(1-\frac{R_{x_1}}{2 \sqrt{Q_5}}\right)\,.
\label{eq:Reg3rodT4}
\end{equation}
The regularity still requires that $R_{x_1} \leq \sqrt{Q_5}$.  Note that, unlike the previous T$^4$ deformation,  the overall length of the configuration is not anymore given by $Q_1 Q_5/R_{y}^2$.  

Moreover, the rods where the $y$ coordinate degenerates carry the same D1 and D5 brane charges \eqref{eq:ChargeAtRodGen}
\begin{equation}
q_{D1}^{(1)}\= q_{D1}^{(3)} \= \frac{Q_1}{2} \,,\qquad q_{D5}^{(1)} \= q_{D5}^{(3)}  \= Q_5\,\frac{\sqrt{Q_5}-R_{x_1}}{2\sqrt{Q_5}-R_{x_1}},
\end{equation}  
while the rod where the $x_1$ degenerates has a D5 brane charge
\begin{equation}
q_{D1}^{(2)} \=0 \,,\qquad q_{D5}^{(2)}  \= \frac{Q_5\, R_{x_1}}{2\sqrt{Q_5}-R_{x_1}}.
\end{equation}

We have depicted the geometries in Fig.\ref{fig:AdS3+T42pic}.  The solutions are smooth and end at $r=0$ as a chain of bolts.  More precisely,  at $r=0$,  we have a S$^3$ that can be decomposed into three regions.  For $0\leq \theta \leq  \theta_c^{(2)}$ and $\theta_c^{(1)} = \pi/2-\theta_c^{(2)} \leq \theta \leq \pi/2$,  the $y$ coordinate degenerates as the usual S$^1$ degeneracy at the center of a global AdS$_3$ spacetime where $\cos^2\theta_c^{(2)} = (2-R_{x_1}/\sqrt{Q_5})^{-1}$.  However,  for $ \theta_c^{(2)} \leq\theta\leq \pi/2-\theta_c^{(2)}$,  this degeneracy has been replaced by the degeneracy of a T$^4$ direction.  Therefore,  the T$^4$ deformation is now centered around the equator of the S$^3$.  Moreover, by allowing a conical defect at one of the rods where the $y$-circle shrinks, the deformation can be centered around any $\theta$ between $0$ and $\pi/2$ (see Appendix \ref{App:GlobalAdS3+T42} for more details).     

Moreover,  for the same arguments as in section \ref{sec:nonBPSness},  the deformation breaks the supersymmetry of the global AdS$_3\times$S$^3\times$T$^4$ solution, and most likely breaks all supersymmetry such that it corresponds to a non-BPS asymptotically-AdS$_3$ solution in type IIB.

As for the previous solution,  we have two interesting limits.  First,  when $R_{x_1} \ll \sqrt{Q_5}$,  the T$^4$ deformation can be treated as a non-BPS smooth perturbation on a global AdS$_3\times$S$^3\times$T$^4$ background.  Indeed,  we have $\ell_2^2 \ll \ell_1^2= \ell_3^2$ and the deformation factors,  $\widetilde{\cF}_1$,  $\widetilde{\cF}_2$ and $\cF_3$ are equal to 1 plus correction of order $\ell_2^2$ as soon as we are not too close to the middle rod.  Since $\theta_c^{(1)} \sim \theta_c^{(2)} \sim \pi/4$ in this limit,  the perturbation is localized at the center of the global AdS$_3$ spacetime, $r=0$, and at the equator of the S$^3$.  The small perturbation has induced a non-trivial degeneracy of the T$^4$ there for which the backreaction has broken the symmetry of the AdS$_3$,  S$^3$, and T$^4$.

\noindent Second,  if one considers $R_{x_1} = \sqrt{Q_5}$,  the two bolts where the S$^1$ degenerates shrink to a point leading to singular D1 brane sources.  The solutions are similar to \eqref{eq:D1SingonD5Bu} but we now have a singular horizon at each pole of the bolt where the $x_1$ circle shrinks.  This limit corresponds to a non-BPS D5 bubble with two singular stacks of D1 branes at its poles.  The singularities can be resolved by considering $R_{x_1} \lesssim \sqrt{Q_5}$ where the point sources undergo a geometric transition into small  bolts where the $y$ coordinate degenerates smoothly.

\subsection{S$^3$ deformation at the center of AdS$_3$}
\label{sec:AdS3+S3}

\begin{figure}[h]
\centering
    \begin{tikzpicture}
%% some definitions

\def\deb{-10} 
\def\inter{0.8} 
\def\ha{2.8} 
\def\zaxisline{5} 
\def\rodsize{1.5} 
\def\numrod{2.5} 

\def\fin{\deb+1+2*\rodsize+\numrod*\rodsize} 

%% 

%\draw (\deb-2,\ha-\zaxisline*0.5*\inter) node{$\Longrightarrow$};

%% Pic title

\draw (\deb+0.5+\rodsize+0.5*\numrod*\rodsize,\ha+1) node{{{\it Global AdS$_3\times$S$^{\,3}\times$T$^{\,4}$ with a S$^{\,3}$ deformation}}}; 

%% Each line black line and names

\draw[black,thin] (\deb+1,\ha) -- (\fin,\ha);
\draw[black,thin] (\deb,\ha-\inter) -- (\fin-1,\ha-\inter);
\draw[black,thin] (\deb,\ha-2*\inter) -- (\fin,\ha-2*\inter);
\draw[black,thin] (\deb,\ha-3*\inter) -- (\fin,\ha-3*\inter);
\draw[black,thin] (\deb,\ha-4*\inter) -- (\fin,\ha-4*\inter);
\draw[black,->, line width=0.3mm] (\deb-0.5,\ha-\zaxisline*\inter) -- (\fin+0.2,\ha-\zaxisline*\inter);

\draw [decorate, 
    decoration = {brace,
        raise=5pt,
        amplitude=5pt},line width=0.2mm,gray] (\deb-0.8,\ha-2.5*\inter+0.05) --  (\deb-0.8,\ha+0.5*\inter-0.05);
\draw [decorate, 
    decoration = {brace,
        raise=5pt,
        amplitude=5pt},line width=0.2mm,gray] (\deb-0.8,\ha-3.5*\inter+0.05) --  (\deb-0.8,\ha-2.5*\inter-0.05);
\draw [decorate, 
    decoration = {brace,
        raise=5pt,
        amplitude=5pt},line width=0.2mm,gray] (\deb-0.8,\ha-4.5*\inter+0.05) --  (\deb-0.8,\ha-3.5*\inter-0.05);
        
\draw[gray] (\deb-1.5,\ha-1*\inter) node{S$^3$};
\draw[gray] (\deb-1.5,\ha-3*\inter) node{S$^1$};
\draw[gray] (\deb-1.5,\ha-4*\inter) node{T$^4$};

\draw (\deb-0.5,\ha) node{$\varphi_1$};
\draw (\deb-0.5,\ha-\inter) node{$\varphi_2$};
\draw (\deb-0.5,\ha-3*\inter) node{$y$};
\draw (\deb-0.5,\ha-2*\inter) node{$\psi$};
\draw (\deb-0.5,\ha-4*\inter) node{$x_a$};

\draw (\fin+0.2,\ha-\zaxisline*\inter-0.3) node{$z$};

%% First two line and their rods

\draw[black, dotted, line width=1mm] (\deb,\ha) -- (\deb+0.5,\ha);
\draw[black,line width=1mm] (\deb+0.5,\ha) -- (\deb+0.5+\rodsize,\ha);
\draw[black,line width=1mm] (\fin-0.5-\rodsize,\ha-\inter) -- (\fin-0.55,\ha-\inter);
\draw[black, dotted,line width=1mm] (\fin-0.5,\ha-\inter) -- (\fin,\ha-\inter);

%% Next lines and their rods

\draw[amaranthred,line width=1mm] (\deb+0.5+\rodsize,\ha-3*\inter) -- (\deb+0.5+2*\rodsize,\ha-3*\inter);
\draw[amaranthred,line width=1mm] (\deb+0.5+2*\rodsize,\ha-3*\inter) -- (\deb+0.5+3*\rodsize,\ha-3*\inter);
\draw[bdazzledblue,line width=1mm] (\deb+0.5+3*\rodsize,\ha-2*\inter) -- (\deb+0.5+3.5*\rodsize,\ha-2*\inter);

\draw[amaranthred,line width=1mm,opacity=0.25] (\deb+0.5+\rodsize,\ha-\zaxisline*\inter) -- (\deb+0.5+3*\rodsize,\ha-\zaxisline*\inter);
\draw[bdazzledblue,line width=1mm,opacity=0.25] (\deb+0.5+3*\rodsize,\ha-\zaxisline*\inter) -- (\deb+0.5+3.5*\rodsize,\ha-\zaxisline*\inter);

%% Vertical lines and coordinates

\draw[gray,dotted,line width=0.2mm] (\deb+0.5+\rodsize,\ha) -- (\deb+0.5+\rodsize,\ha-\zaxisline*\inter);
\draw[gray,dotted,line width=0.2mm] (\deb+0.5+3*\rodsize,\ha) -- (\deb+0.5+3*\rodsize,\ha-\zaxisline*\inter);
\draw[gray,dotted,line width=0.2mm] (\deb+0.5+3.5*\rodsize,\ha) -- (\deb+0.5+3.5*\rodsize,\ha-\zaxisline*\inter);

\draw[line width=0.3mm] (\deb+0.5+3*\rodsize,\ha-\zaxisline*\inter+0.1) -- (\deb+0.5+3*\rodsize,\ha-\zaxisline*\inter-0.1);
\draw[line width=0.3mm] (\deb+0.5+\rodsize,\ha-\zaxisline*\inter+0.1) -- (\deb+0.5+\rodsize,\ha-\zaxisline*\inter-0.1);
\draw[line width=0.3mm] (\deb+0.5+3.5*\rodsize,\ha-\zaxisline*\inter+0.1) -- (\deb+0.5+3.5*\rodsize,\ha-\zaxisline*\inter-0.1);

\draw (\deb+0.5+\rodsize,\ha-\zaxisline*\inter-0.5) node{{\small $0$}};
\draw (\deb+0.5+3*\rodsize,\ha-\zaxisline*\inter-0.5) node{{\small $\frac{\ell_1^2}{4}$}};
\draw (\deb+0.9+3.5*\rodsize,\ha-\zaxisline*\inter-0.5) node{{\small $\frac{\ell_1^2+\ell_2^2}{4}=\frac{\ell^2}{4}$}};

\end{tikzpicture}
\caption{Rod diagram of the shrinking directions on the $z$-axis after sourcing the solutions with two connected rods that force the degeneracy of $y$ and $\psi=k(\varphi_1-\varphi_2)$. }
\label{fig:rodsourceAdS3+S3}
\end{figure}
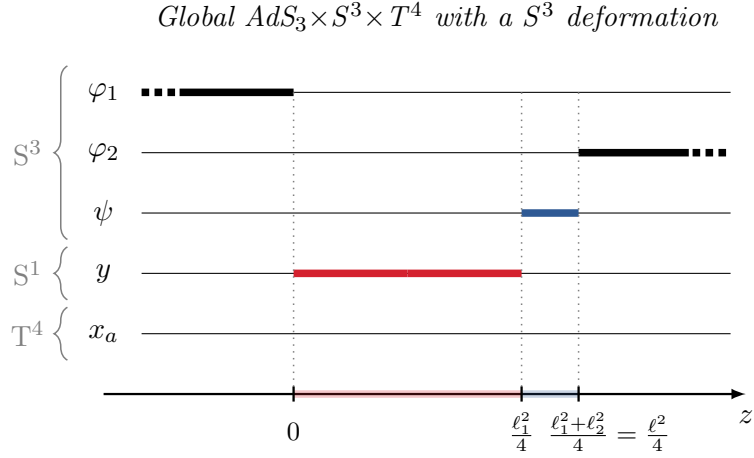

In this section,  we consider two rod sources such that the first rod still induces the degeneracy of the S$^1$ while the second one corresponds to a coordinate degeneracy of the Hopf angle of the S$^3$, $\psi=k(\varphi_1-\varphi_2)$ \eqref{eq:DefHyperspher} (see Fig.\ref{fig:rodsourceAdS3+S3}).  From Table  \ref{tab:internalBC},  this requires fixing the weights of the second rod such that $P_2^{(0)}=1$ while all its other weights are zero.  

\subsubsection{The solution}

We refer the reader interested in the derivation of the type IIB fields to the Appendix \ref{App:GlobalAdS3+S3}. The solution \eqref{eq:TypeIIBAnsatz2}, obtained from \eqref{eq:LinearAdS3} with the rod configuration considered here,  gives\footnote{The metric can be written with the cylindrical Weyl coordinate, $(\rho,z)$, as the main coordinate system by considering \eqref{eq:DefDistanceglobal},  $$ \frac{dr^2}{r^2+\ell^2}+d\theta^2= \frac{4}{\left( r^2+\ell^2\cos^2\theta\right)\left( r^2+\ell^2\sin^2\theta\right)} \left(d\rho^2+dz^2\right) .$$}
\begin{align}
ds_{10}^2 \= &  \frac{1}{\sqrt{Q_1 Q_5}\,\cF_1} \left[-(r^2+\ell^2) \,dt^2 +r^2\,\cF_3 \,dy^2 \right] + \sqrt{\frac{Q_1}{Q_5}}\,\sum_{a=1}^4 dx_a^2 \nn \\
&+k \sqrt{Q_1 Q_5} \, \cF_1 \,\Biggl[ \frac{\cF_2\cF_4}{\cF_3}\,\left(\frac{dr^2}{r^2+\ell^2}+d\theta^2 \right) + \frac{\cF_4}{\cF_3} \cos^2 \theta \sin^2\theta \,d\phi^2 \nn \\
& \hspace{2.5cm} + \frac{1}{4 k^2 \cF_4} \,\left(d\psi+k \left(2 \cF_5 \cos^2\theta-1 \right)\,d\phi \right)^2 \Biggr]\,,\label{eq:met1AdS3+S3}\\\
C^{(2)} \= &\frac{Q_5}{2} \cF_1 \cos^2 \theta \,d\phi \wedge d\psi -\frac{r^2+\ell^2}{Q_1 \cF_1} \,dt\wedge dy \,,\qquad  e^\Phi \= \sqrt{\frac{Q_1}{Q_5}}\,, \nn 
\end{align}
where we used the same deformation factors as in \eqref{eq:FcalDef}, and introduced new ones
\begin{equation}
\cF_4 \equi \frac{\ell^2}{\ell^2+\ell_2^2}\left( 1+\frac{\ell_2^2 (r^2+\ell^2)}{\ell^2 r_2^2} \right)\,,\qquad \cF_5 \equi \frac{\ell^2}{\ell^2+\ell_2^2}\left( 1+\frac{\ell_2^2 r^2}{\ell^2 r_2^2} \right)\,,\label{eq:FcalDef2}
\end{equation}
that all become trivial when the extra rod is turned off, $\ell_2=0$.  We remind that $(\psi,\phi)$ are the angles of the Hopf fibration of the S$^3$ that are related to the spherical angles, $(\varphi_1,\varphi_2)$ in \eqref{eq:DefHyperspher}. Moreover, $(r,\theta)$ are the global spherical coordinates of the three-rod configuration \eqref{eq:DefDistanceglobal}, while $(r_i,\theta_i)$ are the local spherical coordinates centered at the $i^\text{th}$ rod, given in terms of $(\rho,z)$ in \eqref{eq:DefDistance} and of $(r,\theta)$ in \eqref{eq:ri&thetaidef}.

The solution is asymptotic to AdS$_3\times$S$^3/\mathbb{Z}_k\times$T$^4$ since $\cF_I\to 1$ at large $r$.  As we will see later on,  the orbifold action on the S$^3$ is necessary to have smooth geometries.

The extra rod source preserves the rigidity of the T$^4$ but non-trivially deforms the AdS$_3$ part and has broken the symmetry of the S$^3$ by forcing the Hopf angle to degenerate.  As for the previous examples, the rod sources,  located at $\rho=0$ and $0 \leq z \leq\ell^2/4$,  are localized at $r=0$ and $0\leq \theta \leq \pi/2$ in the global spherical coordinate system.  We will study the topology and the regularity condition,  and we will show that the solution corresponds to a regular non-BPS geometry in AdS$_3$ that caps off smoothly at $r=0$ as a chain of two bolts.

\subsubsection{Regularity conditions and topology}

First,  when $r>0$,  one has $r_i>0$ \eqref{eq:ri&thetaidef},  and all $\cF_I$ are finite and positive.  Therefore,  the metric components \eqref{eq:met1AdS3+S3} are finite and the geometries are regular there for $\theta\neq 0,\pi/2$.  The loci $r>0$ and $\theta=0,\pi/2$,  that are the two semi-infinite segments above and below the rod sources on the $z$-axis depicted in Fig.\ref{fig:rodsourceAdS3+S3},  correspond to the North and South poles of the S$^3$ where $\varphi_2$ and $\varphi_1$ degenerate respectively.  One can check that $\cF_2=1$ and $\cF_5=1$ in this region.  Therefore, the poles are regular such that the angles degenerate with the same conical defect as the one imposes asymptotically.  The spacetime is therefore regular outside the rod sources and has a S$^3/\mathbb{Z}_k \times$S$^1\times$T$^4$ topology.

Both rod sources are located at $r=0$ such that the first rod corresponds to $\theta_c\leq \theta \leq \pi/2$ and the second rod is at $0\leq \theta \leq \theta_c$ as given in \eqref{eq:RodLocTheta1} and \eqref{eq:DefThetaCri}.  At the rods,  either the $y$ or the $\psi$ coordinate degenerates defining a bolt.  The local geometries are best described in the local spherical coordinates $(r_i, \theta_i)$ \eqref{eq:LocalSpher1} with $r_i\to 0$.  We find that the time slices of the metric \eqref{eq:met1AdS3+S3} are given by\footnote{We wrote the six-dimensional metric only by omitting the T$^4$ part which is trivially rigid in the whole spacetime.}
\begin{align}
ds_{6}^2 &\sim \frac{\ell^2 \,k\sqrt{Q_1 Q_5}}{\ell_1^2 \left( \ell^2+\ell_2^2\right)}\left[ dr_1^2 + \frac{\ell_1^2 (\ell^2+\ell_2^2)}{\ell^2\,k Q_1 Q_5} r_1^2 \,dy^2 \+ \ell_1^2 \,d\widetilde{\Omega}_3^2 \right]\,, \\
d\widetilde{\Omega}_3^2 & \=  d\theta_1^2 +\frac{\ell_1^2 \sin^2 \theta_1 +\ell_2^2}{\ell^2} \cos^2\theta_1 \,d\phi^2+ \frac{(\ell^2+\ell_2^2)^2\sin^2\theta_1}{4\,\ell^2\,(\ell_1^2 \sin^2 \theta_1 +\ell_2^2)k^2} \left(d\psi+k\frac{\ell_1^2\cos 2\theta_1-2\ell_2^2 }{\ell^2+\ell_2^2}\,d\phi\right)^2, \nn
\end{align}
at the first rod,  $r_1 \to 0$,  while at the second rod,  $r_2 \to 0$,  we have
\begin{align}
ds_{6}^2 &\sim \frac{\ell^2\,k\sqrt{Q_1 Q_5}}{(\ell^2+\ell_2^2)(\ell_1^2+\ell_2^2\cos^2\theta_2 )}\left[ dr_2^2 + \frac{(\ell^2+\ell_2^2)^2}{4\ell^2\ell_2^2 k^2} r_2^2 \left(d\psi+k\frac{\ell_1^2+2\ell_2^2\cos 2\theta_2 }{\ell^2+\ell_2^2}d\phi\right)^2+ \ell_2^2 \,d\widetilde{\Omega}_3^2 \right], \nn\\
d\widetilde{\Omega}_3^2 & \=  d\theta_2^2 +\frac{\ell_1^2 +\ell_2^2 \cos^2 \theta_2}{\ell^2} \left(  \sin^2\theta_2 \,d\phi^2  +\frac{\ell^2+\ell_2^2}{k Q_1 Q_5} \cos^2 \theta_2 \,dy^2 \right)\,.
\end{align}
Therefore the rods correspond to smooth S$^3\times$T$^4$ fibrations over an origin of $\IR^2$ if one imposes
\begin{equation}
R_y \= \frac{\sqrt{k Q_1 Q_5}}{\ell_1}\,\left(1+\frac{\ell_2^2}{\ell^2}\right)^{-\frac{1}{2}} \,\qquad k\=\frac{\ell^2+\ell_2^2}{\ell\,\ell_2}\,,
\end{equation}
where we remind that $\ell^2=\ell_1^2+\ell_2^2$.  One can invert these expressions such that the rod lengths are fixed to be
\begin{equation}
\ell_1^2 \= \frac{k+\sqrt{k^2-4}}{2}\,\frac{Q_1 Q_5}{R_y^2}\,,\quad \ell_2^2 \= \frac{1}{\sqrt{k^2-4}} \,\frac{Q_1 Q_5}{R_y^2}\,,\quad \ell^2 \= \frac{(k+\sqrt{k^2-4})^2}{4\sqrt{k^2-4}}\,\frac{Q_1 Q_5}{R_y^2}\,.
\end{equation}
%Moreover, one can also add potential conical defects on both $\IR^2$. This will introduce two integers, $k_1$ and $k_2$,  and the above expressions will be modified by simply replacing $R_{y} \to k_1 R_{y}/\sqrt{k_2}$ and $k\to k/k_2$ with $k>2 k_2$.

The lengths are well defined if $k\geq 3$.  Therefore, having an orbifolded S$^3/\mathbb{Z}_k$ asymptotically is indeed necessary to have a smooth geometry.  Without this, the solution would have a conical excess at the second rod of order $\frac{\ell^2+\ell_2^2}{\ell \ell_2}>1$. This would correspond to a strut, which is a singular string with negative tension \cite{Costa:2000kf,Bah:2021owp}.

Both rod lengths are generically of order $k Q_1 Q_5/R_y^2$ and bounded by
\begin{equation}
0.87 \,\frac{k Q_1 Q_5}{R_y^2} \,<\, \ell_1^2 \,<\, \frac{k Q_1 Q_5}{R_y^2} \,,\qquad 0\,<\,  \ell_2 \,<\,  0.14 \,\frac{kQ_1 Q_5}{R_y^2}\,.
\end{equation}
The total length, $\ell^2$,  is not strictly equal to $k Q_1 Q_5/R_{y}^2$,  but the ratio between both quantities varies between $1.02$ at $k=3$ and 1 at $k\gg 1$. 

Moreover,  one can check that the three-form flux, $F_3=dC^{(2)}$,  is regular and carries no charges at the second rod while it carries D1 and D5 brane charges,  $Q_1$ and $Q_5$ respectively,  at the first rod \eqref{eq:ChargeAtRodGen}.

%%%%%%%%%%%%%%%%%%%%%%%%%%%%%%%%
\begin{figure}[t]
\centering
\includegraphics[scale=0.7]{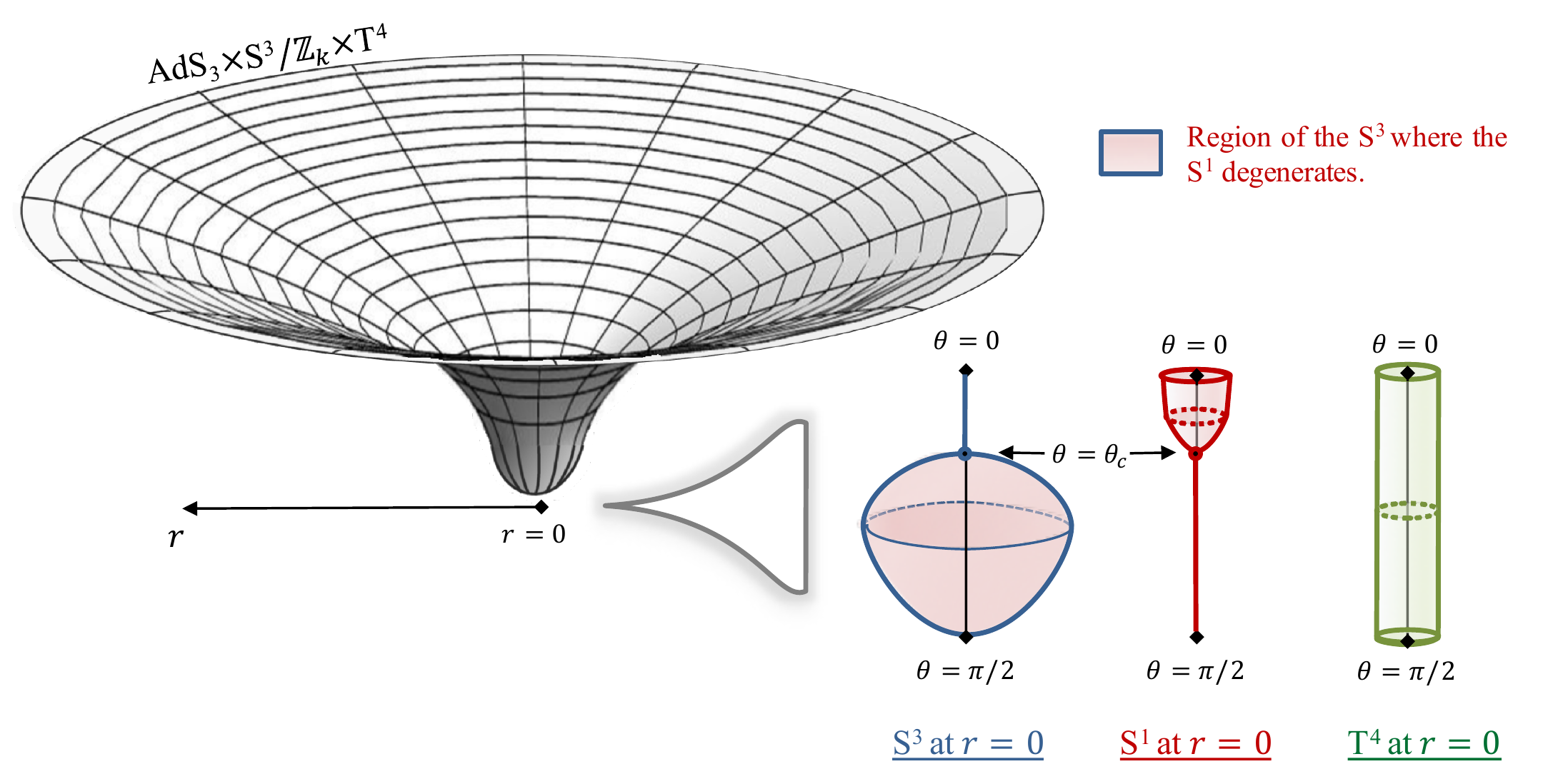}
\caption{Schematic description of the spacetime induced by a chain of two rods,  forcing the degeneracy of the S$^1$ and the S$^3$ Hopf angle $\psi=k(\varphi_1-\varphi_2)$ respectively.  On the left-hand side, we depict the overall geometry in terms of $r$.  On the right-hand side,  we describe the behavior of the S$^3$,  S$^1$, and T$^4$ at $r=0$ and as a function of $\theta$. }
\label{fig:AdS3+S3}
\end{figure}
%%%%%%%%%%%%%%%%%%%%%%%%%%%%%%%%

The geometry has been depicted in Fig.\ref{fig:AdS3+S3} with the same conventions as before.  The spacetime caps off smoothly at $r=0$ where the S$^1$ ($y$) degenerates for $\theta_c\leq \theta\leq \pi/2$ and the S$^3$ ($\psi$) degenerates for $0\leq \theta\leq \theta_c$ such that the critical angle, $\theta_c$ \eqref{eq:DefThetaCri}, is given by
\begin{equation}
\cos \theta_c \= \frac{2}{ k+\sqrt{k^2-4}}\,.
\end{equation}
Thus, the S$^3$ deformation that has replaced the S$^1$ degeneracy at the center of a global AdS$_3\times$S$^3/\mathbb{Z}_k\times$T$^4$ spacetime  is centered on the Northern hemisphere of the S$^3$.  Moreover,  in the  limit $k \gg 1$,  one has $\ell_2 \ll \ell_1$ and the second rod is a small smooth perturbation on a global AdS$_3\times$S$^3/\mathbb{Z}_k\times$T$^4$ spacetime.  Since $\theta_c \sim 0$,  the perturbation is localized at the center of the global AdS$_3$ space and at the North pole of the S$^3$,  and forces the Hopf fibration angle to shrink smoothly here.  The solution is given by \eqref{eq:met1AdS3+S3} such that all deformation factors give $1+\cO(k^{-2})$ that have large values around the source of the perturbation.

Moreover,  the S$^3$ deformation breaks the supersymmetry of the global AdS$_3\times$S$^3\times$T$^4$ solution since there is no null Killing spinor associated with the geometry.   Moreover,  for the same arguments as in section \ref{sec:nonBPSness},  it most likely breaks all supersymmetry such that it corresponds to a non-BPS asymptotically-AdS$_3$ solution in type IIB.  Interestingly,  the T$^4$ remains rigid,  it would be then interesting to have a precise idea of how the degeneracy of the S$^3$ at the center of AdS$_3$ has effectively broken all supersymmetry.  We postpone such an analysis for future projects.

In a similar manner as for the T$^4$ deformations,  nothing forces the S$^3$ deformation to be localized at the North pole of the S$^3$ and one can change its locus by considering a three-rod configuration as in section \ref{sec:AdS3+T42}.  As these solutions are of no particular interest, we ignore them in this paper, and in the next section we construct generic bubbling geometries with an arbitrary number of rods.

%%%%%%%%%%%%%%%%%%%%%%%%%%%%%%%%%%%%%
\section{Generic non-BPS bubbling deformations in AdS$_3\times$S$^3\times$T$^4$}
\label{sec:AdS3Gen}
%%%%%%%%%%%%%%%%%%%%%%%%%%%%%%%%%%%%%

In this section,  we derive more generic bubbling geometries obtained from the linear branch of D1-D5 solutions in section \ref{sec:linearbranchSum}.  Generic solutions are induced by an arbitrary number of rods that force the degeneracy of either the S$^1$ or a T$^4$ direction or the Hopf angle of the S$^3$.  They will correspond to bubbling geometry with a S$^1\times$S$^3/\mathbb{Z}_k\times$T$^4$ topology at $r>0$ that are asymptotic to AdS$_3\times$S$^3/\mathbb{Z}_k\times$T$^4$ and that cap off smoothly at $r=0$.  The solutions terminate in a chain of bolts,  and each bolt spans a region of the S$^3$.

First,  we will focus on solutions with T$^4$ deformations only since it allows for smooth non-BPS geometries that are asymptotic to AdS$_3\times$S$^3\times$T$^4$ without orbifold action on the S$^3$.  Then, we will construct the most generic solutions with S$^3$ deformations too.

\subsection{Conventions}
\label{sec:Conventions}

\begin{figure}[h]
\centering
    \begin{tikzpicture}
%% some definitions

\def\deb{-10} 
\def\inter{0.7} 
\def\ha{2.8} 
\def\zaxisline{8} 
\def\rodsize{1.7} 
\def\numrod{5} 

\def\fin{\deb+1+2*\rodsize+\numrod*\rodsize} 

%% 

%\draw (\deb-2,\ha-\zaxisline*0.5*\inter) node{$\Longrightarrow$};

%% Pic title

\draw (\deb+0.5+\rodsize+0.5*\numrod*\rodsize,\ha+1) node{{{\it Global AdS$_3\times$S$^{\,3}\times$T$^{\,4}$ with T$^{\,4}$ and S$^{\,3}$ deformations}}};

\draw [decorate, 
    decoration = {brace,
        raise=5pt,
        amplitude=5pt},line width=0.2mm,gray] (\deb-0.7,\ha-2.5*\inter+0.05) --  (\deb-0.7,\ha+0.5*\inter-0.05);
\draw [decorate, 
    decoration = {brace,
        raise=5pt,
        amplitude=5pt},line width=0.2mm,gray] (\deb-0.7,\ha-3.5*\inter+0.05) --  (\deb-0.7,\ha-2.5*\inter-0.05);
\draw [decorate, 
    decoration = {brace,
        raise=5pt,
        amplitude=5pt},line width=0.2mm,gray] (\deb-0.7,\ha-7.5*\inter+0.05) --  (\deb-0.7,\ha-3.5*\inter-0.05);
        
\draw[gray] (\deb-1.4,\ha-1*\inter) node{S$^3$};
\draw[gray] (\deb-1.4,\ha-3*\inter) node{S$^1$};
\draw[gray] (\deb-1.4,\ha-5.5*\inter) node{T$^4$};

%% Each line black line and names

\draw[black,thin] (\deb+1,\ha) -- (\deb+0.5+5*\rodsize+0.2,\ha);\draw[black,thin,dotted] (\deb+0.5+5*\rodsize+0.2,\ha) -- (\deb+0.5+5.5*\rodsize,\ha);
\draw[black,thin] (\deb+0.5+5.5*\rodsize,\ha) -- (\fin,\ha);

\draw[black,thin] (\deb,\ha-\inter) -- (\deb+0.5+5*\rodsize+0.2,\ha-\inter);\draw[black,thin,dotted] (\deb+0.5+5*\rodsize+0.2,\ha-\inter) -- (\deb+0.5+5.5*\rodsize,\ha-\inter);
\draw[black,thin] (\deb+0.5+5.5*\rodsize,\ha-\inter) -- (\fin-1,\ha-\inter);

\draw[black,thin] (\deb,\ha-2*\inter) -- (\deb+0.5+5*\rodsize+0.2,\ha-2*\inter);\draw[black,thin,dotted] (\deb+0.5+5*\rodsize+0.2,\ha-2*\inter) -- (\deb+0.5+5.5*\rodsize,\ha-2*\inter);
\draw[black,thin] (\deb+0.5+5.5*\rodsize,\ha-2*\inter) -- (\fin,\ha-2*\inter);

\draw[black,thin] (\deb,\ha-3*\inter) -- (\deb+0.5+5*\rodsize+0.2,\ha-3*\inter);\draw[black,thin,dotted] (\deb+0.5+5*\rodsize+0.2,\ha-3*\inter) -- (\deb+0.5+5.5*\rodsize,\ha-3*\inter);
\draw[black,thin] (\deb+0.5+5.5*\rodsize,\ha-3*\inter) -- (\fin,\ha-3*\inter);

\draw[black,thin] (\deb,\ha-4*\inter) -- (\deb+0.5+5*\rodsize+0.2,\ha-4*\inter);\draw[black,thin,dotted] (\deb+0.5+5*\rodsize+0.2,\ha-4*\inter) -- (\deb+0.5+5.5*\rodsize,\ha-4*\inter);
\draw[black,thin] (\deb+0.5+5.5*\rodsize,\ha-4*\inter) -- (\fin,\ha-4*\inter);

\draw[black,thin] (\deb,\ha-5*\inter) -- (\deb+0.5+5*\rodsize+0.2,\ha-5*\inter);\draw[black,thin,dotted] (\deb+0.5+5*\rodsize+0.2,\ha-5*\inter) -- (\deb+0.5+5.5*\rodsize,\ha-5*\inter);
\draw[black,thin] (\deb+0.5+5.5*\rodsize,\ha-5*\inter) -- (\fin,\ha-5*\inter);

\draw[black,thin] (\deb,\ha-6*\inter) -- (\deb+0.5+5*\rodsize+0.2,\ha-6*\inter);\draw[black,thin,dotted] (\deb+0.5+5*\rodsize+0.2,\ha-6*\inter) -- (\deb+0.5+5.5*\rodsize,\ha-6*\inter);
\draw[black,thin] (\deb+0.5+5.5*\rodsize,\ha-6*\inter) -- (\fin,\ha-6*\inter);

\draw[black,thin] (\deb,\ha-7*\inter) -- (\deb+0.5+5*\rodsize+0.2,\ha-7*\inter);\draw[black,thin,dotted] (\deb+0.5+5*\rodsize+0.2,\ha-7*\inter) -- (\deb+0.5+5.5*\rodsize,\ha-7*\inter);
\draw[black,thin] (\deb+0.5+5.5*\rodsize,\ha-7*\inter) -- (\fin,\ha-7*\inter);

\draw[black,line width=0.3mm] (\deb-0.4,\ha-\zaxisline*\inter) -- (\deb+0.5+5*\rodsize+0.2,\ha-\zaxisline*\inter);\draw[black,line width=0.3mm,dotted] (\deb+0.5+5*\rodsize+0.2,\ha-\zaxisline*\inter) -- (\deb+0.5+5.5*\rodsize,\ha-\zaxisline*\inter);
\draw[black,->, line width=0.3mm] (\deb+0.5+5.5*\rodsize,\ha-\zaxisline*\inter) -- (\fin+0.2,\ha-\zaxisline*\inter);

\draw (\deb-0.4,\ha) node{$\varphi_1$};
\draw (\deb-0.4,\ha-\inter) node{$\varphi_2$};
\draw (\deb-0.4,\ha-2*\inter) node{$\psi$};
\draw (\deb-0.4,\ha-3*\inter) node{$y$};
\draw (\deb-0.4,\ha-4*\inter) node{$x_1$};
\draw (\deb-0.4,\ha-5*\inter) node{$x_2$};
\draw (\deb-0.4,\ha-6*\inter) node{$x_3$};
\draw (\deb-0.4,\ha-7*\inter) node{$x_4$};

\draw (\fin+0.2,\ha-\zaxisline*\inter-0.3) node{$z$};

%% First two line and their rods

\draw[black, dotted, line width=1mm] (\deb,\ha) -- (\deb+0.5,\ha);
\draw[black,line width=1mm] (\deb+0.5,\ha) -- (\deb+0.5+\rodsize-0.3,\ha);
\draw[black,line width=1mm] (\fin-0.5-\rodsize+0.2,\ha-\inter) -- (\fin-0.55,\ha-\inter);
\draw[black, dotted,line width=1mm] (\fin-0.5,\ha-\inter) -- (\fin,\ha-\inter);

%% Next lines and their rods

\draw[byzantine,line width=1mm] (\deb+0.5+\rodsize-0.3,\ha-5*\inter) -- (\deb+0.5+1.5*\rodsize-0.3,\ha-5*\inter);

\draw[absolutezero,line width=1mm] (\deb+0.5+1.5*\rodsize-0.3,\ha-2*\inter) -- (\deb+0.5+2*\rodsize-0.3,\ha-2*\inter);

\draw[amaranthred,line width=1mm] (\deb+0.5+2*\rodsize-0.3,\ha-3*\inter) -- (\deb+0.5+2.75*\rodsize-0.15,\ha-3*\inter);

\draw[turquoise,line width=1mm] (\deb+0.5+2.75*\rodsize-0.15,\ha-7*\inter) -- (\deb+0.5+3.25*\rodsize-0.15,\ha-7*\inter);

\draw[amazon,line width=1mm] (\deb+0.5+3.25*\rodsize-0.15,\ha-4*\inter) -- (\deb+0.5+3.75*\rodsize-0.15,\ha-4*\inter);

\draw[amaranthred,line width=1mm] (\deb+0.5+3.75*\rodsize-0.15,\ha-3*\inter) -- (\deb+0.5+4.5*\rodsize,\ha-3*\inter);

\draw[turquoise,line width=1mm] (\deb+0.5+4.5*\rodsize,\ha-7*\inter) -- (\deb+0.5+5*\rodsize,\ha-7*\inter);

\draw[bitterlemon,line width=1mm] (\deb+0.5+5.5*\rodsize+0.2,\ha-6*\inter) -- (\deb+0.5+6*\rodsize+0.2,\ha-6*\inter);

%% Rods on the z-axis

\draw[byzantine,line width=1mm,opacity=0.25] (\deb+0.5+\rodsize-0.3,\ha-\zaxisline*\inter) -- (\deb+0.5+1.5*\rodsize-0.3,\ha-\zaxisline*\inter);

\draw[absolutezero,line width=1mm,opacity=0.25] (\deb+0.5+1.5*\rodsize-0.3,\ha-\zaxisline*\inter) -- (\deb+0.5+2*\rodsize-0.3,\ha-\zaxisline*\inter);
\draw[amaranthred,line width=1mm,opacity=0.25] (\deb+0.5+2*\rodsize-0.3,\ha-\zaxisline*\inter) -- (\deb+0.5+2.75*\rodsize-0.15,\ha-\zaxisline*\inter);

\draw[turquoise,line width=1mm,opacity=0.25] (\deb+0.5+2.75*\rodsize-0.15,\ha-\zaxisline*\inter) -- (\deb+0.5+3.25*\rodsize-0.15,\ha-\zaxisline*\inter);

\draw[amazon,line width=1mm,opacity=0.25] (\deb+0.5+3.25*\rodsize-0.15,\ha-\zaxisline*\inter) -- (\deb+0.5+3.75*\rodsize-0.15,\ha-\zaxisline*\inter);

\draw[amaranthred,line width=1mm,opacity=0.25] (\deb+0.5+3.75*\rodsize-0.15,\ha-\zaxisline*\inter) -- (\deb+0.5+4.5*\rodsize,\ha-\zaxisline*\inter);
\draw[turquoise,line width=1mm,opacity=0.25] (\deb+0.5+4.5*\rodsize,\ha-\zaxisline*\inter) -- (\deb+0.5+5*\rodsize,\ha-\zaxisline*\inter);

\draw[bitterlemon,line width=1mm,opacity=0.25] (\deb+0.5+5.5*\rodsize+0.2,\ha-\zaxisline*\inter) -- (\deb+0.5+6*\rodsize+0.2,\ha-\zaxisline*\inter);

%% Vertical lines and coordinates

\draw[gray,dotted,line width=0.2mm] (\deb+0.5+\rodsize-0.3,\ha) -- (\deb+0.5+\rodsize-0.3,\ha-\zaxisline*\inter);
\draw[gray,dotted,line width=0.2mm] (\deb+0.5+1.5*\rodsize-0.3,\ha) -- (\deb+0.5+1.5*\rodsize-0.3,\ha-\zaxisline*\inter);
\draw[gray,dotted,line width=0.2mm] (\deb+0.5+2*\rodsize-0.3,\ha) -- (\deb+0.5+2*\rodsize-0.3,\ha-\zaxisline*\inter);
\draw[gray,dotted,line width=0.2mm] (\deb+0.5+2.75*\rodsize-0.15,\ha) -- (\deb+0.5+2.75*\rodsize-0.15,\ha-\zaxisline*\inter);
\draw[gray,dotted,line width=0.2mm] (\deb+0.5+3.25*\rodsize-0.15,\ha) -- (\deb+0.5+3.25*\rodsize-0.15,\ha-\zaxisline*\inter);
\draw[gray,dotted,line width=0.2mm] (\deb+0.5+3.75*\rodsize-0.15,\ha) -- (\deb+0.5+3.75*\rodsize-0.15,\ha-\zaxisline*\inter);
\draw[gray,dotted,line width=0.2mm] (\deb+0.5+4.5*\rodsize,\ha) -- (\deb+0.5+4.5*\rodsize,\ha-\zaxisline*\inter);
\draw[gray,dotted,line width=0.2mm] (\deb+0.5+5*\rodsize,\ha) -- (\deb+0.5+5*\rodsize,\ha-\zaxisline*\inter);
\draw[gray,dotted,line width=0.2mm] (\deb+0.5+5.5*\rodsize+0.2,\ha) -- (\deb+0.5+5.5*\rodsize+0.2,\ha-\zaxisline*\inter);
\draw[gray,dotted,line width=0.2mm] (\deb+0.5+6*\rodsize+0.2,\ha) -- (\deb+0.5+6*\rodsize+0.2,\ha-\zaxisline*\inter);

\draw[line width=0.3mm] (\deb+0.5+\rodsize-0.3,\ha-\zaxisline*\inter+0.1) -- (\deb+0.5+\rodsize-0.3,\ha-\zaxisline*\inter-0.1);
\draw[line width=0.3mm] (\deb+0.5+1.5*\rodsize-0.3,\ha-\zaxisline*\inter+0.1) -- (\deb+0.5+1.5*\rodsize-0.3,\ha-\zaxisline*\inter-0.1);
\draw[line width=0.3mm] (\deb+0.5+2*\rodsize-0.3,\ha-\zaxisline*\inter+0.1) -- (\deb+0.5+2*\rodsize-0.3,\ha-\zaxisline*\inter-0.1);
\draw[line width=0.3mm] (\deb+0.5+2.75*\rodsize-0.15,\ha-\zaxisline*\inter+0.1) -- (\deb+0.5+2.75*\rodsize-0.15,\ha-\zaxisline*\inter-0.1);
\draw[line width=0.3mm] (\deb+0.5+3.25*\rodsize-0.15,\ha-\zaxisline*\inter+0.1) -- (\deb+0.5+3.25*\rodsize-0.15,\ha-\zaxisline*\inter-0.1);
\draw[line width=0.3mm] (\deb+0.5+3.75*\rodsize-0.15,\ha-\zaxisline*\inter+0.1) -- (\deb+0.5+3.75*\rodsize-0.15,\ha-\zaxisline*\inter-0.1);
\draw[line width=0.3mm] (\deb+0.5+4.5*\rodsize,\ha-\zaxisline*\inter+0.1) -- (\deb+0.5+4.5*\rodsize,\ha-\zaxisline*\inter-0.1);
\draw[line width=0.3mm] (\deb+0.5+5*\rodsize,\ha-\zaxisline*\inter+0.1) -- (\deb+0.5+5*\rodsize,\ha-\zaxisline*\inter-0.1);
\draw[line width=0.3mm] (\deb+0.5+5.5*\rodsize+0.2,\ha-\zaxisline*\inter+0.1) -- (\deb+0.5+5.5*\rodsize+0.2,\ha-\zaxisline*\inter-0.1);
\draw[line width=0.3mm] (\deb+0.5+6*\rodsize+0.2,\ha-\zaxisline*\inter+0.1) -- (\deb+0.5+6*\rodsize+0.2,\ha-\zaxisline*\inter-0.1);

\draw (\deb+0.5+1*\rodsize-0.3,\ha-\zaxisline*\inter-0.5) node{{\small $0$}};
\draw (\deb+0.5+1.5*\rodsize-0.3,\ha-\zaxisline*\inter-0.5) node{{\small $\frac{\ell_1^2}{4}$}};
\draw (\deb+0.5+2*\rodsize-0.3,\ha-\zaxisline*\inter-0.5) node{{\small $\frac{\ell_1^2+\ell_2^2}{4}$}};

\draw (\deb+0.5+2.75*\rodsize-0.15,\ha-\zaxisline*\inter-0.5) node{{\tiny $\frac{\ell_1^2+\ell_{2}^2+\ell_{3}^2}{4}$}};
\draw[gray,->,line width=0.1mm] (\deb+0.5+3.25*\rodsize-0.15,\ha-\zaxisline*\inter-0.8) -- (\deb+0.5+3.25*\rodsize-0.15,\ha-\zaxisline*\inter-0.25);
\draw (\deb+0.5+3.25*\rodsize-0.15,\ha-\zaxisline*\inter-1.1) node{{\tiny $\frac{\sum_{i=1}^4\ell_{i}^2}{4}$}};

\draw[black,line width=0.3mm,dotted,double] (\deb+0.5+3.5*\rodsize,\ha-\zaxisline*\inter-0.5) -- (\deb+0.5+4.75*\rodsize,\ha-\zaxisline*\inter-0.5);

\draw (\deb+0.5+5.5*\rodsize+0.2,\ha-\zaxisline*\inter-0.5) node{{\tiny $\frac{\sum_{i=1}^{n-1}\ell_{i}^2}{4}$}};
\draw[gray,->,line width=0.1mm] (\deb+0.5+6*\rodsize+0.2+0.05,\ha-\zaxisline*\inter-0.8) -- (\deb+0.5+6*\rodsize+0.2+0.05,\ha-\zaxisline*\inter-0.25);
\draw (\deb+0.5+6*\rodsize+0.2+0.05,\ha-\zaxisline*\inter-1.1) node{{\tiny $\frac{\sum_{i=1}^n\ell_{i}^2}{4}=\frac{\ell^2}{4}$}};

\end{tikzpicture}
\caption{Rod diagram of the shrinking directions on the $z$-axis after sourcing the solutions with $n$ connected rods that induce the degeneracy of either the S$^1$ or a T$^4$ direction or the S$^3$ Hopf angle.}
\label{fig:rodsourceAdS3+T4+S3}
\end{figure}
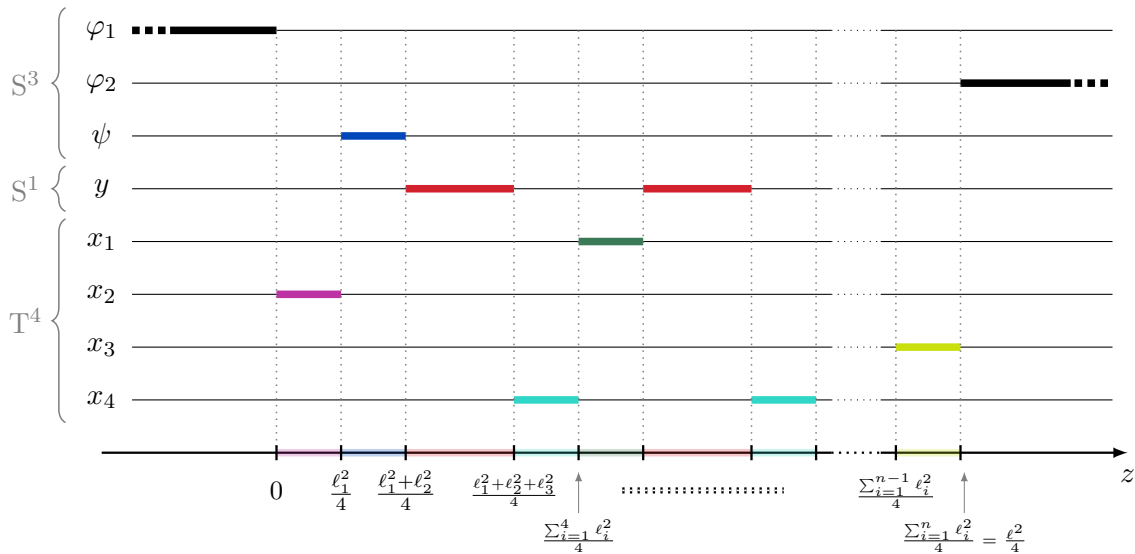  

We consider solutions that are induced by at least one rod that forces the degeneracy of the $y$-circle and by an arbitrary number of connected rods,  forcing either the degeneracy of a T$^4$ direction or the Hopf angle of the S$^3$.  We have depicted a typical rod configuration in Fig.\ref{fig:rodsourceAdS3+T4+S3}.  

The expressions of the type IIB fields from the linear branch of solutions are given by \eqref{eq:LinearAdS3}.  The eight weights at each rod,  $(P_i^{(I)},G_i^{(\Lambda)})$,  are fixed depending on which coordinate shrinks at the rod following Table \ref{tab:internalBC}.  We define six sets of labels,  $U_y$,  $U_{x_a}$ with $a=1,2,3,4$, and $U_\psi$:
\begin{equation}
i \in U_{w} \quad \Rightarrow \quad \text{the }w\text{ direction shrinks smoothly at the }i^\text{th}\text{ rod.}
\label{eq:DefUx}
\end{equation}
For instance,  the example in Fig.\ref{fig:rodsourceAdS3+T4+S3} corresponds to
\begin{align}
U_y &\= \{3,6 , \ldots\}\,, \qquad U_{x_1}=\{5,\ldots\} \,,\qquad U_{x_2}\=\{1,\ldots\} \,,\\
 U_{x_3}& \=\{\ldots,n\} \,,\qquad U_{x_4}=\{4,7,\ldots\} \,,\qquad U_\psi = \{2,\ldots  \}\,. \nn
\end{align}
The weights at the rods can then be read from Table \ref{tab:internalBC}. For instance,
\begin{equation}
i \in U_{x_4}  \quad \Rightarrow \quad P_i^{(0)}=P_i^{(5)}=G_i^{(0)} = 2G_i^{(2)}=2G_i^{(3)}=2G_i^{(4)} =\frac{1}{2}\,,\quad P_i^{(1)} = G_i^{(1)}=0. \nn
\end{equation}
Note that $\alpha_{ij}$,  the exponent in the base warp factor $e^{2\nu}$ \eqref{eq:AlphaDef},  takes simple values for regular rod sources \eqref{eq:AlphaSimple}.  Therefore,  we define the following exponent for the present configurations,  
\begin{equation}
\bar{\alpha}_{ij} = \begin{cases} 
\,\,1 \,,\qquad i,j \in U_{w}\,,\\
\,\,0\,,\qquad i\in U_w\,,\quad j\in U_{w'} \,,\quad w\neq w'.
\label{eq:DefAlphaBar}
\end{cases}
\end{equation}

\subsection{Chain of T$^4$ deformations}

\begin{figure}[h]
\centering
    \begin{tikzpicture}
%% some definitions

\def\deb{-10} 
\def\inter{0.7} 
\def\ha{2.8} 
\def\zaxisline{8} 
\def\rodsize{1.7} 
\def\numrod{5} 

\def\fin{\deb+1+2*\rodsize+\numrod*\rodsize} 

%% 

%\draw (\deb-2,\ha-\zaxisline*0.5*\inter) node{$\Longrightarrow$};

%% Pic title

\draw (\deb+0.5+\rodsize+0.5*\numrod*\rodsize,\ha+1-\inter) node{{{\it Global AdS$_3\times$S$^{\,3}\times$T$^{\,4}$ with T$^{\,4}$ deformations}}};

\draw [decorate, 
    decoration = {brace,
        raise=5pt,
        amplitude=5pt},line width=0.2mm,gray] (\deb-0.7,\ha-2.5*\inter+0.05) --  (\deb-0.7,\ha-0.5*\inter-0.05);
\draw [decorate, 
    decoration = {brace,
        raise=5pt,
        amplitude=5pt},line width=0.2mm,gray] (\deb-0.7,\ha-3.5*\inter+0.05) --  (\deb-0.7,\ha-2.5*\inter-0.05);
\draw [decorate, 
    decoration = {brace,
        raise=5pt,
        amplitude=5pt},line width=0.2mm,gray] (\deb-0.7,\ha-7.5*\inter+0.05) --  (\deb-0.7,\ha-3.5*\inter-0.05);
        
\draw[gray] (\deb-1.4,\ha-1.5*\inter) node{S$^3$};
\draw[gray] (\deb-1.4,\ha-3*\inter) node{S$^1$};
\draw[gray] (\deb-1.4,\ha-5.5*\inter) node{T$^4$};

%% Each line black line and names

\draw[black,thin] (\deb+1,\ha-\inter) -- (\deb+0.5+5*\rodsize+0.2,\ha-\inter);\draw[black,thin,dotted] (\deb+0.5+5*\rodsize+0.2,\ha-\inter) -- (\deb+0.5+5.5*\rodsize,\ha-\inter);
\draw[black,thin] (\deb+0.5+5.5*\rodsize,\ha-\inter) -- (\fin,\ha-\inter);

\draw[black,thin] (\deb,\ha-2*\inter) -- (\deb+0.5+5*\rodsize+0.2,\ha-2*\inter);\draw[black,thin,dotted] (\deb+0.5+5*\rodsize+0.2,\ha-2*\inter) -- (\deb+0.5+5.5*\rodsize,\ha-2*\inter);
\draw[black,thin] (\deb+0.5+5.5*\rodsize,\ha-2*\inter) -- (\fin-1,\ha-2*\inter);

\draw[black,thin] (\deb,\ha-3*\inter) -- (\deb+0.5+5*\rodsize+0.2,\ha-3*\inter);\draw[black,thin,dotted] (\deb+0.5+5*\rodsize+0.2,\ha-3*\inter) -- (\deb+0.5+5.5*\rodsize,\ha-3*\inter);
\draw[black,thin] (\deb+0.5+5.5*\rodsize,\ha-3*\inter) -- (\fin,\ha-3*\inter);

\draw[black,thin] (\deb,\ha-4*\inter) -- (\deb+0.5+5*\rodsize+0.2,\ha-4*\inter);\draw[black,thin,dotted] (\deb+0.5+5*\rodsize+0.2,\ha-4*\inter) -- (\deb+0.5+5.5*\rodsize,\ha-4*\inter);
\draw[black,thin] (\deb+0.5+5.5*\rodsize,\ha-4*\inter) -- (\fin,\ha-4*\inter);

\draw[black,thin] (\deb,\ha-5*\inter) -- (\deb+0.5+5*\rodsize+0.2,\ha-5*\inter);\draw[black,thin,dotted] (\deb+0.5+5*\rodsize+0.2,\ha-5*\inter) -- (\deb+0.5+5.5*\rodsize,\ha-5*\inter);
\draw[black,thin] (\deb+0.5+5.5*\rodsize,\ha-5*\inter) -- (\fin,\ha-5*\inter);

\draw[black,thin] (\deb,\ha-6*\inter) -- (\deb+0.5+5*\rodsize+0.2,\ha-6*\inter);\draw[black,thin,dotted] (\deb+0.5+5*\rodsize+0.2,\ha-6*\inter) -- (\deb+0.5+5.5*\rodsize,\ha-6*\inter);
\draw[black,thin] (\deb+0.5+5.5*\rodsize,\ha-6*\inter) -- (\fin,\ha-6*\inter);

\draw[black,thin] (\deb,\ha-7*\inter) -- (\deb+0.5+5*\rodsize+0.2,\ha-7*\inter);\draw[black,thin,dotted] (\deb+0.5+5*\rodsize+0.2,\ha-7*\inter) -- (\deb+0.5+5.5*\rodsize,\ha-7*\inter);
\draw[black,thin] (\deb+0.5+5.5*\rodsize,\ha-7*\inter) -- (\fin,\ha-7*\inter);

\draw[black,line width=0.3mm] (\deb-0.4,\ha-\zaxisline*\inter) -- (\deb+0.5+5*\rodsize+0.2,\ha-\zaxisline*\inter);\draw[black,line width=0.3mm,dotted] (\deb+0.5+5*\rodsize+0.2,\ha-\zaxisline*\inter) -- (\deb+0.5+5.5*\rodsize,\ha-\zaxisline*\inter);
\draw[black,->, line width=0.3mm] (\deb+0.5+5.5*\rodsize,\ha-\zaxisline*\inter) -- (\fin+0.2,\ha-\zaxisline*\inter);

\draw (\deb-0.4,\ha-\inter) node{$\varphi_1$};
\draw (\deb-0.4,\ha-2*\inter) node{$\varphi_2$};

\draw (\deb-0.4,\ha-3*\inter) node{$y$};
\draw (\deb-0.4,\ha-4*\inter) node{$x_1$};
\draw (\deb-0.4,\ha-5*\inter) node{$x_2$};
\draw (\deb-0.4,\ha-6*\inter) node{$x_3$};
\draw (\deb-0.4,\ha-7*\inter) node{$x_4$};

\draw (\fin+0.2,\ha-\zaxisline*\inter-0.3) node{$z$};

%% First two line and their rods

\draw[black, dotted, line width=1mm] (\deb,\ha-\inter) -- (\deb+0.5,\ha-\inter);
\draw[black,line width=1mm] (\deb+0.5,\ha-\inter) -- (\deb+0.5+\rodsize-0.3,\ha-\inter);
\draw[black,line width=1mm] (\fin-0.5-\rodsize+0.2,\ha-2*\inter) -- (\fin-0.55,\ha-2*\inter);
\draw[black, dotted,line width=1mm] (\fin-0.5,\ha-2*\inter) -- (\fin,\ha-2*\inter);

%% Next lines and their rods

\draw[byzantine,line width=1mm] (\deb+0.5+\rodsize-0.3,\ha-5*\inter) -- (\deb+0.5+1.5*\rodsize-0.3,\ha-5*\inter);

\draw[bitterlemon,line width=1mm] (\deb+0.5+1.5*\rodsize-0.3,\ha-6*\inter) -- (\deb+0.5+2*\rodsize-0.3,\ha-6*\inter);

\draw[amaranthred,line width=1mm] (\deb+0.5+2*\rodsize-0.3,\ha-3*\inter) -- (\deb+0.5+2.75*\rodsize-0.15,\ha-3*\inter);

\draw[turquoise,line width=1mm] (\deb+0.5+2.75*\rodsize-0.15,\ha-7*\inter) -- (\deb+0.5+3.25*\rodsize-0.15,\ha-7*\inter);

\draw[amazon,line width=1mm] (\deb+0.5+3.25*\rodsize-0.15,\ha-4*\inter) -- (\deb+0.5+3.75*\rodsize-0.15,\ha-4*\inter);

\draw[amaranthred,line width=1mm] (\deb+0.5+3.75*\rodsize-0.15,\ha-3*\inter) -- (\deb+0.5+4.5*\rodsize,\ha-3*\inter);

\draw[turquoise,line width=1mm] (\deb+0.5+4.5*\rodsize,\ha-7*\inter) -- (\deb+0.5+5*\rodsize,\ha-7*\inter);

\draw[bitterlemon,line width=1mm] (\deb+0.5+5.5*\rodsize+0.2,\ha-6*\inter) -- (\deb+0.5+6*\rodsize+0.2,\ha-6*\inter);

%% Rods on the z-axis

\draw[byzantine,line width=1mm,opacity=0.25] (\deb+0.5+\rodsize-0.3,\ha-\zaxisline*\inter) -- (\deb+0.5+1.5*\rodsize-0.3,\ha-\zaxisline*\inter);

\draw[bitterlemon,line width=1mm,opacity=0.25] (\deb+0.5+1.5*\rodsize-0.3,\ha-\zaxisline*\inter) -- (\deb+0.5+2*\rodsize-0.3,\ha-\zaxisline*\inter);
\draw[amaranthred,line width=1mm,opacity=0.25] (\deb+0.5+2*\rodsize-0.3,\ha-\zaxisline*\inter) -- (\deb+0.5+2.75*\rodsize-0.15,\ha-\zaxisline*\inter);

\draw[turquoise,line width=1mm,opacity=0.25] (\deb+0.5+2.75*\rodsize-0.15,\ha-\zaxisline*\inter) -- (\deb+0.5+3.25*\rodsize-0.15,\ha-\zaxisline*\inter);

\draw[amazon,line width=1mm,opacity=0.25] (\deb+0.5+3.25*\rodsize-0.15,\ha-\zaxisline*\inter) -- (\deb+0.5+3.75*\rodsize-0.15,\ha-\zaxisline*\inter);

\draw[amaranthred,line width=1mm,opacity=0.25] (\deb+0.5+3.75*\rodsize-0.15,\ha-\zaxisline*\inter) -- (\deb+0.5+4.5*\rodsize,\ha-\zaxisline*\inter);
\draw[turquoise,line width=1mm,opacity=0.25] (\deb+0.5+4.5*\rodsize,\ha-\zaxisline*\inter) -- (\deb+0.5+5*\rodsize,\ha-\zaxisline*\inter);

\draw[bitterlemon,line width=1mm,opacity=0.25] (\deb+0.5+5.5*\rodsize+0.2,\ha-\zaxisline*\inter) -- (\deb+0.5+6*\rodsize+0.2,\ha-\zaxisline*\inter);

%% Vertical lines and coordinates

\draw[gray,dotted,line width=0.2mm] (\deb+0.5+\rodsize-0.3,\ha-\inter) -- (\deb+0.5+\rodsize-0.3,\ha-\zaxisline*\inter);
\draw[gray,dotted,line width=0.2mm] (\deb+0.5+1.5*\rodsize-0.3,\ha-\inter) -- (\deb+0.5+1.5*\rodsize-0.3,\ha-\zaxisline*\inter);
\draw[gray,dotted,line width=0.2mm] (\deb+0.5+2*\rodsize-0.3,\ha-\inter) -- (\deb+0.5+2*\rodsize-0.3,\ha-\zaxisline*\inter);
\draw[gray,dotted,line width=0.2mm] (\deb+0.5+2.75*\rodsize-0.15,\ha-\inter) -- (\deb+0.5+2.75*\rodsize-0.15,\ha-\zaxisline*\inter);
\draw[gray,dotted,line width=0.2mm] (\deb+0.5+3.25*\rodsize-0.15,\ha-\inter) -- (\deb+0.5+3.25*\rodsize-0.15,\ha-\zaxisline*\inter);
\draw[gray,dotted,line width=0.2mm] (\deb+0.5+3.75*\rodsize-0.15,\ha-\inter) -- (\deb+0.5+3.75*\rodsize-0.15,\ha-\zaxisline*\inter);
\draw[gray,dotted,line width=0.2mm] (\deb+0.5+4.5*\rodsize,\ha-\inter) -- (\deb+0.5+4.5*\rodsize,\ha-\zaxisline*\inter);
\draw[gray,dotted,line width=0.2mm] (\deb+0.5+5*\rodsize,\ha-\inter) -- (\deb+0.5+5*\rodsize,\ha-\zaxisline*\inter);
\draw[gray,dotted,line width=0.2mm] (\deb+0.5+5.5*\rodsize+0.2,\ha-\inter) -- (\deb+0.5+5.5*\rodsize+0.2,\ha-\zaxisline*\inter);
\draw[gray,dotted,line width=0.2mm] (\deb+0.5+6*\rodsize+0.2,\ha-\inter) -- (\deb+0.5+6*\rodsize+0.2,\ha-\zaxisline*\inter);

\draw[line width=0.3mm] (\deb+0.5+\rodsize-0.3,\ha-\zaxisline*\inter+0.1) -- (\deb+0.5+\rodsize-0.3,\ha-\zaxisline*\inter-0.1);
\draw[line width=0.3mm] (\deb+0.5+1.5*\rodsize-0.3,\ha-\zaxisline*\inter+0.1) -- (\deb+0.5+1.5*\rodsize-0.3,\ha-\zaxisline*\inter-0.1);
\draw[line width=0.3mm] (\deb+0.5+2*\rodsize-0.3,\ha-\zaxisline*\inter+0.1) -- (\deb+0.5+2*\rodsize-0.3,\ha-\zaxisline*\inter-0.1);
\draw[line width=0.3mm] (\deb+0.5+2.75*\rodsize-0.15,\ha-\zaxisline*\inter+0.1) -- (\deb+0.5+2.75*\rodsize-0.15,\ha-\zaxisline*\inter-0.1);
\draw[line width=0.3mm] (\deb+0.5+3.25*\rodsize-0.15,\ha-\zaxisline*\inter+0.1) -- (\deb+0.5+3.25*\rodsize-0.15,\ha-\zaxisline*\inter-0.1);
\draw[line width=0.3mm] (\deb+0.5+3.75*\rodsize-0.15,\ha-\zaxisline*\inter+0.1) -- (\deb+0.5+3.75*\rodsize-0.15,\ha-\zaxisline*\inter-0.1);
\draw[line width=0.3mm] (\deb+0.5+4.5*\rodsize,\ha-\zaxisline*\inter+0.1) -- (\deb+0.5+4.5*\rodsize,\ha-\zaxisline*\inter-0.1);
\draw[line width=0.3mm] (\deb+0.5+5*\rodsize,\ha-\zaxisline*\inter+0.1) -- (\deb+0.5+5*\rodsize,\ha-\zaxisline*\inter-0.1);
\draw[line width=0.3mm] (\deb+0.5+5.5*\rodsize+0.2,\ha-\zaxisline*\inter+0.1) -- (\deb+0.5+5.5*\rodsize+0.2,\ha-\zaxisline*\inter-0.1);
\draw[line width=0.3mm] (\deb+0.5+6*\rodsize+0.2,\ha-\zaxisline*\inter+0.1) -- (\deb+0.5+6*\rodsize+0.2,\ha-\zaxisline*\inter-0.1);

\draw (\deb+0.5+1*\rodsize-0.3,\ha-\zaxisline*\inter-0.5) node{{\small $0$}};
\draw (\deb+0.5+1.5*\rodsize-0.3,\ha-\zaxisline*\inter-0.5) node{{\small $\frac{\ell_1^2}{4}$}};
\draw (\deb+0.5+2*\rodsize-0.3,\ha-\zaxisline*\inter-0.5) node{{\small $\frac{\ell_1^2+\ell_2^2}{4}$}};

\draw (\deb+0.5+2.75*\rodsize-0.15,\ha-\zaxisline*\inter-0.5) node{{\tiny $\frac{\ell_1^2+\ell_{2}^2+\ell_{3}^2}{4}$}};
\draw[gray,->,line width=0.1mm] (\deb+0.5+3.25*\rodsize-0.15,\ha-\zaxisline*\inter-0.8) -- (\deb+0.5+3.25*\rodsize-0.15,\ha-\zaxisline*\inter-0.25);
\draw (\deb+0.5+3.25*\rodsize-0.15,\ha-\zaxisline*\inter-1.1) node{{\tiny $\frac{\sum_{i=1}^4\ell_{i}^2}{4}$}};

\draw[black,line width=0.3mm,dotted,double] (\deb+0.5+3.5*\rodsize,\ha-\zaxisline*\inter-0.5) -- (\deb+0.5+4.75*\rodsize,\ha-\zaxisline*\inter-0.5);

\draw (\deb+0.5+5.5*\rodsize+0.2,\ha-\zaxisline*\inter-0.5) node{{\tiny $\frac{\sum_{i=1}^{n-1}\ell_{i}^2}{4}$}};
\draw[gray,->,line width=0.1mm] (\deb+0.5+6*\rodsize+0.2+0.05,\ha-\zaxisline*\inter-0.8) -- (\deb+0.5+6*\rodsize+0.2+0.05,\ha-\zaxisline*\inter-0.25);
\draw (\deb+0.5+6*\rodsize+0.2+0.05,\ha-\zaxisline*\inter-1.1) node{{\tiny $\frac{\sum_{i=1}^n\ell_{i}^2}{4}=\frac{\ell^2}{4}$}};

\end{tikzpicture}
\caption{Rod diagram of the shrinking directions on the $z$-axis after sourcing the solutions with $n$ connected rods that force the degeneracy of the S$^1$ or a T$^4$ direction.}
\label{fig:rodsourceAdS3+T4s}
\end{figure}
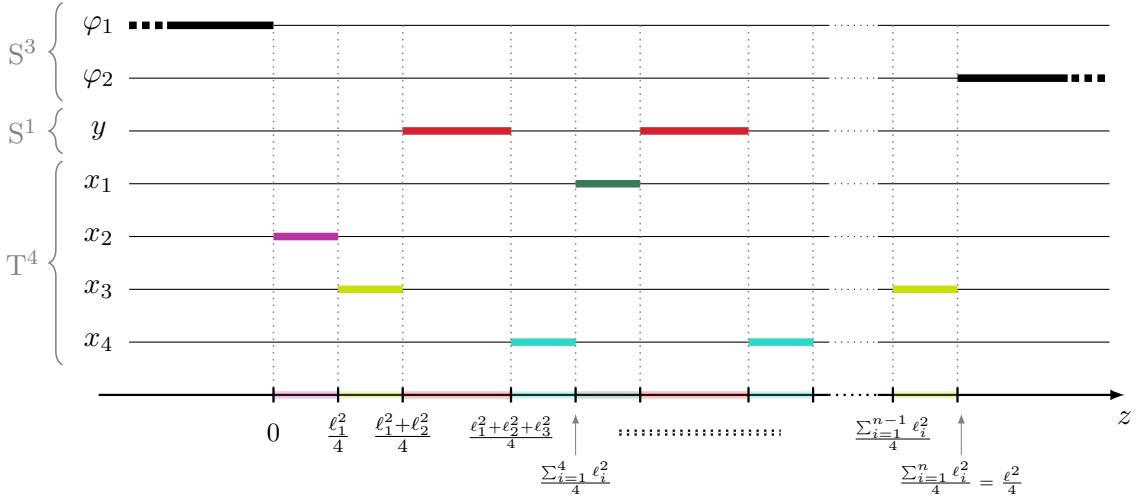

We consider configurations that have no rod sources where the Hopf angle of the S$^3$ degenerates: $U_\psi = \emptyset$.  This allows the geometry to be asymptotic to AdS$_3\times$S$^3\times$T$^4$ without orbifold on the S$^3$.  Therefore,  we consider $k=1$ from now on.  A generic rod configuration has been depicted in Fig.\ref{fig:rodsourceAdS3+T4s}.

\subsubsection{The solutions}

We refer the reader interested in the derivation of the type IIB fields from  \eqref{eq:LinearAdS3} to the Appendix \ref{App:GlobalAdS3+T4s}.  The solutions are given by\footnote{The metric in the Weyl cylindrical coordinate system is obtained by replacing $$ \frac{dr^2}{r^2+\ell^2}+d\theta^2= \frac{4}{\left( r^2+\ell^2\cos^2\theta\right)\left( r^2+\ell^2\sin^2\theta\right)} \left(d\rho^2+dz^2\right),$$
and the component along $y$ can be written as \eqref{eq:SimplRelations2}:
$r^2 \cK_{x_1} \cK_{x_2}\cK_{x_3} \cK_{x_4} = (r^2+\ell^2) \prod_{i\in U_y}\left( 1+\frac{\ell_i^2}{r_i^2}\right)^{-1}.$}
\begin{align}
ds_{10}^2 \= & \frac{1}{\sqrt{Q_1 Q_5\,\cK_1}} \left[-(r^2+\ell^2) dt^2 + r^2 \cK_{x_1}\cK_{x_2}\cK_{x_3}\cK_{x_4}\,dy^2 \right]+\sqrt{\frac{Q_1}{Q_5}\,  \cK_1 }\,\sum_{a=1}^4\frac{dx_a^2}{ \cK_{x_a}} \nn \\
&\+ \sqrt{Q_1 Q_5\,\cK_1} \Biggl[  \cK_2 \left(\frac{dr^2}{r^2+\ell^2}+d\theta^2 \right) + \cos^2 \theta \,d\varphi_1^2 + \sin^2 \theta\, d\varphi_2^2 \Biggr] \,, \label{eq:met1AdS3+T4s}  \\
C^{(2)} \= &Q_5\,\cos^2 \theta\,d\varphi_2 \wedge d\varphi_1 -\frac{r^2+\ell^2}{Q_1\,\cK_1} \,dt\wedge dy \,,\qquad  e^\Phi \= \sqrt{\frac{Q_1}{Q_5}\,\cK_1}\,, \nn 
\end{align}
where we have defined
\begin{align}
\cK_1 &\equi \frac{r^2+\ell^2}{\sum_{i\in U_y} \ell_i^2}\,\left(1-\prod_{i\in U_y} \left(1+\frac{\ell_i^2}{r_i^2} \right)^{-1}\right)\,,\qquad \cK_{x_a}\= \prod_{i\in U_{x_a}} \left(1+\frac{\ell_i^2}{r_i^2} \right)\,, \label{eq:DefDefWarpFac}\\
\cK_2& \equi \prod\limits_{\substack{i,j=1\\j>i}}^n\left(\frac{\left( \left(r_i^2+\ell_i^2 \right) \cos^2\theta_i +  \left(r_j^2+\ell_j^2\right) \sin^2\theta_j \right)\left(r_i^2 \cos^2\theta_i +  r_j^2 \sin^2\theta_j \right)}{\left( \left(r_i^2+\ell_i^2 \right) \cos^2\theta_i + r_j^2 \sin^2\theta_j \right)\left(r_i^2 \cos^2\theta_i +  \left(r_j^2+\ell_j^2\right) \sin^2\theta_j \right)} \right)^{\bar{\alpha}_{ij}-1}\,. \nn
\end{align}
We remind that $(r,\theta)$ are the global spherical coordinates of the $n$-rod configuration \eqref{eq:DefDistanceglobal}, while $(r_i,\theta_i)$ are the local spherical coordinates centered at the $i^\text{th}$ rod, given in terms of $(\rho,z)$ in \eqref{eq:DefDistance} and of $(r,\theta)$ in \eqref{eq:ri&thetaidef}.

The two solutions induced by a single T$^4$ deformation in AdS$_3$ and derived in section \ref{sec:AdS3+T4} and \ref{sec:AdS3+T42}  can be retrieved by considering\footnote{The simplification relations \eqref{eq:SimplRelations2} are required to match the solutions.}
\begin{equation}
\begin{split}
& n=2,\qquad U_y\=\{1\} \qquad U_{x_1}=\{2\}, \qquad U_{x_2}=U_{x_3}=U_{x_4}= \emptyset\,, \\
 &n=3,\qquad U_y\=\{1,3\} \qquad U_{x_1}=\{2\}, \qquad U_{x_2}=U_{x_3}=U_{x_4}= \emptyset\,, 
\end{split}
\end{equation} 

The T$^4$ degeneracies do not only modify the T$^4$ but also deform the S$^3$ and AdS$_3$ spacetime.  From this perspective,  the warp factors  $\cK_I$ are deformation factors that are trivial if one turns off the deformations,  that is $\ell_i \to 0$ for $i\notin U_y$.  In the $(\rho,z)$ Weyl coordinate system,  the deformations are induced by the rod sources $i\notin U_y$ and are located in the $z$-axis,  $\rho=0$ and in the segment $0 \leq z \leq \ell^2/4$.  In the $(r,\theta)$ coordinate system,  the sources are localized at $r=0$ and one moves along the chain of rods by moving along the S$^3$, varying $\theta$ from $0$ to $\pi/2$.

\subsubsection{Regularity conditions and topology}
\label{sec:RegGenSol1}

%One can study the property of the geometries either in the Weyl coordinates or in the global spherical coordinates.  We choose the global spherical coordinates since it gives a more direct comparison to the undeformed global  AdS$_3\times$S$^3\times$T$^4$ background.

At large distance,  $r\to \infty$,  the metric is asymptotic to AdS$_3\times$S$^3\times$T$^4$ as in \eqref{eq:AdS3Asymp} since $\cK_I \to 1$.  At $r>0$ and outside the poles of the S$^3$, $\theta \neq 0,\pi/2$,  the metric components are finite and non-zero,  so the solutions are regular there, and have a S$^1\times$S$^3 \times$T$^4$ topology.

The poles of the S$^3$ correspond to $\theta=0$ and $\pi/2$ at $r>0$ where  the $\varphi_2$ and $\varphi_1$ angles degenerate respectively.  One can check that $\theta_i=0$ and $\theta_i=\pi/2$, $i=1,\ldots ,n$,  at these locii respectively \eqref{eq:ri&thetaidef}.  Thus $\cK_2=1$, and the metric of the S$^3$ at its poles is smooth such that $ds(S^3)^2 \sim d\theta^2 +\cos^2 \theta \,d\varphi_1^2+\sin^2 \theta \,d\varphi_2^2$

%%%%%%%%%%%%%%%%%%%%%%%%%%%%%%%%
\begin{figure}[t]
\centering
\includegraphics[scale=0.7]{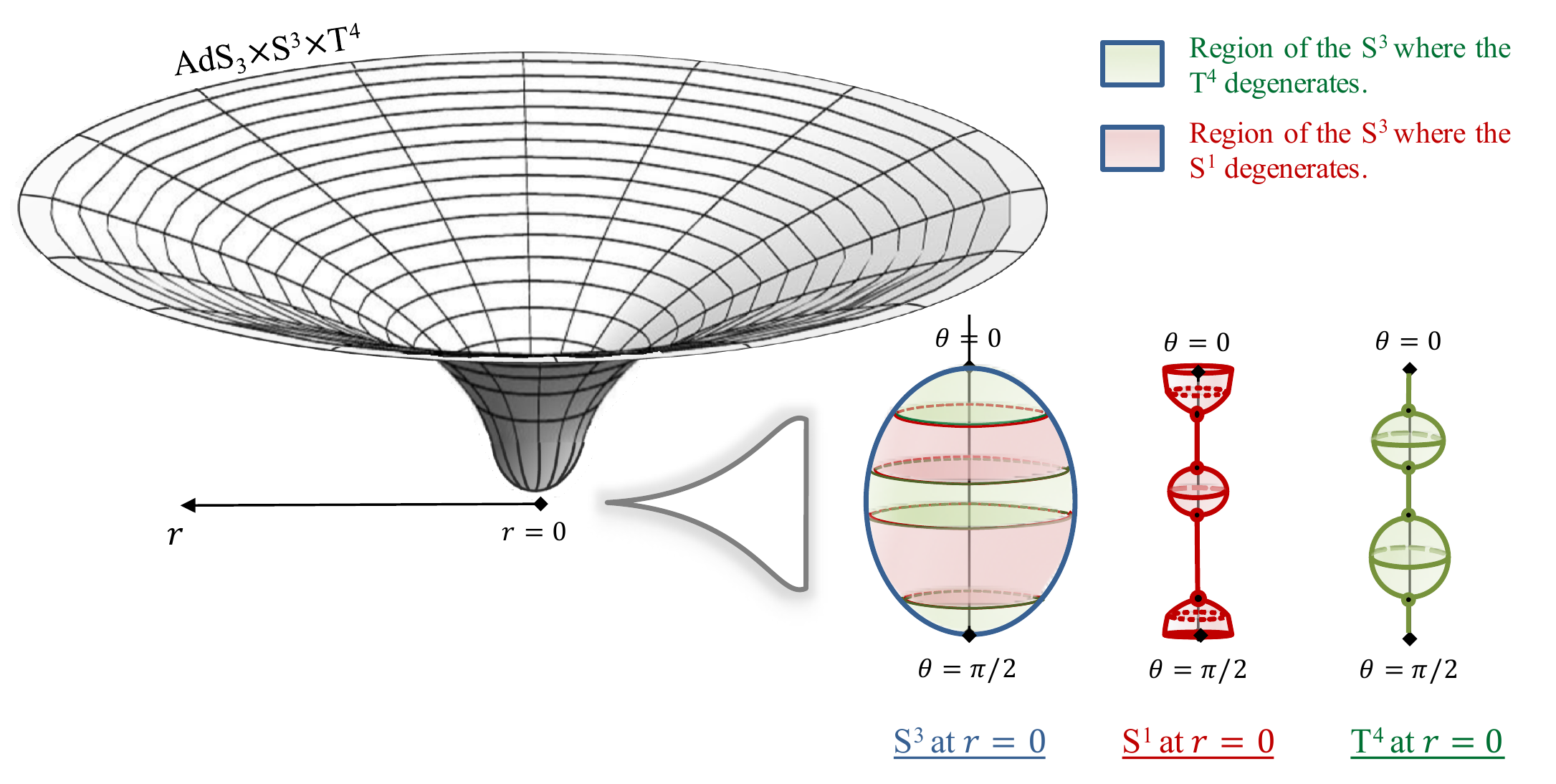}
\caption{Schematic description of the spacetime sourced by $5$ connected rods,  inducing the degeneracy of either the S$^1$ ($y$) or a T$^4$ direction.  On the left-hand side, we depict the overall geometry in terms of $r$.  On the right-hand side,  we describe the behavior of the S$^3$,  S$^1$, and T$^4$ at $r=0$ and as a function of $\theta$,  giving the position along the S$^3$.  At $r=0$,  the spacetime ends smoothly as a coordinate degeneracy of either the S$^1$ or the T$^4$ depending on the position on the S$^3$. }
\label{fig:AdS3+T4spic}
\end{figure}
%%%%%%%%%%%%%%%%%%%%%%%%%%%%%%%%

At $r=0$,  we have a unique $r_i$ that vanishes depending on the value of $\theta$  while all others $r_j$ are non-zero \eqref{eq:ri&thetaidef}.  More precisely we have
\begin{equation}
r=0\,,\qquad \theta_c^{(i)} < \theta < \theta_c^{(i-1)} \quad \Leftrightarrow \quad r_i = 0 \,,\qquad r_j >0, \quad j\neq i\,,
\label{eq:ThetaSection}
\end{equation}
where we have defined the critical angles
\begin{equation}
\cos^2 \theta_c^{(i)} \= \frac{1}{\ell^2} \sum_{j=1}^{i} \ell_j^2\,,\qquad \theta_c^{(n)}=0\,,\qquad \theta_c^{(0)}=\frac{\pi}{2}\,.
\label{eq:DefThetaCrii}
\end{equation}
Thus, we are moving along the chain of rod sources by varying $\theta$ from $0$ to $\pi/2$ at $r=0$.  From the form of the metric \eqref{eq:met1AdS3+T4s},  one can see that a spacelike coordinate degenerates at each section $\theta_c^{(i)} < \theta < \theta_c^{(i-1)}$. The local geometry corresponds to a bolt with a $\IR^2\times \cC_\text{Bubble}$ topology where $\cC_\text{Bubble}$ is a compact space defining the topology of the smooth bubble at the bolt.  Having a regular bolt at each segment $\theta_c^{(i)} < \theta < \theta_c^{(i-1)}$ will impose $n$ \emph{bubble equations} that fix all rod lengths $\ell_i^2$ in terms of the asymptotic quantities.  

Then,  $r=0$ corresponds to a smooth locus where the spacetime ends as a chain of bolts.  Each bolt makes either the S$^1$ or the T$^4$ degenerate smoothly.  It defines a compact bubble that is localized on a specific region of the S$^3$,  given by the critical angles, $\theta_c^{(i)}$.  Moreover,  for the same arguments as in section \ref{sec:nonBPSness},  the deformations break the supersymmetry of the global AdS$_3\times$S$^3\times$T$^4$ solution,  and most likely break all supersymmetry.  Therefore,  the solutions correspond to \emph{asymptotically-AdS$_3$ non-BPS smooth bubbling geometries} without horizon.  We have depicted the profile of the geometries in the same way as the previous examples in Fig.\ref{fig:AdS3+T4spic}.

We divide the regularity analysis depending on whether $i\in U_y$ or $i\in U_{x_a}$.

\begin{itemize}
\item[•] \underline{Regularity at the $i^\text{th}$ rod with $i\in U_y$:}

We consider a segment such that $i\in U_y$,
\begin{equation}
r=0\,,\qquad \theta_c^{(i)} < \theta < \theta_c^{(i-1)} \quad \Leftrightarrow \quad r_i = 0 \,,\qquad 0< \theta_i < \frac{\pi}{2}\,.
\end{equation}
Since $r=0$ and $\cK_{x_a}>0$,   one can check from \eqref{eq:met1AdS3+T4s} that the $y$ coordinate degenerates.  To derive the local geometry at this segment,  one needs to consider $(r_i,\theta_i)$ as the main coordinates and take $r_i\to 0$,  that is to express all other $(r_j,\theta_j)$ and $(r,\theta)$ in terms of $(r_i,\theta_i)$ and expand the metric and fields.\footnote{This is achieved by going first in the Weyl cylindrical coordinates $(\rho,z)$,  using \eqref{eq:DefDistance} and \eqref{eq:DefDistanceglobal}, and then by changing coordinates with  \eqref{eq:LocalSpher1} and \eqref{eq:LocalSpher2}.} We refer the interested reader to previous work of the author \cite{Bah:2020pdz,Bah:2021owp,Heidmann:2021cms} for more details about this derivation.  We find that the time slices of the type IIB metric \eqref{eq:met1AdS3+T4s} converge towards 
\begin{equation}
ds_{10}^2|_{dt=0} \propto dr_i^2 + \frac{r_i^2}{C_i^2}\,dy^2 + ds(\cC^{(i)}_\text{Bubble})^2\,,
\end{equation}
with\footnote{We consider that $\prod_{i=a}^b\ldots =1$ if $a>b$. \label{footnote2}}
\begin{equation}
\begin{split}
C_i^2 \=&\, \frac{Q_1 Q_5\,\ell_i^2}{\ell^2\, \sum_{p\in U_y} \ell_p^2}  \,\prod_{p=1}^{i-1} \prod_{q=i+1}^n \left[\frac{1+ \frac{\ell_q^2}{\sum_{k=p}^{q-1} \ell_k^2}}{1+ \frac{\ell_q^2}{\sum_{k=p+1}^{q-1} \ell_k^2}} \right]^{\bar{\alpha}_{pq}}\\
&\times \prod_{p=1}^{i-1}\left(1+ \frac{\ell_p^2}{\sum_{k=p+1}^{i} \ell_k^2} \right)^{\bar{\alpha}_{ip}} \, \prod_{p=i+1}^{n}\left(1+ \frac{\ell_p^2}{\sum_{k=i}^{p-1} \ell_k^2} \right)^{\bar{\alpha}_{ip}}\,.
\end{split}
\end{equation}
The $(r_i,y)$ subspace describes a smooth origin of a $\IR^2$,  if we impose
\begin{equation}
R_{y}=C_i.
\end{equation}

The line element,  $ds(\cC^{(i)}_\text{Bubble})$,  describes the topology of the bubble at the bolt.  As discussed in \cite{Bah:2020pdz,Bah:2021owp,Heidmann:2021cms},  it can be either a S$^3\times$T$^4$ or a S$^2\times$T$^5$ depending on the near environment of the rod.  More precisely,  if the adjacent rods are of the same category,  let's say they correspond to the degeneracy of the $x_1$ coordinate,  then it is a S$^2\times$T$^5$ where the S$^2$ and T$^5$ are described by $(\theta_i,x_1)$ and $(\varphi_1,\varphi_2,x_2,x_3,x_4)$.  If they are of different nature,  let's say they correspond to the degeneracy of the $x_1$ and $x_2$ coordinates, then we have a S$^3\times$T$^4$ where the S$^3$ and T$^4$ are described by $(\theta_i,x_1,x_2)$ and $(\varphi_1,\varphi_2,x_3,x_4)$.  The rod endpoints,  $\theta_i=0$ or $\pi/2$,  correspond to the poles of either the S$^2$ or the S$^3$.  The regularity at these poles is guaranteed by the regularity at the adjacent rods.\footnote{see \cite{Bah:2020pdz,Bah:2021owp,Heidmann:2021cms} for more details. }

Moreover,  one can show that the three-form field strength,  $F_3 = dC^{(2)}$,  is regular such that the component along $y$ vanishes.  Moreover, the rod carries D1 and D5 charges given by \eqref{eq:ChargeAtRodGen}
\begin{equation}
q_{D1}^{(i)} \= \frac{\ell_i^2}{\sum_{j\in U_y}\ell_j^2}\,Q_1\,,\qquad q_{D5}^{(i)} \= \frac{\ell_i^2}{\ell^2}\,Q_5\,.
\label{eq:LocalChargeT4sS1}
\end{equation}

\item[•] \underline{Regularity at the $i^\text{th}$ rod with $i\in U_{x_a}$:}

We consider a segment such that $i\in U_{x_a}$,
\begin{equation}
r=0\,,\qquad \theta_c^{(i)} < \theta < \theta_c^{(i-1)} \quad \Leftrightarrow \quad r_i = 0 \,,\qquad 0< \theta_i < \frac{\pi}{2}\,.
\end{equation}
The torus direction $x_a$ degenerates.  Indeed, if $i\in U_{x_a}$,  then $\cK_{x_a} \to \infty$ and $r^2 \cK_{x_a}>0$,  so the metric component along $x_a$ vanishes at the rod.  The time slices of the type IIB metric \eqref{eq:met1AdS3+T4s} converge towards 
\begin{equation}
ds_{10}^2|_{dt=0} \propto dr_i^2 + \frac{r_i^2}{C_i^2}\,dx_a^2 + ds(\cC^{(i)}_\text{Bubble})^2\,,
\end{equation}
with
\begin{equation}
\begin{split}
C_i^2 \=&\, \frac{Q_5\,\ell_i^2}{\ell^2}  \,\prod_{p=1}^{i-1} \prod_{q=i+1}^n \left[\frac{1+ \frac{\ell_q^2}{\sum_{k=p}^{q-1} \ell_k^2}}{1+ \frac{\ell_q^2}{\sum_{k=p+1}^{q-1} \ell_k^2}} \right]^{\bar{\alpha}_{pq}}\\
&\times \prod_{p=1}^{i-1}\left(1+ \frac{\ell_p^2}{\sum_{k=p+1}^{i} \ell_k^2} \right)^{\bar{\alpha}_{ip}} \, \prod_{p=i+1}^{n}\left(1+ \frac{\ell_p^2}{\sum_{k=i}^{p-1} \ell_k^2} \right)^{\bar{\alpha}_{ip}}\,.
\end{split}
\end{equation}
The $(r_i,x_a)$ subspace describes a smooth bolt,  if we impose
\begin{equation}
R_{x_a}=C_i.
\end{equation}
As for the rods $i\in U_y$,  $ds(\cC^{(i)}_\text{Bubble})^2$ describes a S$^3\times$T$^4$ or S$^2\times$T$^5$ bubble depending on whether the rod is connected to two rods of the same nature or not.

Moreover,  the rod carries a D5 charge given by \eqref{eq:ChargeAtRodGen}
\begin{equation}
q_{D1}^{(i)} \=0\,,\qquad q_{D5}^{(i)} \= \frac{\ell_i^2}{\ell^2}\,Q_5\,.
\label{eq:LocalChargeT4sT4}
\end{equation}
\end{itemize}

To summarize,  at $r=0$, the solutions correspond to a chain of $n$ bolts where the S$^1$ or a T$^4$ direction smoothly degenerates if $n$ algebraic bubble equations are satisfied:
\begin{align}
\frac{R_y} {\sqrt{Q_1 Q_5}}\=\frac{\ell_i \,d_i}{\ell \sqrt{\sum_{p\in U_y} \ell_p^2}} \,,\quad \text{if } i\in U_y\,,\qquad \qquad 
\frac{R_{x_a}}{\sqrt{Q_5}} \=  \frac{\ell_i \,d_i}{\ell } \,,\quad\text{if }  i\in U_{x_a},\nn
\end{align}
where we have defined the aspect ratios, $d_i$, 
\begin{equation}
d_i \equi \prod_{p=1}^{i-1} \prod_{q=i+1}^n \left[\frac{1+ \frac{\ell_q^2}{\sum_{k=p}^{q-1} \ell_k^2}}{1+ \frac{\ell_q^2}{\sum_{k=p+1}^{q-1} \ell_k^2}} \right]^{\frac{\bar{\alpha}_{pq}}{2}} \,\prod_{p=1}^{i-1}\left(1+ \frac{\ell_p^2}{\sum_{k=p+1}^{i} \ell_k^2} \right)^{\frac{\bar{\alpha}_{ip}}{2}} \, \prod_{p=i+1}^{n}\left(1+ \frac{\ell_p^2}{\sum_{k=i}^{p-1} \ell_k^2} \right)^{\frac{\bar{\alpha}_{ip}}{2}}\,. \label{eq:DefdiAspect}
\end{equation}
These equations fix all rod lengths, $\ell_i^2$,  in terms of the boundary quantities, namely the charges of the D1-D5 branes and the radii of the S$^1$ and T$^4$.  The solutions are therefore completely fixed and the only changeable parameters are the nature of the rods and their total number. The latter can be varied by adding ``deformation quanta'' as we increase $n$.  This will non-trivially modify the geometries by changing the bubble equations.

Remarkably,  the bubble equations can be expressed in terms of the local D1 and D5 brane charges at the rods \eqref{eq:LocalChargeT4sS1} and \eqref{eq:LocalChargeT4sT4} such that
\begin{equation}
R_y \= \frac{\sqrt{q_{D1}^{(i)} q_{D5}^{(i)}}}{\ell_i} \, d_i\,, \quad \text{if } i\in U_y\,,\qquad \qquad 
R_{x_a} \=  \sqrt{q_{D5}^{(i)}}\, d_i\,,\quad\text{if }  i\in U_{x_a}.
\end{equation}
Thus,  the regularity constraint for the rods corresponding to the degeneracy of the S$^1$,  $i\in U_y$,  is similar to the constraint for a global AdS$_3\times$S$^3\times$T$^4$ \eqref{eq:RegCondAdS3} in terms of the local charges. However,  they have an extra deformation factor $d_i$ that accounts for interactions between the rods of the same nature.  Indeed,  $d_i$ depends on the exponent $\bar{\alpha}_{pq}$ which is non-zero only when the $p^\text{th}$ and $q^\text{th}$ rods are of the same kind \eqref{eq:DefAlphaBar}.  Similarly,  the constraints for the rods where a T$^4$ direction shrinks, $i\in U_{x_a}$,  is comparable to the regularity when there is only a single T$^4$ deformation \eqref{eq:RegCondAdS3+T4} with the additional $d_i$ factor.

Moreover,  the bubble equations do not have analytic solutions in general, except for small values of $n$.  However,  an approximation can be performed considering a large number of bolts,  and the equations can be solved at leading order in $n$ \cite{Bah:2021rki}. 

As for the examples constructed in section \ref{sec:AdS3+T4} and \ref{sec:AdS3+T42},  the bolts where a T$^4$ direction degenerates can be considered as small perturbations on top of a global AdS$_3\times$S$^3\times$T$^4$ background if one imposes a hierarchy of scale in between the T$^4$ and the D5 charge.  Indeed,  we have 
\begin{equation}
(R_{x_1},R_{x_2},R_{x_3},R_{x_4}) \,\ll \, \sqrt{Q_5} \quad (\sqrt{\text{Vol}(T^4)} \,\ll Q_5) \quad \Rightarrow \quad \ell_j^2 \,\ll \, \ell_i^2\,,\quad j\in U_{x_a}\,,\, i\in U_y \,.
\label{eq:PerturbCond}
\end{equation}
Then we have $\cK_I = 1+\cO(\ell_j^2)$ as soon as we are not too close to the rod sources and the metric \eqref{eq:met1AdS3+T4s} corresponds to a global AdS$_3\times$S$^3\times$T$^4$ background with small perturbations that break the rigidity of the T$^4$ and the symmetry of the S$^3$.

Note that the bubble equations simplify if we consider that only the first rod forces the degeneracy of the S$^1$ as in section \ref{sec:AdS3+T4},  i.e.  $U_{y}=\{1\}$.  It corresponds to solutions where all the T$^4$ deformations are centered around a pole of the S$^3$.  The bubble equation for the first rod gives
\begin{equation}
\frac{Q_1 Q_5}{R_y^2}\= \ell^2 \= \sum_{i=1}^n \ell_i^2  \,.
\end{equation}
Thus,  we have the same quantization as the undeformed global AdS$_3\times$S$^3\times$T$^4$ background \eqref{eq:RegCondAdS3},  but $\frac{Q_1 Q_5}{R_y^2}$ is now distributed along all rods.

\subsection{Chain of T$^4$ and S$^3$ deformations}
\label{sec:AdS3+T4+S3s}

We now consider generic solutions where the angle of the Hopf fibration of the S$^3$ can also degenerates: $U_\psi \neq \emptyset$.  For such configurations,  one needs to impose the geometries to be asymptotic to AdS$_3\times$S$^3/\mathbb{Z}_k\times$T$^4$ for regularity.  A generic rod configuration has been depicted in Fig.\ref{fig:rodsourceAdS3+T4+S3}.

We refer the reader interested in the derivation of the type IIB fields to Appendix \ref{App:GlobalAdS3+T4+S3s}.  The metric and fields are given by
\begin{align}
ds_{10}^2 \= & \frac{1}{\sqrt{Q_1 Q_5 \cK_1\cK_-}} \left[-(r^2+\ell^2) dt^2 + r^2 \cK_{x_1}\cK_{x_2}\cK_{x_3}\cK_{x_4}\cK_\psi\,dy^2 \right]+\sqrt{\frac{Q_1\,\cK_1}{Q_5\,\cK_-}}  \,\sum_{a=1}^4\frac{dx_a^2}{ \cK_{x_a}} \nn \\
&\+ k \sqrt{Q_1 Q_5 \cK_1 \cK_-}\, \Biggl[  \cK_2  \cK_+ \left(\frac{dr^2}{r^2+\ell^2}+d\theta^2 \right) + \cK_+\cos^2 \theta \sin^2\theta \,d\phi^2 \nn \\
&\hspace{3.3cm} \+ \frac{1}{4k^2 \cK_+ \cK_\psi} \left(d\psi + k \cA_+ d\phi \right)^2 \Biggr] \,, \label{eq:met1AdS3+S3s} \\
C^{(2)} \= &\frac{Q_5}{4}\,\cA_{-} d\phi \wedge d\psi -\frac{r^2+\ell^2}{Q_1\,\cK_1} \,dt\wedge dy \,,\qquad  e^\Phi \= \sqrt{\frac{Q_1\,\cK_1}{Q_5\,\cK_-}}\,, \nn 
\end{align}
where we have defined in addition to the deformation factors \eqref{eq:DefDefWarpFac}
\begin{align}
\cK_\pm &\equi \frac{r^2}{\ell^2\pm \sum_{i\in U_\psi} \ell_i^2}\,\left(1+\frac{\ell^2}{r^2}-\prod_{i\in U_\psi} \left(1+\frac{\ell_i^2}{r_i^2} \right)^{\mp 1}\right)\,,\qquad \cK_{\psi}\= \prod_{i\in U_{\psi}} \left(1+\frac{\ell_i^2}{r_i^2} \right)\,, \nn\\
\cA_\pm & \equi \frac{1}{\ell^2\pm \sum_{i\in U_\psi} \ell_i^2} \,\left(\ell^2 \cos 2\theta \pm  \sum_{i\in U_\psi} \ell_i^2 \cos 2 \theta_i \right)\,.  \label{eq:DefDefWarpFac2}
\end{align}

We will be brief in the analysis of the geometry and the regularity conditions since they are similar to the previous constructions.  First,  the geometries are regular at $r>0$ and $\theta\neq 0 $ or $\pi/2$,  and one can check that the geometry is indeed asymptotic to AdS$_3\times$S$^3/\mathbb{Z}_k\times$T$^4$ since all $\cK_I \to 1$ and $\cA_\pm  \to \cos 2\theta$.

%%%%%%%%%%%%%%%%%%%%%%%%%%%%%%%%
\begin{figure}[t]
\centering
\includegraphics[scale=0.7]{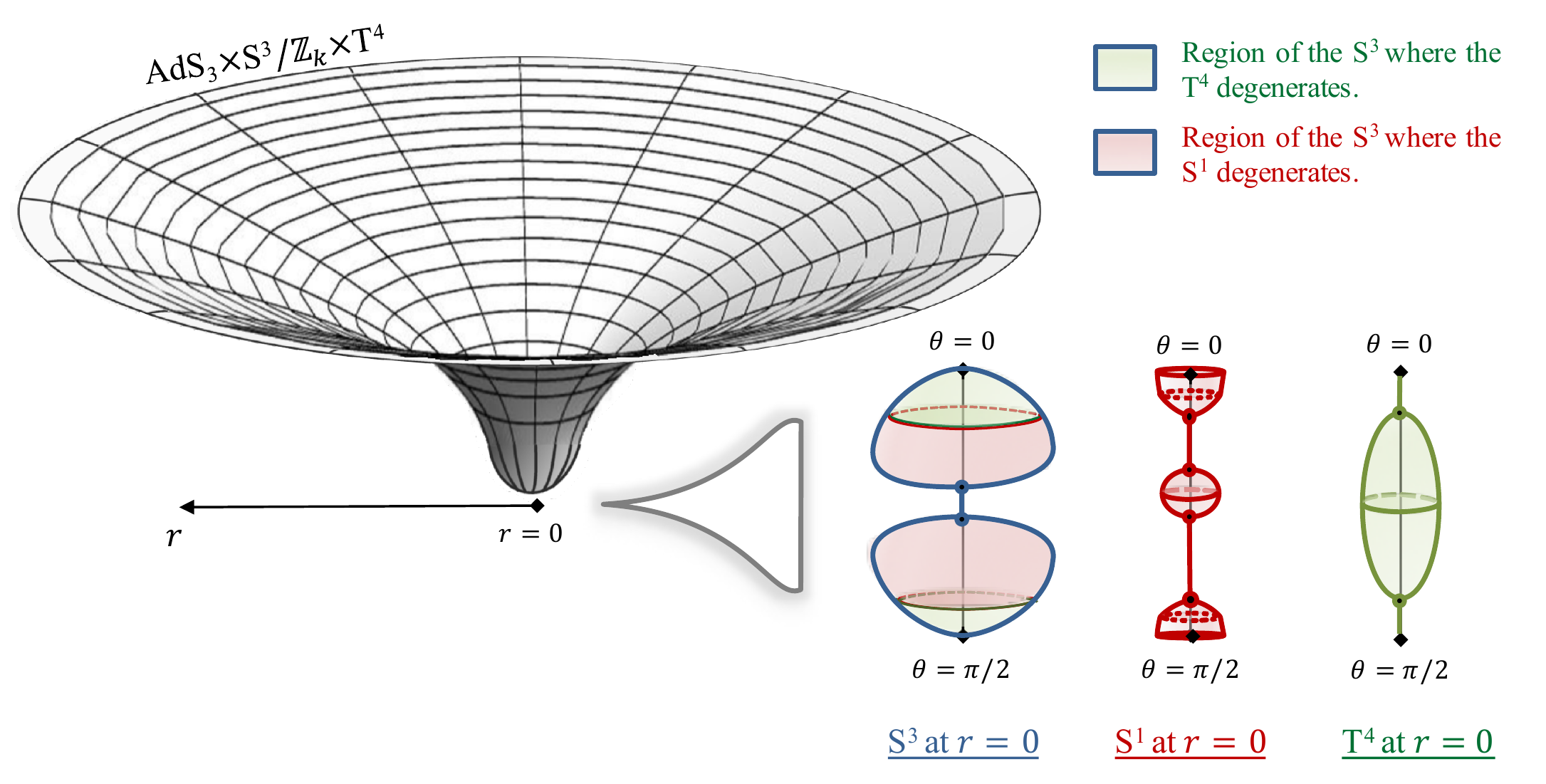}
\caption{Schematic description of the spacetime induced by a chain of $5$ rods,  inducing the degeneracy of either the S$^1$ ($y$) or a T$^4$ direction or the Hopf angle of the S$^3$.  On the left-hand side, we depict the overall geometry in terms of $r$.  On the right-hand side,  we describe the behavior of the S$^3$,  S$^1$, and T$^4$ at $r=0$ and as a function of $\theta$.  At $r=0$,  the spacetime ends smoothly as a coordinate degeneracy of either the S$^1$ or the T$^4$ or the S$^3$. }
\label{fig:AdS3+T4+S3spic}
\end{figure}
%%%%%%%%%%%%%%%%%%%%%%%%%%%%%%%%

The three loci $r=0$, $\theta=0$ and $\theta=\pi/2$ correspond to the $z$-axis where spacelike coordinates degenerate as depicted in Fig.\ref{fig:rodsourceAdS3+T4+S3}.  First,  one can check that the $\varphi_2$ and $\varphi_1$ angle degenerates regularly at $\theta=0$ and $\theta=\pi/2$ respectively  for $r>0$. They define the North and South poles of a S$^3/\mathbb{Z}_k$.  Second,  there is a chain of bolts at the rod sources, $r=0$, where one coordinate degenerates smoothly defining an origin of $\IR^2$ as in \eqref{eq:boltDef}.  The $i^\text{th}$ rod makes either the S$^1$ shrink  if $i\in U_y$ or the T$^4$ if $i\in U_{x_a}$ or the S$^3$ if $i\in U_\psi$.  The regularity at each bolt fixes all rod lengths $\ell_i^2$ in terms of the boundary quantities such as
\begin{align}
\frac{R_y} {\sqrt{k Q_1 Q_5}}\=&\frac{\ell\,\ell_i\,d_i}{\sqrt{\sum_{p\in U_y} \ell_p^2\times \left( \ell^4-\left(\sum_{p\in U_\psi} \ell_p^2\right)^2\right)}} \,,\quad \text{if }i\in U_y\,,\nn\\
\frac{R_{x_a}}{\sqrt{k Q_5}} \= & \frac{\ell \,\ell_i\,d_i}{\sqrt{\ell^4-\left(\sum_{p\in U_\psi} \ell_p^2\right)^2}} \,,\quad \text{if }i\in U_{x_a},\qquad \frac{1}{k}\=  \frac{\ell\,\ell_i\,d_i}{\ell^2+\sum_{i\in U_\psi} \ell_i^2  }\,,\quad \text{if }i\in U_{\psi}.\label{eq:BuEqT4+S3s}
\end{align}
where $d_i$ is defined in \eqref{eq:DefdiAspect}.   One retrieves the solutions of the previous section by simply taking $U_\psi =\emptyset$ or the simple solutions constructed in section \ref{sec:AdS3+S3} by considering $n=2$,  $U_y=\{1\}$ and $U_\psi=\{2\}$.

Once the bubble equations are satisfied  the solutions correspond to \emph{asymptotically-AdS$_3$ smooth bubbling geometries} without horizons that are T$^4$ and S$^3$ deformations of a global AdS$_3\times$S$^3/\mathbb{Z}_k\times$T$^4$ background in type IIB.  Moreover,  for the same arguments as in section \ref{sec:nonBPSness},  the deformations break the supersymmetry of the global AdS$_3\times$S$^3\times$T$^4$ solution,  and most likely break all supersymmetry such that they correspond to non-BPS states.  We depicted generic geometries in the global spherical coordinates in Fig.\ref{fig:AdS3+T4+S3spic}.  The spacetime ends smoothly at $r=0$ as a chain of bolts delimited in sections of $\theta$ as in \eqref{eq:ThetaSection} and where one of the spacelike directions smoothly degenerates.  Depending on their nature,  the bolts may induce D1 and D5 brane charges \eqref{eq:ChargeAtRodGen}.  The bolts where the S$^1$ degenerates have non-zero D1 and D5 charges,  while the bolts where a T$^4$ direction shrinks carry a D5 charge only, and the bolts where $\psi$ degenerates have no D1-D5 charges.  More precisely,  
\begin{equation}
\begin{split}
q_{D1}^{(i)} &\= \frac{\ell_i^2}{\sum_{j\in U_y}\ell_j^2}\,Q_1\,,\qquad q_{D5}^{(i)} \= \frac{\ell_i^2}{\ell^2- \sum_{i\in U_\psi} \ell_i^2}\,Q_5\,,\qquad i\in U_y\,,\\
q_{D1}^{(i)} &\=0\,,\qquad q_{D5}^{(i)} \= \frac{\ell_i^2}{\ell^2- \sum_{i\in U_\psi} \ell_i^2}\,Q_5\,,\qquad i\in U_{x_a}\,, \\
q_{D1}^{(i)}& \= q_{D5}^{(i)} \=0 \,,\qquad i\in U_{\psi}\,.\\
\end{split}
\end{equation}

Moreover,  one can consider the T$^4$ and S$^3$ deformations as small perturbations by assuming \eqref{eq:PerturbCond} and $k \gg 1$.  Then,  the solutions are almost identical to a global AdS$_3\times$S$^3/\mathbb{Z}_k$ $\times$T$^4$ plus corrections that are induced by degeneracies of the T$^4$ and S$^3$ on localized positions of the S$^3$ at $r=0$,  $\theta\sim \theta_c^{(i)}$ \eqref{eq:DefThetaCrii}.

One can rewrite the bubble equations \eqref{eq:BuEqT4+S3s} in terms of the local charges.  They will then be comparable to the constraints obtained for the simple T$^4$ and S$^3$ deformations in section \ref{sec:AdS3} with an extra factor $d_i$ which accounts for interactions in between rods of the same nature.

\section{Regular bound states of non-extremal BTZ black holes}
\label{sec:BTZBS}

In previous sections,  we have restricted the constructions to regular geometries in AdS$_3$ without horizons.  We will now build bound states of non-extremal two-charge black holes using similar techniques.  We will first derive the solution obtained from a single rod inducing the degeneracy of the timelike direction.  It will correspond to a S$^3\times$T$^4$ or a S$^3/\mathbb{Z}_k\times$T$^4$ fibration over a static non-extremal BTZ black hole.  Then, we will construct chains of these two-charge black holes separated by regular bolts.

\subsection{Static BTZ black hole as a single rod solution}
\label{sec:BTZ}

\begin{figure}[h]
\centering
    \begin{tikzpicture}
%% some definitions

\def\deb{-10} 
\def\inter{0.8} 
\def\ha{2.8} 
\def\zaxisline{5} 
\def\rodsize{1.5} 
\def\numrod{2} 

\def\fin{\deb+1+2*\rodsize+\numrod*\rodsize} 

%% 

%\draw (\deb-2,\ha-\zaxisline*0.5*\inter) node{$\Longrightarrow$};

%% Pic title

\draw (\deb+0.5+\rodsize+0.5*\numrod*\rodsize,\ha+1) node{{{\it BTZ$\times$S$^{\,3}\times$T$^{\,4}$}}}; 

%% Each line black line and names

\draw[black,thin] (\deb+1,\ha) -- (\fin,\ha);
\draw[black,thin] (\deb,\ha-\inter) -- (\fin-1,\ha-\inter);
\draw[black,thin] (\deb,\ha-2*\inter) -- (\fin,\ha-2*\inter);
\draw[black,thin] (\deb,\ha-3*\inter) -- (\fin,\ha-3*\inter);
\draw[black,thin] (\deb,\ha-4*\inter) -- (\fin,\ha-4*\inter);
\draw[black,->,line width=0.3mm] (\deb-0.5,\ha-\zaxisline*\inter) -- (\fin+0.2,\ha-\zaxisline*\inter);

\draw [decorate, 
    decoration = {brace,
        raise=5pt,
        amplitude=5pt},line width=0.2mm,gray] (\deb-0.8,\ha-1.5*\inter+0.05) --  (\deb-0.8,\ha+0.5*\inter-0.05);
\draw [decorate, 
    decoration = {brace,
        raise=5pt,
        amplitude=5pt},line width=0.2mm,gray] (\deb-0.8,\ha-2.5*\inter+0.05) --  (\deb-0.8,\ha-1.5*\inter-0.05);
\draw [decorate, 
    decoration = {brace,
        raise=5pt,
        amplitude=5pt},line width=0.2mm,gray] (\deb-0.8,\ha-3.5*\inter+0.05) --  (\deb-0.8,\ha-2.5*\inter-0.05);

\draw[gray] (\deb-1.5,\ha-0.5*\inter) node{S$^3$};
\draw[gray] (\deb-1.5,\ha-2*\inter) node{S$^1$};
\draw[gray] (\deb-1.5,\ha-3*\inter) node{T$^4$};
    
\draw (\deb-0.5,\ha) node{$\varphi_1$};
\draw (\deb-0.5,\ha-\inter) node{$\varphi_2$};
\draw (\deb-0.5,\ha-2*\inter) node{$y$};
\draw (\deb-0.5,\ha-3*\inter) node{$x_a$};
\draw (\deb-0.5,\ha-4*\inter) node{$t$};

\draw (\fin+0.2,\ha-\zaxisline*\inter-0.3) node{$z$};

%% First two line and their rods

\draw[black, dotted, line width=1mm] (\deb,\ha) -- (\deb+0.5,\ha);
\draw[black,line width=1mm] (\deb+0.5,\ha) -- (\deb+0.5+\rodsize,\ha);
\draw[black,line width=1mm] (\fin-0.5-\rodsize,\ha-\inter) -- (\fin-0.55,\ha-\inter);
\draw[black, dotted,line width=1mm] (\fin-0.5,\ha-\inter) -- (\fin,\ha-\inter);

%% Next lines and their rods

\draw[gray,line width=1mm] (\deb+0.5+\rodsize,\ha-4*\inter) -- (\deb+0.5+2*\rodsize,\ha-4*\inter);
\draw[gray,line width=1mm] (\deb+0.5+2*\rodsize,\ha-4*\inter) -- (\deb+0.5+3*\rodsize,\ha-4*\inter);

\draw[gray,line width=1mm,opacity=0.25] (\deb+0.5+\rodsize,\ha-\zaxisline*\inter) -- (\deb+0.5+3*\rodsize,\ha-\zaxisline*\inter);

%% Vertical lines and coordinates

\draw[gray,dotted,line width=0.2mm] (\deb+0.5+\rodsize,\ha) -- (\deb+0.5+\rodsize,\ha-\zaxisline*\inter);
\draw[gray,dotted,line width=0.2mm] (\deb+0.5+3*\rodsize,\ha) -- (\deb+0.5+3*\rodsize,\ha-\zaxisline*\inter);

\draw[line width=0.3mm] (\deb+0.5+3*\rodsize,\ha-\zaxisline*\inter+0.1) -- (\deb+0.5+3*\rodsize,\ha-\zaxisline*\inter-0.1);
\draw[line width=0.3mm] (\deb+0.5+\rodsize,\ha-\zaxisline*\inter+0.1) -- (\deb+0.5+\rodsize,\ha-\zaxisline*\inter-0.1);

\draw (\deb+0.5+\rodsize,\ha-\zaxisline*\inter-0.5) node{{\small $0$}};
\draw (\deb+0.5+3*\rodsize,\ha-\zaxisline*\inter-0.5) node{{\small $\frac{\ell_1^2}{4}$}};

\end{tikzpicture}
\caption{Rod diagram of the shrinking directions on the $z$-axis after sourcing the solutions with a rod that forces the degeneracy of the timelike coordinate. }
\label{fig:rodsourceBTZ}
\end{figure}  

We consider a single rod source, $n=1$,  such that it forces the degeneracy of the timelike coordinate and induces a horizon.  The rod profile has been depicted in Fig.\ref{fig:rodsourceBTZ}.  From Table \ref{tab:internalBC},  only $G_0^{(1)} $ differs from the single-rod solution constructed in section \ref{sec:GlobalAdS3} that led to a global AdS$_3\times$S$^3/\mathbb{Z}_k\times$T$^4$ spacetime.  One can take the same solutions as in \eqref{eq:LinearAdS32} and replace $W_1 \to W_1^{-1}$ or equivalently perform a Wick interchange $(t,y) \to(i\,y, i\,t)$ in \eqref{eq:metGlobalAdS3}.  The type IIB fields for such a rod configuration are then given by
\begin{align}
ds_{10}^2 \= &\frac{1}{\sqrt{Q_1 Q_5}} \left[-r^2\,dt^2 +(r^2+\ell^2) \,dy^2 \right] \+ \sqrt{\frac{Q_1}{Q_5}}\,\sum_{a=1}^4 dx_a^2 \label{eq:metBTZ}\\ 
& \+ k\sqrt{Q_1 Q_5} \left[\frac{dr^2}{r^2+\ell^2} + d\theta^2 + \cos^2 \theta \,d\varphi_1^2+ \sin^2 \theta\,d\varphi_2^2 \right],\nn\\
C^{(2)} \= &k \,Q_5 \cos^2 \theta \,d\varphi_2 \wedge d\varphi_1 -\frac{r^2+\ell^2}{Q_1} \,dt\wedge dy \,,\qquad  e^\Phi \= \sqrt{\frac{Q_1}{Q_5}}\,,\nn
\end{align}
The solution corresponds to \emph{a non-extremal static BTZ black hole with a S$^{\,3}/\mathbb{Z}_k\times$T$^{\,4}$}. The black hole carries $Q_1$ and $Q_5$ D1 and D5 charges,  and the S$^3\times$T$^5$ horizon is located at the rod,  $r=0$. The temperature can be derived from \eqref{eq:ConstraintInternal}
\begin{equation}
T \= \frac{\ell}{2\pi\sqrt{k Q_1 Q_5}}\,,
\label{eq:TemperatureBTZ}
\end{equation}
while the Bekenstein-Hawking entropy is given by\footnotemark
\begin{equation}
S \= \frac{\pi^2 \sqrt{k Q_1 Q_5}\,\ell}{2 G_5}.
\label{eq:EntropyBTZ}
\end{equation}
\footnotetext{The area of the horizon must be derived with the metric in the Einstein frame $ g_E = e^{-\Phi/2} g$ and we have introduced the five-dimensional Newton constant $G_5 = \frac{G_{10}}{(2\pi)^5 R_y R_{x_1}R_{x_2}R_{x_3}R_{x_4}}$. }
One can restrict to solutions where the S$^3$ has no conical defect by simply considering $k=1$ in the above expressions.  The extremal limit is obtained by considering $\ell=0$, that is $T=S=0$.  This corresponds to a BPS D1-D5 black hole that is simply AdS$_3\times$S$^3/\mathbb{Z}_k\times$T$^4$ in the Poincar\'e patch.

\subsection{Black hole bound states without T$^4$ and S$^3$ deformations}

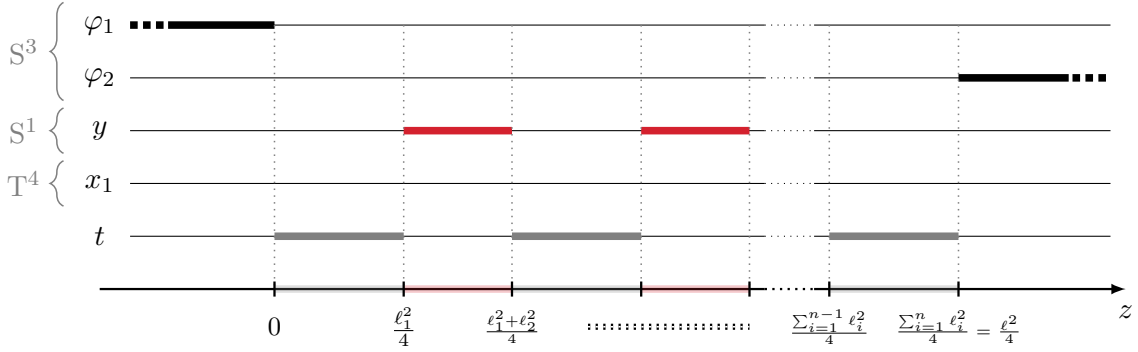
\begin{figure}[h]
\centering
    \begin{tikzpicture}
%% some definitions

\def\deb{-10} 
\def\inter{0.7} 
\def\ha{2.8} 
\def\zaxisline{6} 
\def\rodsize{1.7} 
\def\numrod{5} 

\def\fin{\deb+1+2*\rodsize+\numrod*\rodsize} 

%% 

%\draw (\deb-2,\ha-\zaxisline*0.5*\inter) node{$\Longrightarrow$};

%% Pic title

\draw (\deb+0.5+\rodsize+0.5*\numrod*\rodsize,\ha+1-\inter) node{{{\it Chain of BTZ black holes without T$^4$ and S$^3$ deformations}}};

\draw [decorate, 
    decoration = {brace,
        raise=5pt,
        amplitude=5pt},line width=0.2mm,gray] (\deb-0.7,\ha-2.5*\inter+0.05) --  (\deb-0.7,\ha-0.5*\inter-0.05);
\draw [decorate, 
    decoration = {brace,
        raise=5pt,
        amplitude=5pt},line width=0.2mm,gray] (\deb-0.7,\ha-3.5*\inter+0.05) --  (\deb-0.7,\ha-2.5*\inter-0.05);
\draw [decorate, 
    decoration = {brace,
        raise=5pt,
        amplitude=5pt},line width=0.2mm,gray] (\deb-0.7,\ha-4.5*\inter+0.05) --  (\deb-0.7,\ha-3.5*\inter-0.05);
        
\draw[gray] (\deb-1.4,\ha-1.5*\inter) node{S$^3$};
\draw[gray] (\deb-1.4,\ha-3*\inter) node{S$^1$};
\draw[gray] (\deb-1.4,\ha-4*\inter) node{T$^4$};

%% Each line black line and names

\draw[black,thin] (\deb+1,\ha-\inter) -- (\deb+0.5+4.5*\rodsize+0.2,\ha-\inter);\draw[black,thin,dotted] (\deb+0.5+4.5*\rodsize+0.2,\ha-\inter) -- (\deb+0.5+5*\rodsize,\ha-\inter);
\draw[black,thin] (\deb+0.5+5*\rodsize,\ha-\inter) -- (\fin,\ha-\inter);

\draw[black,thin] (\deb,\ha-2*\inter) -- (\deb+0.5+4.5*\rodsize+0.2,\ha-2*\inter);\draw[black,thin,dotted] (\deb+0.5+4.5*\rodsize+0.2,\ha-2*\inter) -- (\deb+0.5+5*\rodsize,\ha-2*\inter);
\draw[black,thin] (\deb+0.5+5*\rodsize,\ha-2*\inter) -- (\fin-1,\ha-2*\inter);

\draw[black,thin] (\deb,\ha-3*\inter) -- (\deb+0.5+4.5*\rodsize+0.2,\ha-3*\inter);\draw[black,thin,dotted] (\deb+0.5+4.5*\rodsize+0.2,\ha-3*\inter) -- (\deb+0.5+5*\rodsize,\ha-3*\inter);
\draw[black,thin] (\deb+0.5+5*\rodsize,\ha-3*\inter) -- (\fin,\ha-3*\inter);

\draw[black,thin] (\deb,\ha-4*\inter) -- (\deb+0.5+4.5*\rodsize+0.2,\ha-4*\inter);\draw[black,thin,dotted] (\deb+0.5+4.5*\rodsize+0.2,\ha-4*\inter) -- (\deb+0.5+5*\rodsize,\ha-4*\inter);
\draw[black,thin] (\deb+0.5+5*\rodsize,\ha-4*\inter) -- (\fin,\ha-4*\inter);

\draw[black,thin] (\deb,\ha-5*\inter) -- (\deb+0.5+4.5*\rodsize+0.2,\ha-5*\inter);\draw[black,thin,dotted] (\deb+0.5+4.5*\rodsize+0.2,\ha-5*\inter) -- (\deb+0.5+5*\rodsize,\ha-5*\inter);
\draw[black,thin] (\deb+0.5+5*\rodsize,\ha-5*\inter) -- (\fin,\ha-5*\inter);

\draw[black,line width=0.3mm] (\deb-0.4,\ha-\zaxisline*\inter) -- (\deb+0.5+4.5*\rodsize+0.2,\ha-\zaxisline*\inter);\draw[black,line width=0.3mm,dotted] (\deb+0.5+4.5*\rodsize+0.2,\ha-\zaxisline*\inter) -- (\deb+0.5+5*\rodsize,\ha-\zaxisline*\inter);
\draw[black,->, line width=0.3mm] (\deb+0.5+5*\rodsize,\ha-\zaxisline*\inter) -- (\fin+0.2,\ha-\zaxisline*\inter);

\draw (\deb-0.4,\ha-\inter) node{$\varphi_1$};
\draw (\deb-0.4,\ha-2*\inter) node{$\varphi_2$};

\draw (\deb-0.4,\ha-3*\inter) node{$y$};
\draw (\deb-0.4,\ha-4*\inter) node{$x_1$};
\draw (\deb-0.4,\ha-5*\inter) node{$t$};

\draw (\fin+0.2,\ha-\zaxisline*\inter-0.3) node{$z$};

%% First two line and their rods

\draw[black, dotted, line width=1mm] (\deb,\ha-\inter) -- (\deb+0.5,\ha-\inter);
\draw[black,line width=1mm] (\deb+0.5,\ha-\inter) -- (\deb+0.5+\rodsize-0.3,\ha-\inter);
\draw[black,line width=1mm] (\fin-0.5-\rodsize+0.2,\ha-2*\inter) -- (\fin-0.55,\ha-2*\inter);
\draw[black, dotted,line width=1mm] (\fin-0.5,\ha-2*\inter) -- (\fin,\ha-2*\inter);

%% Next lines and their rods

\draw[gray,line width=1mm] (\deb+0.5+\rodsize-0.3,\ha-5*\inter) -- (\deb+0.5+2*\rodsize-0.3,\ha-5*\inter);

\draw[amaranthred,line width=1mm] (\deb+0.5+2*\rodsize-0.3,\ha-3*\inter) -- (\deb+0.5+2.75*\rodsize-0.15,\ha-3*\inter);

\draw[gray,line width=1mm] (\deb+0.5+2.75*\rodsize-0.15,\ha-5*\inter) -- (\deb+0.5+3.75*\rodsize-0.15,\ha-5*\inter);

\draw[amaranthred,line width=1mm] (\deb+0.5+3.75*\rodsize-0.15,\ha-3*\inter) -- (\deb+0.5+4.5*\rodsize,\ha-3*\inter);

\draw[gray,line width=1mm] (\deb+0.5+5*\rodsize+0.2,\ha-5*\inter) -- (\deb+0.5+6*\rodsize+0.2,\ha-5*\inter);

%% Rods on the z-axis

\draw[gray,line width=1mm,opacity=0.25] (\deb+0.5+\rodsize-0.3,\ha-\zaxisline*\inter) -- (\deb+0.5+2*\rodsize-0.3,\ha-\zaxisline*\inter);

\draw[amaranthred,line width=1mm,opacity=0.25] (\deb+0.5+2*\rodsize-0.3,\ha-\zaxisline*\inter) -- (\deb+0.5+2.75*\rodsize-0.15,\ha-\zaxisline*\inter);

\draw[gray,line width=1mm,opacity=0.25] (\deb+0.5+2.75*\rodsize-0.15,\ha-\zaxisline*\inter) -- (\deb+0.5+3.75*\rodsize-0.15,\ha-\zaxisline*\inter);

\draw[amaranthred,line width=1mm,opacity=0.25] (\deb+0.5+3.75*\rodsize-0.15,\ha-\zaxisline*\inter) -- (\deb+0.5+4.5*\rodsize,\ha-\zaxisline*\inter);

\draw[gray,line width=1mm,opacity=0.25] (\deb+0.5+5*\rodsize+0.2,\ha-\zaxisline*\inter) -- (\deb+0.5+6*\rodsize+0.2,\ha-\zaxisline*\inter);

%% Vertical lines and coordinates

\draw[gray,dotted,line width=0.2mm] (\deb+0.5+\rodsize-0.3,\ha-\inter) -- (\deb+0.5+\rodsize-0.3,\ha-\zaxisline*\inter);

\draw[gray,dotted,line width=0.2mm] (\deb+0.5+2*\rodsize-0.3,\ha-\inter) -- (\deb+0.5+2*\rodsize-0.3,\ha-\zaxisline*\inter);
\draw[gray,dotted,line width=0.2mm] (\deb+0.5+2.75*\rodsize-0.15,\ha-\inter) -- (\deb+0.5+2.75*\rodsize-0.15,\ha-\zaxisline*\inter);

\draw[gray,dotted,line width=0.2mm] (\deb+0.5+3.75*\rodsize-0.15,\ha-\inter) -- (\deb+0.5+3.75*\rodsize-0.15,\ha-\zaxisline*\inter);
\draw[gray,dotted,line width=0.2mm] (\deb+0.5+4.5*\rodsize,\ha-\inter) -- (\deb+0.5+4.5*\rodsize,\ha-\zaxisline*\inter);

\draw[gray,dotted,line width=0.2mm] (\deb+0.5+5*\rodsize+0.2,\ha-\inter) -- (\deb+0.5+5*\rodsize+0.2,\ha-\zaxisline*\inter);

\draw[gray,dotted,line width=0.2mm] (\deb+0.5+6*\rodsize+0.2,\ha-\inter) -- (\deb+0.5+6*\rodsize+0.2,\ha-\zaxisline*\inter);

\draw[line width=0.3mm] (\deb+0.5+\rodsize-0.3,\ha-\zaxisline*\inter+0.1) -- (\deb+0.5+\rodsize-0.3,\ha-\zaxisline*\inter-0.1);

\draw[line width=0.3mm] (\deb+0.5+2*\rodsize-0.3,\ha-\zaxisline*\inter+0.1) -- (\deb+0.5+2*\rodsize-0.3,\ha-\zaxisline*\inter-0.1);
\draw[line width=0.3mm] (\deb+0.5+2.75*\rodsize-0.15,\ha-\zaxisline*\inter+0.1) -- (\deb+0.5+2.75*\rodsize-0.15,\ha-\zaxisline*\inter-0.1);

\draw[line width=0.3mm] (\deb+0.5+3.75*\rodsize-0.15,\ha-\zaxisline*\inter+0.1) -- (\deb+0.5+3.75*\rodsize-0.15,\ha-\zaxisline*\inter-0.1);
\draw[line width=0.3mm] (\deb+0.5+4.5*\rodsize,\ha-\zaxisline*\inter+0.1) -- (\deb+0.5+4.5*\rodsize,\ha-\zaxisline*\inter-0.1);

\draw[line width=0.3mm] (\deb+0.5+5*\rodsize+0.2,\ha-\zaxisline*\inter+0.1) -- (\deb+0.5+5*\rodsize+0.2,\ha-\zaxisline*\inter-0.1);

\draw[line width=0.3mm] (\deb+0.5+6*\rodsize+0.2,\ha-\zaxisline*\inter+0.1) -- (\deb+0.5+6*\rodsize+0.2,\ha-\zaxisline*\inter-0.1);

\draw (\deb+0.5+1*\rodsize-0.3,\ha-\zaxisline*\inter-0.5) node{{\small $0$}};

\draw (\deb+0.5+2*\rodsize-0.3,\ha-\zaxisline*\inter-0.5) node{{\small $\frac{\ell_1^2}{4}$}};

\draw (\deb+0.5+2.75*\rodsize-0.15,\ha-\zaxisline*\inter-0.5) node{{\tiny $\frac{\ell_1^2+\ell_{2}^2}{4}$}};

\draw[black,line width=0.3mm,dotted,double] (\deb+0.5+3.25*\rodsize,\ha-\zaxisline*\inter-0.5) -- (\deb+0.5+4.5*\rodsize,\ha-\zaxisline*\inter-0.5);

\draw (\deb+0.5+5*\rodsize+0.2,\ha-\zaxisline*\inter-0.5) node{{\tiny $\frac{\sum_{i=1}^{n-1}\ell_{i}^2}{4}$}};

\draw (\deb+0.5+6*\rodsize+0.2,\ha-\zaxisline*\inter-0.5) node{{\tiny $\frac{\sum_{i=1}^{n}\ell_{i}^2}{4}=\frac{\ell^2}{4}$}};

\end{tikzpicture}
\caption{Rod diagram of the shrinking directions on the $z$-axis after sourcing the solutions with $n$ connected rods that induce alternatively horizons and bolts where the S$^1$ shrinks.}
\label{fig:rodsourceBTZchain1}
\end{figure}  

We construct regular bound states of the non-extremal black holes built in the previous section.  As a first example, we consider a chain of $n$ rods such that the odd rods will correspond to a horizon while the even rod will induce the degeneracy of the $y$ circle.  Moreover,  we consider for simplicity that $n$ is an odd number.  The rod profile has been depicted in Fig.\ref{fig:rodsourceBTZchain1}.  We also restrict to solutions such that the S$^3$ has no conical defect asymptotically, that is $k=1$.

The solution \eqref{eq:TypeIIBAnsatz2}, obtained from \eqref{eq:LinearAdS3} with the rod configuration considered,  gives
\begin{align}
ds_{10}^2 \= &  \frac{1}{\sqrt{Q_1 Q_5}} \left[-\frac{r^2+\ell^2}{\cK_t} \,dt^2 +r^2\,\cK_t \,dy^2 \right] + \sqrt{\frac{Q_1}{Q_5}}\,\sum_{a=1}^4 dx_a^2 \nn \\
&+ \sqrt{Q_1 Q_5} \,\Biggl[ \cK_2\,\left(\frac{dr^2}{r^2+\ell^2}+d\theta^2 \right) + \cos^2 \theta \,d\varphi_1^2+ \sin^2 \theta\,d\varphi_2^2 \Biggr]\,,\label{eq:metAdS3+BTZs} \\
C^{(2)} \= &Q_5 \cos^2 \theta \,d\varphi_2 \wedge d\varphi_1 -\frac{r^2+\ell^2}{Q_1} \,dt\wedge dy \,,\qquad  e^\Phi \= \sqrt{\frac{Q_1}{Q_5}}\,, \nn 
\end{align}
where $\cK_2$ is given in \eqref{eq:DefDefWarpFac} and $\cK_t$ indicates the loci of the horizons by diverging at the odd rods:
\begin{equation}
\cK_t \equi \prod_{\substack{i=1\\i\text{ odd}}}^n \left( 1+\frac{\ell_i^2}{r_i^2}\right)\,.
\end{equation}
\footnotetext{The metric can be written in the Weyl cylindrical coordinate system with \eqref{eq:ChangeMetWeylSp},  and the component along $y$ can be written as 
$r^2 \cK_{t} = (r^2+\ell^2) \prod_{\substack{i=1\\i\text{ even}}}^n\left( 1+\frac{\ell_i^2}{r_i^2}\right)^{-1}$ using \eqref{eq:SimplRelations2}.}
The sources are all localized at $r=0$ and one moves along the chain of rods by changing $\theta$ from $0$ to $\pi/2$.

\subsubsection{Regularity conditions and topology}

The geometry at $r>0$ is regular for the same arguments as given in section \ref{sec:RegGenSol1}.  It has a S$^1\times$S$^3 \times$T$^4$ topology in this region and is asymptotic to AdS$_3\times$S$^3\times$T$^4$.

The rod sources on the $z$-axis are located at $r=0$.  More precisely the $i^\text{th}$ rod is at $r=0$ and in the S$^3$ region,  $\theta_c^{(i)} < \theta < \theta_c^{(i-1)}$ \eqref{eq:DefThetaCrii},  such that $r=r_i=0$ and $ r_j >0$,  $j\neq i$.  
%%%%%%%%%%%%%%%%%%%%%%%%%%%%%%%%
\begin{figure}[t]
\centering
\includegraphics[scale=0.62]{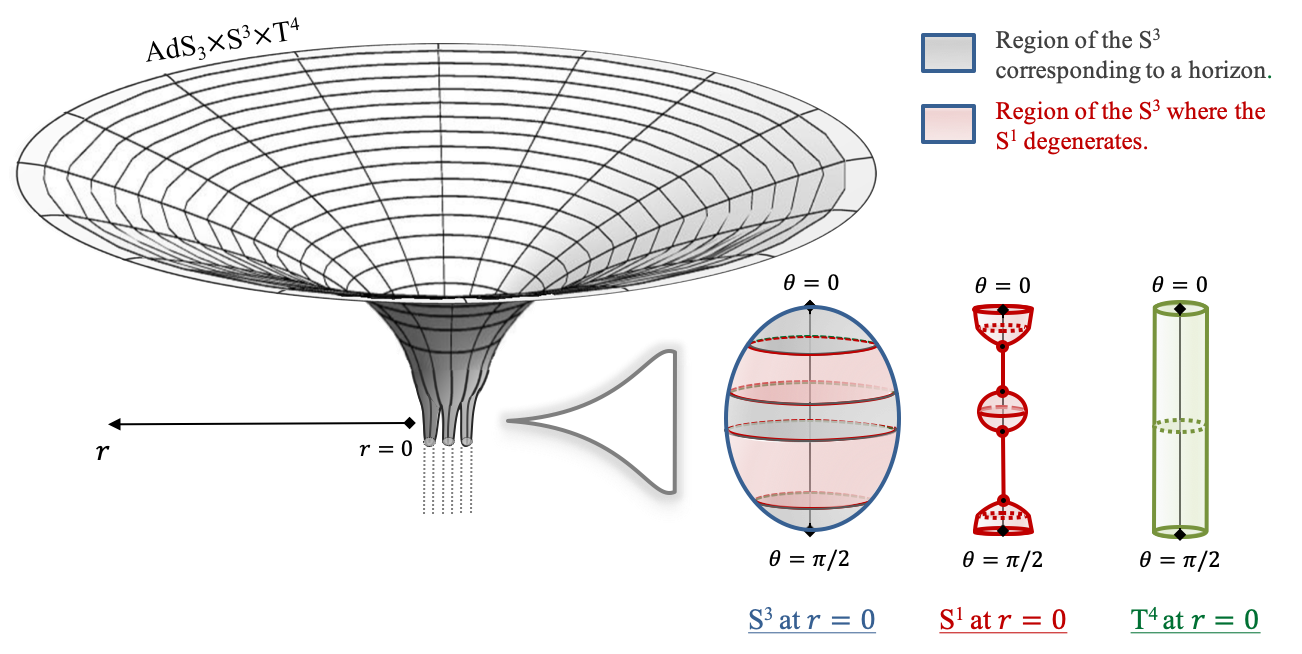}
\caption{Schematic description of the spacetime induced by a chain of $5$ rods,  inducing either horizons or the degeneracy of the S$^1$ ($y$).  On the left-hand side, we depict the overall geometry in terms of $r$.  On the right-hand side,  we describe the behavior of the S$^3$,  S$^1$, and T$^4$ at $r=0$ depending on the position on the S$^3$ given by $\theta$. }
\label{fig:AdS3+BTZpic}
\end{figure}
%%%%%%%%%%%%%%%%%%%%%%%%%%%%%%%%

The S$^1$, parametrized by the $y$ coordinate, degenerates at the even rods,  $r=0$ and $\theta_c^{(2i)} < \theta < \theta_c^{(2i-1)}$,  while the odd rods,  $r=0$ and $\theta_c^{(2i-1)} < \theta < \theta_c^{(2i)}$,  correspond to horizons.  The local geometry at an even rod corresponds to a bolt with a $\IR^2\times \cC_\text{Bubble}$ topology where $\cC_\text{Bubble}$ defines the topology of the smooth bubble at the bolt.  Having regular bolts and black holes in thermal equilibrium will impose $n$ \emph{algebraic bubble and thermal-equilibrium equations} \eqref{eq:ConstraintInternal} that fix all rod lengths $\ell_i^2$ in terms of the asymptotic quantities.  

Then,  $r=0$ will correspond to a chain of bolts and non-extremal black holes, each of them carrying D1-D5 brane charges given by \eqref{eq:ChargeAtRodGen}
\begin{equation}
q_{D1}^{(i)} \= \frac{\ell_i^2}{\ell^2}\,Q_1\,,\qquad q_{D5}^{(i)} \= \frac{\ell_i^2}{\ell^2}\,Q_5\,.
\label{eq:BTZchaincharges1}
\end{equation}
We have depicted the profile of the geometries in Fig.\ref{fig:AdS3+BTZpic}.

We divide the derivation of the bubble and thermal-equilibrium equations depending on whether the rod corresponds to a horizon or a bolt.

\begin{itemize}
\item[•] \underline{Regularity of the $i^\text{th}$ black hole, at the $(2i-1)^\text{th}$ rod:}

We consider the segment
\begin{equation}
r=0\,,\qquad \theta_c^{(2i-1)} < \theta < \theta_c^{(2i)} \quad \Leftrightarrow \quad r=r_{2i-1} = 0 \,,\qquad 0< \theta_{2i-1} < \frac{\pi}{2}\,.
\end{equation}
Since $r=0$ and $r^2 \cK_{t}>0$,   one can check from \eqref{eq:metAdS3+BTZs} that the time coordinate degenerates defining a horizon.  To derive the local geometry at this segment,  one needs to consider $(r_{2i-1},\theta_{2i-1})$ as the main coordinates and take $r_{2i-1}\to 0$. We find that the type IIB metric \eqref{eq:metAdS3+BTZs} converge towards 
\begin{equation}
ds_{10}^2\, \propto\, dr_{2i-1}^2 - \frac{r_{2i-1}^2}{C_{2i-1}^2}\,dt^2 + ds(\cC^{(2i-1)}_\text{Hor})^2\,,
\end{equation}
with
\begin{equation}
\begin{split}
C_{2i-1}^2 \=&\, \frac{Q_1 Q_5\,\ell_{2i-1}^2\,d_{2i-1}^2}{\ell^4} \= \frac{q_{D1}^{(2i-1)} q_{D5}^{(2i-1)}}{\ell_{2i-1}^2}\,d_{2i-1}^2\,\,,
\end{split}
\end{equation}
and $d_{2i-1}$ is the aspect ratio defined in \eqref{eq:DefdiAspect}.  The black hole is in thermal equilibrium if $T=(2\pi C_{2i-1})^{-1}$,  where $T$ is the temperature of the solution. Moreover,  one can show that the three-form field strength is regular such that the component along $t$ vanishes.  Moreover,  since $\ell^2 \cos^2 \theta \sim \ell_{2i}^2 \cos^2 \theta_{2i}+ \sum_{i=1}^{2i-1} \ell_j^2$,  the black hole carries D1-D5 brane charges given by \eqref{eq:BTZchaincharges1}.

The line element,  $ds(\cC^{(2i-1)}_\text{Hor})$,  describes the topology of the horizon.  We have two scenarios.  For the black holes at the extremity of the chain (see Fig.\ref{fig:rodsourceBTZchain1}),  two different spacelike directions shrink at their poles ($\varphi_1$ and $y$ for the first rod and $y$ and $\varphi_2$ for the last), thus defining S$^3\times$T$^5$ horizons.  For the black holes in the middle of the chain,  only $y$ is degenerating at their poles, thereby defining S$^2\times$T$^6$ horizons.  The area of the horizon can be derived following \cite{Bah:2020pdz,Bah:2021owp} and the Bekenstein-Hawking entropy of the $i^\text{th}$ black hole on the chain is given by
\begin{equation}
S_i \= \frac{\pi^2 \sqrt{q_{D1}^{(2i-1)} q_{D5}^{(2i-1)}} \,\ell_{2i-1}\, d_{2i-1}}{2\, G_5} \= \frac{\pi \,\ell_{2i-1}^2}{4T\,G_5}\,,
\end{equation}
where $G_5$ is the five-dimensional Newton constant introduced in \eqref{eq:EntropyBTZ}.

Note that the temperature and entropy resemble the temperature and entropy of a single BTZ black hole \eqref{eq:TemperatureBTZ} and \eqref{eq:EntropyBTZ} in terms of the local D1 and D5 brane charges.  The only difference arises from the aspect ratio $d_{2i-1}$ that accounts for the interaction in between the rods.

\item[•] \underline{Regularity of the $i^\text{th}$ bolt at the $(2i)^\text{th}$ rod:}

We consider the segment
\begin{equation}
r=0\,,\qquad \theta_c^{(2i)} < \theta < \theta_c^{(2i-1)} \quad \Leftrightarrow \quad r=r_{2i} = 0 \,,\qquad 0< \theta_{2i} < \frac{\pi}{2}\,.
\end{equation}
Since $r=0$ and $\cK_{t}>0$,  the $y$ coordinate degenerates.  The time slices of the metric \eqref{eq:metAdS3+BTZs} converge towards 
\begin{equation}
ds_{10}^2|_{dt=0} \propto dr_{2i}^2 + \frac{r_{2i}^2}{C_{2i}^2}\,dy^2 + ds(\cC^{(2i)}_\text{Bubble})^2\,,
\end{equation}
with
\begin{equation}
\begin{split}
C_{2i}^2 \=&\, \frac{Q_1 Q_5\,\ell_{2i}^2\,d_{2i}^2}{\ell^4} \= \frac{q_{D1}^{(2i)} q_{D5}^{(2i)}}{\ell_{2i}^2}\,d_{2i}^2 \,.
\end{split}
\end{equation}
The $(r_{2i},y)$ subspace describes a smooth origin of a $\IR^2$ if we impose $R_{y}=C_{2i}$.

The line element,  $ds(\cC^{(2i)}_\text{Bubble})$,  describes the topology of the bubble at the bolt.  Since the bolt is connected to horizons,  no spacelike direction shrinks at its extremity, $\theta_{2i}=0$ and $\pi/2$.  Then,  $ds(\cC^{(i)}_\text{Bubble})$ describes a T$^7$ space.

Moreover,  the three-form field strength is regular such that the component along $y$ vanishes, and it carries D1-D5 brane charges given by \eqref{eq:BTZchaincharges1}.

\end{itemize}

To summarize,  at $r=0$, the solutions correspond to a static chain of D1-D5 bolts where the S$^1$ direction smoothly degenerates and non-extremal D1-D5 black holes if $n$ algebraic equations are satisfied:
\begin{align}
R_y&\=\frac{\sqrt{Q_1 Q_5}\,\ell_i\,d_i}{\ell^2} =\frac{\sqrt{q_{D1}^{(i)} q_{D5}^{(i)} }}{\ell_i}\,d_i \,,\quad \text{if } i\text{ is even}\,, \\ 
T &\=  \frac{\ell^2}{2\pi\sqrt{Q_1 Q_5} \,\ell_i \,d_i} =\frac{\ell_i}{2\pi \sqrt{q_{D1}^{(i)} q_{D5}^{(i)} }\,d_i}\,,\quad\text{if } i\text{ is odd}.\nn
\end{align}
These equations fix all rod lengths, $\ell_i^2$,  in terms of the boundary quantities, namely the charges,  the radius of the S$^1$, and the temperature.  The solutions are therefore entirely fixed by regularity and the only changeable parameter is the total number of rods.  

The bubble and thermal-equilibrium equations do not have analytic solutions in general, except for small values of $n$ and in the large $n$ limit.  However,  an approximation can be performed considering $R_y T$ small or large.  Indeed, one has $\frac{\ell_{2i}^2}{\ell_{2i-1}^2} = \cO(R_y T)$.  Thus,  in the limit $R_y T\gg 1$,  the rods where the S$^1$ degenerates are much larger than the black hole rods and the solutions can be considered as small black hole perturbations on a global AdS$_3\times$S$^3\times$T$^4$ spacetime that are localized at the center of the global AdS$_3$ and spread along the S$^3$.  Moreover, when $R_y T\ll 1$,  the black hole rods are much larger, and the solutions correspond to small regular perturbations on a static non-extremal BTZ black hole in type IIB that split the horizon into several pieces separated by small bubbles.

The way the horizon splits into different segments where the S$^1$ pinches is reminiscent of the effect of Gregory-Laflamme instability on black strings or black branes \cite{Gregory:1993vy,Gregory:1994bj}. Indeed, one expects the Gregory-Laflamme instability to make a compact circle grow or shrink depending on the position as in \cite{Lehner:2011wc,Emparan:2021ewh}.  Thus, one may wonder whether the present bound state may be a candidate for a final or intermediate state of unstable black strings and black branes.

\subsection{Generic black hole bound states}

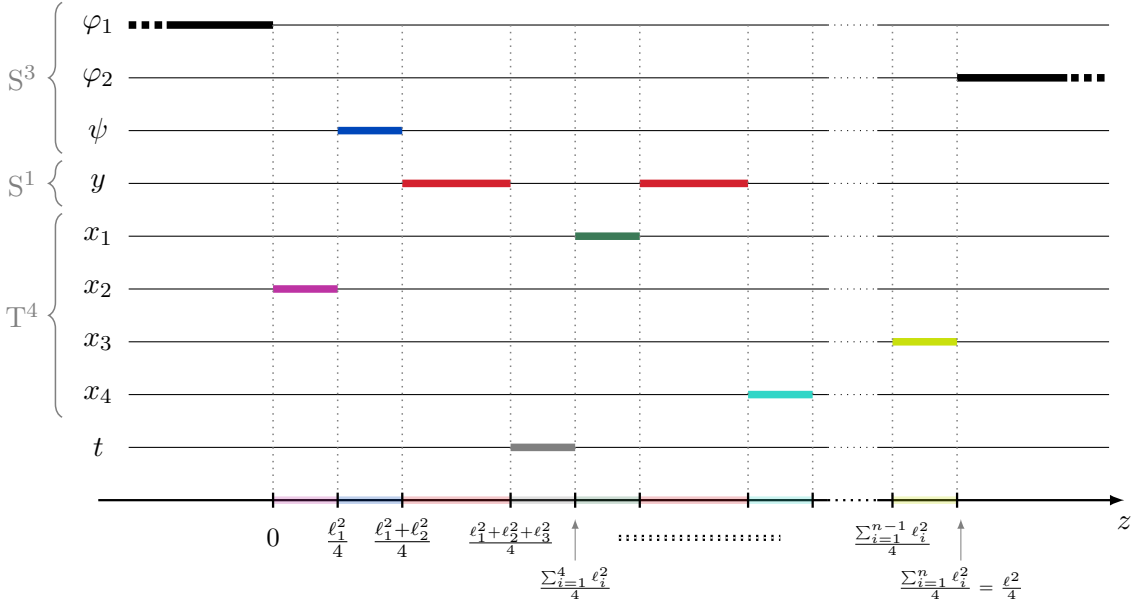
\begin{figure}[h]
\centering
    \begin{tikzpicture}
%% some definitions

\def\deb{-10} 
\def\inter{0.7} 
\def\ha{2.8} 
\def\zaxisline{9} 
\def\rodsize{1.7} 
\def\numrod{5} 

\def\fin{\deb+1+2*\rodsize+\numrod*\rodsize} 

%% 

%\draw (\deb-2,\ha-\zaxisline*0.5*\inter) node{$\Longrightarrow$};

%% Pic title

\draw (\deb+0.5+\rodsize+0.5*\numrod*\rodsize,\ha+1) node{{{\it Generic static axisymmetric solutions in AdS$_3$ with horizons, S$^1$, T$^{\,4}$ and S$^{\,3}$ deformations}}};

\draw [decorate, 
    decoration = {brace,
        raise=5pt,
        amplitude=5pt},line width=0.2mm,gray] (\deb-0.7,\ha-2.5*\inter+0.05) --  (\deb-0.7,\ha+0.5*\inter-0.05);
\draw [decorate, 
    decoration = {brace,
        raise=5pt,
        amplitude=5pt},line width=0.2mm,gray] (\deb-0.7,\ha-3.5*\inter+0.05) --  (\deb-0.7,\ha-2.5*\inter-0.05);
\draw [decorate, 
    decoration = {brace,
        raise=5pt,
        amplitude=5pt},line width=0.2mm,gray] (\deb-0.7,\ha-7.5*\inter+0.05) --  (\deb-0.7,\ha-3.5*\inter-0.05);
        
\draw[gray] (\deb-1.4,\ha-1*\inter) node{S$^3$};
\draw[gray] (\deb-1.4,\ha-3*\inter) node{S$^1$};
\draw[gray] (\deb-1.4,\ha-5.5*\inter) node{T$^4$};

%% Each line black line and names

\draw[black,thin] (\deb+1,\ha) -- (\deb+0.5+5*\rodsize+0.2,\ha);\draw[black,thin,dotted] (\deb+0.5+5*\rodsize+0.2,\ha) -- (\deb+0.5+5.5*\rodsize,\ha);
\draw[black,thin] (\deb+0.5+5.5*\rodsize,\ha) -- (\fin,\ha);

\draw[black,thin] (\deb,\ha-\inter) -- (\deb+0.5+5*\rodsize+0.2,\ha-\inter);\draw[black,thin,dotted] (\deb+0.5+5*\rodsize+0.2,\ha-\inter) -- (\deb+0.5+5.5*\rodsize,\ha-\inter);
\draw[black,thin] (\deb+0.5+5.5*\rodsize,\ha-\inter) -- (\fin-1,\ha-\inter);

\draw[black,thin] (\deb,\ha-2*\inter) -- (\deb+0.5+5*\rodsize+0.2,\ha-2*\inter);\draw[black,thin,dotted] (\deb+0.5+5*\rodsize+0.2,\ha-2*\inter) -- (\deb+0.5+5.5*\rodsize,\ha-2*\inter);
\draw[black,thin] (\deb+0.5+5.5*\rodsize,\ha-2*\inter) -- (\fin,\ha-2*\inter);

\draw[black,thin] (\deb,\ha-3*\inter) -- (\deb+0.5+5*\rodsize+0.2,\ha-3*\inter);\draw[black,thin,dotted] (\deb+0.5+5*\rodsize+0.2,\ha-3*\inter) -- (\deb+0.5+5.5*\rodsize,\ha-3*\inter);
\draw[black,thin] (\deb+0.5+5.5*\rodsize,\ha-3*\inter) -- (\fin,\ha-3*\inter);

\draw[black,thin] (\deb,\ha-4*\inter) -- (\deb+0.5+5*\rodsize+0.2,\ha-4*\inter);\draw[black,thin,dotted] (\deb+0.5+5*\rodsize+0.2,\ha-4*\inter) -- (\deb+0.5+5.5*\rodsize,\ha-4*\inter);
\draw[black,thin] (\deb+0.5+5.5*\rodsize,\ha-4*\inter) -- (\fin,\ha-4*\inter);

\draw[black,thin] (\deb,\ha-5*\inter) -- (\deb+0.5+5*\rodsize+0.2,\ha-5*\inter);\draw[black,thin,dotted] (\deb+0.5+5*\rodsize+0.2,\ha-5*\inter) -- (\deb+0.5+5.5*\rodsize,\ha-5*\inter);
\draw[black,thin] (\deb+0.5+5.5*\rodsize,\ha-5*\inter) -- (\fin,\ha-5*\inter);

\draw[black,thin] (\deb,\ha-6*\inter) -- (\deb+0.5+5*\rodsize+0.2,\ha-6*\inter);\draw[black,thin,dotted] (\deb+0.5+5*\rodsize+0.2,\ha-6*\inter) -- (\deb+0.5+5.5*\rodsize,\ha-6*\inter);
\draw[black,thin] (\deb+0.5+5.5*\rodsize,\ha-6*\inter) -- (\fin,\ha-6*\inter);

\draw[black,thin] (\deb,\ha-7*\inter) -- (\deb+0.5+5*\rodsize+0.2,\ha-7*\inter);\draw[black,thin,dotted] (\deb+0.5+5*\rodsize+0.2,\ha-7*\inter) -- (\deb+0.5+5.5*\rodsize,\ha-7*\inter);
\draw[black,thin] (\deb+0.5+5.5*\rodsize,\ha-7*\inter) -- (\fin,\ha-7*\inter);

\draw[black,thin] (\deb,\ha-8*\inter) -- (\deb+0.5+5*\rodsize+0.2,\ha-8*\inter);\draw[black,thin,dotted] (\deb+0.5+5*\rodsize+0.2,\ha-8*\inter) -- (\deb+0.5+5.5*\rodsize,\ha-8*\inter);
\draw[black,thin] (\deb+0.5+5.5*\rodsize,\ha-8*\inter) -- (\fin,\ha-8*\inter);

\draw[black,line width=0.3mm] (\deb-0.4,\ha-\zaxisline*\inter) -- (\deb+0.5+5*\rodsize+0.2,\ha-\zaxisline*\inter);\draw[black,line width=0.3mm,dotted] (\deb+0.5+5*\rodsize+0.2,\ha-\zaxisline*\inter) -- (\deb+0.5+5.5*\rodsize,\ha-\zaxisline*\inter);
\draw[black,->, line width=0.3mm] (\deb+0.5+5.5*\rodsize,\ha-\zaxisline*\inter) -- (\fin+0.2,\ha-\zaxisline*\inter);

\draw (\deb-0.4,\ha) node{$\varphi_1$};
\draw (\deb-0.4,\ha-\inter) node{$\varphi_2$};
\draw (\deb-0.4,\ha-2*\inter) node{$\psi$};
\draw (\deb-0.4,\ha-3*\inter) node{$y$};
\draw (\deb-0.4,\ha-4*\inter) node{$x_1$};
\draw (\deb-0.4,\ha-5*\inter) node{$x_2$};
\draw (\deb-0.4,\ha-6*\inter) node{$x_3$};
\draw (\deb-0.4,\ha-7*\inter) node{$x_4$};
\draw (\deb-0.4,\ha-8*\inter) node{$t$};

\draw (\fin+0.2,\ha-\zaxisline*\inter-0.3) node{$z$};

%% First two line and their rods

\draw[black, dotted, line width=1mm] (\deb,\ha) -- (\deb+0.5,\ha);
\draw[black,line width=1mm] (\deb+0.5,\ha) -- (\deb+0.5+\rodsize-0.3,\ha);
\draw[black,line width=1mm] (\fin-0.5-\rodsize+0.2,\ha-\inter) -- (\fin-0.55,\ha-\inter);
\draw[black, dotted,line width=1mm] (\fin-0.5,\ha-\inter) -- (\fin,\ha-\inter);

%% Next lines and their rods

\draw[byzantine,line width=1mm] (\deb+0.5+\rodsize-0.3,\ha-5*\inter) -- (\deb+0.5+1.5*\rodsize-0.3,\ha-5*\inter);

\draw[absolutezero,line width=1mm] (\deb+0.5+1.5*\rodsize-0.3,\ha-2*\inter) -- (\deb+0.5+2*\rodsize-0.3,\ha-2*\inter);

\draw[amaranthred,line width=1mm] (\deb+0.5+2*\rodsize-0.3,\ha-3*\inter) -- (\deb+0.5+2.75*\rodsize-0.15,\ha-3*\inter);

\draw[gray,line width=1mm] (\deb+0.5+2.75*\rodsize-0.15,\ha-8*\inter) -- (\deb+0.5+3.25*\rodsize-0.15,\ha-8*\inter);

\draw[amazon,line width=1mm] (\deb+0.5+3.25*\rodsize-0.15,\ha-4*\inter) -- (\deb+0.5+3.75*\rodsize-0.15,\ha-4*\inter);

\draw[amaranthred,line width=1mm] (\deb+0.5+3.75*\rodsize-0.15,\ha-3*\inter) -- (\deb+0.5+4.5*\rodsize,\ha-3*\inter);

\draw[turquoise,line width=1mm] (\deb+0.5+4.5*\rodsize,\ha-7*\inter) -- (\deb+0.5+5*\rodsize,\ha-7*\inter);

\draw[bitterlemon,line width=1mm] (\deb+0.5+5.5*\rodsize+0.2,\ha-6*\inter) -- (\deb+0.5+6*\rodsize+0.2,\ha-6*\inter);

%% Rods on the z-axis

\draw[byzantine,line width=1mm,opacity=0.25] (\deb+0.5+\rodsize-0.3,\ha-\zaxisline*\inter) -- (\deb+0.5+1.5*\rodsize-0.3,\ha-\zaxisline*\inter);

\draw[absolutezero,line width=1mm,opacity=0.25] (\deb+0.5+1.5*\rodsize-0.3,\ha-\zaxisline*\inter) -- (\deb+0.5+2*\rodsize-0.3,\ha-\zaxisline*\inter);
\draw[amaranthred,line width=1mm,opacity=0.25] (\deb+0.5+2*\rodsize-0.3,\ha-\zaxisline*\inter) -- (\deb+0.5+2.75*\rodsize-0.15,\ha-\zaxisline*\inter);

\draw[gray,line width=1mm,opacity=0.25] (\deb+0.5+2.75*\rodsize-0.15,\ha-\zaxisline*\inter) -- (\deb+0.5+3.25*\rodsize-0.15,\ha-\zaxisline*\inter);

\draw[amazon,line width=1mm,opacity=0.25] (\deb+0.5+3.25*\rodsize-0.15,\ha-\zaxisline*\inter) -- (\deb+0.5+3.75*\rodsize-0.15,\ha-\zaxisline*\inter);

\draw[amaranthred,line width=1mm,opacity=0.25] (\deb+0.5+3.75*\rodsize-0.15,\ha-\zaxisline*\inter) -- (\deb+0.5+4.5*\rodsize,\ha-\zaxisline*\inter);
\draw[turquoise,line width=1mm,opacity=0.25] (\deb+0.5+4.5*\rodsize,\ha-\zaxisline*\inter) -- (\deb+0.5+5*\rodsize,\ha-\zaxisline*\inter);

\draw[bitterlemon,line width=1mm,opacity=0.25] (\deb+0.5+5.5*\rodsize+0.2,\ha-\zaxisline*\inter) -- (\deb+0.5+6*\rodsize+0.2,\ha-\zaxisline*\inter);

%% Vertical lines and coordinates

\draw[gray,dotted,line width=0.2mm] (\deb+0.5+\rodsize-0.3,\ha) -- (\deb+0.5+\rodsize-0.3,\ha-\zaxisline*\inter);
\draw[gray,dotted,line width=0.2mm] (\deb+0.5+1.5*\rodsize-0.3,\ha) -- (\deb+0.5+1.5*\rodsize-0.3,\ha-\zaxisline*\inter);
\draw[gray,dotted,line width=0.2mm] (\deb+0.5+2*\rodsize-0.3,\ha) -- (\deb+0.5+2*\rodsize-0.3,\ha-\zaxisline*\inter);
\draw[gray,dotted,line width=0.2mm] (\deb+0.5+2.75*\rodsize-0.15,\ha) -- (\deb+0.5+2.75*\rodsize-0.15,\ha-\zaxisline*\inter);
\draw[gray,dotted,line width=0.2mm] (\deb+0.5+3.25*\rodsize-0.15,\ha) -- (\deb+0.5+3.25*\rodsize-0.15,\ha-\zaxisline*\inter);
\draw[gray,dotted,line width=0.2mm] (\deb+0.5+3.75*\rodsize-0.15,\ha) -- (\deb+0.5+3.75*\rodsize-0.15,\ha-\zaxisline*\inter);
\draw[gray,dotted,line width=0.2mm] (\deb+0.5+4.5*\rodsize,\ha) -- (\deb+0.5+4.5*\rodsize,\ha-\zaxisline*\inter);
\draw[gray,dotted,line width=0.2mm] (\deb+0.5+5*\rodsize,\ha) -- (\deb+0.5+5*\rodsize,\ha-\zaxisline*\inter);
\draw[gray,dotted,line width=0.2mm] (\deb+0.5+5.5*\rodsize+0.2,\ha) -- (\deb+0.5+5.5*\rodsize+0.2,\ha-\zaxisline*\inter);
\draw[gray,dotted,line width=0.2mm] (\deb+0.5+6*\rodsize+0.2,\ha) -- (\deb+0.5+6*\rodsize+0.2,\ha-\zaxisline*\inter);

\draw[line width=0.3mm] (\deb+0.5+\rodsize-0.3,\ha-\zaxisline*\inter+0.1) -- (\deb+0.5+\rodsize-0.3,\ha-\zaxisline*\inter-0.1);
\draw[line width=0.3mm] (\deb+0.5+1.5*\rodsize-0.3,\ha-\zaxisline*\inter+0.1) -- (\deb+0.5+1.5*\rodsize-0.3,\ha-\zaxisline*\inter-0.1);
\draw[line width=0.3mm] (\deb+0.5+2*\rodsize-0.3,\ha-\zaxisline*\inter+0.1) -- (\deb+0.5+2*\rodsize-0.3,\ha-\zaxisline*\inter-0.1);
\draw[line width=0.3mm] (\deb+0.5+2.75*\rodsize-0.15,\ha-\zaxisline*\inter+0.1) -- (\deb+0.5+2.75*\rodsize-0.15,\ha-\zaxisline*\inter-0.1);
\draw[line width=0.3mm] (\deb+0.5+3.25*\rodsize-0.15,\ha-\zaxisline*\inter+0.1) -- (\deb+0.5+3.25*\rodsize-0.15,\ha-\zaxisline*\inter-0.1);
\draw[line width=0.3mm] (\deb+0.5+3.75*\rodsize-0.15,\ha-\zaxisline*\inter+0.1) -- (\deb+0.5+3.75*\rodsize-0.15,\ha-\zaxisline*\inter-0.1);
\draw[line width=0.3mm] (\deb+0.5+4.5*\rodsize,\ha-\zaxisline*\inter+0.1) -- (\deb+0.5+4.5*\rodsize,\ha-\zaxisline*\inter-0.1);
\draw[line width=0.3mm] (\deb+0.5+5*\rodsize,\ha-\zaxisline*\inter+0.1) -- (\deb+0.5+5*\rodsize,\ha-\zaxisline*\inter-0.1);
\draw[line width=0.3mm] (\deb+0.5+5.5*\rodsize+0.2,\ha-\zaxisline*\inter+0.1) -- (\deb+0.5+5.5*\rodsize+0.2,\ha-\zaxisline*\inter-0.1);
\draw[line width=0.3mm] (\deb+0.5+6*\rodsize+0.2,\ha-\zaxisline*\inter+0.1) -- (\deb+0.5+6*\rodsize+0.2,\ha-\zaxisline*\inter-0.1);

\draw (\deb+0.5+1*\rodsize-0.3,\ha-\zaxisline*\inter-0.5) node{{\small $0$}};
\draw (\deb+0.5+1.5*\rodsize-0.3,\ha-\zaxisline*\inter-0.5) node{{\small $\frac{\ell_1^2}{4}$}};
\draw (\deb+0.5+2*\rodsize-0.3,\ha-\zaxisline*\inter-0.5) node{{\small $\frac{\ell_1^2+\ell_2^2}{4}$}};

\draw (\deb+0.5+2.75*\rodsize-0.15,\ha-\zaxisline*\inter-0.5) node{{\tiny $\frac{\ell_1^2+\ell_{2}^2+\ell_{3}^2}{4}$}};
\draw[gray,->,line width=0.1mm] (\deb+0.5+3.25*\rodsize-0.15,\ha-\zaxisline*\inter-0.8) -- (\deb+0.5+3.25*\rodsize-0.15,\ha-\zaxisline*\inter-0.25);
\draw (\deb+0.5+3.25*\rodsize-0.15,\ha-\zaxisline*\inter-1.1) node{{\tiny $\frac{\sum_{i=1}^4\ell_{i}^2}{4}$}};

\draw[black,line width=0.3mm,dotted,double] (\deb+0.5+3.5*\rodsize,\ha-\zaxisline*\inter-0.5) -- (\deb+0.5+4.75*\rodsize,\ha-\zaxisline*\inter-0.5);

\draw (\deb+0.5+5.5*\rodsize+0.2,\ha-\zaxisline*\inter-0.5) node{{\tiny $\frac{\sum_{i=1}^{n-1}\ell_{i}^2}{4}$}};
\draw[gray,->,line width=0.1mm] (\deb+0.5+6*\rodsize+0.2+0.05,\ha-\zaxisline*\inter-0.8) -- (\deb+0.5+6*\rodsize+0.2+0.05,\ha-\zaxisline*\inter-0.25);
\draw (\deb+0.5+6*\rodsize+0.2+0.05,\ha-\zaxisline*\inter-1.1) node{{\tiny $\frac{\sum_{i=1}^n\ell_{i}^2}{4}=\frac{\ell^2}{4}$}};

\end{tikzpicture}
\caption{Rod diagram of the shrinking directions on the $z$-axis after sourcing the solutions with $n$ connected rods that correspond to regular coordinate degeneracies}
\label{fig:rodsourceAdS3+T4+S3+BTZ}
\end{figure}  

In this section,  we construct the most generic static axially-symmetric solutions one can construct within our linear framework in type IIB.  They consist of the same solutions as in section \ref{sec:AdS3+T4+S3s} with additional rods  that force the timelike coordinate to degenerate inducing horizons on the $z$-axis.  A generic rod configuration has been depicted in Fig.\ref{fig:rodsourceAdS3+T4+S3+BTZ}.  We have $n$ connected rod sources,  each one corresponding to one of the seven possible loci detailed

In addition to the conventions of section \ref{sec:Conventions},  we have an extra set of rod labels,  $U_t$,  such that $i\in U_t$ implies that the $i^\text{th}$ rod corresponds to a horizon.  More precisely,  the weights associated with this rod,  which fix the type IIB fields in the linear branch of solutions are given by (see Table \ref{tab:internalBC}):
$P_i^{(0)} = P_i^{(1)} = P_i^{(5)} = G_i^{(1)} = \frac{1}{2}$ and $ G_i^{(0)}= G_i^{(2)}= G_i^{(3)}= G_i^{(4)}= 0\,.$
The type IIB metric and fields \eqref{eq:TypeIIBAnsatz2}, obtained from \eqref{eq:LinearAdS3},  gives\footnotemark
\begin{align}
ds_{10}^2 \= & \frac{1}{\sqrt{Q_1 Q_5 \cK_1\cK_-}} \left[-\frac{r^2+\ell^2}{\cK_t} dt^2 + r^2 \cK_t \cK_{x_1}\cK_{x_2}\cK_{x_3}\cK_{x_4}\cK_\psi\,dy^2 \right]+\sqrt{\frac{Q_1\,\cK_1}{Q_5\,\cK_-}}  \,\sum_{a=1}^4\frac{dx_a^2}{ \cK_{x_a}} \nn \\
&\+ k \sqrt{Q_1 Q_5 \cK_1 \cK_-}\, \Biggl[  \cK_2  \cK_+ \left(\frac{dr^2}{r^2+\ell^2}+d\theta^2 \right) + \cK_+\cos^2 \theta \sin^2\theta \,d\phi^2 \nn \\
&\hspace{3.3cm} \+ \frac{1}{4k^2 \cK_+ \cK_\psi} \left(d\psi + k \cA_+ d\phi \right)^2 \Biggr] \,, \label{eq:met1AdS3+BTZ+S3s} \\
C^{(2)} \= &\frac{Q_5}{4}\,\cA_{-} d\phi \wedge d\psi -\frac{r^2+\ell^2}{Q_1\,\cK_1} \,dt\wedge dy \,,\qquad  e^\Phi \= \sqrt{\frac{Q_1\,\cK_1}{Q_5\,\cK_-}}\,, \hspace{5cm}\nn 
\end{align}
\footnotetext{The metric can be written in the Weyl cylindrical coordinate system with \eqref{eq:ChangeMetWeylSp},  and the component along $y$ can be written as 
$ r^2 \cK_t \cK_{x_1}\cK_{x_2}\cK_{x_3}\cK_{x_4}\cK_\psi = (r^2+\ell^2) \prod_{i\in U_y}\left( 1+\frac{\ell_i^2}{r_i^2}\right)^{-1}$ using \eqref{eq:SimplRelations2}.}
where the deformation factors have been defined in \eqref{eq:DefDefWarpFac} and \eqref{eq:DefDefWarpFac2},  but $\cK_t$ and $\cK_1$ are now given by
\begin{equation}
\cK_1 \= \frac{r^2+\ell^2}{\sum_{i\in U_y \cup U_t} \ell_i^2}\,\left(1-\prod_{i\in U_y \cup U_t} \left(1+\frac{\ell_i^2}{r_i^2} \right)^{-1}\right)\,,\qquad \cK_{t} \= \prod_{i\in U_{t}} \left(1+\frac{\ell_i^2}{r_i^2} \right)\,.
\end{equation}

%%%%%%%%%%%%%%%%%%%%%%%%%%%%%%%%
\begin{figure}[t]
\centering
\includegraphics[scale=0.7]{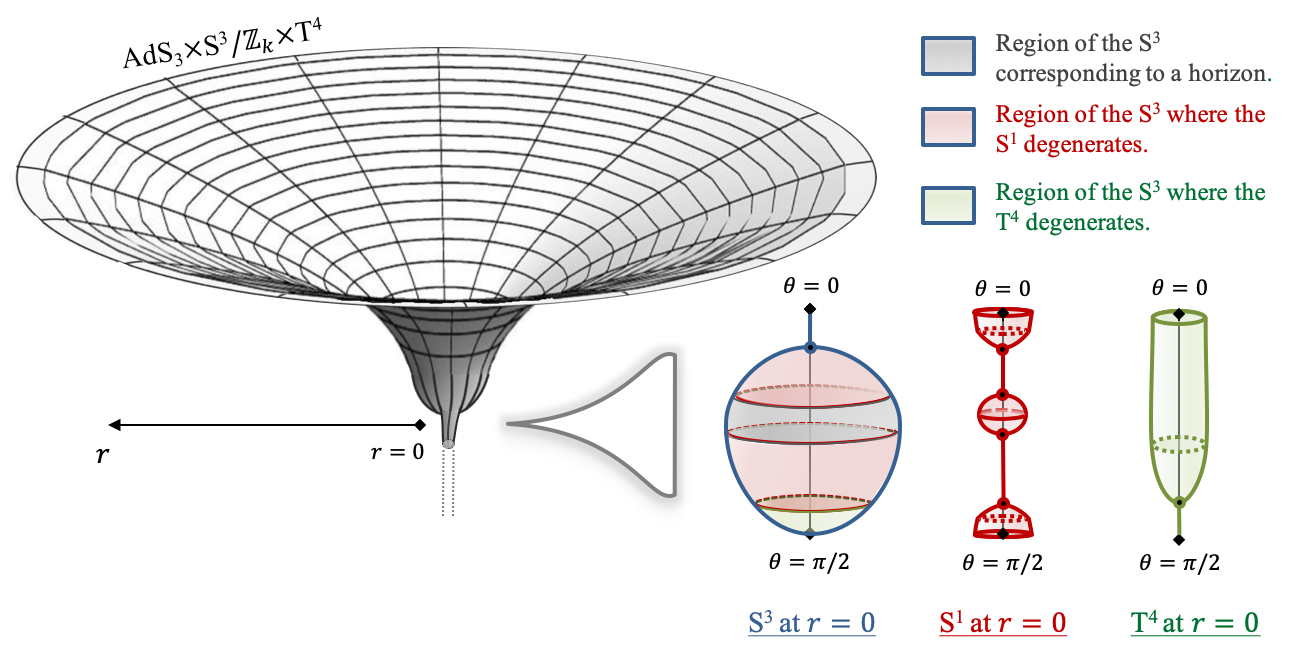}
\caption{Schematic description of the spacetime induced by a chain of $5$ rods,  inducing a horizon or the degeneracy of either the S$^1$ ($y$) or a T$^4$ direction or the Hopf angle of the S$^3$.  On the left-hand side, we depict the overall geometry in terms of $r$.  On the right-hand side,  we describe the behavior of the S$^3$,  S$^1$, and T$^4$ at $r=0$ and as a function of $\theta$. }
\label{fig:AdS3+T4+S3+BTZspic}
\end{figure}
%%%%%%%%%%%%%%%%%%%%%%%%%%%%%%%%

As for the previous geometries,  the solutions are regular for $r>0$ and are asymptotic to AdS$_3\times$S$^3/\mathbb{Z}_k\times$T$^4$. The chain of rods is located at $r=0$ and divided into different sections of $\theta$ delimited by the critical angles $\theta_c^{(i)}$ \eqref{eq:DefThetaCrii}.  The locus of each rod is given by $r=0$ and $\theta_c^{(i-1)}< \theta < \theta_c^{(i)}$ where $r_i = r= 0$ and $r_j>0$ for $j\neq i$.  Depending if the rod induces the degeneracy of the Hopf angle $\psi$,  or a T$^4$ direction or the S$^1$ or the timelike direction,  the regularity condition imposes
\begin{align}
\frac{R_y} {\sqrt{k Q_1 Q_5}}\=&\frac{\ell\,\ell_i\,d_i}{\sqrt{\sum_{p\in U_y\cup U_t} \ell_p^2\times \left( \ell^4-\left(\sum_{p\in U_\psi} \ell_p^2\right)^2\right)}} \,,\quad \text{if }i\in U_y\,,\label{eq:RegGenGen}\\
\sqrt{k Q_1 Q_5}\,T\=&\frac{\sqrt{\sum_{p\in U_y\cup U_t} \ell_p^2\times \left( \ell^4-\left(\sum_{p\in U_\psi} \ell_p^2\right)^2\right)}}{2\pi\,\ell\,\ell_i\,d_i} \,,\quad \text{if }i\in U_t\,,\hspace{5cm}\nn
\end{align}
\begin{align}
\frac{R_{x_a}}{\sqrt{k Q_5}} \= & \frac{\ell \, \ell_i\,d_i}{\sqrt{\ell^4-\left(\sum_{p\in U_\psi} \ell_p^2\right)^2}} \,,\quad \text{if }i\in U_{x_a},\qquad \frac{1}{k}\=  \frac{\ell\,\ell_i\,d_i}{\ell^2+\sum_{i\in U_\psi} \ell_i^2  }\,,\quad \text{if }i\in U_{\psi},\nn
\end{align}
where $d_i$ is defined in \eqref{eq:DefdiAspect}.  We depicted generic geometries in the global spherical coordinates in Fig.\ref{fig:AdS3+T4+S3+BTZspic}.  The spacetime ends at $r=0$ as a chain of bolts and horizons of non-extremal black holes delimited in sections of $\theta$.

Depending on their nature,  the bolts and black holes may carry D1 and D5 brane charges.  The black holes and the bolts where the S$^1$ degenerates have non-zero D1 and D5 charges,  while the bolts where a T$^4$ direction shrinks carry a D5 charge only, and the bolts where $\psi$ degenerates have no charges. More precisely,  we have \eqref{eq:ChargeAtRodGen}
\begin{equation}
\begin{split}
q_{D1}^{(i)} &\= \frac{\ell_i^2}{\sum_{j\in U_y\cup U_t}\ell_j^2}\,Q_1\,,\qquad q_{D5}^{(i)} \= \frac{\ell_i^2}{\ell^2- \sum_{i\in U_\psi} \ell_i^2}\,Q_5\,,\qquad i\in U_y\text{ or } U_t\,,\\
q_{D1}^{(i)} &\=0\,,\qquad q_{D5}^{(i)} \= \frac{\ell_i^2}{\ell^2- \sum_{i\in U_\psi} \ell_i^2}\,Q_5\,,\qquad i\in U_{x_a}\,, \\
q_{D1}^{(i)}& \= q_{D5}^{(i)} \=0 \,,\qquad i\in U_{\psi}\,.\\
\end{split}
\end{equation}

Moreover,  the areas of the horizons in the chain are integrable and we find that the black hole Bekenstein-Hawking entropies are given by
\begin{equation}
S_i \= \frac{\pi^2\,\sqrt{k \,q_{D1}^{(i)} q_{D5}^{(i)}}\,\, \ell_i \ell\,d_i}{2G_5\,\sqrt{\ell^2+\sum_{p\in U_\psi} \ell_p^2}}\=  \frac{\pi \,\ell_{i}^2}{4T\,G_5}\,.
\end{equation}

The regularity constraints \eqref{eq:RegGenGen} are not solvable for generic $n$. However,  assuming large or small values for $R_y T$,  $\frac{R_{x_a}}{\sqrt{k Q_5}}$ and $k$ induces discrepancy in scales in between the rods of different nature,  and the geometries can be considered as small regular perturbations on a simpler asymptotically-AdS$_3$ background.

%%%%%%%%%%%%%%%%%%%%%%%%%%%%%%%%%%%%%
\section*{Acknowledgments}
%%%%%%%%%%%%%%%%%%%%%%%%%%%%%%%%%%%%% 
The work of IB and PH is supported in part by NSF grant PHY-2112699.  The
work of IB is also supported in part by the Simons Collaboration on Global Categorical Symmetries. We are grateful to Iosif Bena,  Bogdan Ganchev,  Gary Horowitz,  Anthony Houppe,  Rodolfo Russo,  Masaki Shigemori,  David Turton,  Nick Warner, and Peter Weck for interesting and stimulating discussions.

\appendix

\section{Einstein-Maxwell equations for the static D1-D5-P system}
\label{App:EOMtypeIIB}

In this section,  we derive the equations of motion for static D1-D5-P backgrounds in type IIB supergravity.  In \cite{Heidmann:2021cms},  the equations of motion for static M2-M2-M2 solutions in M-theory have been derived.  By performing a series of T-dualities (see section 3.3 of \cite{Heidmann:2021cms}),  one can find the equations given in \eqref{eq:EOMVac} and \eqref{eq:EOMMaxwell} after a field redefinition.  However,  here we perform a more direct calculation by working directly in the type IIB framework.

\subsection{Type IIB action and equations}

The bosonic pseudo-action of type IIB supergravity in the Einstein frame is given by \cite{Schwarz:1983qr,Schwarz:1983wa,Fernandez:2008vh,Hamilton:2016ito},
\begin{align}
S_{IIB} &\= S_{NS} \+ S_{RR} \+ S_{CS}\,,\nn\\
(16\pi G_{10})\,S_{NS} &\equi \int R \,\star \mathbb{I} - \frac{1}{2} d\Phi \wedge \star d\Phi - \frac{e^{-\Phi}}{2} H_3 \wedge \star H_3\,,\\
(16\pi G_{10})\,S_{RR} &\equi -\frac{1}{2} \int e^{2\Phi}\,F_1 \wedge \star F_1 + e^{\Phi} F_3 \wedge \star F_3 + \frac{1}{2} F_5 \wedge \star F_5\,,\nn\\
(16\pi G_{10})\,S_{CS} &\equi - \frac{1}{2} \int  C^{(4)} \wedge H_3 \wedge F_3 \,.\nn
\end{align}
where $G_{10}$ is the ten-dimensional Newton constant,  $R$ is the Ricci stalar,  $\Phi$ is the dilaton, $H_3=dB_2$ is the NS-NS three-form field strength,  and $F_p$ are the R-R $p$-form field strengths given in terms of gauge potentials as
\begin{equation}
F_1 \= dC^{(0)} \,,\qquad F_3 \= dC^{(2)} - C_0\,H_3 \,,\qquad F_5 \= dC^{(4)} -\frac{1}{2} H_3 \wedge C^{(2)}-\frac{1}{2} B_2 \wedge dC^{(2)}\,.
\end{equation}
First, the dilaton and Maxwell equations are generically
\begin{align}
d\left(e^{2\Phi} \star F_1\right) &\= - e^{\Phi}\, H_3 \wedge \star F_3\,,\qquad d\star F_5 \= H_3 \wedge F_3\,,\qquad d\left(e^{\Phi} \star F_3\right) \= - H_3 \wedge \star F_5\,, \nn \\
d\left(e^{-\Phi} \star H_3\right)& \=  F_3 \wedge \star F_5 + e^{\Phi} \,F_1 \wedge \star F_3\,, \label{eq:Max&DilatonApp}\\
d\star d \Phi &\=  -\frac{e^{-\Phi}}{2} H_3 \wedge \star H_3 + e^{2\Phi}\,F_1 \wedge \star F_1 + \frac{e^{\Phi}}{2} F_3 \wedge \star F_3 \,.\nn
\end{align}
Moreover,  one has to impose the self-duality equation of $F_5$ and the Bianchi identities:
\begin{equation}
\star F_5 \= F_5 \,,\qquad dH_3 \= dF_1 \= 0\,,\qquad dF_3 \= H_3\wedge F_1 \,,\qquad dF_5 \= H_3 \wedge F_3\,.
\label{eq:BianchiApp}
\end{equation}
Finally,  the Einstein equations are given by
\begin{equation}
\begin{split}
2\,R_{\mu \nu} \=& \left[T(d\Phi)+e^{-\Phi}\, T(H_3)+e^{2\Phi}\, T(F_1) +e^{\Phi}\, T(F_3) +T(F_5)\right]_{\mu \nu} \\
&- \frac{g_{\mu \nu}}{8}  \left[T(d\Phi)+e^{-\Phi}\, T(H_3)+e^{2\Phi}\, T(F_1) +e^{\Phi}\, T(F_3) +T(F_5)\right]_{\sigma}^{\,\,\,\sigma}\,,
\end{split}
\label{ex:EinsteinApp}
\end{equation}
where $T(\cF)$ is the stress tensor of the $p$-form $\cF$ given by
\begin{equation}
T(\mathcal{F})_{\mu \nu}\equi\frac{1}{(p-1) !}\left[\mathcal{F}_{\mu \alpha_2 \ldots \alpha_p} \mathcal{F}_{\nu}^{\,\,\,\alpha_2 \ldots \alpha_p}\-\frac{1}{2 p} g_{\mu \nu} \,\mathcal{F}_{\alpha_1 \ldots \alpha_p} \mathcal{F}^{\alpha_1 \ldots \alpha_{p}}\right]\,.
\end{equation}

\subsection{Equations for static axially-symmetric D1-D5-P backgrounds}

We aim to derive equations of motion for static D1-D5-P solutions with eight commuting Killing vectors.  We therefore consider a T$^4$ and a S$^1$,  parametrized by $(x_1,x_2,x_3,x_4)$ and $y$ respectively.   The T$^4\times$S$^1$ and the S$^1$ are wrapped by D5 branes and D1 branes respectively,  and one allows for P momentum charges along the S$^1$.  The remaining four spacelike directions are expressed as a S$^1$ fibration over a three-dimensional base space.  We parametrize the S$^1$ by an angle $\psi$ while the base is given in cylindrical coordinates, $(\rho,z,\phi)$,  such that all functions depend on $\rho$ and $z$. A generic ansatz of metric and field is therefore given by 
\begin{align}
ds_{10}^2 = &   \frac{-dt^2}{V_t}+ \frac{(dy-T_p dt)^2}{V_y}  + \sum_{i=1}^{4} \frac{dx_i^2}{V_{x_i}}+ V_B \left[\frac{ \left(d\psi+H_0 d\phi\right)^2}{Z_0} +Z_0 \left(e^{2\nu} \left( d\rho^2 +dz^2\right) + \rho^2d\phi^2 \right) \right] \nn\\
C^{(2)} = &H_5 \,d\phi \wedge d\psi -T_1 \,dt\wedge dy \,,\qquad  e^\Phi \= V_\Phi \,,\qquad C^{(0)}=C^{(4)}=B_2=0\,,
\label{eq:TypeIIBAnsatzApp}
\end{align}
where $(T_1,H_5,T_p)$ are the gauge potentials for the D1-D5-P flux, $H_0$ is a KK gauge potential,  and $(V_t,V_y,V_{x_i},V_B,Z_0,e^{2\nu})$ are metric scalars.

\subsubsection{Maxwell and dilaton equations}

First,  we notice that the Bianchi identities \eqref{eq:BianchiApp} are directly satisfied while the Maxwell and dilaton equations \eqref{eq:Max&DilatonApp} lead to
\begin{equation}
\begin{split}
&\partial_a \left( \frac{V_\Phi}{\rho V_t V_y \prod_{i=1}^4 V_{x_i}}\,\partial^a H_5 \right) \= 0\,,\qquad \partial_a \left( \rho V_t V_y V_\Phi\,\partial^a T_1 \right) \= 0\,, \\
& \cG_\Phi \equi \frac{2}{V_\Phi}\,\partial_a \left(\rho \, \partial^a \log V_\Phi \right) + \rho V_t V_y \,(\partial_a T_1)^2 - \frac{1}{\rho V_t V_y \prod_{i=1}^4 V_{x_i}}\, (\partial_a H_5)^2 \=0\,,
\end{split}
\label{eq:Max&DilatonSimApp}
\end{equation}
where the label $a$ runs over the flat two-dimensional $(\rho,z)$ space.

\subsubsection{Ricci tensor}

We derive the Ricci tensor in the tetrad frame by successively computing the tetrad one-forms,  $E^M$,  the spin connection, $dE^M=E^N \wedge \omega^M_{\,\,\, N}$, the curvature two-form, $\mathcal{R}_{\,\,\,N}^M=d \omega_{\,\,\,N}^M+\omega_{\,\,\,O}^M \wedge \omega_{\,\,\,N}^O$, and finally the Ricci tensor,  $R_{MN}= \mathcal{R}_{\,\,\,MON}^O$.  We introduce
\begin{equation}
V_0 \equi \frac{\sqrt{-\det g}}{e^{2\nu} V_B Z_0} \= \frac{\rho V_B}{\sqrt{V_t V_y \prod_{i=1}^4 V_{x_1}}}\,,
\end{equation}
and we find
\begin{align}
2\sqrt{-\det g} \, R_{tt}& \= - \partial_a \left(V_0\, \partial^a \log V_t\right) -\frac{V_0\,V_t}{V_y}\,(\partial_a T_p)^2 \,,\nn\\
2\sqrt{-\det g} \, R_{yy}& \=  \partial_a \left(V_0\, \partial^a \log V_y\right) -\frac{V_0\,V_t}{V_y}\,(\partial_a T_p)^2 \,,\nn\\
2\sqrt{-\det g} \, R_{x_i x_i} &\=  \partial_a \left(V_0\, \partial^a \log V_{x_i}\right) \,,\qquad i=1,2,3,4\,,\\
2\sqrt{-\det g} \, R_{\psi \psi}& \=  -\partial_a \left(V_0\, \partial^a \log \frac{V_B}{Z_0}\right) + \frac{V_0}{\rho^2 Z_0^2}\,(\partial_a H_0)^2 \,,\nn\\
2\sqrt{-\det g} \, R_{\phi \phi}& \=  -\partial_a \left(V_0\, \partial^a \log (\rho^2 V_B Z_0)\right) -\frac{V_0}{\rho^2 Z_0^2}\,(\partial_a H_0)^2 \,,\nn\\
2\sqrt{-\det g} \, R_{t y}& \= \sqrt{\frac{V_y}{V_t}} \, \partial_a \left( \frac{V_0\,V_t}{V_y}\,\partial^a T_p\right)\,, \qquad 2\sqrt{-\det g} \, R_{\phi \psi} \=-\rho Z_0 \, \partial_a \left( \frac{V_0}{\rho^2 Z_0^2}\,\partial^a H_0\right)\,, \nn 
\end{align}
\begin{align}
2 e^{2\nu} V_B Z_0 \,R_{ab}&\=-2 \partial_a \partial_b \log V_0 - \frac{1}{2} \sum_{w=t,y,x_i} \partial_a \log V_w \,\partial_b \log V_w - \frac{1}{2} \partial_a \log\frac{V_B}{Z_0} \, \partial_b \log\frac{V_B}{Z_0} \nn \\
&- \frac{1}{2} \partial_a \log(\rho^2 V_B Z_0) \partial_b \log(\rho^2 V_B Z_0)- \frac{1}{\rho^2 Z_0^2} \,\partial_a H_0 \,\partial_b  H_0 + \frac{V_t}{V_y} \,\partial_a T_p ,\partial_b  T_p \nn \\
&+\partial_{(a} \log (e^{2\nu} V_B Z_0) \, \partial_{b)} \log V_0 - \delta_{ab} \left[ \partial_c \partial^c (e^{2\nu} V_B Z_0) +\partial_{c} \log (e^{2\nu} V_B Z_0) \, \partial^c \log V_0\right] \,, \nn
\end{align}
where the labels $a,b,c$ run over the flat two-dimensional $(\rho,z)$ space.

\subsubsection{Stress tensors}

We compute the stress tensors of the dilaton and the three-form field strength in the tetrad frame and find
\begin{align}
2\sqrt{-\det g}\,\,T(d\Phi)_{MN} &\= -V_0 \left(\partial_a \log V_\Phi\,\partial^a \log V_\Phi \, \,  \eta_{MN}-2 \partial^a \log V_\Phi\,\partial^b \log V_\Phi \,\delta_{aM} \delta_{bN} \right) \,,\nn\\
2\sqrt{-\det g}\,\,T(F_3)_{MN} &\= V_0 \Biggl[V_t V_y \,\partial_a T_1 \, \partial^a T_1 \, \left( \eta_{MN} + 2\delta_{M t}\delta_{N t}-2\delta_{M y}\delta_{N y}\right)  \\
&\hspace{1.1cm} - \frac{1}{\rho^2 V_B^2} \,\partial_a H_5 \, \partial^a H_5 \, \left( \eta_{MN} - 2\delta_{M \phi}\delta_{N \phi}-2\delta_{M \psi}\delta_{N \psi}\right)  \nn \\
&\hspace{1.1cm} -2 \left(V_t V_y \,\partial^a T_1 \,\partial^b T_1 -  \frac{1}{\rho^2 V_B^2}\,\partial^a H_5 \,\partial^b H_5\right)\,\delta_{aM} \delta_{bN}\Biggr]\,. \nn
\end{align}

\subsubsection{Einstein equations and field redefinition}

We now simplify the equations obtained from \eqref{ex:EinsteinApp}.  For that purpose, we introduce
\begin{equation}
\cG_{MN} \equi \sqrt{-\det g}\left( R_{MN} \- \frac{1}{2} \left[T(d\Phi)+e^{\Phi}\, T(F_3)\right]_{MN} + \frac{g_{MN}}{16}  \left[T(d\Phi)+e^{\Phi}\, T(F_3) \right]_{P}^{\,\,\,P}\right)\,,
\end{equation}
which vanish for on-shell solutions.  First, we notice that 
\begin{equation}
 \cG_{tt}-\cG_{yy}-\cG_{\phi\phi}-\cG_{\psi\psi}-\sum_{i=1}^4 \cG_{x_i x_i} \= \partial_a \partial^a V_0 \=0\,.
\end{equation}
As explained in \cite{Emparan:2001wk},  one can consider $V_0= \rho$ without restriction.  This is equivalent to performing a conformal transformation of the $(\rho,z)$ coordinate system. These new coordinates are referred to as Weyl's canonical coordinates and imply:
\begin{equation}
V_B \= \sqrt{V_t V_y \prod_{i=1}^4 V_{x_i}}\,.
\end{equation}
Thus all components in the Ricci tensor $\partial^a (V_0\, \partial^a \ldots)$ give $\rho \,\Delta (\ldots)$ where $\Delta$ is the Laplacian operator of a flat three-dimensional base for axisymmetric functions \eqref{eq:Laplacian}.

Then,  by mixing the Einstein equations with the dilaton equation \eqref{eq:Max&DilatonSimApp},  we find 4 harmonic constraints such that
\begin{equation}
\frac{V_\Phi\, \cG_\Phi }{2}+4 \cG_{x_1 x_1} \=\rho\, \Delta \log (V_\Phi V_{x_1}^2) =0\,,\quad 2( \cG_{x_1 x_1} - \cG_{x_i x_i}  ) = \rho\, \Delta \log \frac{V_{x_i}}{V_{x_1}} =0 \,,\quad i=2,3,4\,.
\end{equation}
We have four additional combinations that encompass the interaction between the metric and the gauge potentials:
\begin{align}
&-\frac{1}{\rho}(\cG_{tt}+\cG_{yy} )\=\Delta \log \sqrt{\frac{V_t}{V_y} } +\frac{V_t}{V_y} \,\partial_a T_p \,\partial^a T_p \=0\,,  \nn\\
&\frac{1}{\rho}\left(-\cG_{tt}+\cG_{yy}+ \frac{V_\Phi\, \cG_\Phi }{4} \right) \= \Delta \sqrt{V_\Phi V_t V_y } + V_\Phi V_t V_y  \,\partial_a T_1 \,\partial^a T_1 \=0\,,  \nn\\
&\frac{1}{\rho}(\cG_{tt} -\cG_{yy} - \sum_{i=1}^4 \cG_{x_i x_i}-\cG_{\psi \psi}) \= \Delta \log Z_0 + \frac{1}{\rho^2 Z_0^2} \,\partial_a H_0 \,\partial^a H_0 \=0\,,\nn\\
&\frac{1}{\rho} \left(\cG_{yy}-\cG_{tt} + \sum_{i=1}^4 \cG_{x_i x_i} - \frac{V_\Phi\, \cG_\Phi }{4} \right) = \Delta \log \sqrt{\frac{V_t V_y \prod_{i=1}^4 V_{x_i}}{V_\Phi}} + \frac{V_\Phi}{\rho^2\,V_t V_y \prod_{i=1}^4 V_{x_i}} \,\partial_a H_5 \,\partial^a H_5 =0.\nn
\end{align}
They are accompanied by the Maxwell equations for $(H_5,T_1)$ \eqref{eq:Max&DilatonSimApp} and we find for $(H_0,T_p)$:
\begin{equation}
\begin{split}
&\frac{-2}{\rho Z_0}\cG_{\phi\psi} \= \partial_a \left( \frac{1}{\rho Z_0^2}\,\partial^a H_0 \right) \= 0\,,\qquad 2 \sqrt{\frac{V_t}{V_y}}\,\cG_{ty}\= \partial_a \left( \rho\frac{V_t}{V_y} \,\partial^a T_p \right) \= 0\,, 
\end{split}
\end{equation}
At this level,  the seven metric functions,  $(V_t,V_y,V_{x_i},Z_0)$, the dilaton function $V_\Phi$,  and the four gauge potentials $(H_0,H_5,T_1,T_p)$ are entirely constrained by the twelve equations given above.  Three equations,  obtained from $(\cG_{\rho\rho},\cG_{\rho z},\cG_{z z})$,  fix the base warp factor $e^{2\nu}$ by constraining $\partial_{\rho} \nu$ and $\partial_z \nu$.  However,  before simplifying those last equations,  one can redefine appropriately the metric functions:
\begin{equation}
\begin{split}
V_t& \= Z_p\,\left(\frac{Z_1^3 Z_5}{W_0}\right)^\frac{1}{4}\,,\qquad V_y \= \frac{1}{Z_p}\,\left(\frac{Z_1^3 Z_5}{W_0}\right)^\frac{1}{4}\,,\qquad V_{x_i} \= \frac{1}{W_{i+1}}\,\left(\frac{Z_5 W_0}{Z_1}\right)^\frac{1}{4}\,,\quad i=1,2,3\,, \\
V_{x_4}&\= W_2 W_3 W_4 \left(\frac{Z_5 W_0}{Z_1}\right)^\frac{1}{4}\,,\qquad V_\Phi \= \sqrt{\frac{Z_1 W_0}{Z_5}}\,.
\end{split}
\end{equation}
By plugging into the equations above we find that the logarithms of $W_\Lambda$ are harmonic functions, $$\Delta \log W_\Lambda\=0,\qquad \Lambda=0,2,3,4\,,$$
and the $Z_I$ couples to the gauge potentials such that
\begin{align}
&\Delta \log Z_I + Z_I^2 \,\partial_a T_I \,\partial^a T_I \=0\,,\qquad \partial_a \left(\rho Z_I\,\partial^a T_I \right) \=0\,,\qquad I=1,p\,, \\
&\Delta \log Z_I + \frac{1}{\rho^2 Z_I^2} \,\partial_a H_I \,\partial^a H_I \=0\,,\qquad \partial_a \left(\frac{1}{\rho Z_I}\,\partial^a H_I \right) \=0\,.\qquad I=0,5\,,.\nn
\end{align}
Moreover,  the ansatz of metric and dilaton \eqref{eq:TypeIIBAnsatzApp} becomes 
\begin{equation}
\begin{split}
ds_{10}^2 &=  e^{-\frac{\Phi}{2}}\,\Biggl[\sqrt{\frac{W_0}{Z_1 Z_5}} \,\left[ - Z_p \,dt^2+ \frac{(dy-T_p dt)^2}{Z_p}\right]  \+ \sqrt{\frac{Z_1}{Z_5}}\,\left[\sum_{i=1}^{3} W_{a+1}\,dx_a^2+\frac{dx_4^2}{W_2 W_3 W_4}\right] \\
&+ \sqrt{W_0\,Z_1 Z_5} \left[\frac{1}{Z_0} \left(d\psi+H_0 \,d\phi\right)^2 +Z_0 \,\left(e^{2\nu} \left( d\rho^2 +dz^2\right) + \rho^2d\phi^2 \right) \right]\Biggr] \,,\quad e^\Phi \= \sqrt{\frac{Z_1}{Z_5}\,W_0}\,, \nn
\end{split}
\end{equation}
and the metric in the string frame is obtained by multiplying by $e^\frac{\Phi}{2}$.  We recover the D1-D5 ansatz \eqref{eq:TypeIIBAnsatz} by imposing $T_p=0$ and $Z_p=\frac{1}{W_1}$,  and we obtain the same set of equations as \eqref{eq:EOMVac} and \eqref{eq:EOMMaxwell} by considering the electric duals of the magnetic gauge potentials \eqref{eq:ElecDual}.

Finally,  three equations remain from $(\cG_{\rho\rho},\cG_{\rho z},\cG_{z z})$. However,  they are not independent and we derive the equations for $\nu$ by simplifying $\cG_{\rho \rho}-\cG_{zz}$ and $\cG_{\rho z}$ and by splitting $\nu$ into different contributions as in \eqref{eq:ElecDual}:
\begin{equation}
\nu \= \nu_{Z_1}+\nu_{Z_5} +\nu_{Z_p}+\nu_{Z_0}+\nu_{W_0}+\sum_{i=2}^5 \nu_{W_i}\,.
\end{equation}
The equations for each contribution are identical to \eqref{eq:EOMVac} and \eqref{eq:EOMMaxwell} with $W_5=(W_2 W_3 W_4)^{-1}$.

\section{Non-BPS bubbling deformations in AdS$^3\times$S$^3\times$T$^4$}

We detail the derivations of the smooth geometries constructed in the sections \ref{sec:AdS3} and \ref{sec:AdS3Gen}.  They are derived from the linear branch \eqref{eq:LinearAdS3} of type IIB asymptotically-AdS$_3$ solutions \eqref{eq:TypeIIBAnsatz2} and are sourced by connected rods of different nature.

\subsection{Single T$^4$ deformation}

We discuss the solutions of the sections \ref{sec:AdS3+T4} and \ref{sec:AdS3+T42}. They are obtained from adding sources that only force the T$^4$ to degenerate on a global AdS$_3\times$S$^3\times$T$^4$ background.  

\subsubsection{Single deformation at the pole of the S$^3$}
\label{App:GlobalAdS3+T4}

We consider two connected rod sources on the $z$-axis as depicted in Fig.\ref{fig:rodsourceAdS3+T4}.  The first induces the degeneracy of the $y$-circle while the other forces the $x_1$-circle to shrink.  From Table \ref{tab:internalBC},  the associated weights that fix the type IIB fields are
\begin{equation}
\begin{split}
P_1^{(0)}&=P_1^{(1)}= P_1^{(5)}= -G_1^{(1)} =1/2 \,,\qquad G_1^{(0)}=G_1^{(2)}=G_1^{(3)}=G_1^{(4)} = 0\,,\\
P_2^{(0)}&=P_2^{(5)}=G_2^{(0)} = -\frac{2}{3}G_2^{(2)}=2G_2^{(3)}=2G_2^{(4)} =\frac{1}{2}\,,\qquad P_2^{(1)}=G_2^{(1)}=0\,.
\end{split}
\end{equation}
By plugging into the expressions \eqref{eq:LinearAdS3},  we find
\begin{align}
Z_0 &\= \frac{4k}{r \sqrt{r^2+\ell^2}},\qquad Z_1 \= \frac{Q_1\,\cF_1}{r \sqrt{r^2+\ell^2}\,\sqrt{\cF_3}},\qquad Z_5 \= \frac{Q_5}{r \sqrt{r^2+\ell^2}}\,, \nn\\
 W_0& \= W_2^{-\frac{2}{3}}\=W_3^{2} \=W_4^{2} \=\sqrt{\cF_3}\,,\qquad W_1 \= \frac{r\,\sqrt{\cF_3}}{\sqrt{r^2+\ell^2}}\,, \qquad T_1 \= \frac{r^2+\ell^2}{Q_1 \cF_1}- \frac{\ell^2-\ell_2^2}{2Q_1},\nn\\
e^{2\nu} &\=\frac{r^2 (r^2+\ell^2)\,\cF_2}{\left( r^2+\ell^2\cos^2\theta\right)\left( r^2+\ell^2\sin^2\theta\right)} \,,\quad  H_0 \= k \,\cos 2\theta \,,\quad H_5 \= \frac{Q_5}{4} \cos 2\theta \,,  \label{eq:LinearglobalAdS3+T4App} 
\end{align}
where we have used the deformation factors $\cF_I$ \eqref{eq:FcalDef} and have introduced the global spherical coordinates, $(r,\theta)$ \eqref{eq:DefDistanceglobal},  using \eqref{eq:SimplRelations2}. The type IIB metric and fields \eqref{eq:TypeIIBAnsatz2} are given in  \eqref{eq:met1AdS3+T4} after having performed a gauge transformation to simplify the gauge field and having considered that the S$^3$ has no conical defect asymptotically $k=1$.

Note that one has
\begin{equation}
\frac{1}{Z_0}(d\psi+H_0 d\phi)^2 + \rho^2 Z_0 d\phi^2 \= r \sqrt{r^2+\ell^2} \left( \cos^2 \theta \,d\varphi_1^2 +\sin^2 \theta \,d\varphi_2^2 \right)\,,
\end{equation}
which allows going simply from the Hopf coordinates of the S$^3$ to the spherical coordinates \eqref{eq:DefHyperspher}.

\subsubsection{Single deformation with an arbitrary position on the S$^3$}
\label{App:GlobalAdS3+T42}

We now consider the three-rod configuration in Fig.\ref{fig:rodsourceAdS3+T42}.  The associated weights are
\begin{equation}
\begin{split}
P_i^{(0)}&=P_i^{(1)}= P_i^{(5)}= -G_i^{(1)} =1/2 \,,\qquad G_i^{(0)}=G_i^{(2)}=G_i^{(3)}=G_i^{(4)} = 0\,,\quad i=1,3,\\
P_2^{(0)}&=P_2^{(5)}=G_2^{(0)} = -\frac{2}{3}G_2^{(2)}=2G_2^{(3)}=2G_2^{(4)} =\frac{1}{2}\,,\qquad P_2^{(1)}=G_2^{(1)}=0\,.
\end{split}
\end{equation}
Thus,  the linear branch of solutions \eqref{eq:LinearAdS3} gives the same expressions as in \eqref{eq:LinearglobalAdS3+T4App} except that $\ell^2=\ell_1^2+\ell_2^2+\ell_3^2$, the functions $(r_i,\theta_i)$ have not the same dependence in $(\rho,z)$ \eqref{eq:DefDistance} and $(r,\theta)$ \eqref{eq:ri&thetaidef} due to the third rod,  and $e^{2\nu}$ is now given by
\begin{equation}
e^{2\nu} \=\frac{r^2 (r^2+\ell^2)\,\widetilde{\cF}_2}{\left( r^2+\ell^2\cos^2\theta\right)\left( r^2+\ell^2\sin^2\theta\right)} \,,
\end{equation}
where $\widetilde{\cF}_2$ is defined in \eqref{eq:FcalDefbis}.

\begin{itemize}
\item[•] \underline{Regularity and topology:}
\end{itemize}

The type IIB solutions,  given by \eqref{eq:metAdS3+T42},  are regular at $r>0$ since $r_i>0$ there \eqref{eq:ri&thetaidef} and all $\cF_I$ are finite and positive.  Moreover,  we have $\widetilde{\cF}_2=1$ at $r>0$ and $\theta=0,\pi/2$.  Thus, the poles of the S$^3$ are regular and the solutions have a S$^3\times$S$^1\times$T$^4$ topology outside the sources at $r=0$.

At $r=0$,  one of the $r_i$ vanishes depending on the range of $\theta$ given by \eqref{eq:RodLocTheta2}.  This locus corresponds to the three rod sources that are located at $\rho=0$ and $0\leq z\leq \ell^2/4$ in the $(\rho,z)$ coordinate system.  To study the local geometry at each rod,  one needs to change to the associated local coordinates $(r_i,\theta_i)$ \eqref{eq:LocalSpher1} and take the limit $r_i \to 0$.

At the first rod,  $r_1 \to 0$,  we have
\begin{equation}
\begin{split}
r_2^2 &\sim r_3^2 - \ell_2^2 \sim \ell_1^2 \sin^2 \theta_1 \,,\qquad \ell^2 \cos^2 \theta \sim \ell_1^2 \cos^2 \theta_1 \,,\qquad r^2 \sim \frac{\ell^2\sin^2 \theta}{\ell^2-\ell_1^2 \cos^2 \theta_1}\,r_1^2\,,\\
\cos^2 \theta_2 &\sim \frac{r_1^2\,\cos^2\theta_1}{\ell_2^2+\ell_1^2 \sin^2\theta_1}\,,\qquad \cos^2 \theta_3 \sim \frac{r_1^2 \,\ell_1^2 \,\cos^2\theta_1\,\sin^2\theta_1}{(\ell^2-\ell_1^2 \cos^2\theta_1)(\ell_2^2+\ell_1^2 \sin^2\theta_1)}\,.
\end{split}
\end{equation}
Thus,  the time slices of the type IIB metric \eqref{eq:metAdS3+T42} converge towards
\begin{align}
ds_{10} |_{dt=0}\sim & \frac{\ell\,  \sqrt{Q_1 Q_5} \,(\ell_2^2+\ell_1^2 \sin^2\theta_1)}{(\ell^2-\ell_3^2)\sqrt{\ell^2-\ell_2^2}(\ell^2-\ell_1^2 \cos^2\theta_1)} \left[ dr_1^2 + \frac{(\ell^2-\ell_3^2)(\ell^2-\ell_2^2)}{Q_1 Q_5 \,\ell_1^2} \, r_1^2 \,dy^2 +\ell_1^2 \, d{\Omega_3^{(1)}}^2\right] \nn\\
& + \frac{\ell}{\sqrt{\ell^2-\ell_2^2}} \,\sqrt{\frac{Q_1}{Q_5}} \left(dx_2^2+dx_3^2+dx_4^2 \right) +\frac{\sqrt{Q_1 Q_5}\,(\ell^2-\ell_1^2 \cos^2\theta_1)}{\ell \,\sqrt{\ell^2-\ell_2^2}}\,d\varphi_2^2\,,\\
d{\Omega_3^{(1)}}^2 \equi & d\theta_1^2 +\frac{(\ell^2-\ell_3^2)(\ell^2-\ell_1^2 \cos^2\theta_1)}{\ell^2 (\ell_2^2+\ell_1^2 \sin^2\theta_1)} \left[\cos^2\theta_1 \,d\varphi_1^2 + \frac{\ell^2\,\sin^2\theta_1}{Q_5 \,(\ell_2^2+\ell_1^2 \sin^2\theta_1)} \,dx_1^2 \right]\,.  \nn
\end{align}

At the second rod,  $r_2 \to 0$,  we have
\begin{equation}
\begin{split}
r_1^2 &\sim \ell_2^2 \cos^2\theta_2,\quad  r_3^2  \sim \ell_2^2 \sin^2 \theta_2 \,,\quad \ell^2 \cos^2 \theta \sim \ell_1^2+\ell_2^2 \cos^2 \theta_2 \,,\quad \sin^2 \theta_1 \sim \frac{r_2^2\, \sin^2\theta_2}{\ell_1^2+\ell_2^2 \cos^2 \theta_2}\,,\\
\cos^2 \theta_3 &\sim \frac{r_2^2 \,\cos^2\theta_2}{\ell_3^2+\ell_2^2 \sin^2\theta_2}\,,\qquad r^2 \sim \frac{r_2^2 \,\ell^2 \ell_2^2\,\cos^2\theta_2\,\sin^2\theta_2}{(\ell_3^2+\ell_2^2 \sin^2\theta_2)(\ell_1^2+\ell_2^2 \cos^2 \theta_2)}\,,
\end{split}
\end{equation}
and the time slices of the type IIB metric \eqref{eq:metAdS3+T42} converge towards
\begin{align}
ds_{10} |_{dt=0}\propto & \frac{\ell\ell_2^2\,  \sqrt{Q_1 Q_5}}{(\ell^2-\ell_3^2)(\ell^2-\ell_1^2)\sqrt{\ell^2-\ell_2^2}} \left[ dr_2^2 + \frac{(\ell^2-\ell_3^2)(\ell^2-\ell_1^2)}{Q_5 \,\ell_2^4} \, r_2^2 \,dx_1^2 +\ell_2^2 \, d{\Omega_2^{(2)}}^2\right] \nn\\
& + \frac{\ell}{\sqrt{\ell^2-\ell_2^2}} \,\sqrt{\frac{Q_1}{Q_5}} \left(dx_2^2+dx_3^2+dx_4^2 \right) \nn \\
&+\frac{\sqrt{Q_1 Q_5}}{\ell \,\sqrt{\ell^2-\ell_2^2}}\,\left[(\ell_1^2+\ell_2^2\cos^2\theta_2) d\varphi_1^2 +(\ell_3^2+\ell_2^2\sin^2\theta_2) d\varphi_2^2  \right]\,,\\
d{\Omega_2^{(2)}}^2 \equi & d\theta_2^2 +\frac{(\ell^2-\ell_1^2)(\ell^2-\ell_2^2)(\ell^2-\ell_3^2)}{4 Q_1 Q_5(\ell_1^2\ell_3^2 +\ell_2^2(\ell_3^2\cos^2\theta_2+\ell_1^2\sin^2\theta_2))}\,\sin^2(2\theta_2) \,dy^2\,.  \nn
\end{align}

Finally,  at the third rod,  $r_3 \to 0$,  we have
\begin{equation}
\begin{split}
r_2^2 &\sim r_1^2 - \ell_2^2 \sim \ell_3^2 \cos^2 \theta_3 \,,\qquad \ell^2 \sin^2 \theta \sim \ell_3^2 \sin^2 \theta_3 \,,\qquad r^2 \sim \frac{\ell^2\cos^2 \theta_3}{\ell^2-\ell_3^2 \sin^2 \theta_3}\,r_3^2\,,\\
\sin^2 \theta_2 &\sim \frac{r_3^2 \,\sin^2\theta_3}{\ell_2^2+\ell_3^2 \cos^2\theta_3}\,,\qquad \sin^2 \theta_1 \sim \frac{r_3^2\,\ell_3^2 \,\cos^2\theta_3\,\sin^2\theta_3}{(\ell^2-\ell_3^2 \sin^2\theta_3)(\ell_2^2+\ell_3^2 \cos^2\theta_3)}\,.
\end{split}
\end{equation}
Thus,  the time slices of the type IIB metric \eqref{eq:metAdS3+T42} converge towards
\begin{align}
ds_{10} |_{dt=0}\sim & \frac{\ell\,  \sqrt{Q_1 Q_5} \,(\ell_2^2+\ell_3^2 \cos^2\theta_3)}{(\ell^2-\ell_1^2)\sqrt{\ell^2-\ell_2^2}(\ell^2-\ell_3^2 \sin^2\theta_3)} \left[ dr_3^2 + \frac{(\ell^2-\ell_1^2)(\ell^2-\ell_2^2)}{Q_1 Q_5 \,\ell_3^2} \, r_3^2 \,dy^2 +\ell_3^2 \, d{\Omega_3^{(3)}}^2\right] \nn\\
& + \frac{\ell}{\sqrt{\ell^2-\ell_2^2}} \,\sqrt{\frac{Q_1}{Q_5}} \left(dx_2^2+dx_3^2+dx_4^2 \right) +\frac{\sqrt{Q_1 Q_5}\,(\ell^2-\ell_3^2 \sin^2\theta_3)}{\ell \,\sqrt{\ell^2-\ell_2^2}}\,d\varphi_1^2\,,\\
d{\Omega_3^{(3)}}^2 \equi & d\theta_3^2 +\frac{(\ell^2-\ell_1^2)(\ell^2-\ell_3^2 \sin^2\theta_3)}{\ell^2 (\ell_2^2+\ell_3^2 \cos^2\theta_3)} \left[\sin^2\theta_3 \,d\varphi_2^2 + \frac{\ell^2\,\cos^2\theta_3}{Q_5 \,(\ell_2^2+\ell_3^2 \cos^2\theta_3)} \,dx_1^2 \right]\,.  \nn
\end{align}

Therefore,  the local geometries correspond to regular S$^3\times$T$^4$ or S$^2\times$T$^5$ fibration over an origin of $\IR^2$ if one imposes
\begin{equation}
R_y^2 \=\frac{Q_1 Q_5 \,\ell_3^2} {(\ell^2-\ell_1^2)(\ell^2-\ell_2^2)}\=\frac{Q_1 Q_5 \,\ell_1^2} {(\ell^2-\ell_3^2)(\ell^2-\ell_2^2)}\,,\qquad R_{x_1}^2 \=  \frac{Q_5 \,\ell_2^4}{(\ell^2-\ell_3^2)(\ell^2-\ell_1^2)}\,,
\end{equation}
which gives \eqref{eq:Reg3rodT4} when we invert these constraints in terms of $(\ell_1^2,\ell_2^2,\ell_3^2)$.  Moreover,  one can also show that the gauge field is regular there such that the components of the field strength along the shrinking direction vanish.

Thus,  the locus $r=0$ corresponds to a smooth end to spacetime where either the $y$-circle or the $x_1$-circle degenerates depending on the value of $\theta$, and the overall geometries are depicted in Fig.\ref{fig:AdS3+T42pic}.

Moreover,  one can introduce conical defects at the $\IR^2$ by replacing $R_y \to k_1 R_y$,  $R_y \to k_3 R_y$ and $R_{x_1} \to k_2 R_{x_1}$ in the three constraints above with $k_i\in \mathbb{N}$.  This allows changing the position where the T$^4$ direction shrinks.  Without defect,  this region is localized around the equator of the S$^3$ as depicted in Fig.\ref{fig:AdS3+T42pic}.  Let us assume now that $k_1=k_2=1$ but $k_3$ is non-trivial.  Then, we have
\begin{equation}
\begin{split}
\ell_1^2 &\= \frac{Q_1 Q_5}{2R_y^2} \left( 1- \frac{k_3^2-1+\frac{2R_{x_1}^2}{Q_5}}{\sqrt{(k_3^2-1)^2+\frac{4 k_3^2 R_{x_1}^2}{Q_5}}}\right)\,,\qquad \ell_2^2 \= \frac{Q_1 R_{x_1}^2}{R_y^2\,\sqrt{(k_3^2-1)^2+\frac{4 k_3^2 R_{x_1}^2}{Q_5}}} \,,\\
\ell_3^2 &\= \frac{Q_1 Q_5}{2R_y^2} \left( 1+ \frac{k_3^2-1-\frac{2R_{x_1}^2}{Q_5}}{\sqrt{(k_3^2-1)^2+\frac{4 k_3^2 R_{x_1}^2}{Q_5}}}\right)\,,
\end{split}
\end{equation}
which is well-defined as soon as $Q_5\geq R_{x_1}^2$.

One can check that the zone of the S$^3$ where the T$^4$ degenerates,  delimited by the critical angles, $\theta_c^{(1)}$ and $\theta_c^{(2)}$ \eqref{eq:DefThetaCri2},  is indeed centered around the equator for $k_3=1$ such that $\theta_c^{(1)} \= \pi/2 -\theta_c^{(2)}$ with $\cos^2\theta_c^{(2)} = (2-R_{x_1}/\sqrt{Q_5})^{-1}$.  However,  both angles increase with $k_3$ such that
\begin{equation}
\cos^2 \theta_c^{(1)} \,\sim\,  \frac{R_{x_1}^2 (Q_5-R_{x_1}^2)}{k_3^2 Q_5^2}\,,\qquad \cos^2 \theta_c^{(2)} \,\sim\,  \frac{R_{x_1}^2}{ Q_5}\,,\qquad k_3\gg 1\,.
\end{equation}
Thus,  the region of the S$^3$ where the T$^4$ degenerates can indeed move from the equator to one of the hemispheres by forcing conical defects at the rods.

\subsection{Single S$^3$ deformation}
\label{App:GlobalAdS3+S3}

We consider two connected rod sources on the $z$-axis as depicted in Fig.\ref{fig:rodsourceAdS3+S3}.  One induces the degeneracy of the $y$-circle while the other forces the $\psi$-circle to shrink.  From Table \ref{tab:internalBC},  the associated weights that fix the type IIB fields are
\begin{equation}
\begin{split}
P_1^{(0)}&=P_1^{(1)}= P_1^{(5)}= -G_1^{(1)} =1/2 \,,\qquad G_1^{(0)}=G_1^{(2)}=G_1^{(3)}=G_1^{(4)} = 0\,,\\
P_2^{(0)}& =1\,,\qquad P_2^{(1)}=P_2^{(5)}=G_2^{(0)}=G_2^{(1)} = G_2^{(2)}=G_2^{(3)}=G_2^{(4)}=0\,.
\end{split}
\end{equation}
Using the generic expressions \eqref{eq:LinearAdS3},  we find
\begin{align}
Z_0 &\= \frac{4k\,\cF_4}{r \sqrt{r^2+\ell^2}\,\sqrt{\cF_3}},\qquad \frac{Z_1}{Q_1}\= \frac{Z_5}{Q_5}\= \= \frac{\cF_1}{r \sqrt{r^2+\ell^2}\,\sqrt{\cF_3}}\,, \label{eq:LinearglobalAdS3+S3App} \\
 W_0& \= W_2\=W_3 \=W_4 \=1\,,\quad W_1 \= \frac{r\,\sqrt{\cF_3}}{\sqrt{r^2+\ell^2}}\,, \quad e^{2\nu} \=\frac{r^2 (r^2+\ell^2)\,\cF_2}{\left( r^2+\ell^2\cos^2\theta\right)\left( r^2+\ell^2\sin^2\theta\right)} ,\nn\\
T_1 &\= \frac{r^2+\ell^2}{Q_1 \cF_1}- \frac{\ell^2-\ell_2^2}{2Q_1} \,,\quad  H_0 \= k \left(2 \cF_5 \cos^2\theta-1 \right) \,,\quad H_5 \= \frac{Q_5}{4}  \left(2 \cF_1 \cos^2\theta-1 \right) \,,   \nn
\end{align}
where we have used the deformation factors $\cF_I$ \eqref{eq:FcalDef} and \eqref{eq:FcalDef2} and have introduced the global spherical coordinates, $(r,\theta)$ \eqref{eq:DefDistanceglobal},  using \eqref{eq:SimplRelations2}. The type IIB metric and fields \eqref{eq:TypeIIBAnsatz2} are given in \eqref{eq:met1AdS3+T4} after having performed a gauge transformation to simplify the gauge field.

\subsection{Solutions with an arbitrary number of T$^4$ deformations}
\label{App:GlobalAdS3+T4s}

We consider $n$ connected rod sources on the $z$-axis,  as depicted in Fig.\ref{fig:rodsourceAdS3+T4s},  that induce the degeneracy of either the $y$-circle or a T$^4$ direction.  From Table \ref{tab:internalBC},  the associated weights that fix the type IIB fields are
\begin{equation}
\begin{split}
P_i^{(0)}&=P_i^{(1)}= P_i^{(5)}= -G_i^{(1)} =1/2 \,,\qquad G_i^{(0)}=G_i^{(2)}=G_i^{(3)}=G_i^{(4)} = 0\,,\quad i\in U_y\,,\\
P_i^{(0)}&=P_i^{(5)}=G_i^{(0)} = -\frac{2}{3}G_i^{(2)}=2G_i^{(3)}=2G_i^{(4)} =\frac{1}{2}\,,\qquad P_i^{(1)}=G_i^{(1)}=0\,, \quad i\in U_{x_1}\,,\\
P_i^{(0)}&=P_i^{(5)}=G_i^{(0)} = -\frac{2}{3}G_i^{(3)}=2G_i^{(2)}=2G_i^{(4)} =\frac{1}{2}\,,\qquad P_i^{(1)}=G_i^{(1)}=0\,, \quad i\in U_{x_2}\,,\\
P_i^{(0)}&=P_i^{(5)}=G_i^{(0)} = -\frac{2}{3}G_i^{(4)}=2G_i^{(2)}=2G_i^{(3)} =\frac{1}{2}\,,\qquad P_i^{(1)}=G_i^{(1)}=0\,, \quad i\in U_{x_3}\,,\\
P_i^{(0)}&=P_i^{(5)}=G_i^{(0)} = 2G_i^{(2)}=2G_i^{(3)}=2G_i^{(4)} =\frac{1}{2}\,,\qquad P_i^{(1)}=G_i^{(1)}=0\,, \quad i\in U_{x_4}\,, 
\end{split}
\label{eq:WeightsAdS3+T4sApp} 
\end{equation}
where $U_y$ and $U_{x_a}$ are the set of labels defined in \eqref{eq:DefUx}.
Using the generic expressions \eqref{eq:LinearAdS3},  we find
\begin{align}
Z_0 &\= \frac{4k}{r \sqrt{r^2+\ell^2}},\qquad Z_1 \= \frac{Q_1\,\cK_1}{r \sqrt{r^2+\ell^2}\,\sqrt{\prod_{a=1}^4\cK_{x_a}}},\qquad Z_5 \= \frac{Q_5}{r \sqrt{r^2+\ell^2}}\,, \nn\\
 W_0& \= \cK_{x_1}^2 W_2^2  \= \cK_{x_2}^2 W_3^2  \= \cK_{x_3}^2 W_4^2 \=\sqrt{\prod_{a=1}^4\cK_{x_a}}\,,\quad  W_1 \= \frac{r\,\sqrt{\prod_{a=1}^4\cK_{x_a}}}{\sqrt{r^2+\ell^2}}\,,\nn\\
e^{2\nu} &\=\frac{r^2 (r^2+\ell^2)\,\cK_2}{\left( r^2+\ell^2\cos^2\theta\right)\left( r^2+\ell^2\sin^2\theta\right)} \,,\\
 T_1& \= \frac{r^2+\ell^2}{Q_1 \cK_1}- \frac{\sum_{i\in U_y}\ell_i^2}{2Q_1},\qquad  H_0 \= k \cos 2\theta \,,\quad H_5 \= \frac{Q_5}{4} \cos 2\theta \,,  \nn \label{eq:LinearglobalAdS3+T4sApp} 
\end{align}
where we have used the deformation factors $\cK_I$ \eqref{eq:DefDefWarpFac} and have introduced the global spherical coordinates, $(r,\theta)$ \eqref{eq:DefDistanceglobal},  using \eqref{eq:SimplRelations2}. The type IIB metric and fields \eqref{eq:TypeIIBAnsatz2} are given in \eqref{eq:met1AdS3+T4s} after having performed a gauge transformation to simplify the gauge field and having considered that the S$^3$ has no conical defect asymptotically $k=1$.

Note that one has
\begin{equation}
\frac{1}{Z_0}(d\psi+H_0 d\phi)^2 + \rho^2 Z_0 d\phi^2 \= r \sqrt{r^2+\ell^2} \left( \cos^2 \theta \,d\varphi_1^2 +\sin^2 \theta \,d\varphi_2^2 \right)\,,
\end{equation}
which allows going simply from the Hopf coordinates of the S$^3$ to the spherical coordinates \eqref{eq:DefHyperspher}.

\subsection{Solutions with an arbitrary number of T$^4$ and S$^3$ deformations}
\label{App:GlobalAdS3+T4+S3s}

We consider the same rod configuration as the previous section but we allow some rods to force the degeneracy of the Hopf angle of the S$^3$, $\psi$.  A generic configuration has been depicted in Fig.\ref{fig:rodsourceAdS3+T4+S3}. In addition to \eqref{eq:WeightsAdS3+T4sApp},  we have
\begin{equation}
\begin{split}
P_i^{(0)}& =1\,,\qquad P_i^{(1)}=P_i^{(5)}=G_i^{(0)}=G_i^{(1)} = G_i^{(2)}=G_i^{(3)}=G_i^{(4)}=0\,, \qquad i\in U_\psi\,.
\end{split}
\end{equation}
Using the generic expressions \eqref{eq:LinearAdS3},  we find 
\begin{align}
Z_0 &\= \frac{4k\,\sqrt{\cK_\psi}\,\cK_+}{r \sqrt{r^2+\ell^2}},\quad Z_1 \= \frac{Q_1\,\cK_1}{r \sqrt{r^2+\ell^2}\,\sqrt{\cK_\psi \,\prod_{a=1}^4\cK_{x_a}}},\quad Z_5 \= \frac{Q_5\, \cK_-}{r \sqrt{r^2+\ell^2}\,\sqrt{\cK_\psi}}\,, \nn\\
 W_0& \= \cK_{x_1}^2 W_2^2  \= \cK_{x_2}^2 W_3^2  \= \cK_{x_3}^2 W_4^2 \=\sqrt{\prod_{a=1}^4\cK_{x_a}}\,,\quad  W_1 \= \frac{r\,\sqrt{\cK_{\psi}\, \prod_{a=1}^4\cK_{x_a}}}{\sqrt{r^2+\ell^2}}\,,\nn\\
e^{2\nu} &\=\frac{r^2 (r^2+\ell^2)\,\cK_2}{\left( r^2+\ell^2\cos^2\theta\right)\left( r^2+\ell^2\sin^2\theta\right)} \,,\\
 T_1& \= \frac{r^2+\ell^2}{Q_1 \cK_1}- \frac{\sum_{i\in U_y}\ell_i^2}{2Q_1},\qquad  H_0 \= k \cA_+\,,\quad H_5 \= \frac{Q_5}{4} \cA_- \,,  \nn \label{eq:LinearglobalAdS3+T4sApp} 
\end{align}
where we have used the deformation factors and the gauge potentials $\cK_I$ and $\cA_\pm$ from \eqref{eq:DefDefWarpFac} and \eqref{eq:DefDefWarpFac2},  and have introduced the global spherical coordinates, $(r,\theta)$ \eqref{eq:DefDistanceglobal},  using \eqref{eq:SimplRelations2}. The type IIB metric and fields \eqref{eq:TypeIIBAnsatz2} are given in \eqref{eq:met1AdS3+S3s} after having performed a gauge transformation to simplify the gauge field.

\bibliographystyle{utphys}      

\bibliography{microstates}       % calls file "microstates.bib"

\providecommand{\href}[2]{#2}\begingroup\raggedright\begin{thebibliography}{10}

\bibitem{Bah:2020ogh}
I.~Bah and P.~Heidmann, ``{Topological Stars and Black Holes},''
  \href{http://arxiv.org/abs/2011.08851}{{\ttfamily arXiv:2011.08851
  [hep-th]}}.

\bibitem{Bah:2020pdz}
I.~Bah and P.~Heidmann, ``{Topological stars, black holes and generalized
  charged Weyl solutions},''
  \href{http://dx.doi.org/10.1007/JHEP09(2021)147}{{\em JHEP} {\bfseries 09}
  (2021) 147}, \href{http://arxiv.org/abs/2012.13407}{{\ttfamily
  arXiv:2012.13407 [hep-th]}}.

\bibitem{Bah:2021owp}
I.~Bah and P.~Heidmann, ``{Smooth bubbling geometries without supersymmetry},''
  \href{http://dx.doi.org/10.1007/JHEP09(2021)128}{{\em JHEP} {\bfseries 09}
  (2021) 128}, \href{http://arxiv.org/abs/2106.05118}{{\ttfamily
  arXiv:2106.05118 [hep-th]}}.

\bibitem{Bah:2021rki}
I.~Bah and P.~Heidmann, ``{Bubble bag end: a bubbly resolution of curvature
  singularity},'' \href{http://dx.doi.org/10.1007/JHEP10(2021)165}{{\em JHEP}
  {\bfseries 10} (2021) 165}, \href{http://arxiv.org/abs/2107.13551}{{\ttfamily
  arXiv:2107.13551 [hep-th]}}.

\bibitem{Heidmann:2021cms}
P.~Heidmann, ``{Non-BPS floating branes and bubbling geometries},''
  \href{http://dx.doi.org/10.1007/JHEP02(2022)162}{{\em JHEP} {\bfseries 02}
  (2022) 162}, \href{http://arxiv.org/abs/2112.03279}{{\ttfamily
  arXiv:2112.03279 [hep-th]}}.

\bibitem{Bah:2022yji}
I.~Bah, P.~Heidmann, and P.~Weck, ``{Schwarzschild-like Topological
  Solitons},'' \href{http://arxiv.org/abs/2203.12625}{{\ttfamily
  arXiv:2203.12625 [hep-th]}}.

\bibitem{Gibbons:2013tqa}
G.~Gibbons and N.~Warner, ``{Global structure of five-dimensional fuzzballs},''
  \href{http://dx.doi.org/10.1088/0264-9381/31/2/025016}{{\em
  Class.Quant.Grav.} {\bfseries 31} (2014) 025016},
\href{http://arxiv.org/abs/1305.0957}{{\ttfamily arXiv:1305.0957 [hep-th]}}.
%%CITATION = ARXIV:1305.0957;%%.

\bibitem{Bah:2021irr}
I.~Bah, A.~Dey, and P.~Heidmann, ``{Stability of topological solitons, and
  black string to bubble transition},''
  \href{http://arxiv.org/abs/2112.11474}{{\ttfamily arXiv:2112.11474
  [hep-th]}}.

\bibitem{Lin:2004nb}
H.~Lin, O.~Lunin, and J.~M. Maldacena, ``{Bubbling AdS space and 1/2 BPS
  geometries},'' {\em JHEP} {\bfseries 10} (2004) 025,
\href{http://arxiv.org/abs/hep-th/0409174}{{\ttfamily arXiv:hep-th/0409174}}.
%%CITATION = HEP-TH/0409174;%%.

\bibitem{Lunin:2002iz}
O.~Lunin, J.~M. Maldacena, and L.~Maoz, ``{Gravity solutions for the D1-D5
  system with angular momentum},''
\href{http://arxiv.org/abs/hep-th/0212210}{{\ttfamily arXiv:hep-th/0212210}}.
%%CITATION = HEP-TH/0212210;%%.

\bibitem{Kanitscheider:2007wq}
I.~Kanitscheider, K.~Skenderis, and M.~Taylor, ``{Fuzzballs with internal
  excitations},'' {\em JHEP} {\bfseries 06} (2007) 056,
\href{http://arxiv.org/abs/0704.0690}{{\ttfamily arXiv:0704.0690 [hep-th]}}.
%%CITATION = 0704.0690;%%.

\bibitem{Giusto:2012yz}
S.~Giusto, O.~Lunin, S.~D. Mathur, and D.~Turton, ``{D1-D5-P microstates at the
  cap},'' \href{http://dx.doi.org/10.1007/JHEP02(2013)050}{{\em JHEP}
  {\bfseries 02} (2013) 050}, \href{http://arxiv.org/abs/1211.0306}{{\ttfamily
  arXiv:1211.0306 [hep-th]}}.

\bibitem{Giusto2015}
S.~Giusto, E.~Moscato, and R.~Russo, ``{AdS3 Holography for 1/4 and 1/8 BPS
  geometries},''
\href{http://arxiv.org/abs/1507.00945}{{\ttfamily arXiv:1507.00945 [hep-th]}}.
%%CITATION = ARXIV:1507.00945;%%.

\bibitem{Bena:2015bea}
I.~Bena, S.~Giusto, R.~Russo, M.~Shigemori, and N.~P. Warner, ``{Habemus
  Superstratum! A constructive proof of the existence of superstrata},''
  \href{http://dx.doi.org/10.1007/JHEP05(2015)110}{{\em JHEP} {\bfseries 05}
  (2015) 110},
\href{http://arxiv.org/abs/1503.01463}{{\ttfamily arXiv:1503.01463 [hep-th]}}.
%%CITATION = ARXIV:1503.01463;%%.

\bibitem{Shigemori:2020yuo}
M.~Shigemori, ``{Superstrata},''
  \href{http://dx.doi.org/10.1007/s10714-020-02698-8}{{\em Gen. Rel. Grav.}
  {\bfseries 52} no.~5, (2020) 51},
  \href{http://arxiv.org/abs/2002.01592}{{\ttfamily arXiv:2002.01592
  [hep-th]}}.

\bibitem{Bena:2018bbd}
I.~Bena, P.~Heidmann, and D.~Turton, ``{AdS$_{2}$ holography: mind the cap},''
  \href{http://dx.doi.org/10.1007/JHEP12(2018)028}{{\em JHEP} {\bfseries 12}
  (2018) 028},
\href{http://arxiv.org/abs/1806.02834}{{\ttfamily arXiv:1806.02834 [hep-th]}}.
%%CITATION = ARXIV:1806.02834;%%.

\bibitem{Jejjala:2005yu}
V.~Jejjala, O.~Madden, S.~F. Ross, and G.~Titchener, ``{Non-supersymmetric
  smooth geometries and D1-D5-P bound states},''
  \href{http://dx.doi.org/10.1103/PhysRevD.71.124030}{{\em Phys. Rev.}
  {\bfseries D71} (2005) 124030},
\href{http://arxiv.org/abs/hep-th/0504181}{{\ttfamily arXiv:hep-th/0504181}}.
%%CITATION = HEP-TH/0504181;%%.

\bibitem{Bena:2009en}
I.~Bena, S.~Giusto, C.~Ruef, and N.~P. Warner, ``{Multi-Center non-BPS Black
  Holes: the Solution},''
  \href{http://dx.doi.org/10.1088/1126-6708/2009/11/032}{{\em JHEP} {\bfseries
  11} (2009) 032}, \href{http://arxiv.org/abs/0908.2121}{{\ttfamily
  arXiv:0908.2121 [hep-th]}}.

\bibitem{DallAgata:2010srl}
G.~Dall'Agata, S.~Giusto, and C.~Ruef, ``{U-duality and non-BPS solutions},''
  \href{http://dx.doi.org/10.1007/JHEP02(2011)074}{{\em JHEP} {\bfseries 02}
  (2011) 074},
\href{http://arxiv.org/abs/1012.4803}{{\ttfamily arXiv:1012.4803 [hep-th]}}.
%%CITATION = ARXIV:1012.4803;%%.

\bibitem{Bossard:2014yta}
G.~Bossard and S.~Katmadas, ``{A bubbling bolt},''
  \href{http://dx.doi.org/10.1007/JHEP07(2014)118}{{\em JHEP} {\bfseries 1407}
  (2014) 118},
\href{http://arxiv.org/abs/1405.4325}{{\ttfamily arXiv:1405.4325 [hep-th]}}.
%%CITATION = ARXIV:1405.4325;%%.

\bibitem{Bossard:2014ola}
G.~Bossard and S.~Katmadas, ``{Floating JMaRT},''
  \href{http://dx.doi.org/10.1007/JHEP04(2015)067}{{\em JHEP} {\bfseries 04}
  (2015) 067},
\href{http://arxiv.org/abs/1412.5217}{{\ttfamily arXiv:1412.5217 [hep-th]}}.
%%CITATION = ARXIV:1412.5217;%%.

\bibitem{Bena:2015drs}
I.~Bena, G.~Bossard, S.~Katmadas, and D.~Turton, ``{Non-BPS multi-bubble
  microstate geometries},''
  \href{http://dx.doi.org/10.1007/JHEP02(2016)073}{{\em JHEP} {\bfseries 02}
  (2016) 073},
\href{http://arxiv.org/abs/1511.03669}{{\ttfamily arXiv:1511.03669 [hep-th]}}.
%%CITATION = ARXIV:1511.03669;%%.

\bibitem{Bossard:2017vii}
G.~Bossard, S.~Katmadas, and D.~Turton, ``{Two Kissing Bolts},''
\href{http://arxiv.org/abs/1711.04784}{{\ttfamily arXiv:1711.04784 [hep-th]}}.
%%CITATION = ARXIV:1711.04784;%%.

\bibitem{Edery:2020kof}
A.~Edery, ``{Non-singular vortices with positive mass in 2+1 dimensional
  Einstein gravity with AdS$_3$ and Minkowski background},''
  \href{http://dx.doi.org/10.1007/JHEP01(2021)166}{{\em JHEP} {\bfseries 01}
  (2021) 166}, \href{http://arxiv.org/abs/2004.09295}{{\ttfamily
  arXiv:2004.09295 [hep-th]}}.

\bibitem{Edery:2022crs}
A.~Edery, ``{Nonminimally coupled gravitating vortex: Phase transition at
  critical coupling \ensuremath{\xi}c in AdS3},''
  \href{http://dx.doi.org/10.1103/PhysRevD.106.065017}{{\em Phys. Rev. D}
  {\bfseries 106} no.~6, (2022) 065017},
  \href{http://arxiv.org/abs/2205.12175}{{\ttfamily arXiv:2205.12175
  [hep-th]}}.

\bibitem{Mayerson:2020tcl}
D.~R. Mayerson, R.~A. Walker, and N.~P. Warner, ``{Microstate Geometries from
  Gauged Supergravity in Three Dimensions},''
  \href{http://dx.doi.org/10.1007/JHEP10(2020)030}{{\em JHEP} {\bfseries 10}
  (2020) 030}, \href{http://arxiv.org/abs/2004.13031}{{\ttfamily
  arXiv:2004.13031 [hep-th]}}.

\bibitem{Houppe:2020oqp}
A.~Houppe and N.~P. Warner, ``{Supersymmetry and superstrata in three
  dimensions},'' \href{http://dx.doi.org/10.1007/JHEP08(2021)133}{{\em JHEP}
  {\bfseries 08} (2021) 133}, \href{http://arxiv.org/abs/2012.07850}{{\ttfamily
  arXiv:2012.07850 [hep-th]}}.

\bibitem{Ganchev:2021pgs}
B.~Ganchev, A.~Houppe, and N.~P. Warner, ``{Q-balls meet fuzzballs: non-BPS
  microstate geometries},''
  \href{http://dx.doi.org/10.1007/JHEP11(2021)028}{{\em JHEP} {\bfseries 11}
  (2021) 028}, \href{http://arxiv.org/abs/2107.09677}{{\ttfamily
  arXiv:2107.09677 [hep-th]}}.

\bibitem{Chakrabarty:2015foa}
B.~Chakrabarty, D.~Turton, and A.~Virmani, ``{Holographic description of
  non-supersymmetric orbifolded D1-D5-P solutions},''
  \href{http://dx.doi.org/10.1007/JHEP11(2015)063}{{\em JHEP} {\bfseries 11}
  (2015) 063},
\href{http://arxiv.org/abs/1508.01231}{{\ttfamily arXiv:1508.01231 [hep-th]}}.
%%CITATION = ARXIV:1508.01231;%%.

\bibitem{Ganchev:2021ewa}
B.~Ganchev, S.~Giusto, A.~Houppe, and R.~Russo, ``{$\hbox {AdS}_3$ holography
  for non-BPS geometries},''
  \href{http://dx.doi.org/10.1140/epjc/s10052-022-10133-2}{{\em Eur. Phys. J.
  C} {\bfseries 82} no.~3, (2022) 217},
  \href{http://arxiv.org/abs/2112.03287}{{\ttfamily arXiv:2112.03287
  [hep-th]}}.

\bibitem{NonBPSAdS2}
P.~Heidmann and A.~Houppe, ``{Solitonic Excitations in AdS$_2$},''
  \href{http://arxiv.org/abs/2212.05065}{{\ttfamily arXiv:2212.05065
  [hep-th]}}.

\bibitem{Belinski:2001ph}
V.~Belinski and E.~Verdaguer,
  \href{http://dx.doi.org/10.1017/CBO9780511535253}{{\em {Gravitational
  solitons}}}.
\newblock Cambridge Monographs on Mathematical Physics. Cambridge University
  Press, 2005.

\bibitem{Belinsky:1971nt}
V.~A. Belinsky and V.~E. Zakharov, ``{Integration of the Einstein Equations by
  the Inverse Scattering Problem Technique and the Calculation of the Exact
  Soliton Solutions},'' {\em Sov. Phys. JETP} {\bfseries 48} (1978) 985--994.

\bibitem{Belinsky:1979mh}
V.~A. Belinsky and V.~E. Sakharov, ``{Stationary Gravitational Solitons with
  Axial Symmetry},'' {\em Sov. Phys. JETP} {\bfseries 50} (1979) 1--9.

\bibitem{PhysRevLett.41.1197}
B.~K. Harrison, ``B\"acklund transformation for the ernst equation of general
  relativity,'' \href{http://dx.doi.org/10.1103/PhysRevLett.41.1197}{{\em Phys.
  Rev. Lett.} {\bfseries 41} (Oct, 1978) 1197--1200}.
  \url{https://link.aps.org/doi/10.1103/PhysRevLett.41.1197}.

\bibitem{Alekseev:1999kj}
G.~A. Alekseev,
  \href{http://dx.doi.org/10.1142/9789812817587_0002}{``{Monodromy transform
  approach to solution of some field equations in general relativity and string
  theory},''} in {\em {Nonlinearity, Integrability and All That: Twenty Years
  after NEEDS 79}}.
\newblock 9, 1999.
\newblock \href{http://arxiv.org/abs/gr-qc/9911045}{{\ttfamily
  arXiv:gr-qc/9911045}}.

\bibitem{Alekseev:1999bv}
G.~A. Alekseev, ``{Gravitational solitons and monodromy transform approach to
  solution of integrable reductions of Einstein equations},''
  \href{http://dx.doi.org/10.1016/S0167-2789(01)00162-2}{{\em Physica D}
  {\bfseries 152} (2001) 97--103},
  \href{http://arxiv.org/abs/gr-qc/0001012}{{\ttfamily arXiv:gr-qc/0001012}}.

\bibitem{Stephani:2003tm}
H.~Stephani, D.~Kramer, M.~A.~H. MacCallum, C.~Hoenselaers, and E.~Herlt,
  \href{http://dx.doi.org/10.1017/CBO9780511535185}{{\em {Exact solutions of
  Einstein's field equations}}}.
\newblock Cambridge Monographs on Mathematical Physics. Cambridge Univ. Press,
  Cambridge, 2003.

\bibitem{Geroch:1970nt}
R.~P. Geroch, ``{A Method for generating solutions of Einstein's equations},''
  \href{http://dx.doi.org/10.1063/1.1665681}{{\em J. Math. Phys.} {\bfseries
  12} (1971) 918--924}.

\bibitem{Geroch:1972yt}
R.~P. Geroch, ``{A Method for generating new solutions of Einstein's equation.
  2},'' \href{http://dx.doi.org/10.1063/1.1665990}{{\em J. Math. Phys.}
  {\bfseries 13} (1972) 394--404}.

\bibitem{Lehner:2011wc}
L.~Lehner and F.~Pretorius, {\em {Final state of Gregory\textendash{}Laflamme
  instability}}, pp.~44--68.
\newblock 2012.
\newblock \href{http://arxiv.org/abs/1106.5184}{{\ttfamily arXiv:1106.5184
  [gr-qc]}}.

\bibitem{Emparan:2021ewh}
R.~Emparan, D.~Licht, R.~Suzuki, M.~Toma\v{s}evi\'c, and B.~Way, ``{Black
  tsunamis and naked singularities in AdS},''
  \href{http://dx.doi.org/10.1007/JHEP02(2022)090}{{\em JHEP} {\bfseries 02}
  (2022) 090}, \href{http://arxiv.org/abs/2112.07967}{{\ttfamily
  arXiv:2112.07967 [hep-th]}}.

\bibitem{Weyl:book}
H.~Weyl {\em Ann. Phys. (Leipzig)} {\bfseries 54} (1917) 117.

\bibitem{Emparan:2001wk}
R.~Emparan and H.~S. Reall, ``{Generalized Weyl solutions},''
  \href{http://dx.doi.org/10.1103/PhysRevD.65.084025}{{\em Phys. Rev. D}
  {\bfseries 65} (2002) 084025},
  \href{http://arxiv.org/abs/hep-th/0110258}{{\ttfamily arXiv:hep-th/0110258}}.

\bibitem{Gauntlett:2002nw}
J.~P. Gauntlett, J.~B. Gutowski, C.~M. Hull, S.~Pakis, and H.~S. Reall, ``{All
  supersymmetric solutions of minimal supergravity in five- dimensions},''
  \href{http://dx.doi.org/10.1088/0264-9381/20/21/005}{{\em Class.Quant.Grav.}
  {\bfseries 20} (2003) 4587--4634},
\href{http://arxiv.org/abs/hep-th/0209114}{{\ttfamily arXiv:hep-th/0209114
  [hep-th]}}.
%%CITATION = HEP-TH/0209114;%%.

\bibitem{Bena:2005va}
I.~Bena and N.~P. Warner, ``{Bubbling supertubes and foaming black holes},''
  \href{http://dx.doi.org/10.1103/PhysRevD.74.066001}{{\em Phys. Rev.}
  {\bfseries D74} (2006) 066001},
\href{http://arxiv.org/abs/hep-th/0505166}{{\ttfamily arXiv:hep-th/0505166}}.
%%CITATION = HEP-TH/0505166;%%.

\bibitem{Bena:2007kg}
I.~Bena and N.~P. Warner, ``{Black holes, black rings and their microstates},''
  \href{http://dx.doi.org/10.1007/978-3-540-79523-0}{{\em Lect. Notes Phys.}
  {\bfseries 755} (2008) 1--92},
\href{http://arxiv.org/abs/hep-th/0701216}{{\ttfamily arXiv:hep-th/0701216}}.
%%CITATION = HEP-TH/0701216;%%.

\bibitem{Heidmann:2017cxt}
P.~Heidmann, ``{Four-center bubbled BPS solutions with a Gibbons-Hawking
  base},'' \href{http://dx.doi.org/10.1007/JHEP10(2017)009}{{\em JHEP}
  {\bfseries 10} (2017) 009},
\href{http://arxiv.org/abs/1703.10095}{{\ttfamily arXiv:1703.10095 [hep-th]}}.
%%CITATION = ARXIV:1703.10095;%%.

\bibitem{Bena:2017fvm}
I.~Bena, P.~Heidmann, and P.~F. Ramirez, ``{A systematic construction of
  microstate geometries with low angular momentum},''
  \href{http://dx.doi.org/10.1007/JHEP10(2017)217}{{\em JHEP} {\bfseries 10}
  (2017) 217},
\href{http://arxiv.org/abs/1709.02812}{{\ttfamily arXiv:1709.02812 [hep-th]}}.
%%CITATION = ARXIV:1709.02812;%%.

\bibitem{Heidmann:2018vky}
P.~Heidmann and S.~Mondal, ``{The full space of BPS multicenter states with
  pure D-brane charges},''
  \href{http://dx.doi.org/10.1007/JHEP06(2019)011}{{\em JHEP} {\bfseries 06}
  (2019) 011}, \href{http://arxiv.org/abs/1810.10019}{{\ttfamily
  arXiv:1810.10019 [hep-th]}}.

\bibitem{Elvang:2002br}
H.~Elvang and G.~T. Horowitz, ``{When black holes meet Kaluza-Klein bubbles},''
  \href{http://dx.doi.org/10.1103/PhysRevD.67.044015}{{\em Phys. Rev. D}
  {\bfseries 67} (2003) 044015},
  \href{http://arxiv.org/abs/hep-th/0210303}{{\ttfamily arXiv:hep-th/0210303}}.

\bibitem{Astorino:2022fge}
M.~Astorino, R.~Emparan, and A.~Vigan\`o, ``{Bubbles of nothing in binary black
  holes and black rings, and viceversa},''
  \href{http://dx.doi.org/10.1007/JHEP07(2022)007}{{\em JHEP} {\bfseries 07}
  (2022) 007}, \href{http://arxiv.org/abs/2204.09690}{{\ttfamily
  arXiv:2204.09690 [hep-th]}}.

\bibitem{Tomasiello:2011eb}
A.~Tomasiello, ``{Generalized structures of ten-dimensional supersymmetric
  solutions},'' \href{http://dx.doi.org/10.1007/JHEP03(2012)073}{{\em JHEP}
  {\bfseries 03} (2012) 073}, \href{http://arxiv.org/abs/1109.2603}{{\ttfamily
  arXiv:1109.2603 [hep-th]}}.

\bibitem{Giusto:2013rxa}
S.~Giusto, L.~Martucci, M.~Petrini, and R.~Russo, ``{6D microstate geometries
  from 10D structures},''
  \href{http://dx.doi.org/10.1016/j.nuclphysb.2013.08.018}{{\em Nucl.Phys.}
  {\bfseries B876} (2013) 509--555},
\href{http://arxiv.org/abs/1306.1745}{{\ttfamily arXiv:1306.1745 [hep-th]}}.
%%CITATION = ARXIV:1306.1745;%%.

\bibitem{Bena:2016agb}
I.~Bena, E.~Martinec, D.~Turton, and N.~P. Warner, ``{Momentum Fractionation on
  Superstrata},'' \href{http://dx.doi.org/10.1007/JHEP05(2016)064}{{\em JHEP}
  {\bfseries 05} (2016) 064},
\href{http://arxiv.org/abs/1601.05805}{{\ttfamily arXiv:1601.05805 [hep-th]}}.
%%CITATION = ARXIV:1601.05805;%%.

\bibitem{Bena:2017xbt}
I.~Bena, S.~Giusto, E.~J. Martinec, R.~Russo, M.~Shigemori, D.~Turton, and
  N.~P. Warner, ``{Asymptotically-flat supergravity solutions deep inside the
  black-hole regime},'' \href{http://dx.doi.org/10.1007/JHEP02(2018)014}{{\em
  JHEP} {\bfseries 02} (2018) 014},
\href{http://arxiv.org/abs/1711.10474}{{\ttfamily arXiv:1711.10474 [hep-th]}}.
%%CITATION = ARXIV:1711.10474;%%.

\bibitem{Ceplak:2018pws}
N.~Ceplak, R.~Russo, and M.~Shigemori, ``{Supercharging Superstrata},''
\href{http://arxiv.org/abs/1812.08761}{{\ttfamily arXiv:1812.08761 [hep-th]}}.
%%CITATION = ARXIV:1812.08761;%%.

\bibitem{Heidmann:2019zws}
P.~Heidmann and N.~P. Warner, ``{Superstratum Symbiosis},''
\href{http://arxiv.org/abs/1903.07631}{{\ttfamily arXiv:1903.07631 [hep-th]}}.
%%CITATION = ARXIV:1903.07631;%%.

\bibitem{Heidmann:2019xrd}
P.~Heidmann, D.~R. Mayerson, R.~Walker, and N.~P. Warner, ``{Holomorphic Waves
  of Black Hole Microstructure},''
  \href{http://dx.doi.org/10.1007/JHEP02(2020)192}{{\em JHEP} {\bfseries 02}
  (2020) 192}, \href{http://arxiv.org/abs/1910.10714}{{\ttfamily
  arXiv:1910.10714 [hep-th]}}.

\bibitem{Bellorin:2006yr}
J.~Bellorin, P.~Meessen, and T.~Ortin, ``{All the supersymmetric solutions of
  N=1,d=5 ungauged supergravity},''
  \href{http://dx.doi.org/10.1088/1126-6708/2007/01/020}{{\em JHEP} {\bfseries
  01} (2007) 020}, \href{http://arxiv.org/abs/hep-th/0610196}{{\ttfamily
  arXiv:hep-th/0610196}}.

\bibitem{Costa:2000kf}
M.~S. Costa and M.~J. Perry, ``{Interacting black holes},''
  \href{http://dx.doi.org/10.1016/S0550-3213(00)00577-0}{{\em Nucl. Phys. B}
  {\bfseries 591} (2000) 469--487},
  \href{http://arxiv.org/abs/hep-th/0008106}{{\ttfamily arXiv:hep-th/0008106}}.

\bibitem{Regge:1961px}
T.~Regge, ``{General Relativity without coordinates},''
  \href{http://dx.doi.org/10.1007/BF02733251}{{\em Nuovo Cim.} {\bfseries 19}
  (1961) 558--571}.

\bibitem{Gregory:1993vy}
R.~Gregory and R.~Laflamme, ``{Black strings and p-branes are unstable},''
  \href{http://dx.doi.org/10.1103/PhysRevLett.70.2837}{{\em Phys. Rev. Lett.}
  {\bfseries 70} (1993) 2837--2840},
  \href{http://arxiv.org/abs/hep-th/9301052}{{\ttfamily arXiv:hep-th/9301052}}.

\bibitem{Gregory:1994bj}
R.~Gregory and R.~Laflamme, ``{The Instability of charged black strings and
  p-branes},'' \href{http://dx.doi.org/10.1016/0550-3213(94)90206-2}{{\em Nucl.
  Phys. B} {\bfseries 428} (1994) 399--434},
  \href{http://arxiv.org/abs/hep-th/9404071}{{\ttfamily arXiv:hep-th/9404071}}.

\bibitem{Schwarz:1983qr}
J.~H. Schwarz, ``{Covariant Field Equations of Chiral N=2 D=10 Supergravity},''
  \href{http://dx.doi.org/10.1016/0550-3213(83)90192-X}{{\em Nucl. Phys. B}
  {\bfseries 226} (1983) 269}.

\bibitem{Schwarz:1983wa}
J.~H. Schwarz and P.~C. West, ``{Symmetries and Transformations of Chiral N=2
  D=10 Supergravity},''
  \href{http://dx.doi.org/10.1016/0370-2693(83)90168-5}{{\em Phys. Lett. B}
  {\bfseries 126} (1983) 301--304}.

\bibitem{Fernandez:2008vh}
F.~C. Fernandez, {\em {D-branes in Supersymmetric Backgrounds}}.
\newblock PhD thesis, Santiago de Compostela U., 2008.
\newblock \href{http://arxiv.org/abs/0804.4878}{{\ttfamily arXiv:0804.4878
  [hep-th]}}.

\bibitem{Hamilton:2016ito}
M.~J.~D. Hamilton, ``{The field and Killing spinor equations of M-theory and
  type IIA/IIB supergravity in coordinate-free notation},''
  \href{http://arxiv.org/abs/1607.00327}{{\ttfamily arXiv:1607.00327
  [math.DG]}}.

\end{thebibliography}\endgroup

%\end{adjustwidth}
%%%%%%%%%%%%%%%%%%%%%%%%%%%%%%%%%%%%%%%%%%%%%%%%%%%%%

%%%%%%%%%%%%%%%%%%%%%%%%%%%%%%%%%%%%%%%%%%%%%%%%%%%%%
\end{document}